\documentclass[rmp,amsmath,amssymb,amsfonts,apsref]{revtex4}

\usepackage{graphicx}
\usepackage{bm}
\usepackage[latin1]{inputenc}

\newtheorem{theorem}{Theorem}[section]

\newtheorem{remark}{Remark} 
 
\newtheorem{lemma}[theorem]{Lemma}
\newtheorem{proposition}{Proposition}
\newtheorem{corollary}{Corollary} 
\newtheorem{conjecture}{Conjecture}
\newtheorem{assumption}{Assumption}

\newtheorem{condition}{Condition}

\newcommand{\bd}{
\begin{document}} 
\newcommand{\ed}{\end{document}}
\newcommand{\beq}{\begin{equation}} 
\newcommand{\eeq}{\end{equation}}
\newcommand{\baR}{\begin{array}}
\newcommand{\eaR}{\end{array}}
\newcommand{\bef}{\begin{figure}} 
\newcommand{\enf}{\end{figure}}
\newcommand{\bea}{\begin{eqnarray}} 
\newcommand{\eea}{\end{eqnarray}}
\newcommand{\bth}{\begin{theorem}} 
\newcommand{\enth}{\end{theorem}}
\newcommand{\bhyp}{\begin{assumption}}
\newcommand{\ehyp}{\end{assumption}}
\newcommand{\bp}{\begin{proposition}}
\newcommand{\ep}{\end{proposition}}
\newcommand{\bco}{\begin{corollary}}
\newcommand{\eco}{\end{corollary}}
\newcommand{\bconj}{\begin{conjecture}}
\newcommand{\econj}{\end{conjecture}}
\newcommand{\ble}{\begin{lemma}} 
\newcommand{\ele}{\end{lemma}}
\newcommand{\bR}{\begin{remark}} 
\newcommand{\eR}{\end{remark}}
\newcommand{\bC}{\begin{condition}} 
\newcommand{\eC}{\end{condition}}
\newcommand{\bc}{\begin{center}} 
\newcommand{\ec}{\end{center}}
\newcommand{\ben}{\begin{enumerate}}
\newcommand{\een}{\end{enumerate}}
\newcommand{\bit}{\begin{itemize}} 
\newcommand{\eit}{\end{itemize}}
\newcommand{\su}{\section} 
\newcommand{\ssu}{\subsection}
\newcommand{\sssu}{\subsubsection} 
\newcommand{\nid}{\noindent}
\newcommand{\nnb}{\nonumber}

\newcommand\bbbr{{\sf I\!R}} 
\newcommand\bbbc{{\sf I\!\!C}}
\newcommand\bbbn{{\sf I\!N}} 
\newcommand\bbbh{{\sf I\!H}}
\newcommand\bbbz{{\sf Z\!\!Z}}

\newcommand\bxi{\bm{\xi}}
\newcommand\Bth{\bm{\theta}}
\newcommand\bt{\bar{\theta}}
\newcommand\st{\sigma_\theta}
\newcommand\sdt{\sigma^2_\theta}
\newcommand\sxi{{\sigma_\xi}}

\newcommand\cA{{\cal A}} 
\newcommand\cB{{\cal B}} 
\newcommand\cC{{\cal C}} 
\newcommand\cD{{\cal D}} 
\newcommand\cE{{\cal E}}
\newcommand\cF{{\cal F}} 
\newcommand\cG{{\cal G}} 
\newcommand\cH{{\cal H}} 
\newcommand\cI{{\cal I}} 
\newcommand\cJ{{\cal J}}
\newcommand\cK{{\cal K}}
\newcommand\cL{{\cal L}} 
\newcommand\cM{{\cal M}} 
\newcommand\cO{{\cal O}}
\newcommand\bcM{{\bar{\cal M}}} 
\newcommand\cN{{\cal N}} 
\newcommand\cP{{\cal P}} 
\newcommand\tcP{\tilde{\cal P}} 
\newcommand\cQ{{\cal Q}}
\newcommand\cR{{\cal R}} 
\newcommand\cS{{\cal S}} 
\newcommand\cW{{\cal W}} 
\newcommand\cU{{\cal U}} 
\newcommand\cV{{\cal V}}
\newcommand\cT{{\cal T}} 
\newcommand\cX{{\cal X}} 
\newcommand\cZ{{\cal Z}} 
\newcommand\cl{{\cal l}} 
\newcommand\cn{{\cal n}}

\newcommand\hX{\hat{\bX}} 
\newcommand\hY{\hat{\bY}} 
\newcommand\hF{\hat{\bF}}
\newcommand\hG{\hat{\bG}}
\newcommand\hL{\hat{\bL}}
\newcommand\bE{{\bf E}} 
\newcommand\bF{{\bf F}} 
\newcommand\Bf{{\bf f}} 
\newcommand\bG{{\bf G}} 
\newcommand\bg{{\bf g}} 
\newcommand\bH{{\bf H}} 
\newcommand\bJ{{\bar{J}}}
\newcommand\bL{{\bf L}}
\newcommand\bX{{\bf X}} 
\newcommand\bY{{\bf Y}} 
\newcommand\bU{{\bf U}}
\newcommand\bZ{{\bf Z}} 
\newcommand\ba{{\bf a}} 
\newcommand\bb{{\bf b}} 
\newcommand\be{{\bf e}}
\newcommand\bk{{\bf k}}
\newcommand\bh{{\bf h}} 
\newcommand\bi{{\bf i}} 
\newcommand\bl{{\bf l}} 
\newcommand\bn{{\bf n}} 
\newcommand\bq{{\bf q}} 
\newcommand\bu{{\bf u}}
\newcommand\bv{{\bf v}}
\newcommand\bx{{\bf x}} 
\newcommand\bxs{{\bx^\ast}} 
\newcommand\bXs{{\bX^\ast}} 
\newcommand\by{{\bf y}} 
\newcommand\bz{{\bf z}} 
\newcommand\bw{{\bf w}}
\newcommand\btX{{\bf \tilde{X}}} 

\newcommand\va{{\vec a}} 
\newcommand\vb{{\vec b}} 
\newcommand\vu{{\vec u}} 
\newcommand\vv{{\vec v}} 

\newcommand\ta{{\tilde a}}
\newcommand\tb{{\tilde b}}
\newcommand\ti{\tilde{\i}}
\newcommand\tx{\tilde{x}}
\newcommand\txT{\tilde{x}_T}
\newcommand\txTk{\tilde{x}_{T,k}}
\newcommand\tbx{\tilde{\bx}}
\newcommand\tbxT{\tilde{\bx}_T}
\newcommand\ctbxT{\left[\tbx\right]_T}
\newcommand\tu{\tilde{u}}
\newcommand\tuT{\tilde{u}_T}
\newcommand\tbeta{\tilde{\beta}} 
\newcommand\tom{\tilde{\omega}} 
\newcommand\bom{\bar{\nu}} 
\newcommand\tml{\tilde{\mu}_L} 
\newcommand\ml{m_L} 
\newcommand\hml{\hat{\mu}_L} 
\newcommand\bel{\bar{e}_L} 
\newcommand\bxl{\bar{X}_L} 
\newcommand\bpxl{\bar{X}^+_L} 
\newcommand\brl{\bar{r}_L} 
\newcommand\boml{\bar{\nu}_L} 
\newcommand\tbu{\tilde{\bu}}
\newcommand\tbG{{\tilde{\bf G}}}
\newcommand\obs{A}
\newcommand\hchi{\hat\chi}
\newcommand\tow{\stackrel{w}{\to}}

\newcommand\bcm{\bar{\cM}} \newcommand\cm{\cM}

\newcommand{\deq}{\stackrel {\rm def}{=}}

\newcommand{\sep}{\; \,} 
\newcommand{\D}{\displaystyle}
\newcommand{\T}{\textstyle} 
\newcommand{\etc}{etc $\dots$}
\newcommand{\etal}{etc $\dots$} 
\newcommand\dL{^\partial \Lambda}

\voffset=1truecm

\def\Appendix{\section*{APPENDIX}}

\bd

\title{Chapter 1: From Neuron to Neural Networks dynamics.}
\author{B. Cessac}
\affiliation{Institut Non Lin\'eaire de Nice, 1361 Route des Lucioles, 06560 Valbonne, France}
\author{M. Samuelides}
\affiliation{Ecole Nationale Sup\'erieure de l'A\'eronautique et de l'espace and ONERA/DTIM,
2 Av. E. Belin, 31055 Toulouse, France.}

\begin{abstract}
This paper presents an overview of some techniques and concepts coming from  dynamical system theory and  used for the analysis
of dynamical neural networks models. In a first section, we describe the dynamics of the neuron, starting from the Hodgkin-Huxley description, which
is somehow the canonical description for the ``biological neuron''. We discuss some models reducing the Hodgkin-Huxley model to
a two dimensional dynamical system, keeping one of the main feature of the neuron: its excitability.  We present then  examples of
phase diagram and bifurcation analysis for the Hodgin-Huxley equations. Finally, we end this section by a dynamical system analysis for the
nervous flux propagation along the axon. We then consider neuron couplings, with a brief description of synapses,
synaptic plasticiy and learning, in a second section. We also briefly discuss the delicate issue of causal action from
one neuron to another when complex feedback effects and non linear dynamics are involved. The third section presents the limit of weak
coupling and the use of normal forms technics to handle this situation. We consider then several examples of recurrent models
with different type of synaptic interactions (symmetric, cooperative, random). We introduce various techniques
coming from statistical physics and dynamical systems theory. A last section is devoted to a detailed example
of recurrent model where we go in deep in the analysis of the dynamics and discuss the effect of learning on the
neuron dynamics. We also present recent methods allowing the analysis of the non linear effects of the neural dynamics
on signal propagation and causal action. An appendix, presenting the main notions of dynamical systems theory
useful for the comprehension of the chapter, has been added for the convenience of the reader.
\end{abstract}

\maketitle 

\tableofcontents
\vfill
\hfill

\pagebreak

\su{Introduction.}

The present chapter aims to give an outlook of the various dynamical
systems notions and techniques that are used while modeling  Neural Network dynamics. 
Actually, there are a lot of such models.
One reason is that there are several levels of description and 
abstraction in this context : from a biologically realistic 
modeling of a neuron to neurons with a binary state; from an isolated neuron
to Neural Networks, composed by several functional parts, each of them
constituted by many neurons, and interacting in complex fashion, etc ....
Another reason is that the Neural Network community is wide :
 from biologists, neurophysiologists, pharmacologists,
to mathematician, theoretical physicists, including engineers, computer scientists, robot designers,
etc... Clearly the motivations and questions are different.  Models that are designed
to tackle a given problem may have very different structure and properties.
It follows from these remarks that any attempt to ''give an outlook of the various dynamical
systems notions and techniques used while modeling  Neural Network dynamics''
 is necessarily partial, biased et includes arbitrary and subjective
choices. For sure, this chapter is subject to these restrictions.\\

\begin{figure}
\begin{center}
\includegraphics[height=10cm,width=14cm,clip=false]{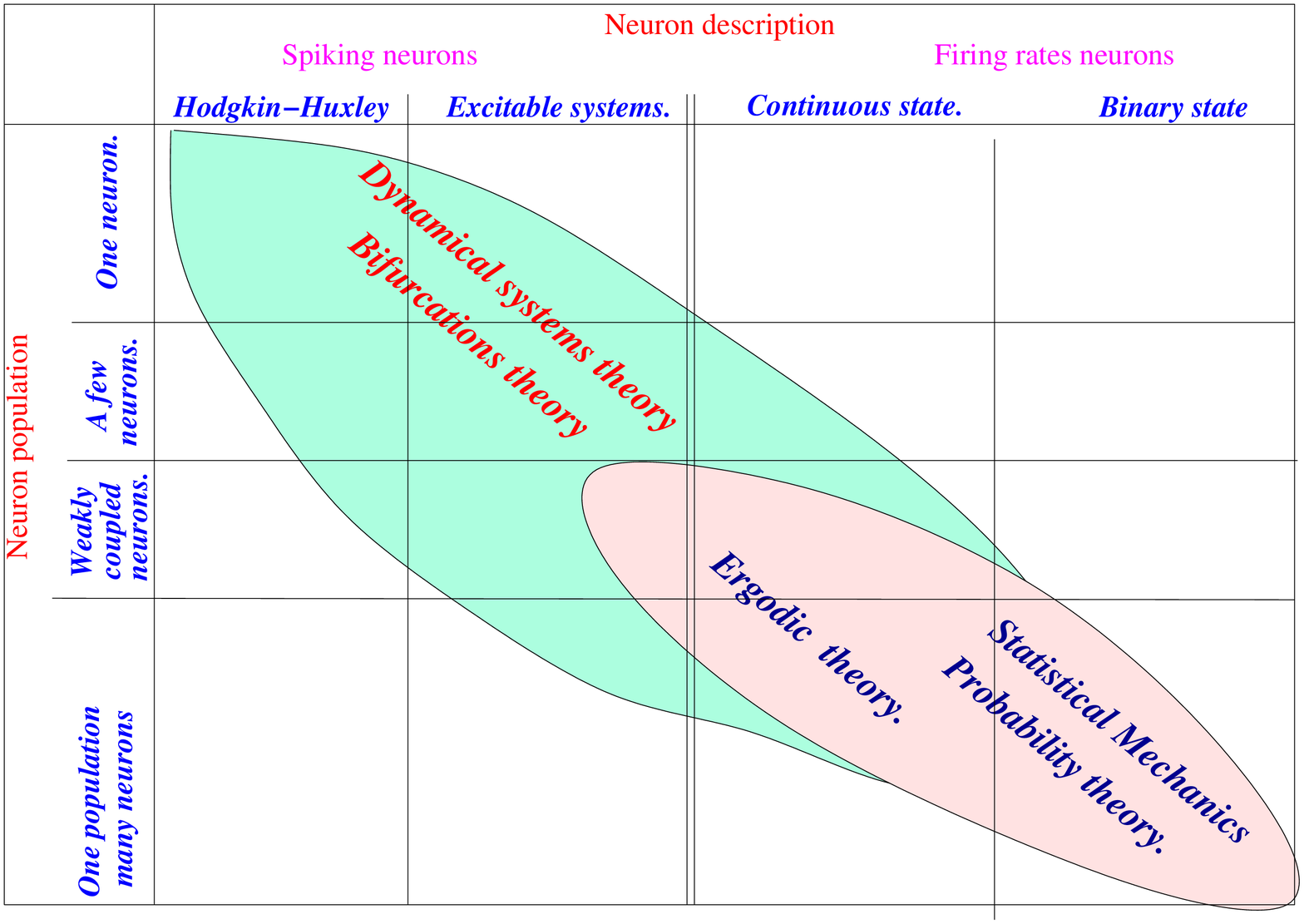}
\vspace{0.5cm}
\caption{Different levels of description of the neuron/network and techniques used to handle
the dynamics of the models presented in this chapter.\label{TabNN}}
\end{center}
\end{figure}

\begin{figure}[ht]
\begin{center}
\includegraphics[height=3cm,width=6cm,clip=false]{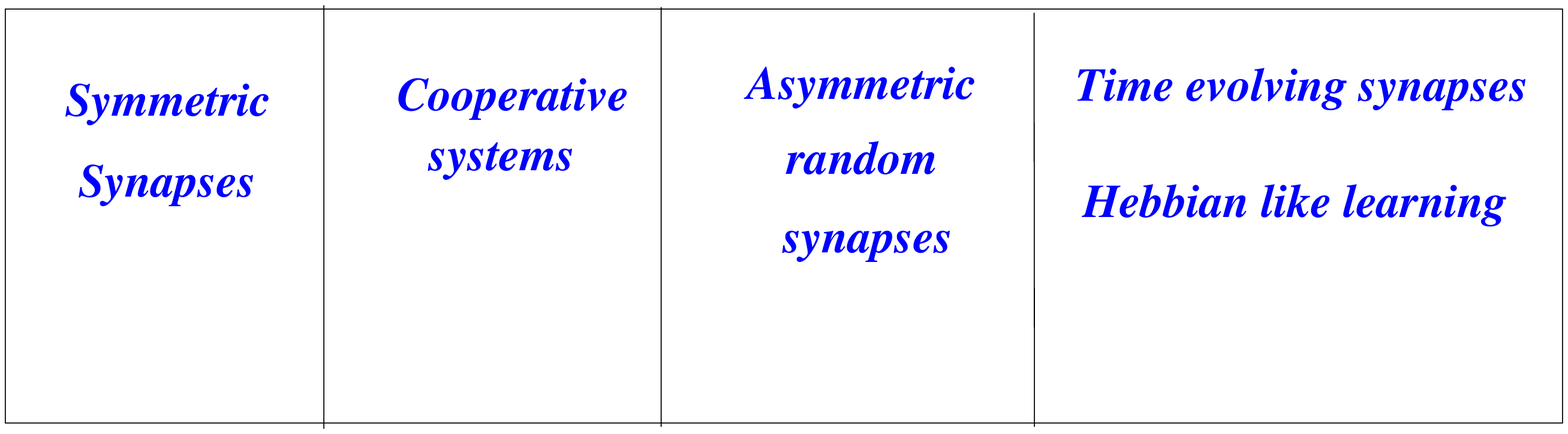}
\vspace{0.5cm}
\caption{Different type of (formal) synapses considered in this chapter (section \ref{Rec}).\label{TabSyn}}
\end{center}
\end{figure}

With this idea in mind we made the choice to explore the world of Neural Networks according
to a specific map, represented in the Figures \ref{TabNN},\ref{TabSyn}, localizing the various
models studied in this chapter in a 3 dimensional space. We started from the obvious remark
that a Neural Network is roughly made of neurons and synapses. But
there are different levels of complexity and accuracy in the description of
neurons  and synapses. For neurons we used a
model categorization along two axis. The first axis is relative to the proximity
to biology. In this hierarchy,
the Hodgkin-Huxley model is at the first rank (Section \ref{SHH}).
 The Hodgkin-Huxley equations are derived in the section  \ref{SHH} and some aspects of their dynamical properties
are briefly described in the sections \ref{SBifHH} (examples of bifurcations occurring in the Hodgkin-Huxley model when
a control parameter such as the external current is applied) and \ref{SProp} (propagation of a spike along the axon).
Before this, one remarks that  the Hodgkin-Huxley equations 
can  be
reduced to a two dimensional dynamical system, taking various forms according to the modeling,
but retaining in particular one of the main feature of the neuron:
the property of \textit{excitability}. The two dimensional excitable dynamical system obtained
by reducing the Hodgkin-Huxley equations are easy to understand and provide fairly pedagogical examples.
The excitable systems come therefore next in our hierarchy 
 (section \ref{SRHH}). They allow one to capture some important dynamical aspects in neuronal behavior, such as spike generation,
refractory period, threshold, and they exhibit various dynamical regimes observed in the experiments. 
After presenting the general structure of models for excitable membranes (section \ref{Sexc}) we discuss several canonical
examples in neuron modeling.
The first example is the  Fitzhugh-Nagumo model (Section \ref{SFN}), then we briefly present the Morris-Lecar 
model (Section \ref{SML}), and  Integrate and Fire models (Section \ref{SIF}). 
%
%
%
%
%
%

These sections essentially deal with ``spiking'' neurons, namely the activity of the
neuron is manifested by emission of action potential or ``spikes'' according to various pattern
(individuals spikes, periodic spiking, bursting, \etc).  On biological grounds, this is certainly a
fundamental aspect in neuronal dynamics. However, another description of the neuron can be
made in  terms of ``firing rates''. The firing rate is the frequency of the spikes 
occurring during a certain time window of length $T$ (typically, $T \sim 100ms$). It  plays certainly
also an important role in a certain number of neurological processes. For example, it is known
since a long time \cite{Adrian1},\cite{Adrian2} 
that the firing rate of stretch receptor neurons in the
muscles is related to the force applied to the muscle. However, during recent years, experimental evidences have suggested that this concept may be too 
simplistic to describe brain activity. It neglects indeed important aspects such as the information
possibly contained in the exact timing of the spikes \cite{Bialek,Abeles1,Shadlen,Hopfield3,Softky,Rieke,Oram}.
Also, the reaction times in behavioral experiments are often too short to allow long temporal averages
(see for example the experiments by S. Thorpe \cite{Thorpe} on the vision).
  
Nevertheless, firing rate models play an important role in the Neural Network community since
they have been often used to model the collective activity of a neural assembly \cite{Amari1},\cite{Amari2},\cite{CG},
 and also
to perform recognition tasks \cite{Hopfield1}. Henceforth, we have included them in our table,
and we have placed them after the spiking neurons in our rough hierarchy. In the examples
described in the sections \ref{Rec},\ref{NotreModele}, corresponding to 
 recurrent neural networks, the neuron is basically considered as an entity having an input and an output
with a non linear transfer function (typically a sigmoid). This nonlinearity has several deep effects on the
dynamics and a detailed example is described in section \ref{NotreModele}. 

Finally, if one makes the further 
approximation that the slope of the sigmoid function is infinite, one ends up with a binary state neuron
(or Mac Cullogh and Pitts neuron \cite{MCP}). Neural Networks with such binary ``spin'' like neurons had
a great  success \cite{Hopfield1} but we shall not discuss them in this chapter.\\

The second axis of the table \ref{TabNN} takes into account the collective aspects of Neural Networks. 
We establish a hierarchy ordering the models by increasing complexity in 
the neural population: one neuron, then a few neurons, then one population of
weakly coupled neurons, then one population with arbitrary couplings (one could also 
consider  several populations interacting with each others, but we do not consider this case in this chapter). 

If one observes this space and asks which analytical methods allow us to describe
the dynamics, one obtains the Table  \ref{TabNN}. A simple glance 
 reveals that the methods discussed in this chapter essentially belong to three different domains of mathematics and physics:
Dynamical systems theory, statistical mechanics and probability theory, and, at the intersection, ergodic
theory. Also, one can remark that we essentially deal
with the diagonal of this Table. As a matter of fact, when one moves away 
from the diagonal one meets, on one side, more and more trivial
models (e.g. an isolated binary neuron), and, on the other side,
more and more complex cases (a big  population of many Hodgkin-Huxley neurons).
In the first case, there is almost nothing non trivial to say, and in
the second one, very little is known at least from the analytical point of view. In this chapter, we therefore choose some examples on
the diagonal and we analyze the corresponding dynamics. \\

There is actually, behind this choice, a fundamental aspect in modeling and analyzing Neural Networks, and more
generally,  modeling and analyzing the so-called ``complex systems''. Complex systems are often composed by elementary units (in our case, neurons), having
their own intrinsic characteristic dynamics and interacting with each others in a complicated way (nonlinear, non symmetric,
with delays,  \etc). The intrinsic dynamics of the units can already be quite a bit complex (see, for example, the
section \ref{SBifHH}) so one may expect the collective dynamics to be even more complex. This is certainly true, but
coupling the units give usually rise to a collective  \textit{emergent} behavior that one may characterize by the
sentence: ``The system as a whole is not reducible to the superposition of its elementary components''.  This is
usually due to non linear effects but this can also result from large numbers effects. Nevertheless, when one builds a dynamical system
by coupling entities, each of them described by a lower dimensional dynamical system, the wisdom acquired when observing
individual units is usually not sufficient to handle the collective behavior. The coupled system inherits characteristics that cannot
be inferred from the knowledge of the uncoupled one. Also, some characteristics of the individual units may be hidden
or may become irrelevant in the collective dynamics. 
These emergent effects can arise even if the coupling is weak. Starting from isolated neurons and ``switching on'' an interaction
(synapses) between them, with an increasing intensity controlled by some parameter, the coupled system may, in some
situations, exhibit a sharp, drastic change in its dynamics even if the parameter is small. This change usually corresponds
to a bifurcation and it has often some analogies with phase transitions in statistical mechanics. Some prominent
examples are presented in section \ref{WCNN}. 

The existence of \textit{emergence} has two consequences. 
Firstly, this justifies somehow the simplifications inherent to \textit{modeling}. If one
desires to understand some emergent properties resulting from coupling neurons it might not be necessary to integrate all the 
features of the isolated neuron. It is often possible to drop some feature (preventing, for example, an analytic computation)
and to capture nevertheless some important collective aspect. This outlines one important feature of the diagonal in table  
\ref{TabNN}. When going from one ``level of complexity'' (detailed description of the neuron dynamics) to another
level (coupling neurons) one often simplifies the characteristics of the neuron in order to have a tractable model.
This is in some sense what we do when going from spiking Hodgkin-Huxley neurons to firing rate neurons. 
However another consequence results from the modeling process aiming to capture some characteristics considered as ``relevant'' and
eliminate others considered as ``details''. The mathematical structure and properties of
the coupled model might be drastically different from the uncoupled one. This means that the tools, techniques or even
philosophy adopted to handle the dynamics may change from one level of complexity to another. As we shall see, for example,
the normal forms theory is quite a bit useful to handle dynamical changes in isolated neurons or in weakly
coupled neural networks (provided some necessary assumptions are made), but it is of little help in randomly coupled
recurrent neural networks, at least before any prior treatment (such as the dynamic mean field equations of
section \ref{MFT}).    
 
It results from these remarks that there is, currently, no general strategy to study Neural Networks dynamics.
 Nevertheless, as we shall see, dynamical systems theory, probability theory,
statistical physics and ergodic theory can sometimes be used and combined to give partial solutions and can be
tailored to build new tools and methods. A few examples are given in this chapter.\\ 

Now, a few words about Table \ref{TabSyn} below. It defines a third dimension in our classification space, where
we define several levels of description for the ``synapses'' (interactions between neurons). 
The detailed physiology of the synapse is complex and, actually, there exists 
different types of synapses: chemical or electrical (gap junction). However, in most \textit{models} the
mathematical description is rough and, quite often, synapses
are basically modeled in a way allowing to store information \textit{in the network}, this information being extracted from the dynamical
evolution of the neurons. Depending on the modeling chosen for the synapses, the dynamics can be very different,
and their modification can induce drastic dynamical changes. In this chapter, we essentially give one example of the
changes induced when one considers the different types of synapses presented in table 2, for
recurrent networks (section \ref{Rec}). We discuss first the convergence properties of the Cohen-Grossberg model
when the synapses are symmetric (section \ref{SCG}). Then we discuss the case of cooperative networks. The main result is a
convergence theorem
from Hirsch \cite{Hirsch} which had recently some extensions in the field of genetic networks \cite{Gouze,Soule}.
We also discuss in this section the notion of frustration resulting from the competition of excitatory/inhibitory 
effects. The section \ref{NotreModele} is devoted to the complete
analysis of a recurrent model with asymmetric interactions, exhibiting complex regimes such as chaos. One
can indeed go quite a bit deep in the description of the dynamics, by combining dynamical systems theory, statistical
mechanics and ergodic theory (sections
\ref{Mod},\ref{ResGen},\ref{KO},\ref{MFT}).
This model exhibits interesting properties when submitted to Hebbian learning (section \ref{App}). We also present new developments 
characterizing the ability of such a network to
transmit a signal. The basic tools is a linear response theory recently developed by Ruelle \cite{Ruelle2}
 (section \ref{RepLin}).    \\      
%
%
%
%
%
%
%
%
%
%
%
%

To conclude this introduction we would like to point out an important aspect. Many techniques described
here have been developed out of the field of Neural Networks. But, in many cases they have been tailored
or adapted to tackle specific problems in this field, and new methods have emerged. The interesting remark
is that some of these techniques have now applications in other fields such as genetic networks, communication networks,
or more generally non linear dynamical systems on non regular graphs\footnote{
In this way, the wisdom coming from the field of Neural Network is different (and complementary)
from the knowledge acquired in parallel fields, such as coupled map lattices.}, with a large number of degree of freedom
(but \textit{finite}).  Some examples of applications to other
fields are discussed in this chapter.

\su{Spiking neurons and excitable systems.}\label{Spiking}

\begin{figure}[ht]
\begin{center}
\includegraphics[height=5cm,width=7cm,clip=false]{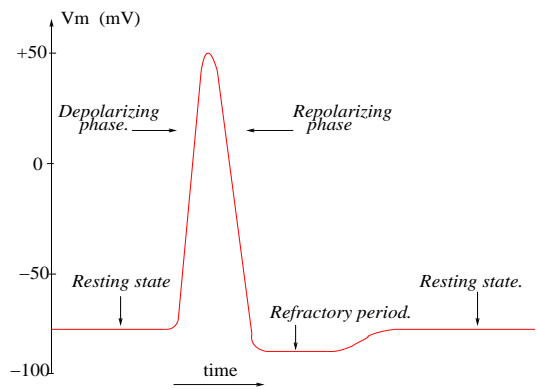}
\caption{Typical action potential of a neuron.
\label{PA}}
\end{center}
\end{figure}

The activity of a neuron occurs by the emission of \textit{action potentials} (or \textit{spikes})
 (see Fig. \ref{PA}).
In the simplest cases, they are controlled
by ions (mainly Sodium (Na+) and
Potassium (K+)) and their concentration around the nerve cell (see section \ref{SHH}).
 An external stimulus causes Na-selective ion channel to open
causing an influx of Sodium in the nerve cell. If the corresponding potential
exceeds a threshold value (depolarization threshold) an action
potential is generated. The action potential 
propagates then along the axon (section \ref{SProp}).  After the cell depolarizes, it must repolarize to its
resting potential before it can depolarize again. This repolarization
phase is controlled by an efflux of Potassium (repolarization phase). 
This phase is followed by a refractory period where
the neuron cannot be excited. The initial balance between Sodium
and Potassium is restored by  ionic pumps.
%
%
%
%
%
Different models accounting for action potential generation exist and some of them are described
below. But, the core of all these models is certainly the Hodgkin-Huxley's that
we describe in the next section.

\ssu{Hodgkin-Huxley neurons.}\label{SHH}

The classical description of neuronal spiking dates back to Hodgkin and Huxley
\cite{HH}. After extensive experimental studies these authors were able
to propose a model for the dynamics of the giant axon of the squid. This 
constituted a significant breakthrough in the description of action potential.
 At the time of their experiments (1952), the modern concept of ion-selective channels
controlling the flow of current through the membrane was only
one hypothesis among several competing others. Their model ruled out
alternative ideas and gave correct predictive results of experiments
that were not used in formulating the model. It reproduces and explain a remarkable
range of data from squid axon, including the shape and propagation of the action
potential, its sharp threshold, refractory period, anode-break excitation, accommodation
and sub-threshold oscillations.  Hodgkin \& Huxley also proposed a set of  equations
modeling spikes propagation along the axon (see section \ref{SProp}). They were
in particular able to predict the propagation rate of spikes with a remarkable accuracy.
The  Hodgkin-Huxley modeling is
generic, tractable and gave rise to new techniques and concepts. Consequently, 
the actual models of neural excitability are greatly influenced 
this work which resulted in a Nobel price (1961) for the authors. 
There is a large number of papers and books dealing with Hodgkin-Huxley model. Our
main references are \cite{Cronin,Gerstner,Hille,Keener,Koch,Nelson}\\

In their work, Hodgkin and Huxley start from the idea that the action potential results
from transmembrane currents mainly constituted by Sodium ($Na^+$) and Potassium
($K^+$) ions. Consider a neuron at rest in its natural environment, namely in the intra cellular
fluid where the Sodium and Chloride concentration is similar to sea water. One observes that,
at rest, the $Na^+$ concentration is about $10$ times higher outside the neuron than inside,
while the $K^+$ concentration is about $5$ times higher inside than outside. Assuming that
the system is locally at thermal equilibrium with a temperature $T$, the difference in 
concentration between the inside and the outside, for the ionic species $X$, results
in a potential difference $E_X \deq V_{in}[X] - V_{out}[X]$ called the \textit{Nernst potential}
and given by: 
\beq \label{Nernst}
E_X=\frac{RT}{F}\log\left(
\frac{\left[X\right]_{out}}{\left[X\right]_{in}} \right)
\eeq
\nid where $R=\cN k=8.315 J/K$ is the ideal gas constant, ($\cN=6.02 \times 10^{23}$ is
the Avogadro number, $k=1.38 \times 10^{-23} J/K$ the Boltzmann constant), $F=\cN e =96500$ C is the Faraday number
($e=1.602 \times 10^{-19} C$ is the charge of the proton), and $\left[X\right]_{out}$ (resp. $\left[X\right]_{in}$)
is the concentration of $X$ outside  (resp. inside) the neuron. With this convention, for positive
ions, the effective
electric force has the same direction as the force induced by the concentration gradient. For the
giant axon of the skid and  for a temperature $T=6.3 \ ^\circ C$, the Nernst potential
for Sodium and Potassium are respectively $E_{Na} \sim 56 mV$, $E_{K} \sim -77 mV$. Moreover,
taking into account the respective concentration of all ionic species the membrane potential
is about $-70 mV$ at rest. Were the membrane to be permeable to ions, would one observe
ionic currents through the membrane. These currents are not observed at rest, but arise
during an action potential. Consequently, the ionic permeability of the membrane (conductance) depends
on the neuron state (i.e. its membrane potential). \\

In Hodgkin-Huxley modeling the (macroscopic) membrane conductances are determined
by the combined effects of a large number of microscopic
\textit{ionic channels} located in the membrane. One considers a channel
as an ensemble of\textit{ independent gates} (that can be of different type) 
with a binary open-closed state. Denote  by
$p_i\equiv p_i(V)$ the probability that a a gate of type $i$ is open.
Then the conductivity $G_X$ for
channels of ionic species $X$, with gates of type $i=1 \dots N$, is proportional to
the product of the probabilities $p_i$ that the gate $i$ is open :
$G_X =g_X \prod_{i=1}^Np_i$, where $g_X$ is the maximal conductance for
channels of type $X$.
Each $p_i$ depends on the potential
$V$ and on the fraction of open ($p_i$) and closed ($1-p_i$)
gates.  In the Hodgkin-Huxley model the time dependence of the $p_i$'s is given by a master equation:

\beq \label{GatesHH}
\frac{dp_i}{dt}=\alpha_i(V)(1-p_i) - \beta_i(V)p_i= \frac{p_i^\infty(V)-p_i}{\tau_i(V)}
\eeq

\nid where $\alpha_i$ (resp. $\beta_i$) are the transition rates from close to 
open (resp. open to close) or  \textit{gate inactivation} (resp. \textit{activation}).
They have been empirically determined by  Hodgkin and Huxley
for each ion species. 
They are function of the membrane potential $V$ (see eq. (\ref{HHparam}) below).
 In the second equality one introduces the natural quantities:

\beq 
\tau_i(V)=\frac{1}{\alpha_i(V)+\beta_i(V)} ; \quad p_i^\infty(V)=\frac{\alpha_i(V)}{\alpha_i(V)+\beta_i(V)}
\eeq

\nid where $\tau_i$ is a characteristic time constant and
$p_i^\infty$ is called the \textit{steady state activation}.
This is the value reached by $p_i$ when it is held at a potential
$V$ for a long period (say larger than the characteristic time $\tau_i$).  
The solution of (\ref{GatesHH}) is obviously:

\beq 
p_i(t)=p_i^\infty(V)-(p_i^\infty(V)-p_i^0)e^{-\frac{t}{\tau_i(V)}}
\eeq 

\nid Consequently, for a fixed $V$, $p_i$ has a simple exponential time dependence governed by
$\tau_i$.

From their experiments Hodgkin-Huxley proposed
 to model the $K$ conductance with an equation of the form:

\beq\label{GK}
G_K=g_K n^4
\eeq 

\nid where $g_K$ is the maximum Potassium conductance. This corresponds to have 
a $K$ channel  with four independent gates of type $n$. 
The probability  $n$ is called the \textit{$K$ activation variable}.

A similar equation can be written
for the sodium:

\beq\label{GNa}
G_{Na}=g_{Na} m^3 h
\eeq 
 
\nid This corresponds
to model a $Na^+$ channel with  three gates of type ``$m$'' and one gate of type ``$h$''.
 $m$ is the \textit{$Na$ activation variable},
 and $h$ is called the \textit{$Na$ inactivation variable}. The $Na^+$
ions can penetrate in the cell only if the $m$ and $h$ gates are both open (see Fig. \ref{gateNa}).\\

The membrane potential $V$ is now given by Kirchhoff law 
\beq \label{Kirch}
C_m \frac{dV}{dt}+I_{Na} + I_{K} +I_L = I_{ext}
\eeq
\nid where $I_{Na},I_{K}$ are the sodium and potassium ionic currents
through the cell membrane, $I_L$ the leakage current (mainly composed
by $Cl^-$ ions)
 and $I_{ext}$ is some external current
(for example applied during an experiment). $C_m$ is the membrane
capacity ($\sim 1 \mu F/cm^2$).  The currents are given
by the Ohm's law $I_i = G_i(V-E_i)$ where $E_i$ is the Nernst
potential of the species $i=Na,K,L$.

Finally, the ionic currents  are given by :
\bea 
C_m \frac{dV}{dt}&=& - g_{Na} m^3 h (V - E_{Na}) - g_K n^4(V - E_{K}) - g_L(V - E_L)+ I_{ext} \label{HHV}\\
\frac{1}{\gamma(T)}\frac{dn}{dt}&=&\alpha_n(V)(1-n) - \beta_n(V)n= \frac{n^\infty(V)-n}{\tau_n(V)}\label{HHn}\\
\frac{1}{\gamma(T)}\frac{dm}{dt}&=&\alpha_m(V)(1-m) - \beta_m(V)m= \frac{m^\infty(V)-m}{\tau_m(V)}\label{HHm}\\
\frac{1}{\gamma(T)}\frac{dh}{dt}&=&\alpha_h(V)(1-h) - \beta_h(V)h= \frac{h^\infty(V)-h}{\tau_h(V)}\label{HHh}
\eea
The dynamical system (\ref{HHV}-\ref{HHh})
constitutes the complete Hodgkin-Huxley system. It involves a temperature dependent factor :
\beq
\gamma(T) = 3^{\frac{(T-6.3)}{10}}
\eeq
This factor has the only effect of modifying the time constants in the equations
for the activation/inactivation  variables\footnote{
For a recent numerical work on the effects of temperature on the dynamics of a network composed 
by Hodgkin-Huxley neurons, coupled with gap junctions, see \cite{Yoshioka}.}. In the sequel we shall forget it
and assume that the temperature is $T=6.3 \ ^\circ C$ ($\gamma(T)=1$).\\

The $V$ dependence of the parameters $\alpha_n,\beta_n,\alpha_m,\beta_m,\alpha_h,\beta_h$  
was determined empirically by Hodgkin and Huxley.
They found \footnote{\label{Norigin}In the literature one may find different forms for
these equations depending on the zero of the potential. Here we have chosen it
such that the membrane potential at rest is $V_{rest}=-70mV$. One can
also choose it such that  $V=0$ at rest.}:
\bea \label{HHparam}
\alpha_m(V)= \Psi\left(\frac{-(V+45)}{10}\right); \quad && \quad \beta_m(V)=4e^{\frac{-(V+70)}{18}}\\
\alpha_n(V)= 0.1\Psi\left(\frac{-(V+60)}{10}\right); \quad && \quad \beta_n(V)=0.125e^{\frac{-(V+70)}{80}}\\
\alpha_h(V)= 0.07 e^{\frac{-(V+70)}{20}}; \quad && \quad \beta_h(V)=\frac{1}{1+e^{\frac{(-(V+40))}{10}}}
\eea
\nid with:
\beq
\Psi(x)=\left\{
\baR{ccc}
\frac{x}{e^x-1} \quad && \quad \mbox{if $x\neq 0$}\\
1  \quad && \quad \mbox{if $x=0$}
\eaR
\right.
\eeq
In  Fig. \ref{FigHHparam}a have we drawn the time constants $\tau_n,\tau_h,\tau_m$
deduced from eq. (\ref{HHparam}) as functions of $V$, while in fig. \ref{FigHHparam} b the  steady state
values $n^\infty,m^\infty,h^\infty$ as functions of $V$ are shown.
One notes in particular that the time constant for the Na activation variable
is about  \textit{one order of magnitude less than for the Na inactivation and the K activation,
through the entire range}. This means that the response in the $m$ variable
is quite a bit faster than the other variables. Consequently, during an action potential,
when the voltage is high and $m$ is large, it will take a while for $h$ to decrease
and for $n$ to increase and contribute to the opposite $K$ current.
%
%
%
%
%
\begin{figure}[ht]
\begin{center}
\includegraphics[height=5cm,width=5cm,clip=false]{hhact1}
\hspace{1cm}
\includegraphics[height=5cm,width=5cm,clip=false]{hhact2}
\caption{Time constants  $\tau_n,\tau_h,\tau_m$ and  steady state
values $n^\infty,m^\infty,h^\infty$ as functions of $V$.
\label{FigHHparam}}
\ec
\enf
The mechanism of action potential emission is then the following.  
 In the resting phase (a) the $m,n$ gates are closed
while the $h$ gate is open. Therefore, sodium and potassium are neither leaving
nor entering the cell (fig \ref{gateNa}a).  During depolarization, the $m$
gates open fast allowing sodium to diffuse inside the cell, 
following the concentration gradient, while the $n$ gates are still closed (fig \ref{gateNa}b).
This increases the membrane potential.
Then $n$ increases slowly, more and more $K$ gates are open,
 generating an opposite $K$ current. In the same time, $h$ decreases
and more and more $h$ gates close, preventing sodium from coming
into the cell (fig \ref{gateNa}c). This corresponds to the repolarization phase. In the refractory period
 the $m$ gates close, the $h$ gates stay closed and
the $n$ gates stay open. It is not possible to excite the neuron in this phase (fig \ref{gateNa}d).
Finally, the $h$ gates open, the ionic balance is restored by ionic pumps, and the resting state is once again achieved.
If one draws the membrane potential versus time one obtains a picture similar to  figure \ref{PA}.
The action potential is then propagated along the axon. The propagation equations
are studied in section \ref{SProp}.
%
%
%
%
%
\begin{figure}[ht]
\begin{center}
\includegraphics[height=5cm,width=8cm,clip=false]{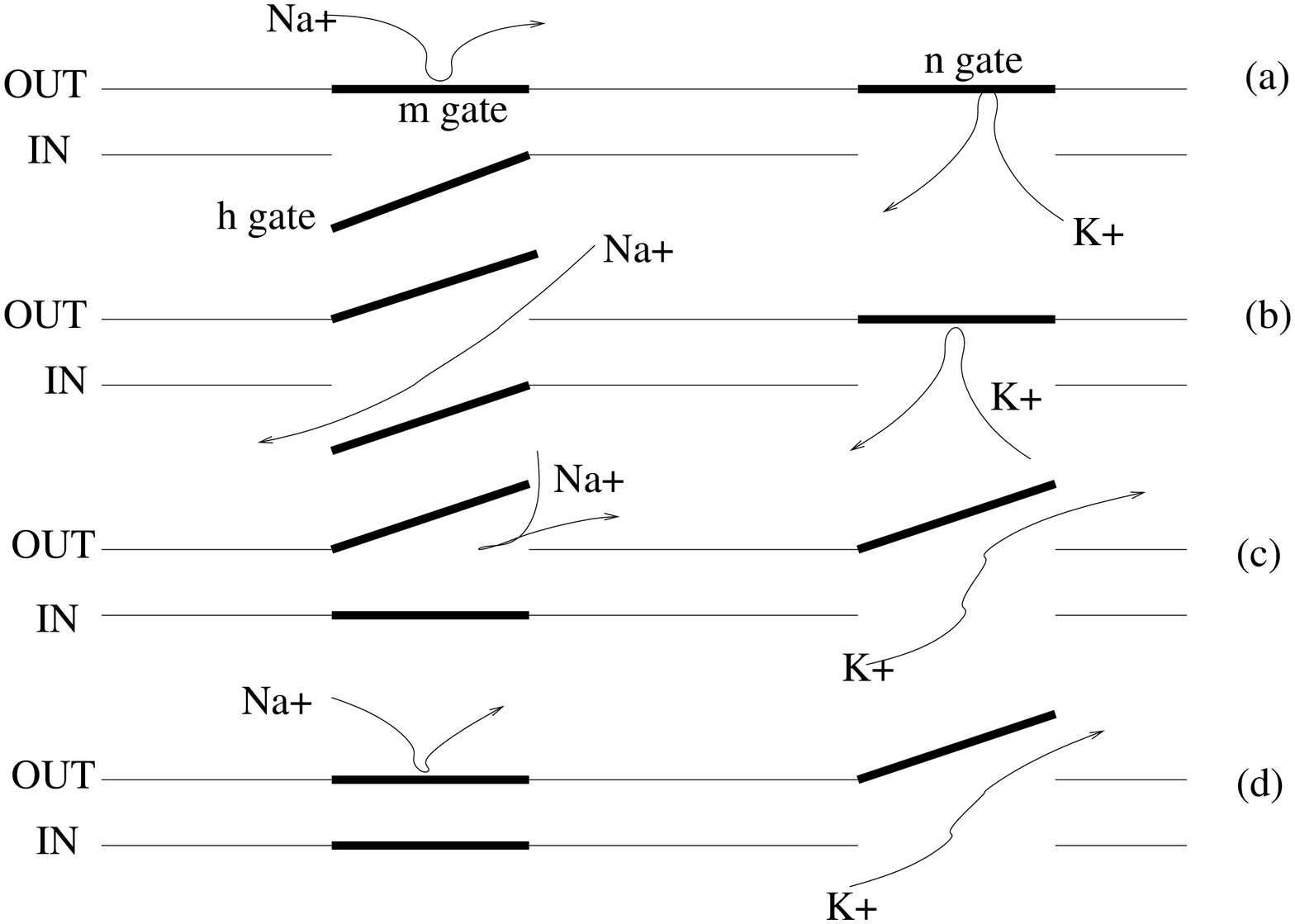}
\hspace{1cm}
\caption{The various phase of the action potential
in terms of the Hodgkin-Huxley equations.
\label{gateNa}}
\ec
\enf

The preceding analysis is only qualitative but deeper mathematical investigations can be done
(see section \ref{SBifHH}) and numerical simulations can be performed. One observes
spike generations but also periodic spiking, bursting etc... 
The Hodgkin-Huxley equations describe therefore the neural dynamics with a fantastic accuracy
accounting of the wide variability in neuron activity.  In particular,
one predicts various situations observed in experiments.   On the other
hand  they equations can be simplified  giving rise to many models of formal neural networks. 
Despite this simplification (that can be quite a bit rough) it is still possible to obtain a huge quantity of information
 about the neural dynamics. In the next section we present a few  models derived from Hodgkin-Huxley equation and capturing
one of the main feature of the biological neuron: \textit{excitability}.

\ssu{Reducing the Hodgkin-Huxley equations.} \label{SRHH}

\sssu{General structure of excitable membrane.} \label{Sexc}

Most models for excitable membrane retain the general Hodgkin-Huxley 
structure (eq. (\ref{HHV})-(\ref{HHh})) and can be written in the form.

\bea\label{NGen}
C_m\frac{dV}{dt}&=&-I_{ion}(V,X_1, \dots, X_n)+I_{ext} = -\sum_{k=1}^N I_k(V,X_1, \dots, X_n)
 + I_{ext}, \  \label{Kirchg}\\
I_k&=&g_k\sigma_k(V,X_1, \dots, X_n)(V - E_k), \qquad  k=1 \dots N, \label{Ohm} \\
\frac{dp_i}{dt}&=&\frac{p_i^\infty(V)-p_i}{\tau_i(V)}, \qquad i=1 \dots l, \label{ions} 
\eea

\nid where $V$ denotes membrane potential,
$C_m$ the membrane capacity, $I_{ion}$ is the sum  of ionic currents, $I_{ext}$
an external or applied current. The variables $p_i$ are used to describe the fraction
of open channels of type $i$. $\tau_i$ is the characteristic time that the
ions of type $i$ need to reach the rest state $p_i^\infty(V)$. In the Ohm's law
(\ref{Ohm}), $I_k$ is the current for the $k$ th ion species,
$g_k$ is the maximal conductivity for the ions channels of type
$k$, $\sigma_k$ is the product  of gate $k$-channels activity, and $E_k$ is the Nernst
equilibrium potential. 
In some situations it is fundamental to have an accurate
models of the neuron excitability, if one seeks, for example, to account
for rather detailed aspects of spike shape, dependence upon many pharmacological
agents, etc .... However,  in many cases a rough description is enough to capture the
main qualitative and quantitative aspects of the dynamics of excitability.
Consequently, one can reduce the complexity of the set of equations (\ref{Kirchg},\ref{Ohm},
\ref{ions})
in order to obtain an analytically tractable model.
Henceforth, many models of neuronal dynamics are reduction of these general equations.

\sssu{The FitzHugh-Nagumo model} \label{SFN}

In this spirit  FitzHugh \cite{FitzHugh} and independently Nagumo, Arimoto et Yoshizawa \cite{Nagumo},
 considered reductions of the
Hodgkin-Huxley model and introduced an analytically tractable
two variables model.  

The basic observation is the \textit{time scale separation} between the variables
$V,m,n,h$ in eq. ((\ref{HHV})-(\ref{HHh})). According to Fig. \ref{FigHHparam} the characteristic time
for Sodium activation is so fast compared to the other variables that one may consider
$m$ essentially as a constant. This eliminates the variable $m$. Also, FitzHugh
observed that $h+n$ is essentially a constant $\sim 0.8$ during the action potential.
Consequently, one can eliminate one more variable. One finally obtains a model of the
form (for the detailed reduction see e.g. \cite{AK},\cite{Kall},\cite{Keener},\cite{Gerstner},\cite{Rinzel}):
\bea
\epsilon \frac{dv}{dt}&=& f_\lambda(v,w) \label{SystExca}\\
\frac{dw}{dt}&=& g_\lambda(v,w)\label{SystExcb}
\eea
\nid where $\D{\epsilon = \frac{C_m}{\max_V \tau_n(V)}}$ is typically \textit{small}. 
The index $\lambda$ refers to the control parameters of the system. In the FitzHugh-Nagumo model 
$f_\lambda(v,w)=v-v^3-w+I$ is a cubic polynomial in $v$ and is linear in $w$,
while $g_\lambda(v,w)=(v-a-bw)$. The parameters $\lambda=(a,b,I)$ are deduced from 
the physiological characteristics of the neuron.
It can also
be useful to consider the dynamical system
\bea
\frac{dv}{dt}&=& f_\lambda(v,w)\label{SystExc1a} \\
\frac{dw}{dt}&=& \epsilon g_\lambda(v,w)\label{SystExc1b}
\eea
\nid obtained from (\ref{SystExca},\ref{SystExcb}) by a time rescaling
$t \to \frac{t}{\epsilon}$.

 The system of equations (\ref{SystExca},\ref{SystExcb}) is the canonical form
for  \textit{excitable systems}. That is why we used the ``generic'' variables namely $v,w$
instead of $V,n$. They are usually called \textit{excitation} and \textit{recovery} variables.
The excitation variable governs the rise to the excited state while the recovery variable causes
the return to the steady state. 
Since $\epsilon$ is typically a small parameter, there is a separation
of time scales between the two variables.\\

On technical grounds, the analysis is simplified
by the two dimensional geometry of the phase space.
Indeed, in the phase plane, the slope of the trajectory of a given point
is $\frac{dw}{dv}=\frac{g_\lambda(v,w)}{f_\lambda(v,w)}$ and consequently the phase portrait
can easily been drawn. In particular a trajectory is vertical (resp. horizontal)
at the points such that $f_\lambda(v,w)=0$ (resp. $g_\lambda(v,w)=0$). The set
of points $N_v \deq \lbrace (v,w) \ | \ f_\lambda(v,w)=0 \rbrace$
 (resp. $N_w \deq \lbrace (v,w) \ | \ g_\lambda(v,w)=0 \rbrace$) is composed
by a union of curves called the 
$v$-\textit{nullclines} (resp. $w$-nullclines). Thus, the fixed points of (\ref{SystExca},\ref{SystExcb})
are at the intersection of nullclines. More generally, the shape of
the nullclines gives strong informations on the dynamics. As shown below
the nullclines shape changes when the parameters $\lambda$ are varying, leading to bifurcations
for some values of $\lambda$.\\

When $\epsilon$ is small one  uses an additional property to analyze the dynamical
system (\ref{SystExca},\ref{SystExcb}). Setting $\epsilon=0$ in (\ref{SystExca},\ref{SystExcb}) one obtains
$f_\lambda(v,w)=0;\frac{dw}{dt}=g_\lambda(v,w)$.
 This means that, whenever it is possible, $v$ is adjusted rapidly to maintain a pseudo-equilibrium
corresponding to $f(v,w)=0$ and plays the role of an implicit parameter
in the evolution of $w$. In other words, the point $(v,w)$ moves slowly along the (stable) branches of the
$v$ nullclines.  These branches compose the so-called ``slow manifold'':
it is only ``on'' (or very near) this curve that the motion of the solution curves is not very fast in a nearly
horizontal direction (see e.g. Fig. \ref{FToy}).
 
On the other hand, away from the $N_w$ nullcline, the vector field
is essentially horizontal and one has a fast motion of $v$. Indeed, a time rescaling $t \to \frac{t}{\epsilon}$
gives the system (\ref{SystExc1a},\ref{SystExc1b}). Then, setting $\epsilon=0$   one can approximate the (regular)
trajectories of the system (\ref{SystExca},\ref{SystExcb}) by the (non regular) trajectories
of the \textit{degenerated system}:  
\bea\label{SystSingExc}
\frac{dv}{dt}&=& f_\lambda(v,w) \\
\frac{dw}{dt}&=& 0
\eea
\nid where the vector field is horizontal with a norm $f_\lambda(v,w)$.

The trajectories of the real system are composed by  pieces coming from these two approximations. 
There are theorems controlling how the real trajectories of (\ref{SystExca},\ref{SystExcb}) are close to
the piecewise trajectories, for a sufficiently small $\epsilon$ allowing to obtain the characteristic trajectories 
of the initial system from the solutions of the degenerated system. This is the essence of the
singular perturbation theory developed by  Mischenko \& Rozov \cite{MR}.\\

To illustrate this, let us start we a simple example used as a preliminary
step to analyze later on the FitzHugh-Nagumo equations:

\bea\label{Toy}
\epsilon \frac{dv}{dt}&=&v - v^3 - w \\
\frac{dw}{dt}&=&v - a
\eea

The $v$-nullcline is given by $w=v - v^3$
while the $w$-nullcline is the vertical line $v=a$.
The nullclines and the flow of (\ref{Toy}) are depicted fig. \ref{FToy}.  
Due to the smallness of the parameter
$\epsilon$, the flow is essentially horizontal\footnote{Note that, strictly speaking, the vector field on the $y$ axis is not
horizontal (its component are $(-\frac{w}{\epsilon},-a)$). However, for simplicity, we are drawn
it horizontal, assuming a very small $\epsilon$ value.} ($\frac{dw}{dt} \sim 0$)
except close to the $v$-nullcline. Crossing the $v$-nullcline
(resp. the $w$-nullcline) 
makes the $v$ component of the flow (resp. the $w$ component) changing its sign. 
The $v$ nullcline has two ``stable'' branches  denoted by $N_v^\pm$.
Namely the flow is attracted in a neighborhood of these branches and
stays a long time in this neighborhood, moving slowly upward for the $+$ branch and downward for the branch $-$. In the case
of the $+$ branch the flow finally  reaches the extremum.
Then it moves fast to the other branch. The middle branch is called the \textit{unstable} branch. As discussed below it acts
(roughly) as a threshold for spike generation.

%
%
%
%
\begin{figure}[ht]
\begin{center}
\includegraphics[height=6cm,width=8cm,clip=false]{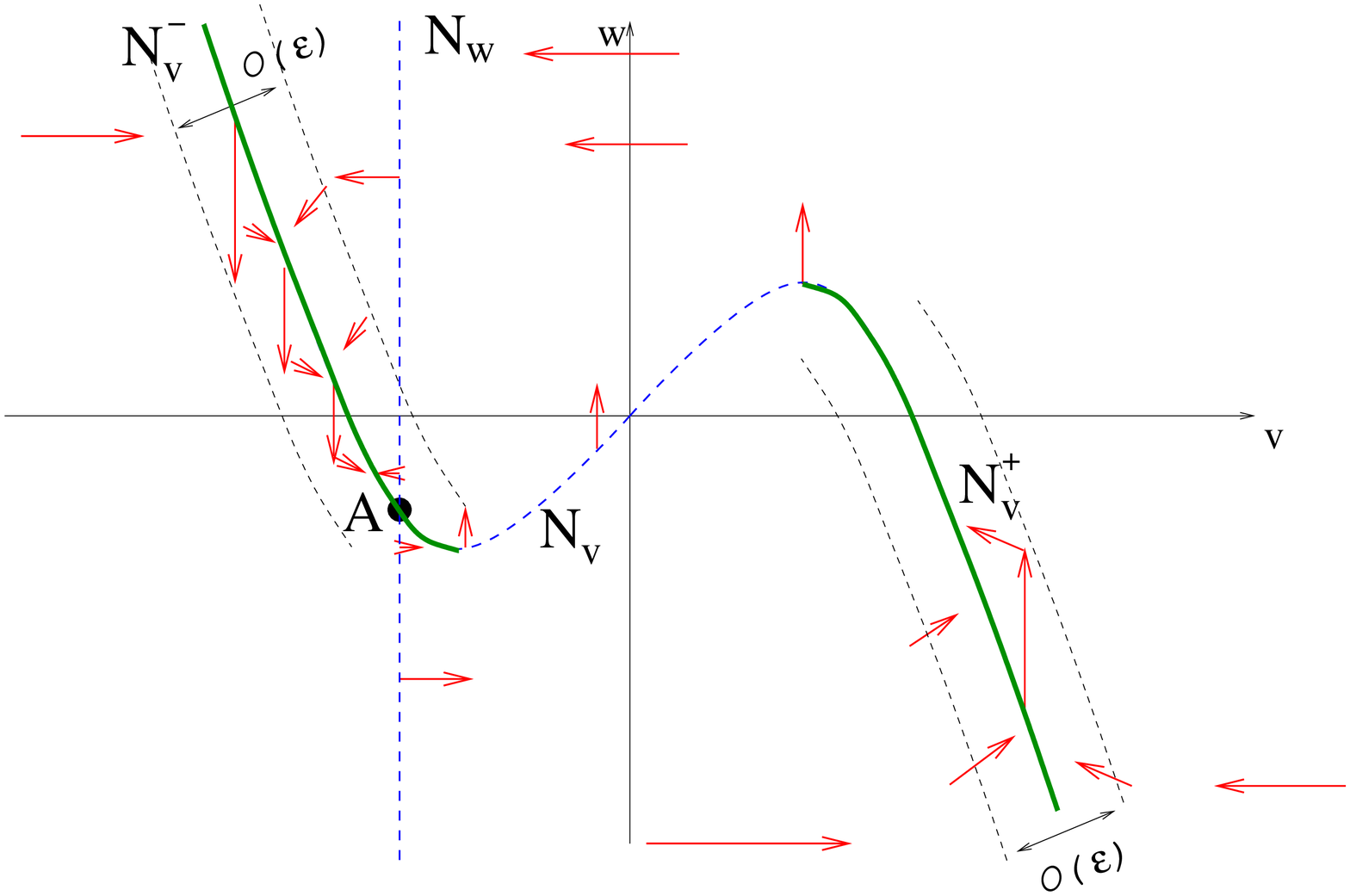}
\vspace{0.5cm}
\caption{Nullclines and vector field for the toy model (\ref{Toy}). This a qualitative
drawing and the phase portrait has been drawn ``by hand''. Consequently, the arrows representing
 the vector field are drawn as indicators. The picture is \textit{not} scaled. In particular, the vicinity of
the slow manifolds (in green) is of order $\epsilon$. Practically, the trajectories near the slow manifold can 
essentially be considered as being ``on'' the slow manifold.  \label{FToy}}
\end{center}
\end{figure}
%
%
%
%
%
The point $A=\left(v_A=a,w_A=-a+a^3\right)$, where the nullclines intersect, is a fixed
point. The eigenvalues of the corresponding Jacobian matrix $DF_A$ are
$\lambda_{1,2} = \frac{1-3a^2 \pm \sqrt{(1-3a^2)^2 - 4\epsilon}}{2}$.
Consequently, the eigenvalues are complex for $a \in ]\frac{-1+2\sqrt(\epsilon)}{3},
-\frac{1-2\sqrt(\epsilon)}{3}[ \cup ]\frac{1-2\sqrt(\epsilon)}{3},
\frac{1+2\sqrt(\epsilon)}{3}[$
and real otherwise. Moreover, $A$ is stable when $|a| > \frac{1}{\sqrt{3}}$
and unstable otherwise. More precisely, this is a sink 
($\lambda_1, \lambda_2 <0$) for $a \in
]-\infty, -\sqrt{\frac{1+2\sqrt{\epsilon}}{3}}] \cup 
[\sqrt{\frac{1+2\sqrt{\epsilon}}{3}},+\infty[$, 
a stable focus ($\Re(\lambda_{1,2})<0$) for 
$a \in
]-\sqrt{\frac{1+2\sqrt{\epsilon}}{3}}, -\frac{1}{\sqrt{3}}[ \cup 
]\frac{1}{\sqrt{3}},\sqrt{\frac{1+2\sqrt{\epsilon}}{3}}[$, 
a center ($\Re(\lambda_{1,2})=0$) for $|a|=\frac{1}{\sqrt{3}}$, 
an unstable focus ($\Re(\lambda_{1,2})>0$) for
$a \in ] -\frac{1}{\sqrt{3}},-\sqrt{\frac{1-2\sqrt{\epsilon}}{3}}[ \cup 
]\sqrt{\frac{1-2\sqrt{\epsilon}}{3}},\frac{1}{\sqrt{3}}[$,
and a source ($\lambda_1, \lambda_2 >0$) for
$a \in
[-\sqrt{\frac{1-2\sqrt{\epsilon}}{3}},\sqrt{\frac{1-2\sqrt{\epsilon}}{3}}]$ (see the appendix
for more details about the classification of fixed points).\\

Assume now that we are in the situation depicted in Fig. \ref{FigToy}a, with $a < -\frac{1}{\sqrt{3}}$. 
The system is at rest
in $A$. Now, we excite it moving $A$ to $B=(v_B,w_B$. There are two possibilities.
Either $w_B > -\frac{2}{3\sqrt(3)}$, then the excitation relaxes down to the rest
state
(Fig. \ref{FigToy}a). Or $w_B < -\frac{2}{3\sqrt(3)}$. Then we have the situation depicted
in fig. \ref{FigToy}b. The trajectory  flows rapidly parallel to $v$ until it approaches 
the $v$-nullcline and crosses it in $C$. Then it follows slowly the stable branch $(C,D)$.
At this point, the $v$ flow is zero while the $w$ flow is positive.
Consequently, the trajectory leaves the nullclines, and is fast driven
by the flow until the point $E$. It follows then the stable branch $(E,A)$ until
the rest state $A$. The corresponding trajectory of $v$ is depicted in the
inset of 
Fig.\ref{FigToy}b. It has a spike shape where one recognizes the equivalent of the
depolarizing phase $(B,C)$,  the repolarizing phase ($C,E$), and the 
refractory period ($E,A$) of the figure \ref{PA}. 
Consequently, this simplified model
gives already a fairly good example of an excitable dynamical system.

Note that the dynamical system (the neuron) is more sensitive to excitation
when the fixed point $A$ is closer to the local extremum $M_1=(-\frac{1}{\sqrt{3}},
-\frac{2}{3\sqrt(3)})$ of the nullcline (resp. $M_2=(\frac{1}{\sqrt{3}},
\frac{2}{3\sqrt(3)})$), namely when the control parameter
$a$ is close to the bifurcation value $a=-\frac{1}{\sqrt{3}}$ (resp. $a=\frac{1}{\sqrt{3}}$).
In this way, one may consider that excitable neurons are dynamical systems
close to a bifurcation point.  This idea is further developed in section \ref{WCNN}.
This dynamical system has moreover an additional feature which makes it
relevant to neuronal dynamics. Assume now that $|a| < \frac{1}{\sqrt{3}}$.
Then the rest state $A$ is unstable. If we slightly perturb $A$ one generates
a \textit{periodic activity} depicted in fig. \ref{FigToy}c.\\
%
%
%
%
\begin{figure}[ht]
\begin{center}
\includegraphics[width=5cm,clip=false]{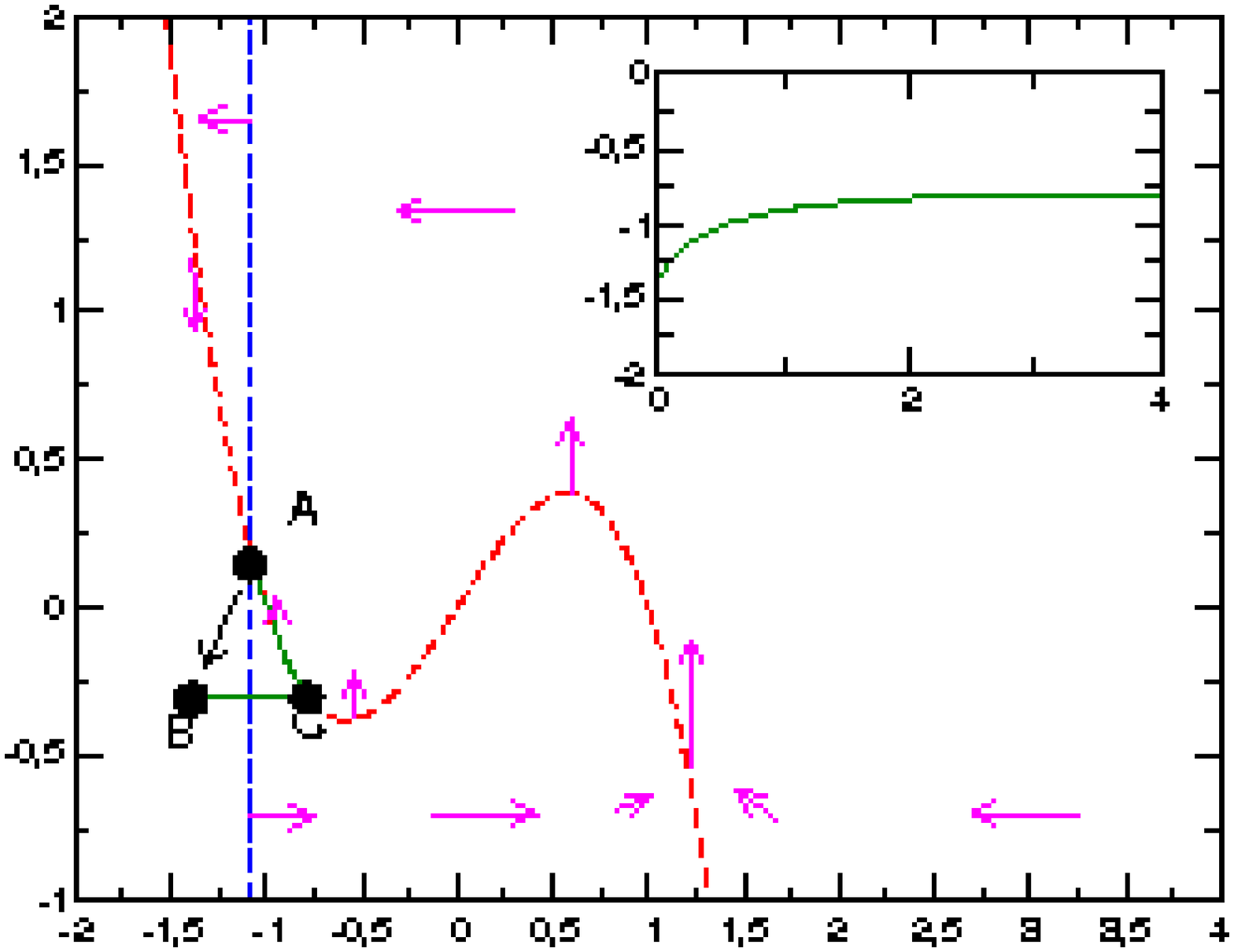}
\hspace{0.5cm}
\includegraphics[width=5cm,clip=false]{Toy_exc_XY}
\hspace{0.5cm}
\includegraphics[width=5cm,clip=false]{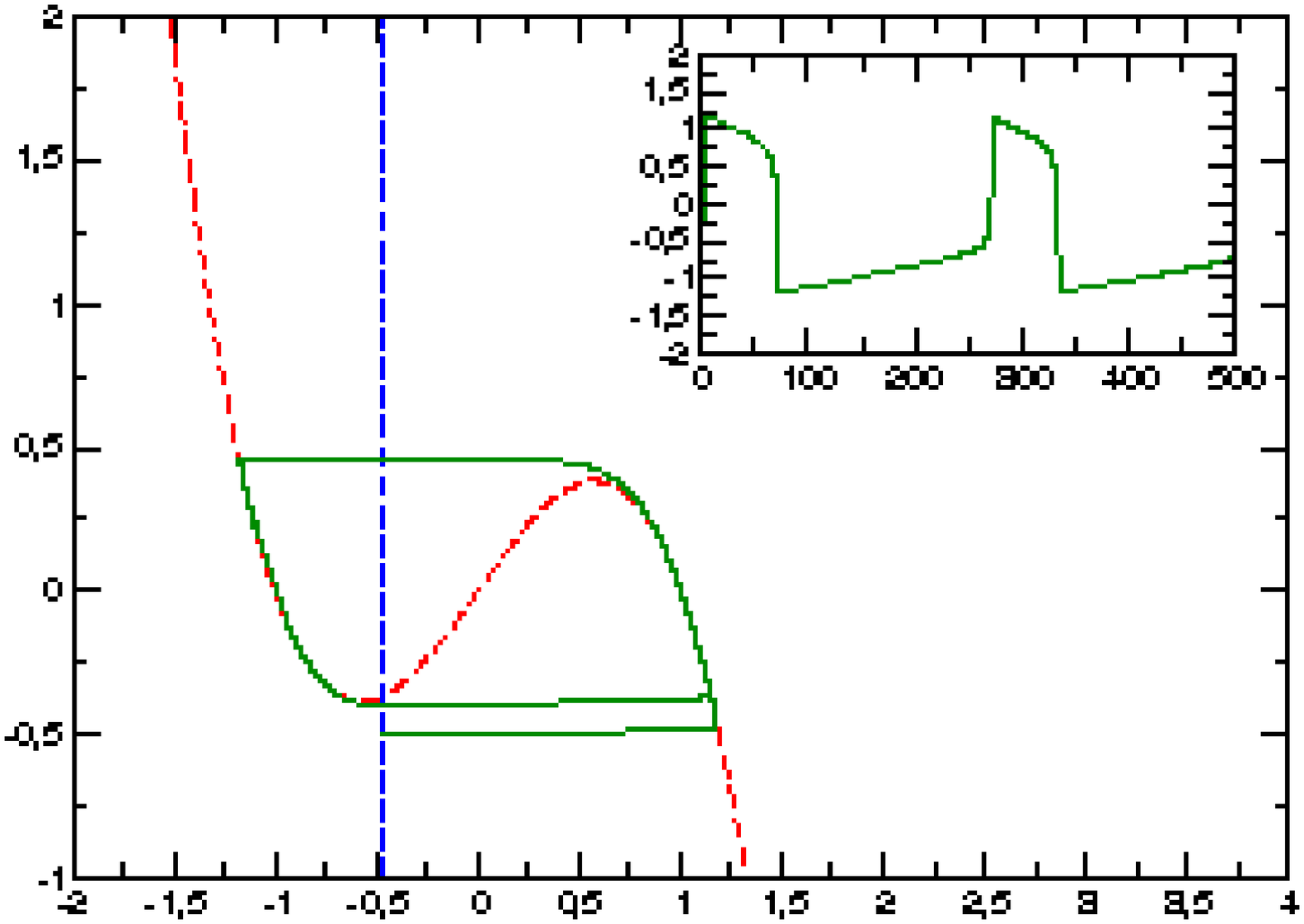}
\vspace{0.5cm}
\caption{Examples of possible behaviors for the equation (\ref{Toy})
in response to a perturbation of the rest state $A$.
Fig. \ref{FigToy}a. Relaxation to the rest state $A$.
Fig. \ref{FigToy}b. Spike emission.
Fig. \ref{FigToy}c. Periodic spikes train emission.
\label{FigToy}}
\end{center}
\end{figure}
%
%
%
%
%

For general systems of the form (\ref{SystExca},\ref{SystExcb})  the nullclines 
have a more complex shape and the dynamics is richer. 
It is an interesting exercise,
illustrating the spirit of dynamical systems theory, to start from the system
(\ref{Toy}), and to ask what are the changes induced in the dynamics by deformations
of the nullclines. Let us do this for the FitzHugh-Nagumo model.
\bea\label{FN}
\frac{dv}{dt}&=&v - v^3 - w +I \\
\frac{dw}{dt}&=&\epsilon(v - a -bw)
\eea
\nid It is deduced from the system (\ref{Toy}) by translating the $v$-nullclines with
a vertical displacement $I$ and by tilting the $w$ nullclines which becomes
the straight line $w=\frac{v-a}{b}$, for $b\neq 0$. From a qualitative
point of view one can figure out without any computation which type of novelties will be induced
by these changes. As shown in Fig. \ref{FigSNFN}, \ref{FigSNinv} 
we can for example have appearance/coalescence of pairs of fixed points by \textit{saddle-node bifurcations} and \textit{bistability}.
%
%
%
%
\begin{figure}[ht]
\begin{center}
\includegraphics[height=5cm,width=5cm,clip=false]{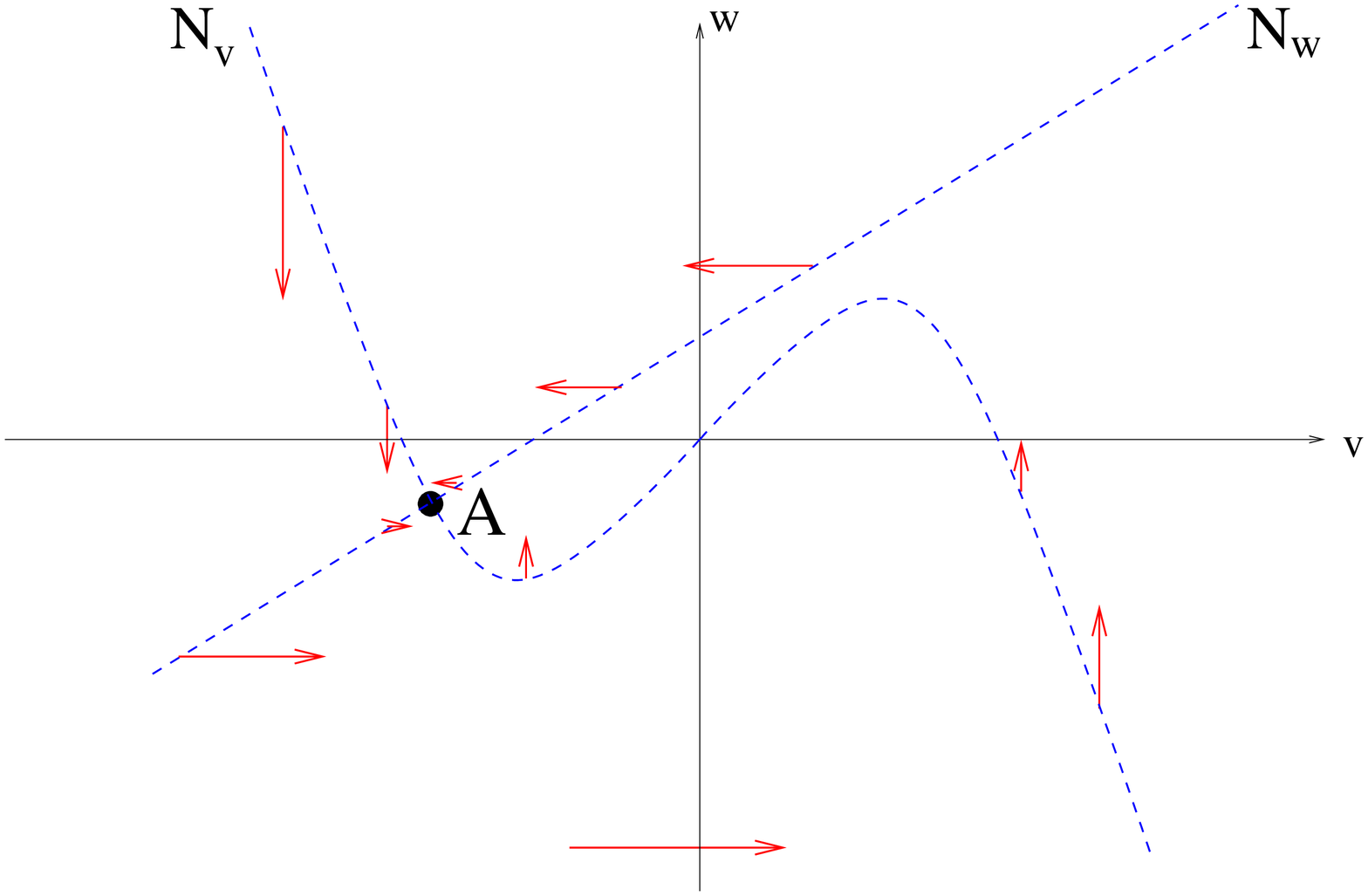}
\hspace{0.5cm}
\includegraphics[height=5cm,width=5cm,clip=false]{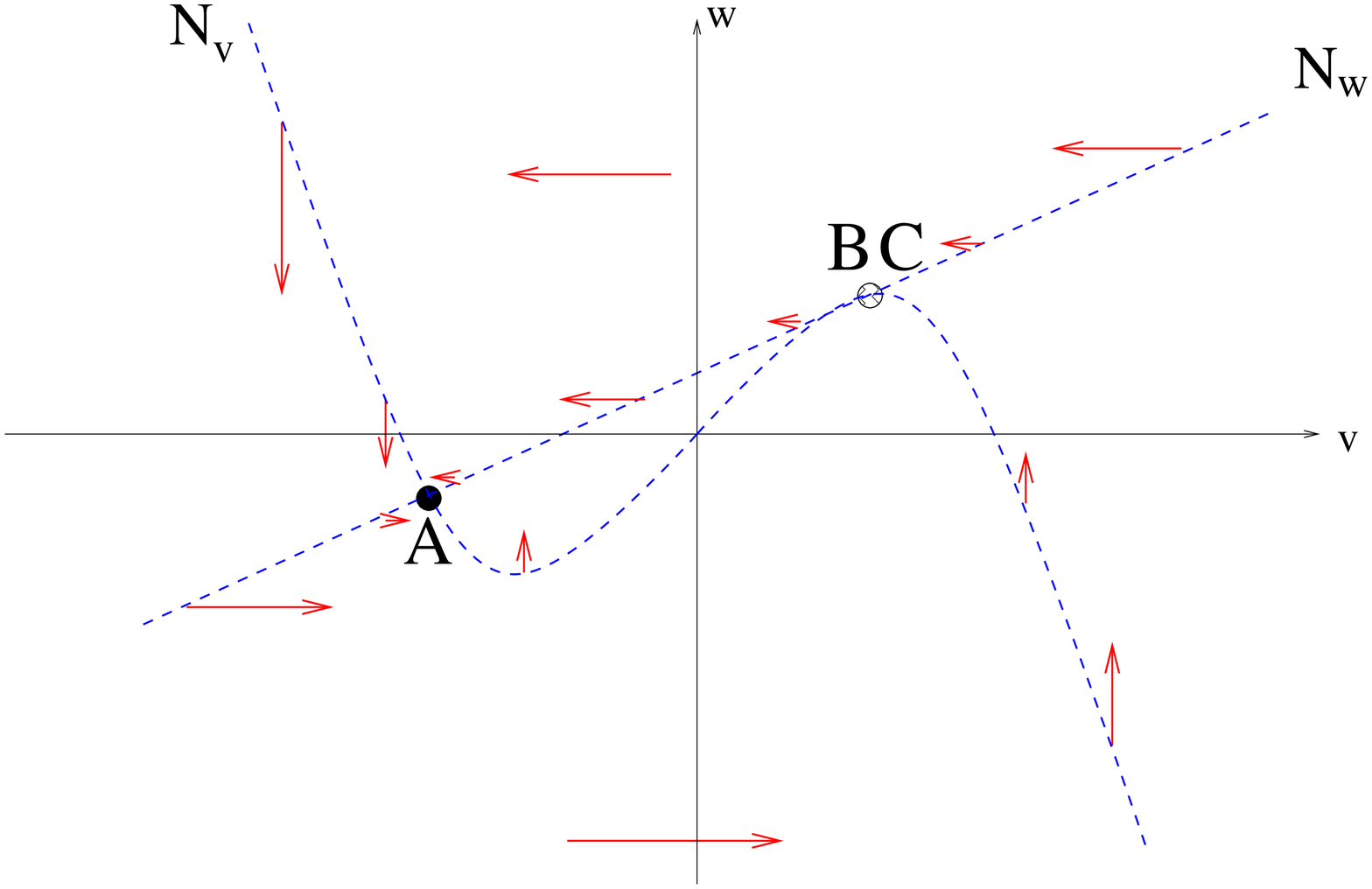}
\hspace{0.5cm}
\includegraphics[height=5cm,width=5cm,clip=false]{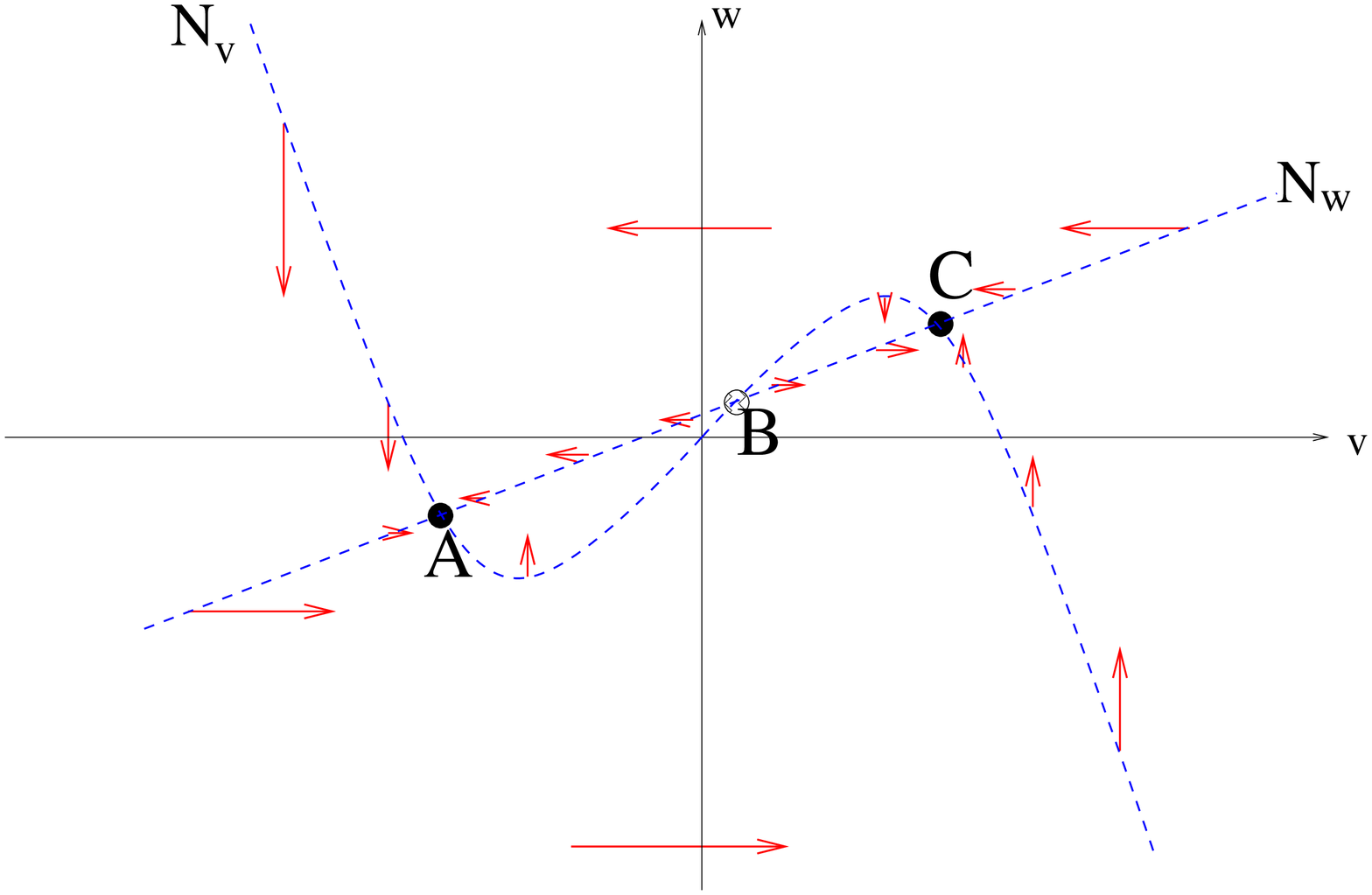}
\vspace{0.5cm}
\caption{Saddle node bifurcation and bistability in the FitzHugh-Nagumo model (\ref{FN}) when
the parameter $b$ increases. Note that the slope of the $w$ nullcline is $\frac{1}{b}$. The
same remarks as in Fig. \ref{FToy} holds for the scaling of the arrows.
\label{FigSNFN}}
\end{center}
\end{figure}
\begin{figure}[ht]
\begin{center}
\includegraphics[height=3cm,width=5cm,clip=false]{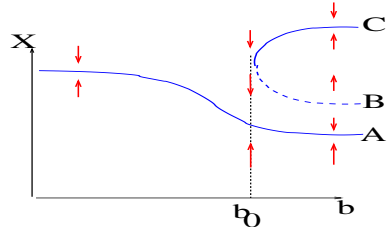}
\vspace{0.5cm}
\caption{Bifurcation diagram corresponding to Fig.\ref{FigSNFN}. $X$ corresponds to the projection on the $N_y$ nullcline.
$b_0$ is the critical point.
\label{FigSNinv}}
\end{center}
\end{figure}

On more general biological grounds, and though the FitzHugh-Nagumo equations are a simplification of the Hodgkin-Huxley
equations, they exhibit some typical behavior of the real neuron. Let us
list a few examples.\\

\bit

\item \textit{Action potential emission and threshold.} The first observation
is that a suitable input current can generate an action potential. Consider
the case depicted in Fig. \ref{FigPAVFN}. There is a unique stable fixed point $A$.
Consider now the line labeled by $S$. This line is called the threshold separatrix
since it separates solution curves that represent action potentials from
those that do not represent action potentials \cite{Cronin}. This curve is not sharply defined here
(see the discussion of type I excitability for a definition)
but it is very close to the unstable branch and, between the minimum and the maximum of the  $v$ nullclines,
 it essentially corresponds
to the set $\left\{(x,y) \ | \ f_v(v,w)=f_w(v,w) \right\}$,
where the vector
field makes an angle of $45 \ ^\circ$ with the $v$ axis. Let us now consider the situations
corresponding to the case $1$ and $2$ in Fig. \ref{FigPAVFN}. One perturbs the rest state by changing the membrane
potential such that $v$ 
is close to $S$, but in the case $1$ the perturbed point is ``above'' $S$ and
in the case $2$ it is ``below'' $S$. Even if these two points are close to each other,
the vector fields have a different orientation since the
angle of the vector field with the $v$ axis is, in the case $1$ larger than $45 \ ^\circ$
and in the case $2$ it is smaller. This has the following consequence.
In the case $1$  the neuron returns to equilibrium
without emitting a spike. On the other hand, in the case $2$
the trajectory has to make a big excursion before returning to the rest state:
  there is a spike emission. The horizontal distance from $A$ to $S$ corresponds therefore to
a threshold value $\theta$. Note however that the concept of threshold, corresponding to a sharp
transition, is questionable, in the Hodgkin-Huxley model, since there is no real clear cut firing threshold
(see \cite{KBD,RE}; see also the discussion below about type I and type II models of excitability). \\

%
%
%
%
\begin{figure}[ht]
\begin{center}
\includegraphics[height=5cm,width=7cm,clip=false]{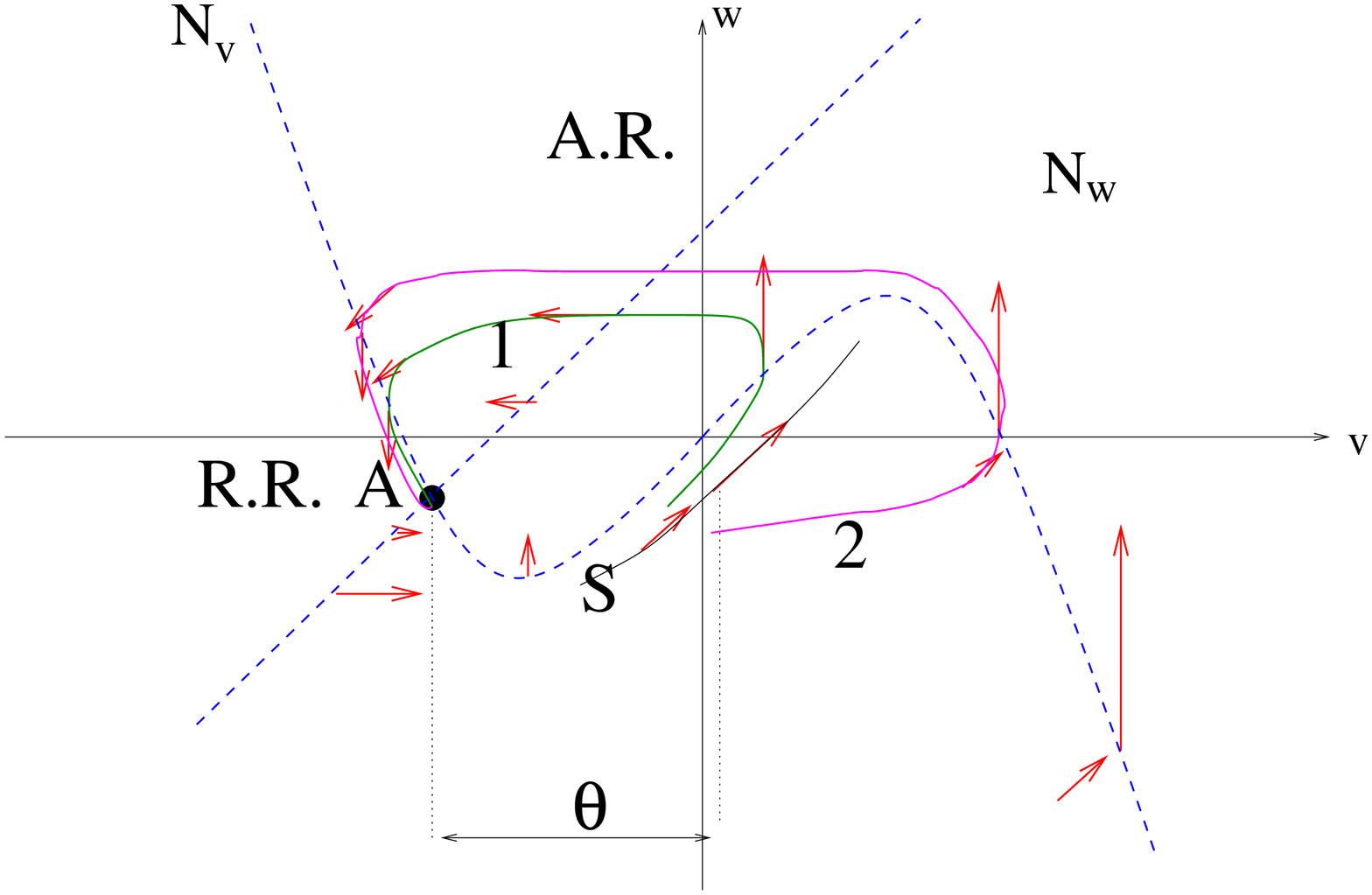}
\hspace{2cm}
\includegraphics[height=5cm,width=7cm,clip=false]{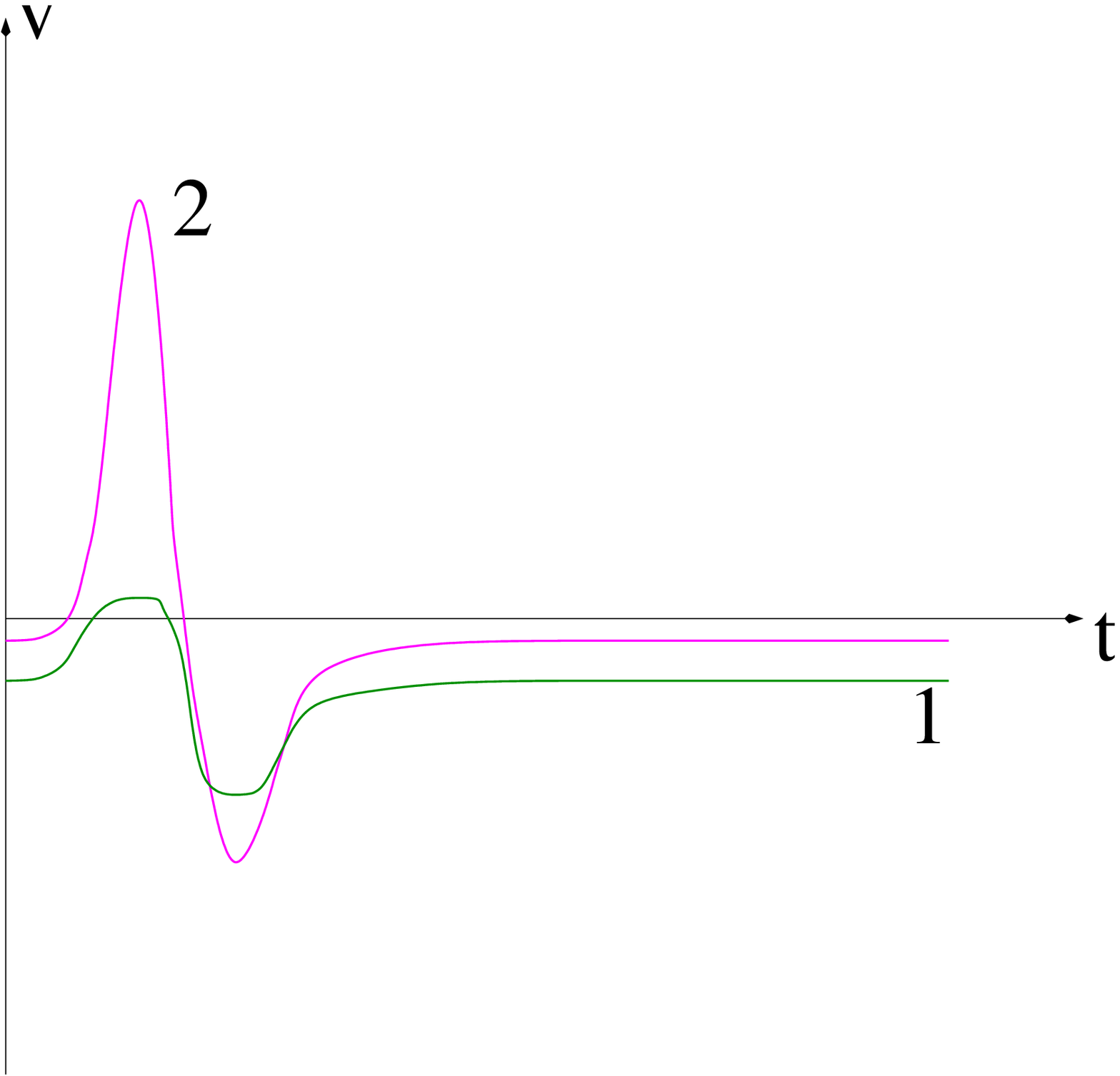}
\vspace{0.5cm}
\caption{Spike emission in the FitzHugh-Nagumo model.\label{FigPAVFN}}
\end{center}
\end{figure}
%
%
%
%
%

\item \textit{Existence of a refractory period.} The Figure \ref{FigPAVFN} also exhibits two regions
labeled by AR for ``Absolute Refractory'', and RR for ``Relative Refractory''. These regions are defined
as follows. Assume that the
neuron is spiking. If the corresponding point in the phase space is in
the region AR, any further positive increase in the membrane potential will not be able to generate a new spike.
On the other hand, in the region RR a a spike can be generated provided the clamped potential
is strong enough.\\

\item \textit{Anodal break excitation.} Assume that an action potential is generated and, during this, an external potential
(anodal shock) is applied at the instant where the system is the point $P$ in Fig.
\ref{FigABEFN}, with the effect to move $P$ to $P'$. If the shock is large enough such
that $P'$ is on the left of the threshold separatrix, the action potential is abolished
by the anodal shock. This phenomenon has been observed experimentally (see \cite{Cronin} and references
therein). \\

%
%
%
%
\begin{figure}[ht]
\begin{center}
\includegraphics[height=5cm,width=7cm,clip=false]{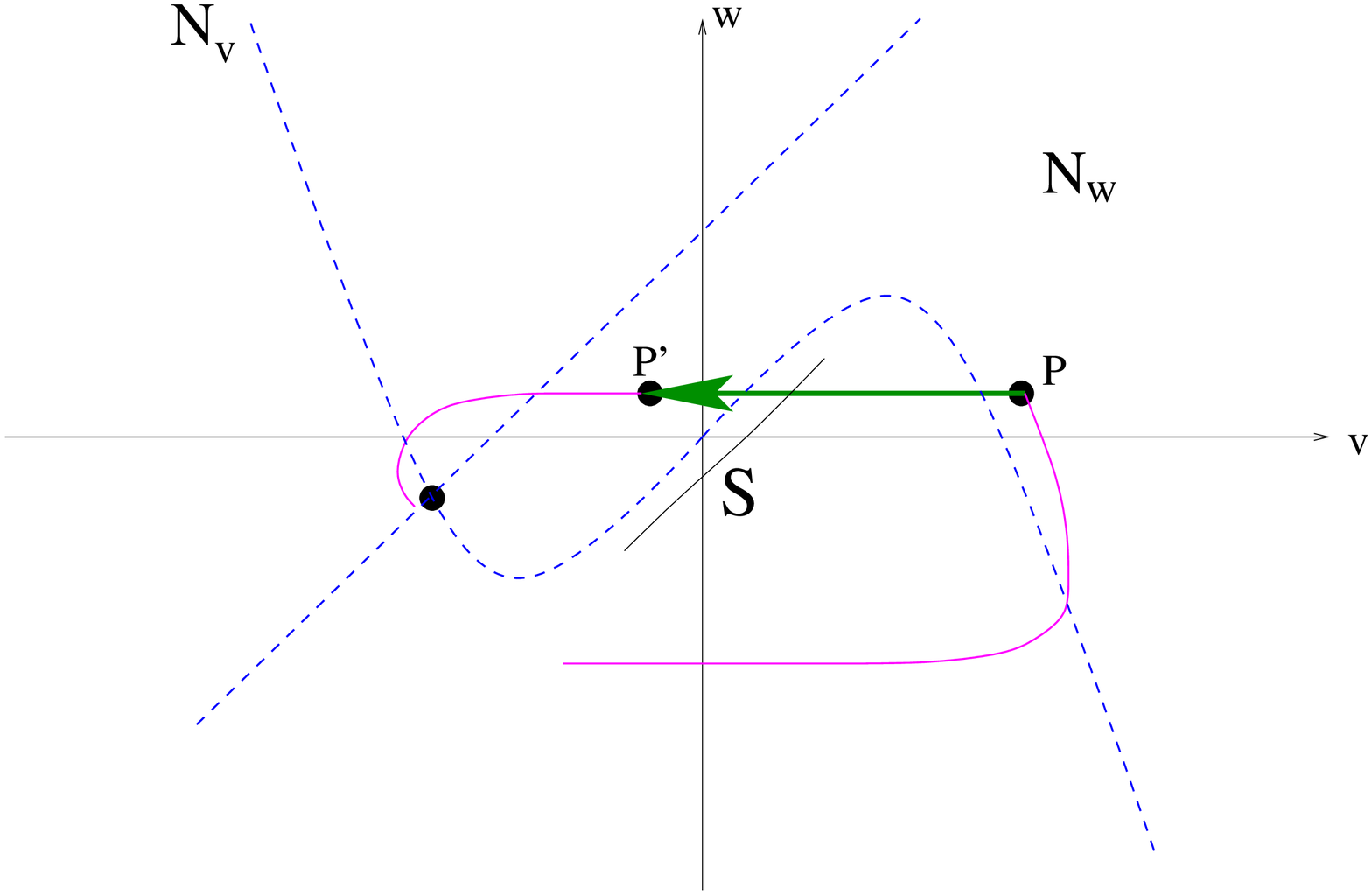}
\hspace{2cm}
\includegraphics[height=5cm,width=7cm,clip=false]{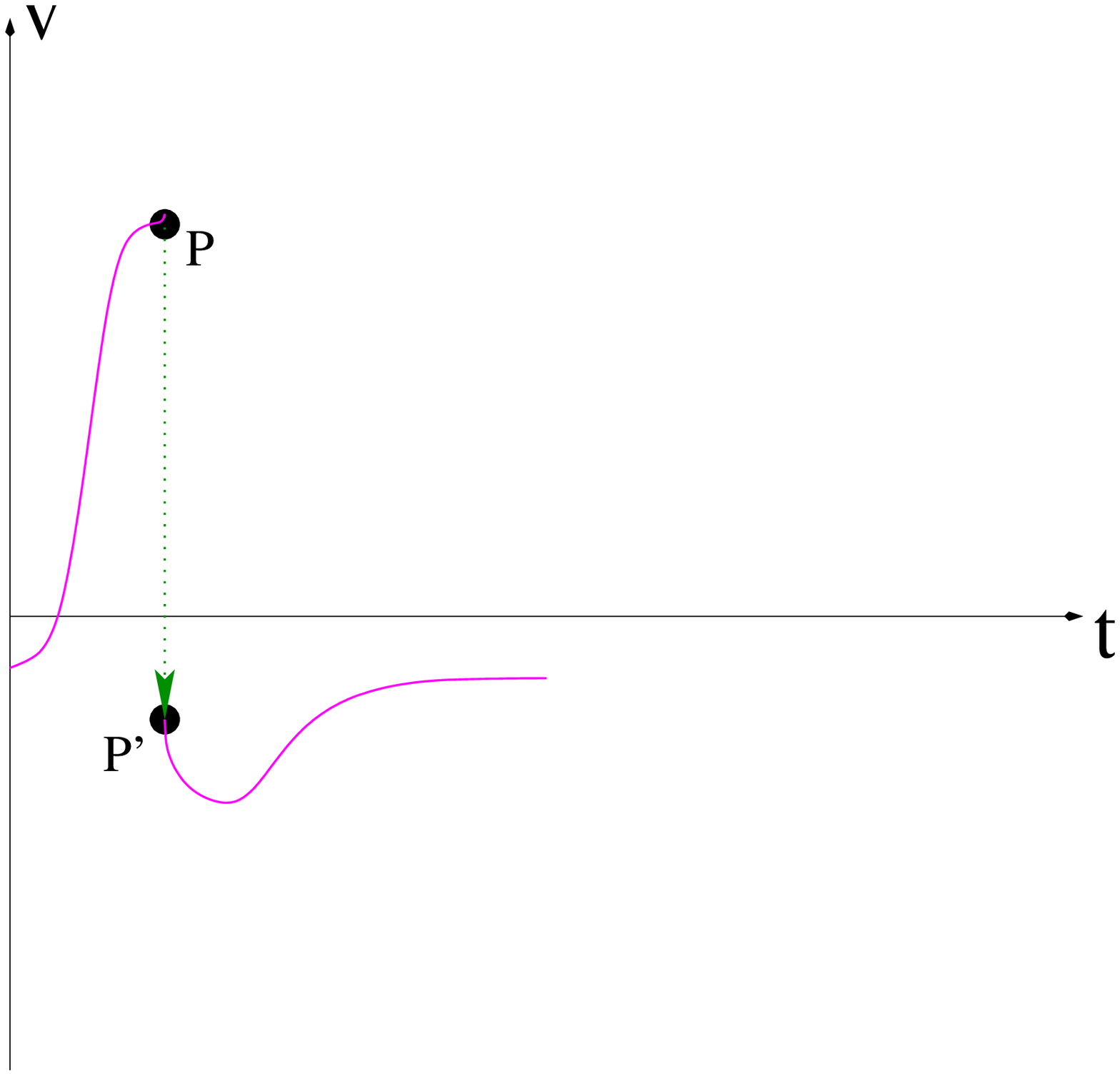}
\vspace{0.5cm}
\caption{Anodal break excitation in the FitzHugh-Nagumo model. \label{FigABEFN}}
\end{center}
\end{figure}
%
%
%
%
%

\item \textit{Spike emission by hyperpolarization.} Assume now that we apply
a negative current $I<0$ in the situation where the system is initially at rest, with
a stable fixed point $A$ (Fig. \ref{FigSEHFN}). The cubic nullcline moves downward and
$A$ moves to $A'$. If we removes the current, the cubic moves upward. But then $A'$
is no more a fixed point. Its trajectory  is described in Fig. \ref{FigSEHFN}. This
corresponds to a spike emission.\\

%
%
%
%
\begin{figure}[ht]
\begin{center}
\includegraphics[height=6cm,width=8cm,clip=false]{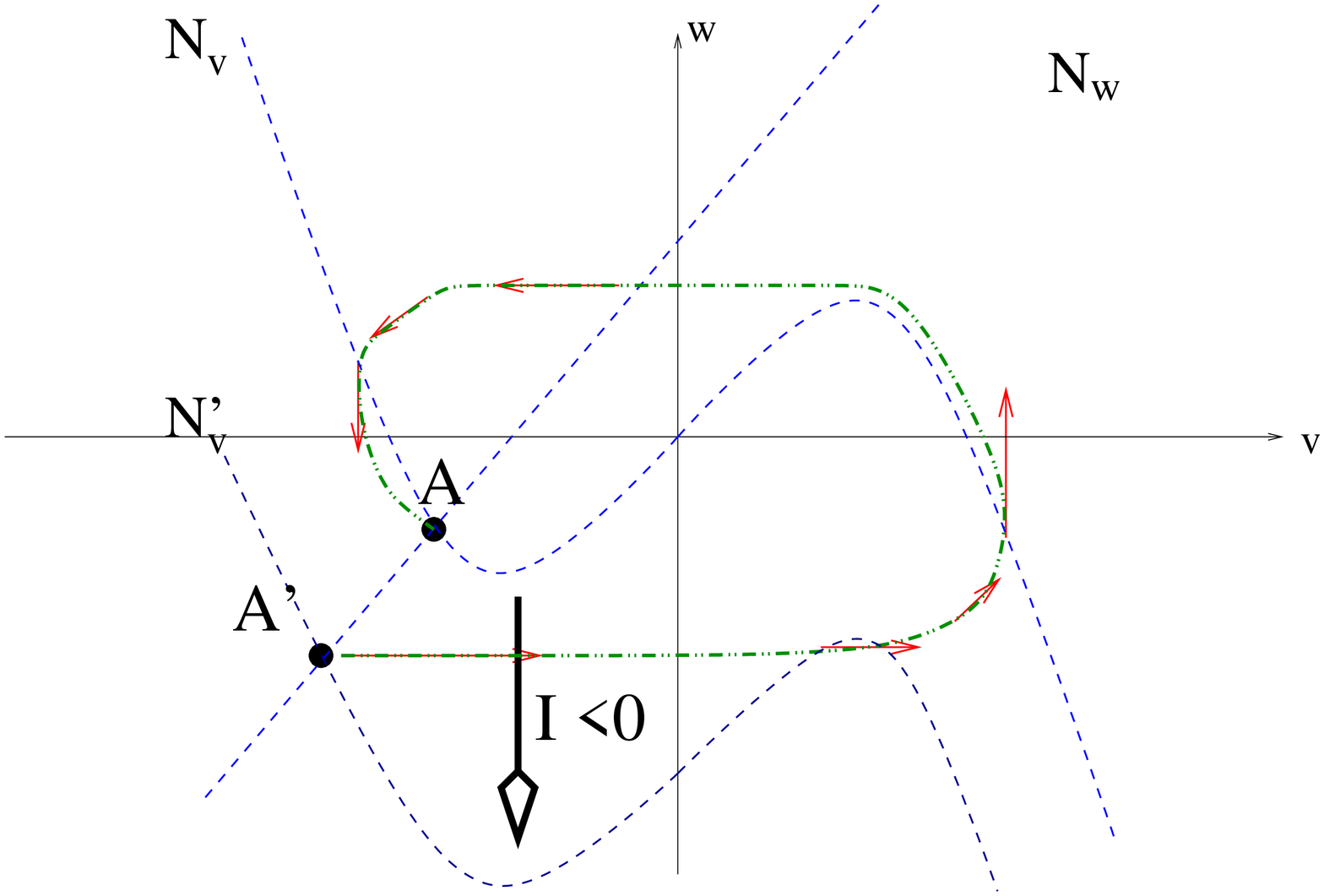}
\vspace{0.5cm}
\caption{Spike emission by hyperpolarization in the FitzHugh-Nagumo model.\label{FigSEHFN}}
\end{center}
\end{figure}
%
%
%
%
%

\item \textit{Periodic sequences of action potential.} Assume now that we apply
a \textit{positive current $I>0$}. For sufficiently high $I$ $A$ becomes unstable.
Then the slightest excitation generates a \textit{periodic emission of spikes}.
It is indeed possible to show rigorously, using Mischenko and Rozov theorems
combined with the Poincar\'e-Bendixon theorem \cite{GH}, that there exists a stable limit
cycle (depicted Fig.  \ref{FigPAIFN}).\\
%
%
%
%
\begin{figure}[ht]
\begin{center}
\includegraphics[height=6cm,width=8cm,clip=false]{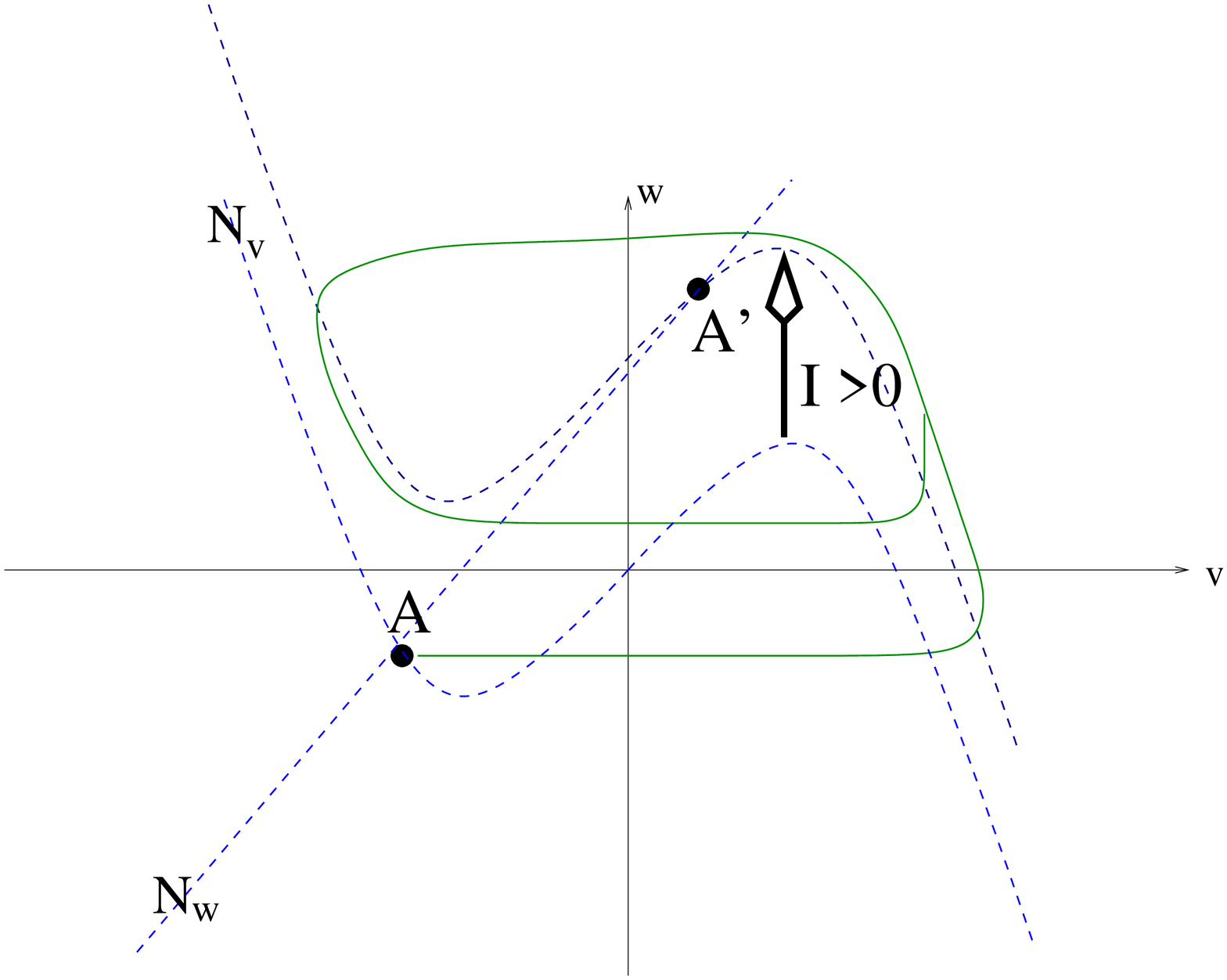}
\vspace{0.5cm}
\caption{Periodic sequences of spikes in the FitzHugh-Nagumo model.\label{FigPAIFN}}
\end{center}
\end{figure}
%
%
%
%
%
\eit

What happens now if we go on deforming the nullclines ? For example,
one can bend the line corresponding to the $w$ nullcline
 transforming it into a parabola: this is the deformation of lowest non linear order.
It is quite interesting to remark that this leads to a system exhibiting
neural excitability\footnote{Note that type II excitability exists already in the
previous case.} of type I and II. Indeed, the response of a neuron to permanent
current stimulus can generate a periodic train of spikes with a determined
frequency. In this case, one distinguishes two types of such excitability (this classification
was proposed by Hodgkin in 1948). 

\bit
\item \textit{Type I excitability}. The spike train is generated with an arbitrary small frequency, depending
on the applied current (Fig. \ref{Typefreq}). From a dynamical point of view, such type
of excitability can be generated by the scenario depicted Fig. \ref{TypeI}a,b,c. The variation 
of a control parameter (here the applied current) moves the $v$ nullcline such that a \textit{saddle-node} bifurcation
on a limit cycle occurs. For a critical value $I=I_c$ there is an homoclinic connexion
on the fixed point $A$. Consequently, the period is infinite (and the frequency is zero). Note
that the amplitude of the cycle is independent of $I$.

In figure \ref{TypeI}a we have also qualitatively plotted the separatrix $S$ which is here the \textit{stable manifold} of $B$. Clearly, a perturbation
to the left of $S$ does not generate a spike, while a perturbation to the right corresponds to a trajectory making
a big excursion around the unstable fixed point $C$, before returning to the rest state: this corresponds to a spike. \\

%
%
%
%
\begin{figure}[ht]
\begin{center}
\includegraphics[height=5cm,width=5cm,clip=false]{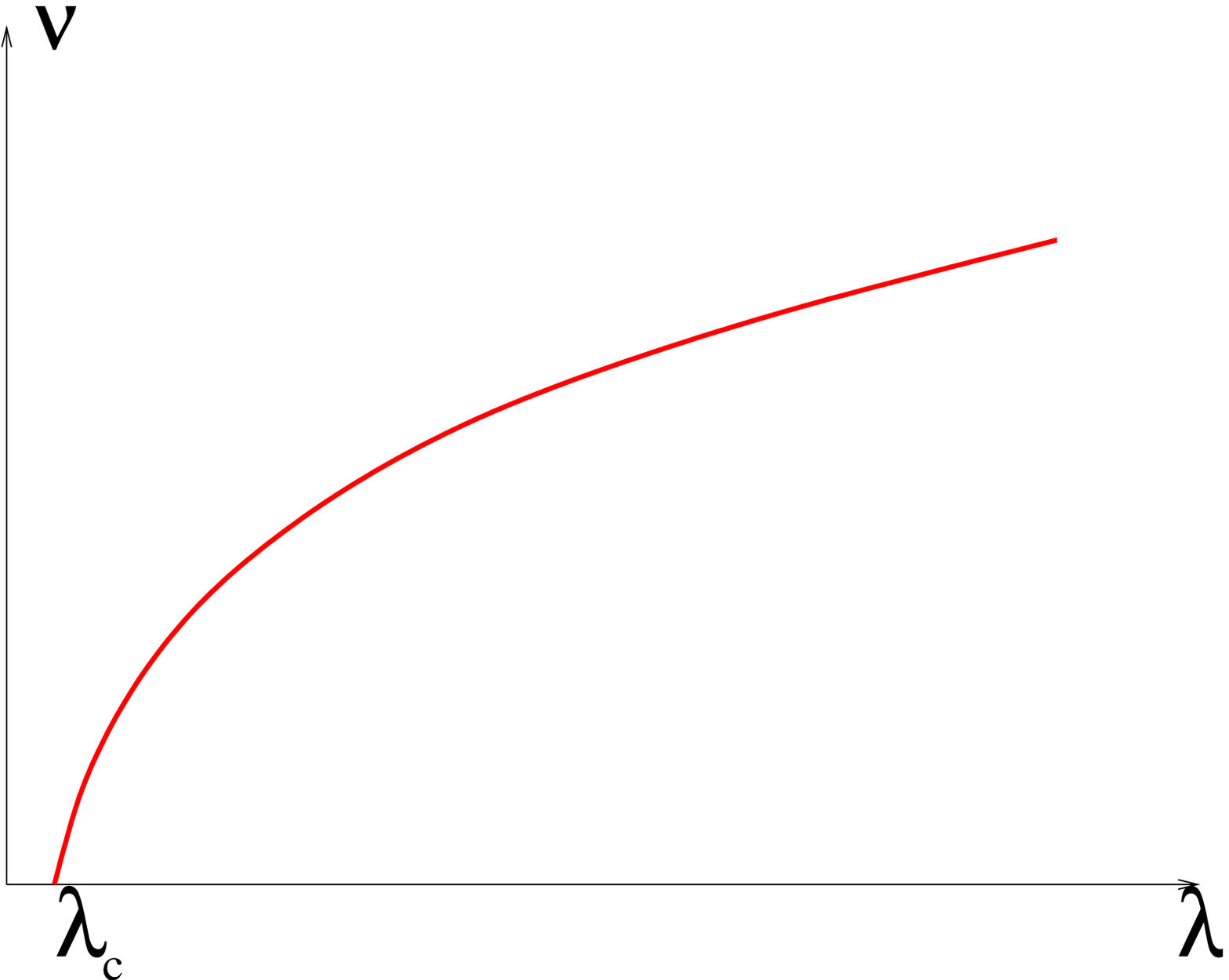}
\hspace{0.5cm}
\includegraphics[height=5cm,width=5cm,clip=false]{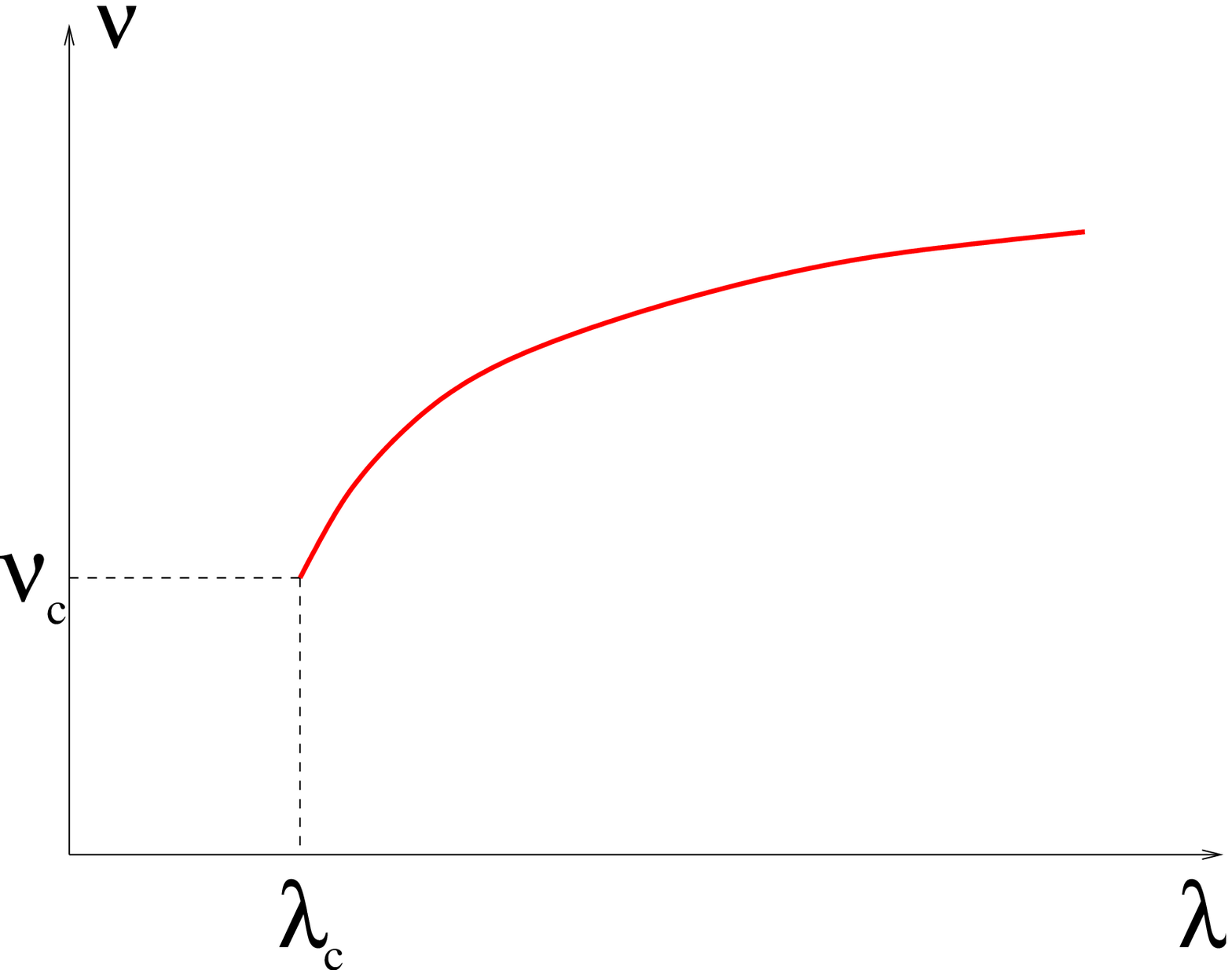}
\vspace{0.5cm}
\caption{Variation of the spikes frequency with the control parameter
(applied current in Fig. \ref{TypeI}a,b,c, \ref{TypeII}a,b,c). Fig. \ref{Typefreq}a.  Type I excitability.  )
Fig. \ref{Typefreq}b. Type II excitability.).
\label{Typefreq}}
\end{center}
\end{figure}
\begin{figure}[ht]
\begin{center}
\includegraphics[height=5cm,width=5cm,clip=false]{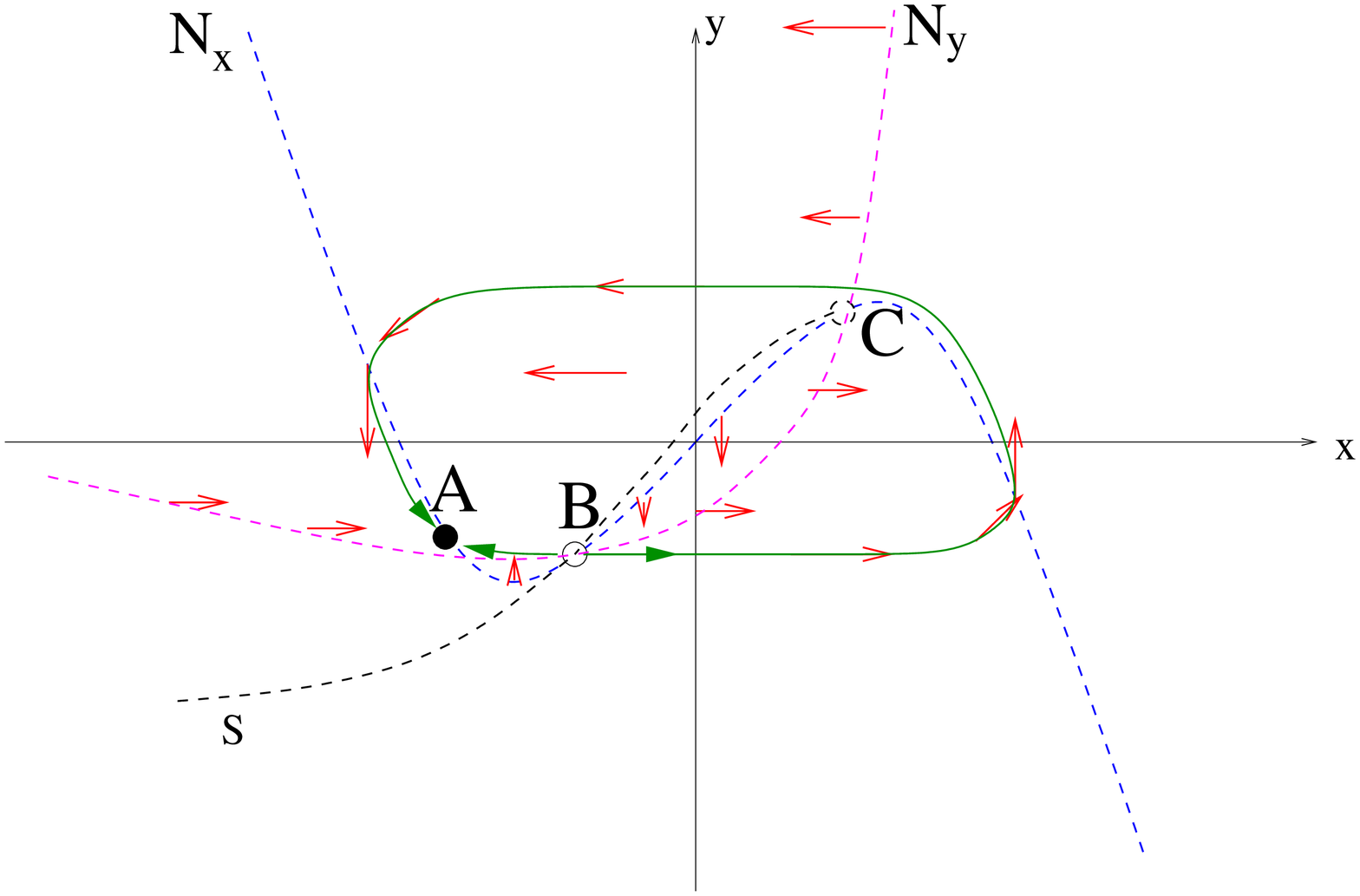}
\hspace{0.5cm}
\includegraphics[height=5cm,width=5cm,clip=false]{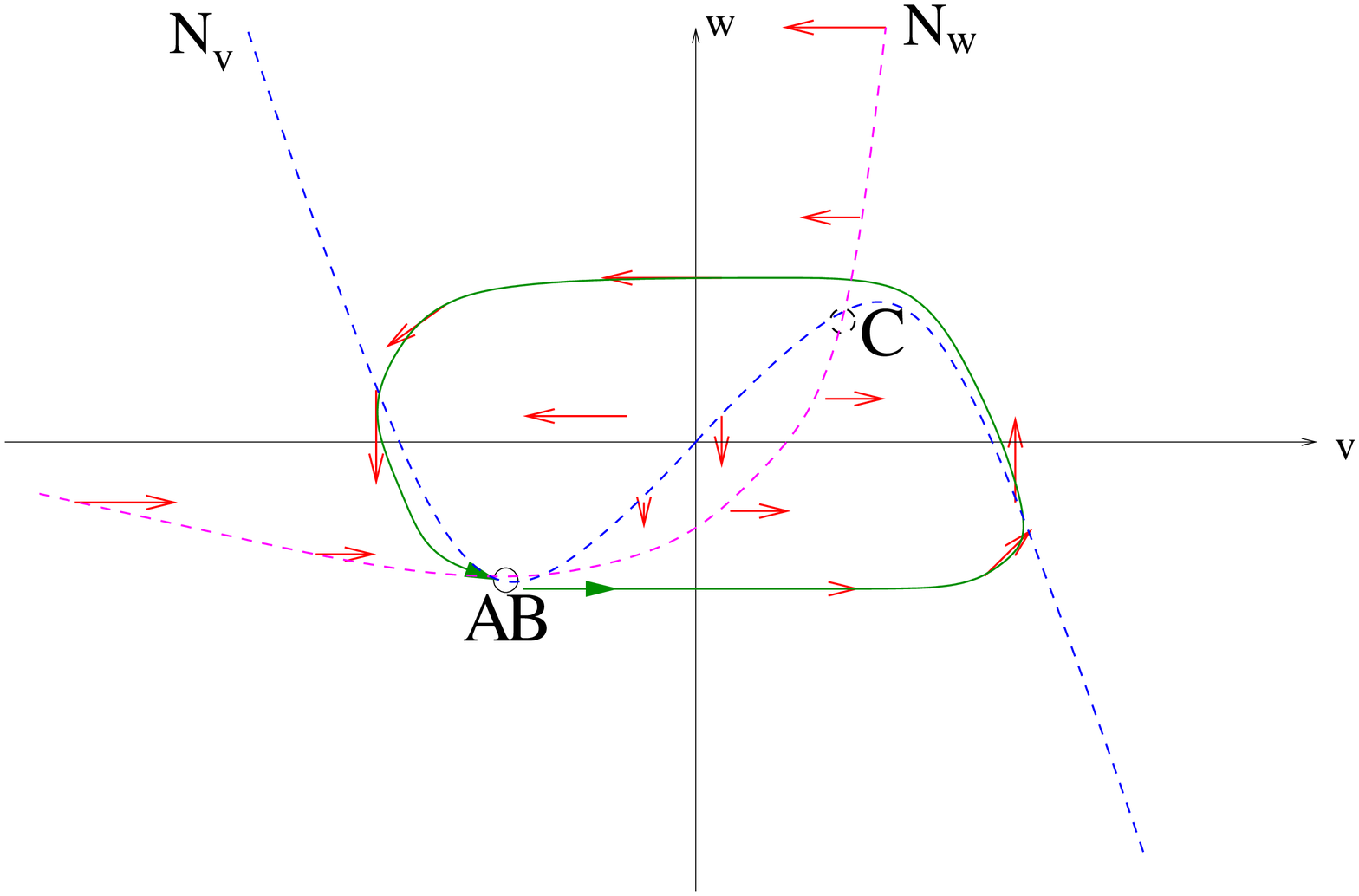}
\hspace{0.5cm}
\includegraphics[height=5cm,width=5cm,clip=false]{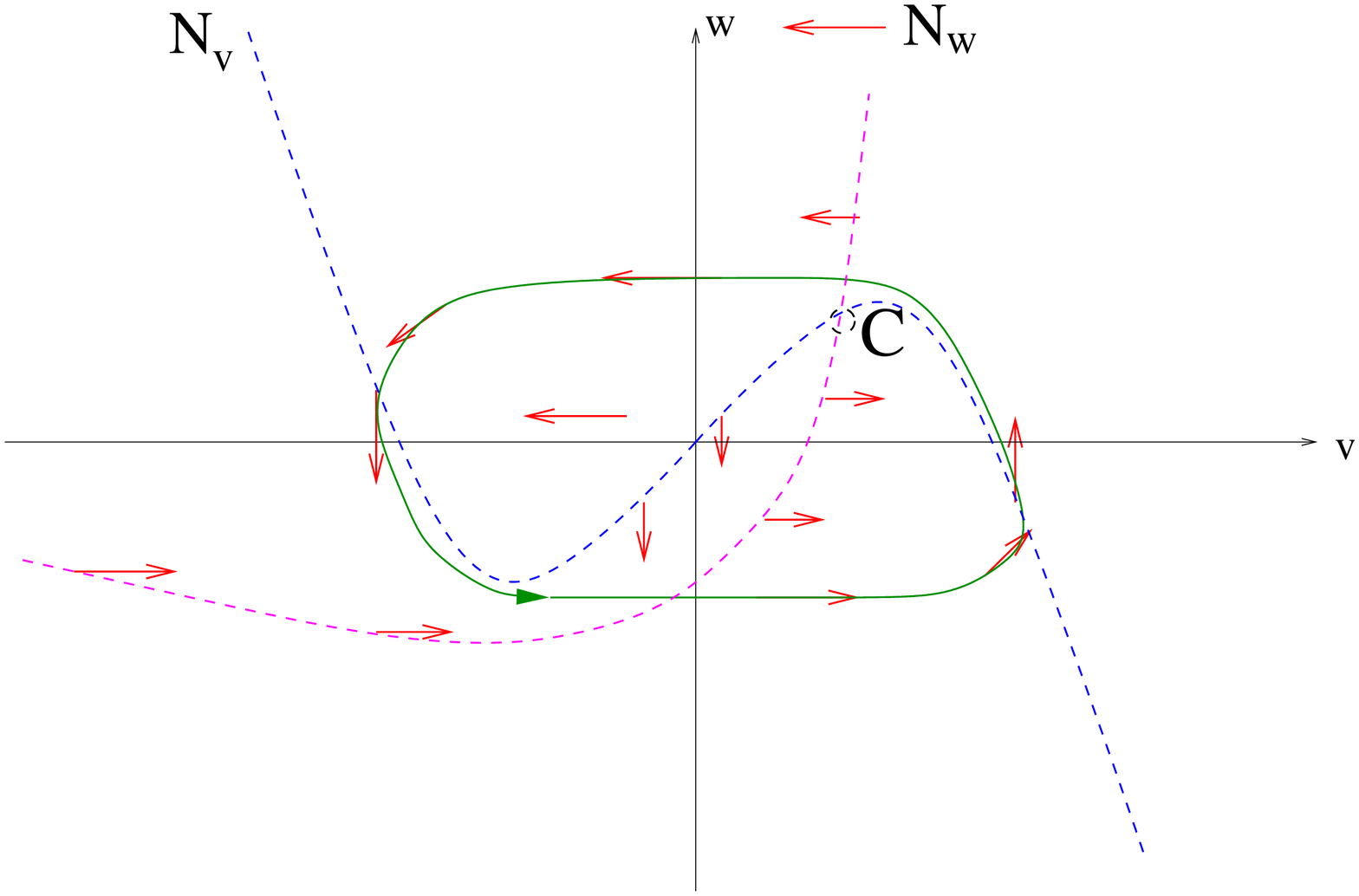}
\vspace{0.5cm}
\caption{Type I excitability. Schematic example of a dynamical system exhibiting type I excitability.
\label{TypeI}}
\end{center}
\end{figure}
%
%
%
%
%

\item\textit{Type II excitability}. The spike train is generated with a frequency staying a specific domain
 (Fig. \ref{Typefreq}b). From a dynamical point of view, such type
of excitability can be generated by the scenario depicted Fig. \ref{TypeII}. The variation 
of the applied current moves the $x$ nullcline such that a \textit{Hopf} bifurcation
occurs. The frequency depends slightly on $I$ and the amplitude increases like the square root of
the parameter distance to the critical value, as long as one stays close to the bifurcation point.
Note that the example depicted in Fig. \ref{TypeII} does not use the fact that $N_y$ has
a quadratic shape. Actually, the same is obtained with a straight line. 
%
%
%
%
\begin{figure}[ht]
\begin{center}
\includegraphics[height=5cm,width=5cm,clip=false]{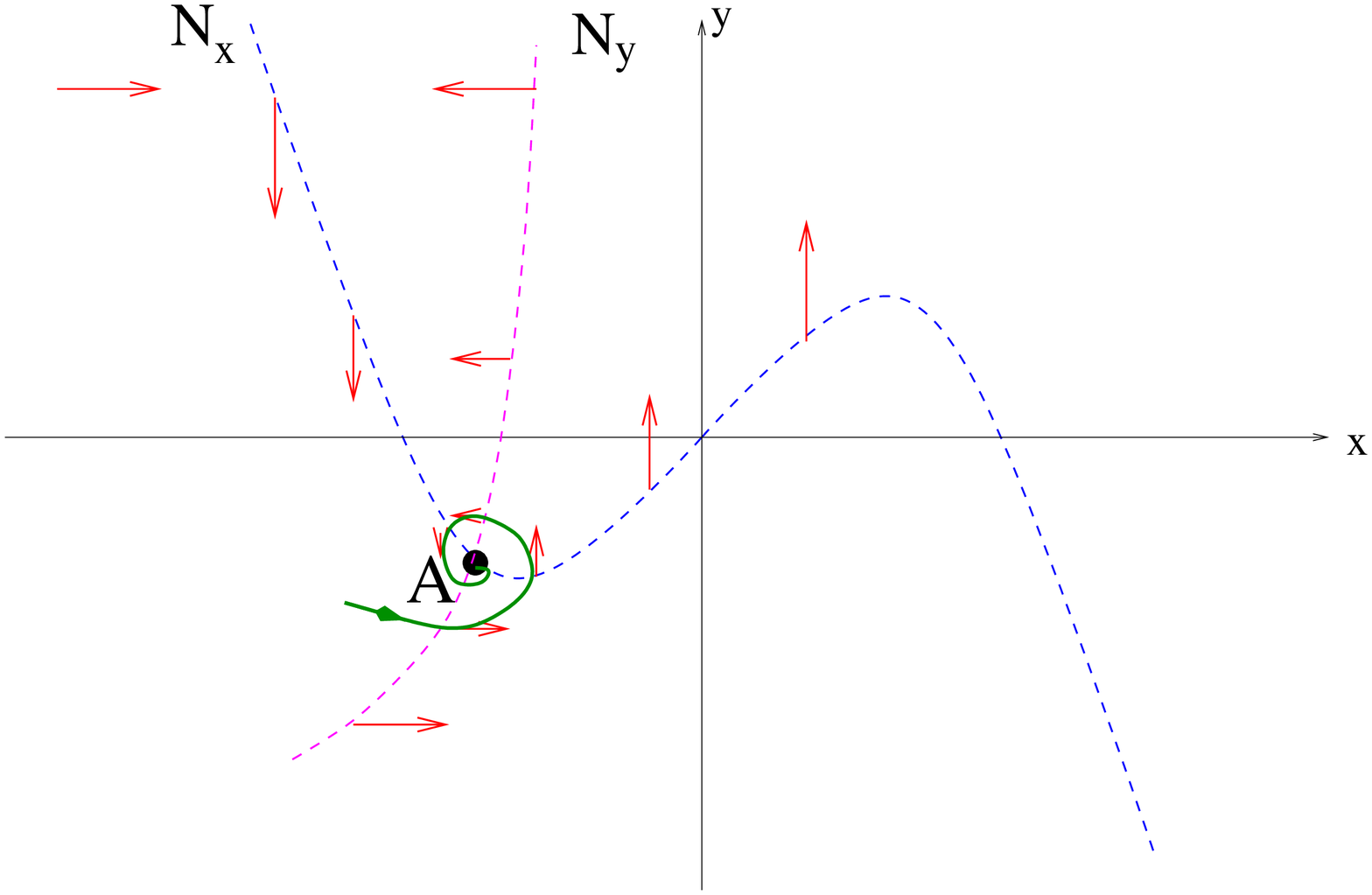}
\hspace{0.5cm}
\includegraphics[height=5cm,width=5cm,clip=false]{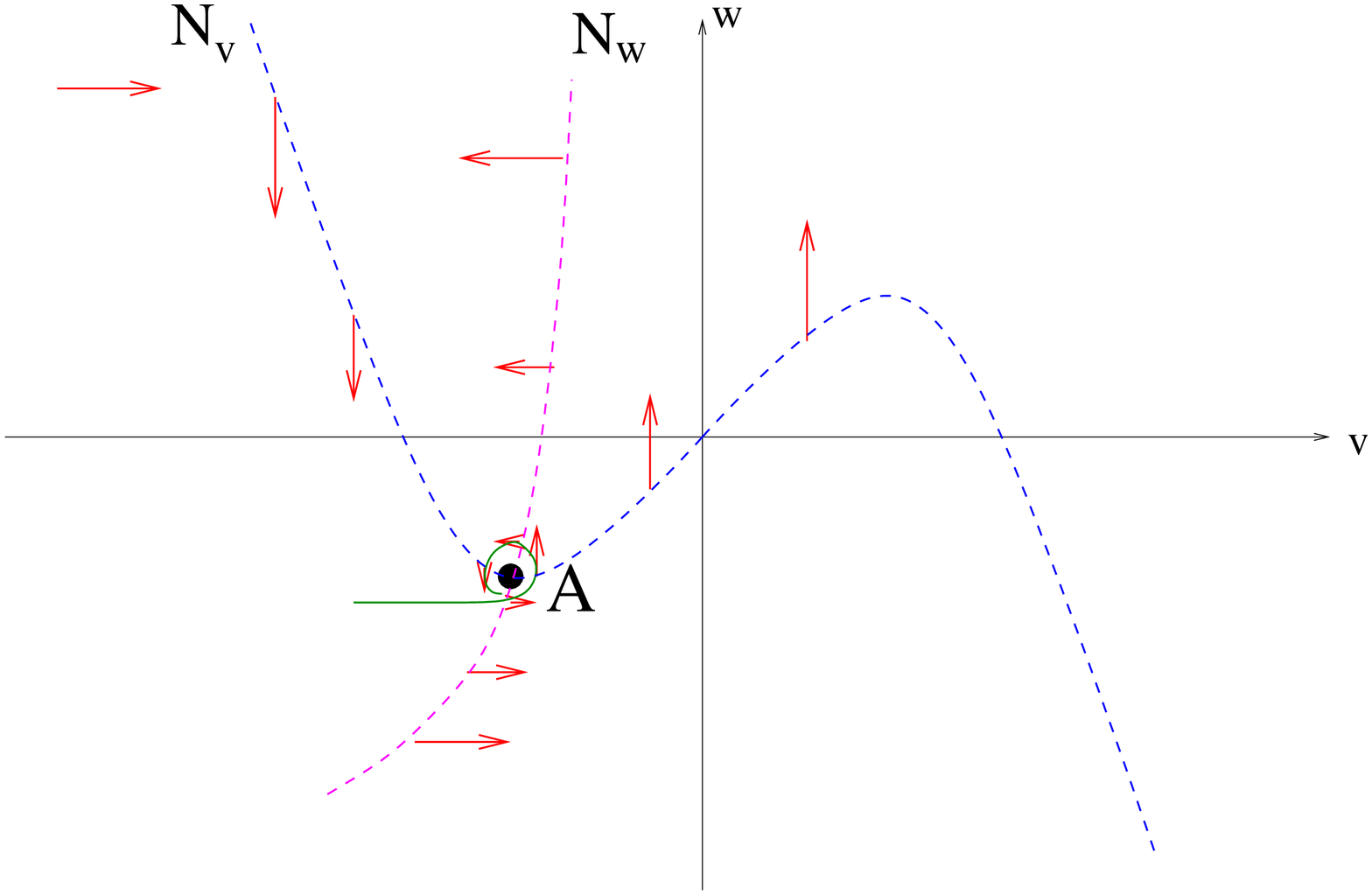}
\hspace{0.5cm}
\includegraphics[height=5cm,width=5cm,clip=false]{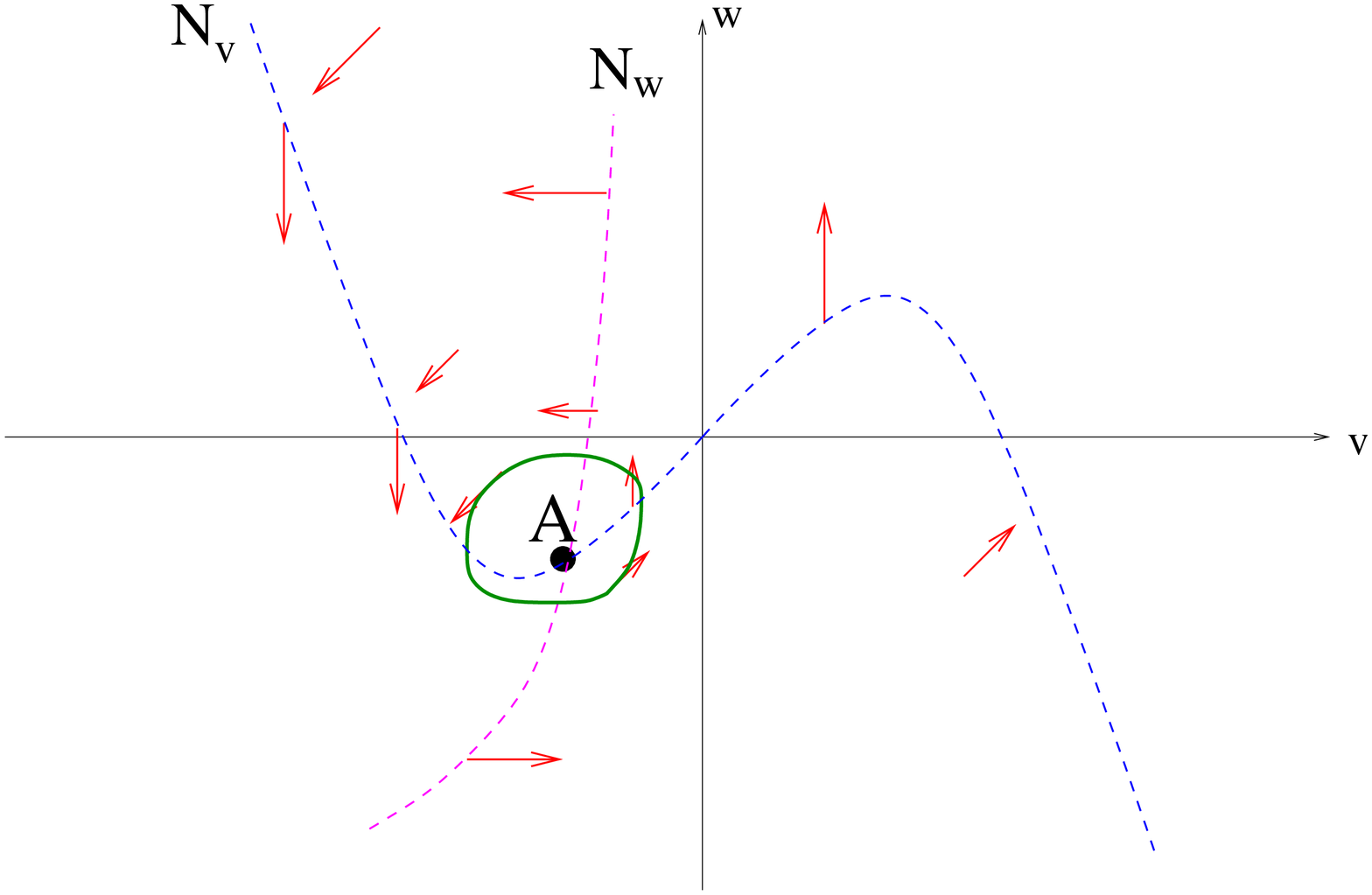}
\vspace{0.5cm}
\caption{Schematic example of a dynamical system exhibiting type II excitability.\label{TypeII}}
\end{center}
\end{figure}
%
%
%
%
%
\eit  
\sssu{The Morris Lecar model}\label{SML}

The previous examples may look quite abstract since we deformed the nullclines
freely, without paying  much attention to the biological relevance of this operation.
Actually, there exist biologically plausible models exhibiting the behaviors presented above.
 An example is  the Morris and Lecar model \cite{ML,EK} which was formulated 
in the context of the electrical activity of the barnacle muscle
fiber. The Sodium channels are replaced by Calcium channels. One calls
$m$ the activation variable. The Calcium conductance is given by $G_{Ca}=g_{Ca}m(V)$.
There is no inactivation variable $h$. The dynamics is given by:
\bea
C_m\frac{dV}{dt}&=&-g_{Ca}m_\infty(V)(V-E_{Ca}) - g_K w(V-E_K) - g_L(V-E_L) + I \\
\frac{dw}{dt}&=&\epsilon\frac{[w_\infty(V)-w]}{\tau_w(V)}
\eea
\nid where :
\bea
m_\infty(V) &=& \frac{1}{2}\left[1+\tanh\left(\frac{V - V_1}{V_2}\right) \right]\\
w_\infty(V) &=& \frac{1}{2}\left[1+\tanh\left(\frac{V - V_3}{V_4}\right) \right]\\
\tau_w(V) &=& \frac{1}{\cosh\left(\frac{V - V_3}{2 V_4}\right)}\\
\eea
\nid $w$ is the fraction of open  $K^+$ channels. 
This set of equations as a large number of parameter that one may vary in order
to study the behavior of the neuron when physical characteristics, such has $V_1,V_2,V_3,V_4$,
 are varying. However, from an experimentalist point of view, the only free parameter
is the external current $I$. 

The $V$ nullcline corresponds to a situation where the applied current exactly
cancels the ionic current. It is given by 
$$I=g_{Ca}w_\infty(V)
(V-E_{Ca}) - g_K w(V-E_K) - g_L(V-E_L)$$
\nid It has a cubic  shape  
and a variation of $I$ as simply the effect of translating
it parallel to the $V$ axis. The $w$ nullcline is  the activation curve
$w=w_\infty(V)$.
This model displays a wide variety of dynamics such as spikes, oscillations
emerging with zero or non-zero frequency  and  bistability.

\sssu{Integrate and Fire models.} \label{SIF}

A convenient and simple model producing spikes is the so called leaky integrate and fire model.
Consider the circuit drawn in Fig. \ref{FIF}. The device $D$ is conducting when the potential is
above a threshold $\theta$ and has an infinite resistance otherwise. It acts therefore as a potential
dependent switch. The total current is $I=I_R+I_C = \frac{u}{R}+C\frac{du}{dt}$. Using the time constant
$\tau_m=RC$ one obtains the equation of the leaky integrate and fire model:
\beq\label{EIF}
\tau_m \frac{du}{dt}=-u(t)+RI(t)
\eeq
\nid with the additional condition that $u$ cannot increase above $\theta$.
Starting, say, from a zero
potential $u$, $u(t)$ increases until it reaches the threshold value $\theta$. Then $D$ switches on and the capacity
unloads. Consequently, the potential $u$ decreases exponentially fast. $u$ is interpreted as a membrane potential and 
$\tau_m$ as the membrane time constant of the neuron.

%
%
%
%
\begin{figure}[ht]
\begin{center}
\includegraphics[height=3cm,width=6cm,clip=false]{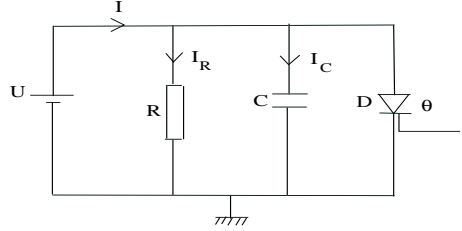}
\vspace{0.5cm}
\caption{Schematic circuit of the integrate and fire model.\label{FIF}}
\end{center}
\end{figure}
%
%
%
%
%
In integrate and fire models, the form of the action potential is not  explicitly described. Instead, one models the situation
above by saying that, when the potential $u$ reach the value $\theta$, at some time $t_f$, it is immediately reset
to a new value $u_r \deq u(t_f^+) < \theta$ while a spike is emitted. Then, the membrane potential keeps
the value $u_r$ for a time $\tau_a$ corresponding to the refractory period. In this sense, spikes are formal events characterized by
the ``firing time'' $t_f$.      

A more general version of (\ref{EIF}) is a non linear integrate and fire model \cite{AVV}:
\beq\label{ENLIF}
\tau_m \frac{du}{dt}=F(u(t))+G(u(t))I(t)
\eeq

\nid where $F,G$ are non linear functions of $u$.\\

Though the integrate and fire model  is a rough modeling of a spiking neuron it has several advantages.
Firstly, the linear model (\ref{EIF}) is exactly solvable. The potential $u(t)$ resulting from an excitation
with a time dependent current is easily found. For example, the current after a spike arising at time $t_1$ and before
the next spike ($u(t_2)=\theta$) is given by:
$$u(t)=u_r e^{-\left(\frac{t-t_1}{\tau_m}\right)}+\frac{1}{C} \int_{0}^{t-t_1} e^{\left(-\frac{s}{\tau_m}\right)}I(t-s)ds; \qquad t \in [t_1,t_2[ $$

Also, it is easy to model a network of integrate and fire neurons \footnote{Note however that the equation (\ref{CIF}) holds
in a more general setting.}. In this framework the neuron $i$ receives
the spikes coming from other neurons, and the total current $I_i(t)$ is the sum of  spikes coming from each neuron $j$
weighted by a quantity $J_{ij}$ roughly modeling the synaptic connexion between $j$ and $i$:
\beq \label{CIF}
I_i(t)=\sum_j J_{ij} \sum_{n(j)=1}^{n_{max}(j)} \alpha(t-t_n(j))
\eeq
\nid where $t_n(j)$ is the $n$-th time of firing of the neuron $j$, $\alpha$ is a function modeling the spike,
and the sum $\sum_{n(j)=1}^{n_{max}(j)}$ corresponds to an integration over a small time window. The spike
function $\alpha$ can have different forms, but the simplest one is a Dirac $\delta$ distribution, corresponding
to have an ``instantaneous'' spike.

Note that an equation with the form ( \ref{CIF}) is particularly well suited for a stochastic
approach, where the firing times are randomly distributed e.g. according to a Poisson process.
An example of this is given in Chapter II. Most of the analysis use a stochastic
approach. However the evolution can also be investigated in a deterministic context,
where the firing condition is determined e.g. by an Heaviside function. Then
one has to handle a deterministic dynamical system with singularities.  
 
\ssu{Qualitative analysis of the Hodgkin-Huxley equations.} \label{SBifHH}

We now return back to the Hodgkin-Huxley equations. The analysis made in section  \ref{SHH}
was only quantitative. But it has allowed us to understand the spike generation,
by simple arguments on the characteristic times of the variables $m,n,h$, and their
interpretation in terms of probabilities that a gate of a given ionic species is open or closed.
A further analysis requires however to consider the complete non linear  dynamical system (\ref{HHV}-\ref{HHh})
and its dependence in  control parameters such as the external current $I$. Actually,
the simplifications made in section \ref{SRHH} lead us to find several situations
having a correspondence with experiments on real neurons. Since the equations
(\ref{FN}) are a simplification of the Hodgkin-Huxley system, one expects to observe similar effects
in the dynamical system (\ref{HHV}-\ref{HHh}).  However, the reduced systems were two dimensional
while the Hodgkin-Huxley system has four dimensions. Therefore, bifurcations and dynamical regimes (such as
chaos) occurring in phase space having more than two dimensions are not observed in reduced systems like
(\ref{FN}). Rinzel and Miller \cite{RM} first gave evidence of this. Doi and Kumagai \cite{DK} recently showed the existence
of chaotic attractors  in a modified
Hodgkin-Huxley model that changes the time constant of one of the current
by a factor $100$,  and, more recently, Guckenheimer and Oliva \cite{GO} showed rigorously the existence of a Smale horseshoe
(hence of chaos) in the Hodgkin-Huxley model with its original parameters. 
Finally, the reduction performed to obtain the equations (\ref{FN}) used several simplifications
that can be discussed and that may bring some exogenous properties, not present in 
the initial model.    

For all these reasons, there is a clear need to perform an analysis of the Hodgkin-Huxley system.
Obviously, it is always possible to make numerical simulations of this dynamical system and
many papers have been written on the subject (see for example \cite{Mascagni} and reference therein).
 Also,  there exists currently a lot of ``on line'' simulators on the Internet \cite{Genesis,Hippo}.
However, analytical results are also useful since they allow in particular to locate
bifurcations points. This is certainly useful because this permits
to reduce the explored area in the (huge) parameters space and to locate small regions
that could be missed by a discrete sampling in a numerical simulation.
 In this section we present one example of such an investigation, due to Guckenheimer \& Labouriau \cite{GL}. This
paper presents actually an approach combining
rigorous methods from dynamical systems theory with numerical tools 
of formal calculus (for more details on this type of approach see also \cite{GW}). This allows the authors to draw a bifurcation diagram
in a two dimensional parameter space corresponding to the potassium reversal potential\footnote{In the paper,
the variable  $\nu$ corresponds to a clamped potential with the opposite convention as in the section
\ref{SHH} (see note \ref{Norigin}) $\nu=V_{rest}-V$ where $V$ is the membrane potential.}
$\nu_K=V_{rest}-E_K$  and to the current $I$ (the reversal potential of Sodium and Potassium can indeed
be controlled experimentally \cite{HH0,Jack}). Consequently, the bifurcations presented
are generic codimension one and two bifurcations. Actually, the bifurcation diagram presented 
in Fig. \ref{FGL} presents an overwhelming richness of dynamical behaviors in a rather small parameter
space region. This is a ``zoo'' in which one meets basically all species described in standard
textbooks about bifurcation theory \cite{GH,RuelleBif} (see the appendix) plus some more 
``exotic'' individuals such as the twisted saddle loop bifurcations. This is one reason why
we have chosen this example: it shows how deep the dynamical systems analysis can go and how
rich are the Hodgkin-Huxley equations.  Additional references are \cite{Hassard,Labouriau,DK}. \\

Let us start from elementary remarks. It is easy to show that the asymptotic
solutions of eq. (\ref{HHV})-(\ref{HHn}) are contained in the set $\left\{m,h,n \in [0,1]^3 \times \nu \in [\nu_- - r,\nu_++r \right\}$,
for some $r>0$, and where $\nu_- = \min(\nu_{Na},\nu_K,\nu_L)$ and $\nu_+ = \max(\nu_{Na},\nu_K,\nu_L)$.
Fortunately $m,h,n$ stay dynamically in $[0,1]^3$ (these are probabilities !!). Indeed,
if $m$ (resp. $h,n$) is equal to zero $\frac{dm}{dt} > 0$ and if $m=1$, $\frac{dm}{dt} < 0$.
Also, if $\nu > \nu_+$, $\frac{d\nu}{dt} <0$ and if $\nu < \nu_-$, $\frac{d\nu}{dt} >0$. 
As $t \to \infty$, $m,h,n \to m_\infty,h_\infty,n_\infty$. Consequently, if $\nu^\ast$
is a equilibrium of eq. (\ref{HHV}) then $(\nu^\ast,m_\infty(\nu^\ast),h_\infty(\nu^\ast),n_\infty(\nu^\ast)$
is an equilibrium of eq. ((\ref{HHV})-(\ref{HHh})). Also, $\frac{d\nu}{dt}=0 \Rightarrow G(\nu^\ast,m_\infty(\nu^\ast),h_\infty(\nu^\ast),n_\infty(\nu^\ast))
\deq f(\nu^\ast)=I$.
Consequently, there exists a unique value of $I$ for which $A=(\nu^\ast,m_\infty(\nu^\ast),h_\infty(\nu^\ast),n_\infty(\nu^\ast)$
is an equilibrium. When $\nu_K$ has the value found by Hodgkin-Huxley, $f$ is monotonic and
(\ref{HHV}-\ref{HHn}) has a unique equilibrium for each value of $I$. For fixed \textit{lower} values of $\nu_K$ there are
two saddle node bifurcations as $I$ is varied, creating a region with three equilibria and corresponding
to multistability (as in the example depicted in the previous section, Fig. \ref{FigSNFN}). The two
curves of saddle node terminate at a cusp point. These curves are obtained by varying
$\nu^\ast$, considered as a parameter and taking into account the transversality conditions
TSN1,TSN2 in the appendix. In particular the determinant of the Jacobian matrix has to vanish.
Given the equilibrium point, one also obtains the parameters value where Hopf bifurcations
occur. Hopf bifurcation requires that two complex conjugate eigenvalues appear or disappear.
This corresponds to conditions on the coefficient of the characteristic polynomial
of the Jacobian matrix. This polynomial has the form $x^4+c_3 x^3 + c_2 x^2 + c_1 x +c_0$.
Considering $\nu^\ast$ as a control parameter and solving simultaneously the fixed point equations
and the transversality conditions (TH1,TH2) in the appendix) one finds the set of parameters $\nu_K,I$ where a Hopf
bifurcation occurs. At the intersection of the Hopf bifurcation line and the saddle-node bifurcation
line, a Bogdanov-Takens bifurcation occurs (see the appendix). 
%
%
%
%
\begin{figure}[ht]
\begin{center}
\includegraphics[height=13cm,width=13cm,clip=false]{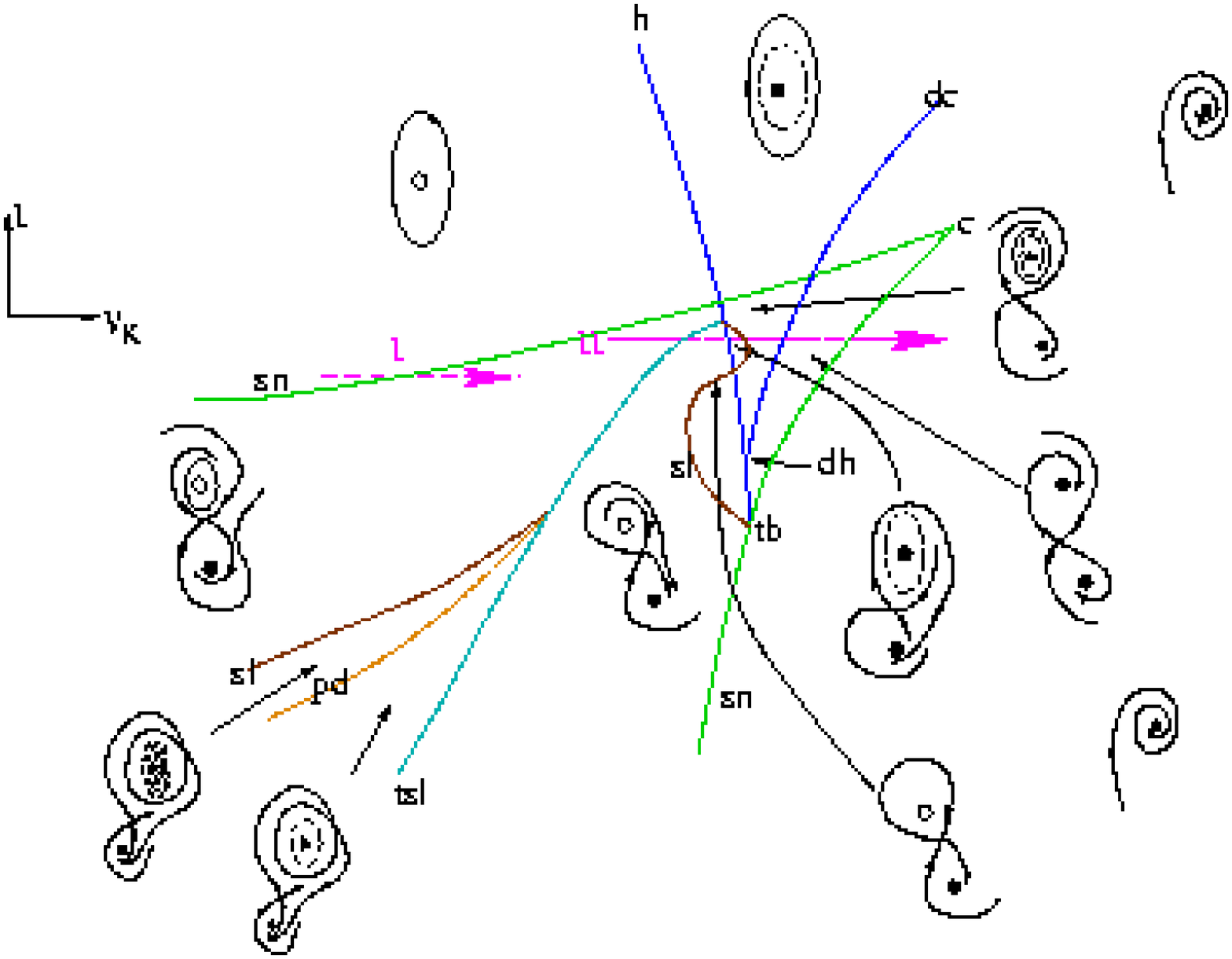}
\vspace{1cm}
\caption{
Bifurcation diagram of the Hodgkin-Huxley equations when varying the parameters $I,\nu_K$.
This figure has been drawn ``by hand'' from the Figure 1 in \cite{GL}.
 Stable equilibrium points are shown as black
dots, unstable focus as white dots, stable limit cycles are closed curves with solid lines and unstable
periodic orbits are dashed lines. One dimensional unstable manifolds of equilibrium points are shown together
with curves of the ``weak stable manifolds'' of equilibrium points with three dimensional stable
manifolds (see e.g. in the ``tsl'' and ``pd'' regions).
\label{FGL}
}
\end{center}
\end{figure}
%
%
%
%
%
One observes also global bifurcations: collapse of two limit cycles, homoclinic connexion at the
Bogdanov-Takens point, twisted saddle loop, degenerate Hopf bifurcation, etc ... 
The various bifurcations are depicted in Fig. \ref{FGL}. We used the following nomenclature
(from \cite{GL}). For a description of the corresponding bifurcations see the appendix.

\bigskip

\textit{Codimension one bifurcations.}

\smallskip

\bit
\item \textbf{sn:} Saddle-node bifurcation: two fixed point coalesce and disappear (resp. appear), see Fig. \ref{FSN} in the appendix.\\

\item \textbf{h:} Hopf bifurcation. A  fixed point changes its stability  and a  limit cycle appear with a radius
increasing with the control parameter (resp. a limit cycle decreases until it is reduced to a point and disappear
while the point at the center changes its stability), see Fig \ref{FH} in the appendix. As discussed above this corresponds
to type II excitability.\\

\item \textbf{sl:} Saddle-loop or homoclinic bifurcation. The amplitude of a periodic orbit increases until it captures a saddle point and 
disappears, its period tending to infinity when the control parameter tends to the critical value. As discussed above this corresponds
to type I excitability.\\

\item \textbf{tsl:} Twisted saddle-loop bifurcation. In dimension larger than two an orientation reversal
along a homoclinic may occur. The homoclinic orbit
is a two dimensional ribbon which is invariant under the flow with tangents in the directions of the weakest
contraction at the saddle point. A twisted saddle loop occurs if the ribbon is not orientable.This bifurcation is also met in physical experiments
about Rayleigh-Benard convection in a small geometry (see \cite{IC} for a mathematical analysis). 
Note that this bifurcation is usually related to period doubling. Also, for any $n$ value, $n$ integer, there exists a dynamical system, arbitrary close to
the bifurcating system, having homoclinic connexions with loops of order $n$ (see \cite{IW}). The dynamics can therefore be quite
complex in the vicinity of this bifurcation.\\

%
%
%
%
\begin{figure}[ht]
\begin{center}
\includegraphics[height=4cm,width=10cm,clip=false]{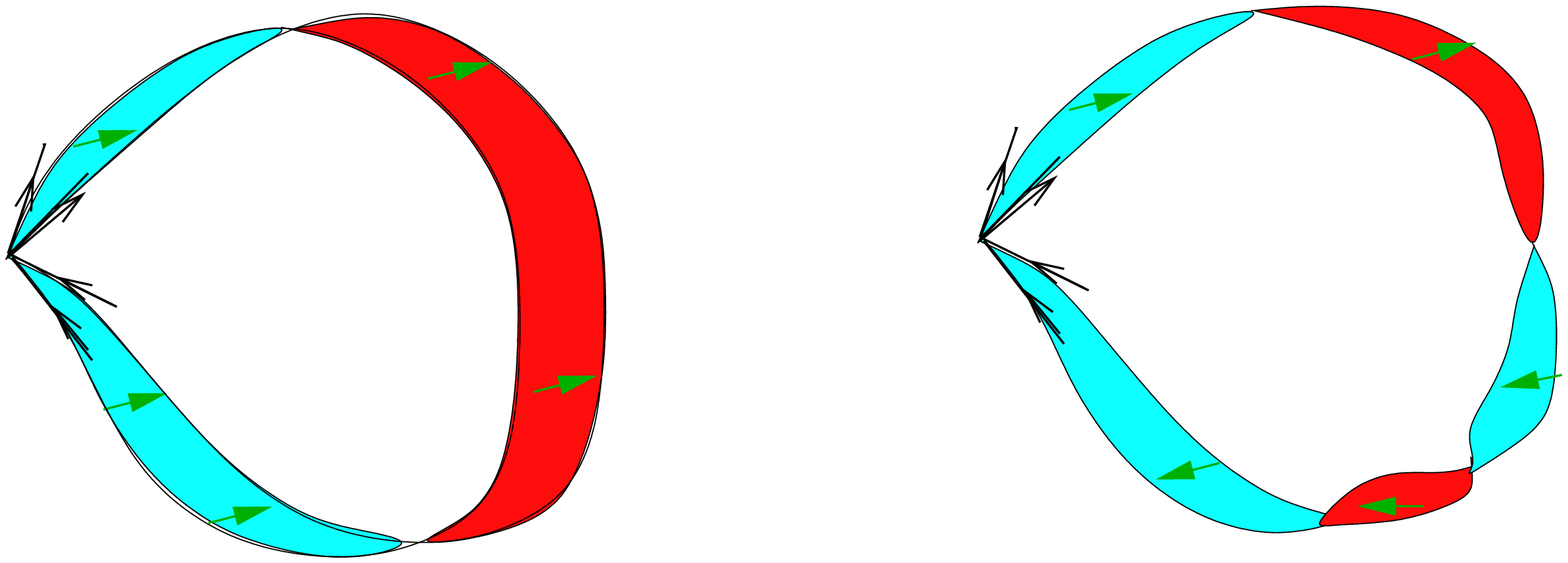}
\vspace{1cm}
\caption{(a) Untwisted saddle loop. (b) Twisted saddle loop.).
\label{FTSL}
}
\end{center}
\end{figure}
%
%
%
%
\item \textbf{dc:} Double cycle or saddle-node bifurcation of cycles. Two periodic orbit coalesce and disappear.\\

\item \textbf{pd:} Period doubling bifurcation. A periodic orbit changes its stability, while a periodic orbit of twice
its period coalesce with the bifurcating periodic orbit.\\
\eit

\textit{Codimension two bifurcations.}
\bit
\item \textbf{c:} Cusp. Three equilibria coalesce into one (see Fig. \ref{FSN} in the appendix).\\

\item \textbf{tb:} Takens-Bogdanov bifurcation (see appendix, Fig. \ref{FBT}).\\

\item \textbf{nsl:} Neutral saddle-loop or homoclinic bifurcation. A periodic orbit changes its stability
in a saddle loop at a point where the sum of the eigenvalues of the Jacobian matrix is zero. \\

\item \textbf{tnsl:} Twisted neutral saddle-loop bifurcation.\\

\item \textbf{snl:} Saddle-node loop.\\

\item \textbf{dh:} Degenerate Hopf bifurcation.
\eit

\bigskip

We have also represented some qualitative changes in the dynamics arising when varying
the Nernst potential $V_K$ in Fig. \ref{FPathGL} a,b,c.
%
%
%
%
\begin{figure}[ht]
\begin{center}
\includegraphics[height=4cm,width=3cm,clip=false]{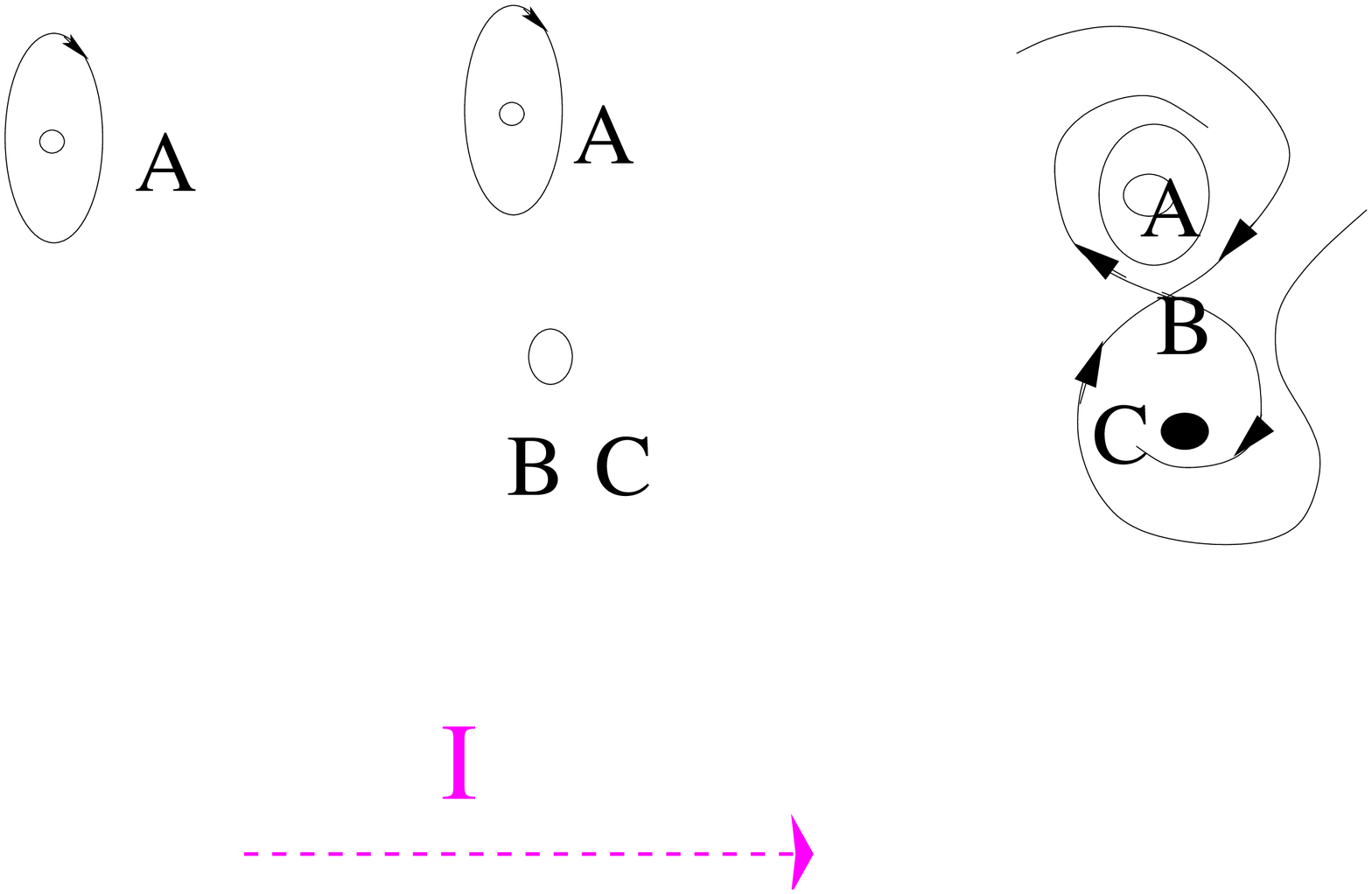}
\hspace{1.5cm}
\includegraphics[height=6cm,width=4cm,clip=false]{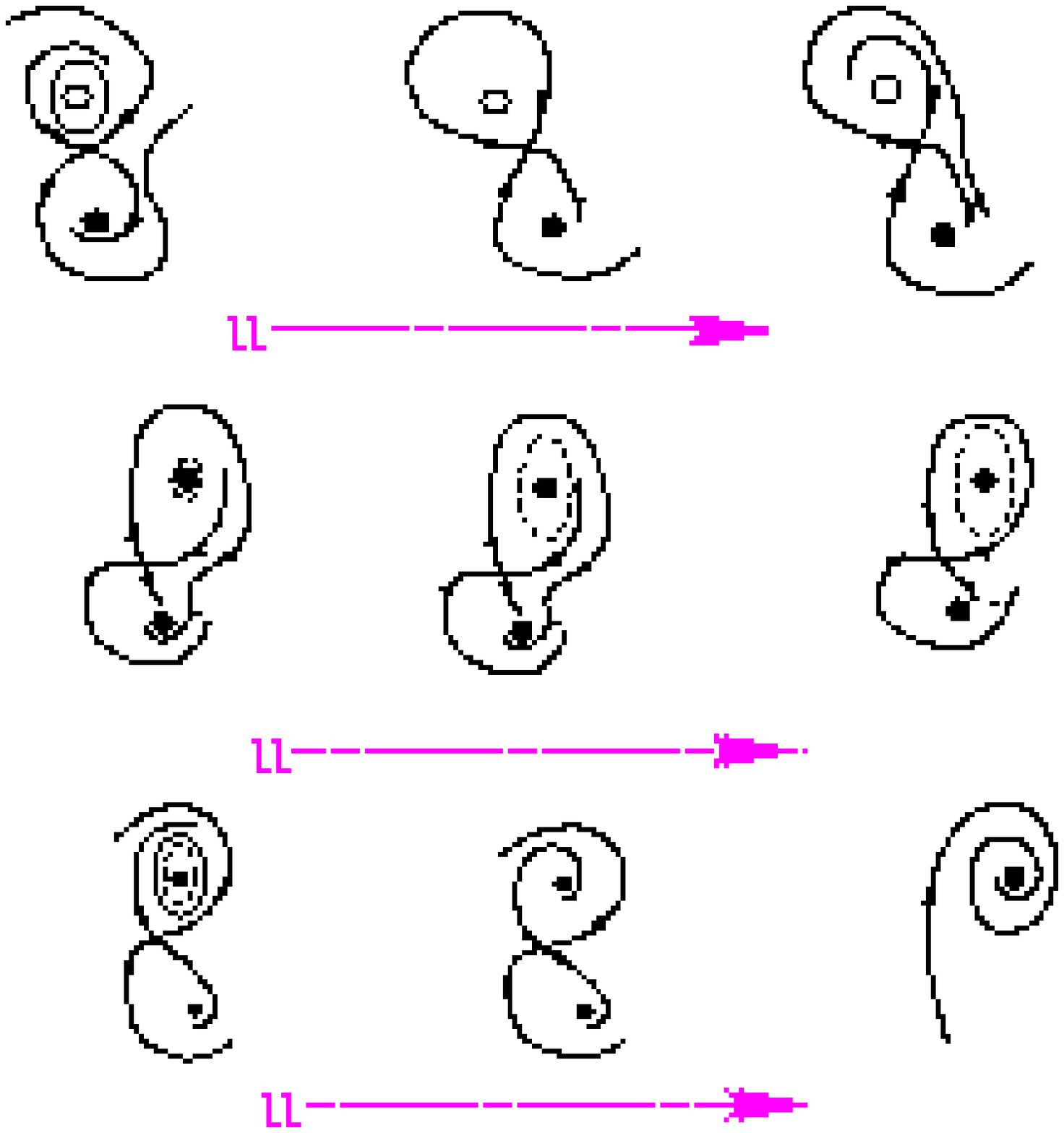}
\hspace{1.5cm}
\vspace{1cm}
\caption{Bifurcations occurring when following the paths I (Fig.\ref{FPathGL} a),II
(Fig.\ref{FPathGL} b) drawn in figure \ref{FGL}. The corresponding values
for the potential $V_K$ range from $-5.155$ to $-5.129$ mV.
 \label{FPathGL}}
\end{center}
\end{figure}
%
%
%
%
%

 These results illustrate the complexity of the dynamics occurring in the Hodgkin-Huxley
equations. There are many possibilities for the spiking patterns when the parameters
are changed. One may however ask about the biological relevance of these results.
Note that in Fig. \ref{FGL} the usual value of $V_K \sim 10 mV$ is far on the right of the graph
and does not appear in a scaled figure. Indeed, the region corresponding to the path II
ranges from $-5.155$ to $-5.129$ mV. Thus its width is of order $20 \mu V$ ... and
the potential is negative... Thus, some of these regimes may be difficult to find experimentally,
since they correspond to very tiny regions in the parameters space and quite unusual
value of parameters\footnote{Note that the Authors of \cite{GL}  also  explored  the changes induced by a variation
of the Potassium conductance $g_K$ but we do not discuss this here.
}. Another related question is:
what happens when coupling such neurons ? For example, do the regions $sl,pd,tsl$,
exhibiting a complex behavior, still exist when considering a neural network of
Hodgkin-Huxley neurons ? We shall see in this chapter that coupling neurons
with complex dynamics does not necessarily imply that the coupled dynamical
system will have a complex dynamics. On the opposite, coupling neuron
models with a simple evolution may lead to a complex evolution.

\ssu{Axon propagation.}\label{SProp}

The Hodgkin-Huxley equations (\ref{HHV}-\ref{HHh}) describe the behavior of a small piece
of neuron membrane. From the fundamental laws of Physics, one can use them 
to obtain an equation describing the propagation of the action potential
along the axon. One can in particular obtain the propagation speed. In this section
we derive the propagation equation. We then discuss the existence
of propagating solutions in a simplified version of the propagation equations, based on the FitzHugh-Nagumo model.\\

Let $V$ be the local membrane potential and $R$ the resistance per unit length
(as discussed in section \ref{SHH} it depends on $V$). For simplicity, we shall use in this section the convention
where $V=0$ at rest and we shall set $V_X = V_{rest}-E_X$ where $X=Na,K,L$ and
$E_X$ is the Nernst potential. Denote by $x$ the coordinate
longitudinal to the axon. One decomposes the current
in the membrane into an longitudinal  part ($i_a$) and a transverse part $i_m$. From
local charge conservation one has: $i_a(x+dx)=i_a(x)-i_m(x) \Rightarrow \frac{\partial i_a}{\partial x}=-i_m(x)$, 
while the Ohm's law writes: $V(x+dx)-V(x)=-Ri_a(x) \Rightarrow \frac{\partial V}{\partial x}=-R i_a(x)$. Consequently:

\beq
\frac{\partial^2 V}{\partial x^2}=R i_m(x)
\eeq

The local transmembrane current is given by the Hodgkin-Huxley system (\ref{HHV}-\ref{HHh}):

\beq
i_m dx=dx(C_m \frac{\partial V}{\partial x} + I_{ion})
= C_m \frac{\partial V}{\partial x}dx + S(x)\left[ g_{Na} m^3 h (V - V_{Na}) + g_K n^4(V - V_{K}) + g_L(V - V_L)\right]
\eeq

\nid where $S(x)=2\pi r(x)dx$ is the membrane surface per unit length and $r(x)$ the axon radius at $x$.
Finally, the equations describing the spike propagation along the axon are:

\bea \label{HHProp}
\frac{1}{R}\frac{\partial^2 V}{\partial x^2} 
&=& C_m \frac{dV}{dt}+2\pi r(x)\left[g_{Na} m^3 h (V - V_{Na}) + g_K n^4(V - V_{K}) + g_L(V - V_L)\right]\\
\frac{dn}{dt}&=&\alpha_n(V)(1-n) - \beta_n(V)n= \frac{n^\infty(V)-n}{\tau_n(V)}\\
\frac{dm}{dt}&=&\alpha_m(V)(1-m) - \beta_m(V)m= \frac{m^\infty(V)-m}{\tau_m(V)}\\
\frac{dh}{dt}&=&\alpha_h(V)(1-h) - \beta_h(V)h= \frac{h^\infty(V)-h}{\tau_h(V)}
\eea

Since we are interested in traveling solutions, it is natural to seek
solutions of type $V(x,t)=\cU(x-c t) \equiv \cU(\xi)$, where $c$ is the propagation speed.
To avoid boundary conditions problems, one may assume that the neuron is infinite.
Moreover the neuron is at rest at infinity, namely we are looking for solutions
such that :

\beq\label{BCHHProp}
\lim_{\xi \pm \infty}\cU(\xi)=0
\eeq 

The variable change $\xi=x-ct$ allows us to convert the partial differential equation
above in an ordinary differential equation where $\xi$ plays the role of a formal time:

\bea 
\frac{1}{R}\frac{d^2 \cU}{d \xi^2} 
&=& -c C_m \frac{d\cU}{d\xi}+\left[g_{Na} m^3 h (\cU - V_{Na}) + g_K n^4(\cU - V_{K}) + g_L(\cU - V_L)\right]\label{HHPropxia}\\
\frac{dn}{d\xi}&=&\alpha_n(\cU)(1-n) - \beta_n(\cU)n= \frac{n^\infty(\cU)-n}{\tau_n(\cU)}\label{HHPropxib}\\
\frac{dm}{d\xi}&=&\alpha_m(\cU)(1-m) - \beta_m(\cU)m= \frac{m^\infty(\cU)-m}{\tau_m(\cU)}\label{HHPropxic}\\
\frac{dh}{d\xi}&=&\alpha_h(\cU)(1-h) - \beta_h(\cU)h= \frac{h^\infty(\cU)-h}{\tau_h(\cU)}\label{HHPropxid}
\eea

\nid where we assumed for simplicity that $2\pi r(x)=1, \forall x$.\\

Instead of solving these equations we shall study the corresponding equation for the FitzHugh-Nagumo model.
They are indeed simpler and they allow us to figure out why  traveling
wave with a determined speed $c$ are selected. The equivalent of the equations 
(\ref{HHPropxia},\ref{HHPropxib},\ref{HHPropxic},\ref{HHPropxid}) for
the FitzHugh-Nagumo model (\ref{FN}) are:

\bea 
\epsilon^2\ddot{v}+ \epsilon c\dot{v}+f(v,w)&=&0 \label{FNPropv}\\
c \dot{w}+g(v,w)&=&0 \label{FNPropw}
\eea
\nid where $f(v,w)$ has a cubic shape (e.g. $f(v,w)=v-v^3-w$)
and $g(v,w)$ is linear (e.g. $g(v,w)=(v - a -bw)$). More specifically we shall assume that we are in the situation
of the Fig. \ref{FigPAVFN} where only one fixed point exists for the  model (\ref{FN}).
In eq. (\ref{FNPropv},\ref{FNPropw}) we forgot $C_m$ and $R$ which play no relevant
role in the mechanism described below. Since $\xi$ plays the role of a formal
time we used the notation $\frac{d u}{d \xi}=\dot{u}, \frac{d^2 u}{d \xi^2}=\ddot{u}$
Note that the variable $v$, representing the local membrane potential,
 is spatially coupled by the diffusion term, while $w$, representing a slow ionic current or gating variable, is not.

We describe the spike propagation by using the singular perturbation theory. If we set  $\epsilon=0$
in the equations (\ref{FNPropv},\ref{FNPropw}) we obtain the system of equations (called ``outer equations'' \cite{Keener}):

\bea 
f(v,w)&=&0 \label{Outv}\\
c \dot{w}+g(v,w)&=&0 \label{Outw}
\eea
As in section \ref{SFN} the solution of (\ref{Outv}) is the $v$ nullcline and $v$ depends parametrically on $w$.
The trajectory moves slowly on the stable branch  $N_v^+$ (resp.$N_v^-$ ) and this motion corresponds to the
excited phase (resp. recovery phase) of the pulse (see Fig. \ref{FPropPulse}).  

The pulse appears then as a trajectory connecting the two branches. To characterize the dynamics
between the two branches, it is convenient to rescale the variable $\xi$ as $\frac{\xi}{\epsilon}$
and  to write (\ref{FNPropv},\ref{FNPropw}) in the following form:

\bea \label{FNPropxiV}
\ddot{v}&=& -c\dot{v}-\frac{\partial \cV}{\partial v} \label{FNPropuV}\\
\dot{w} &=&-\frac{\epsilon}{c}(v-a-bw)
\eea

\nid where  we have introduced the ``potential'':

\beq
\cV(v,w)=\frac{v^2}{2}-\frac{v^4}{4}-wv
\eeq

Indeed, introducing $\cV$ allows us to interpret the equation (\ref{FNPropuV}) as the formal
equivalent of the motion of a particle moving in a potential well with a shape $\cV$,
with a ``friction coefficient'' $c$ and where $\xi$ plays the role of time. This picture
is especially useful to understand intuitively the mechanism at work. 
The potential $\cV$ depends parametrically on $w$ and has the typical shape depicted in Fig. \ref{FVFN}.\\
%
%
%
%
\begin{figure}[ht]
\begin{center}
\includegraphics[height=5cm,width=5cm,clip=false]{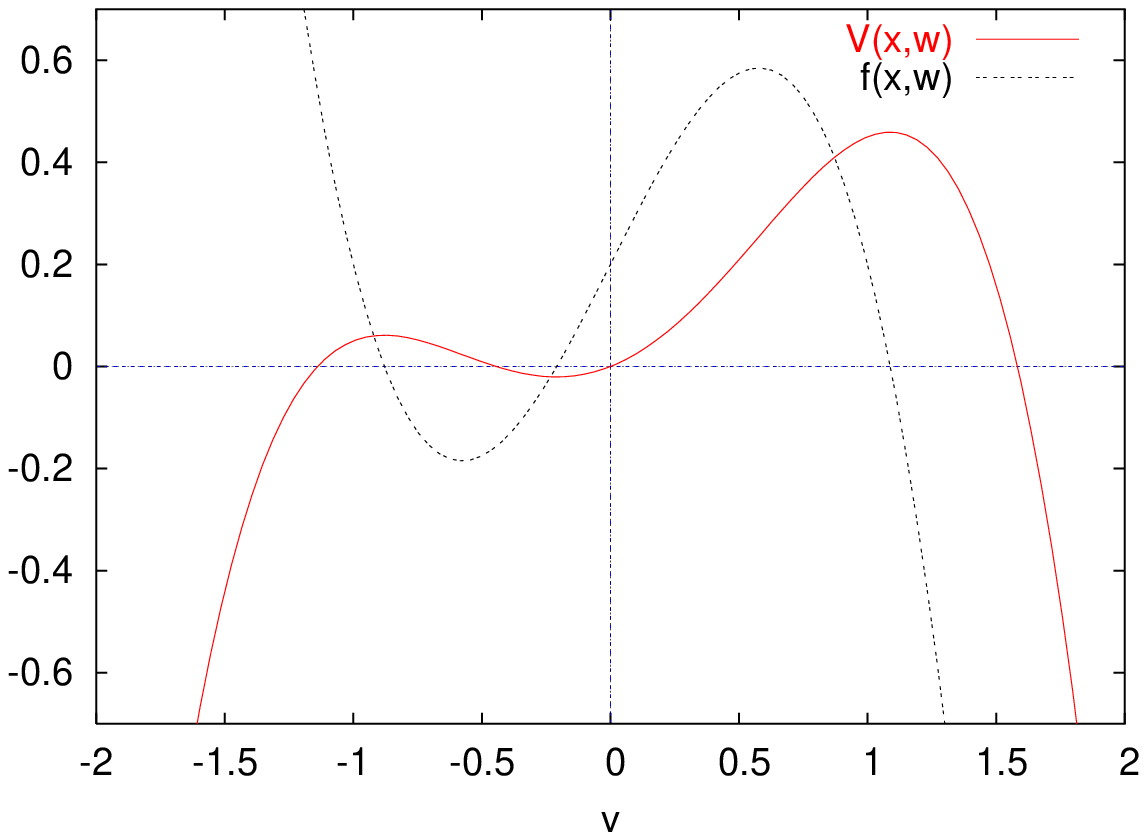}
\hspace{0.5cm}
\includegraphics[height=5cm,width=5cm,clip=false]{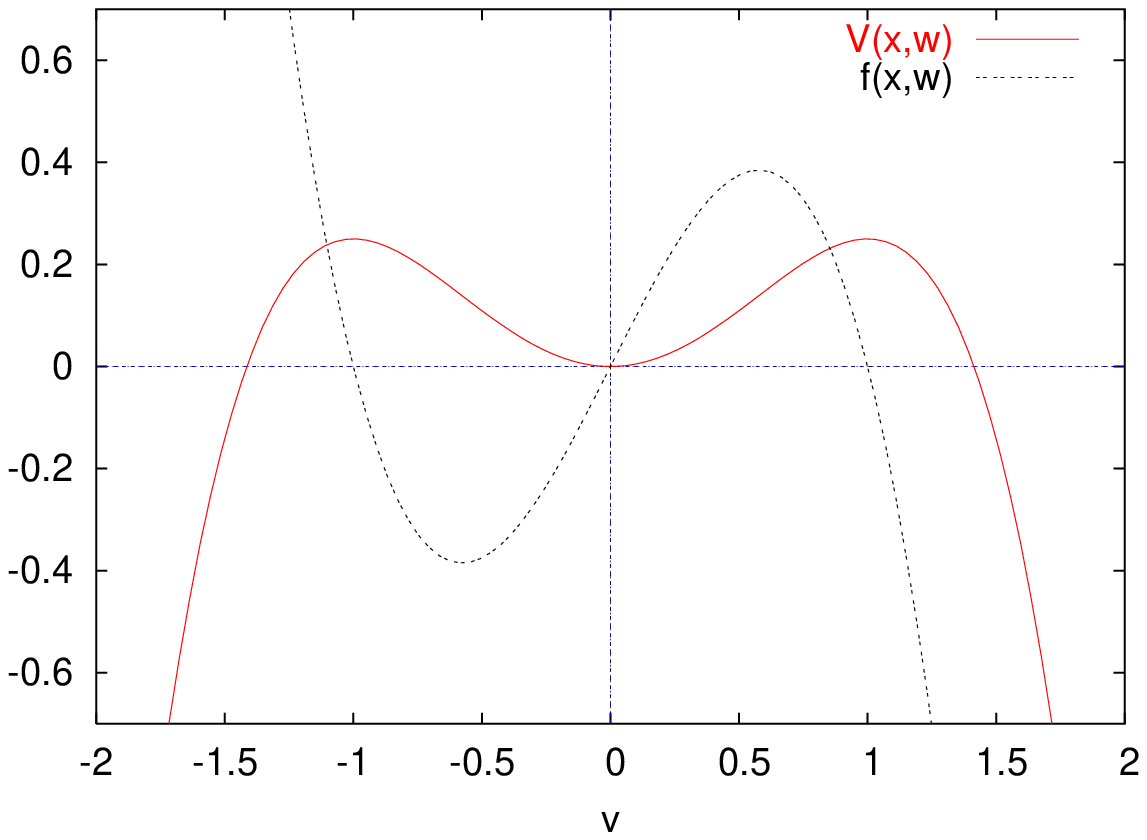}
\hspace{0.5cm}
\includegraphics[height=5cm,width=5cm,clip=false]{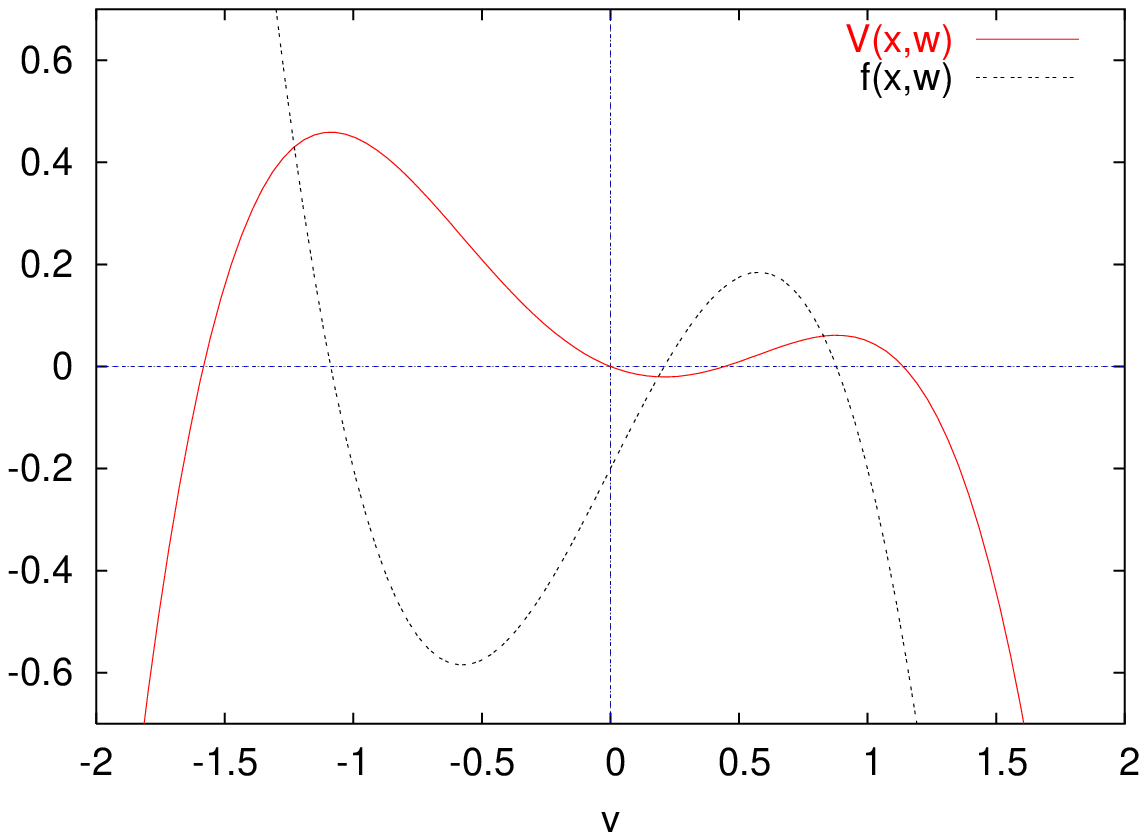}
\vspace{0.5cm}
\caption{Potential $\cV$ of eq. (\ref{FNPropv}) for : Fig. \ref{FVFN}a : $w < 0$;
Fig. \ref{FVFN}b : $w=0$;Fig. \ref{FVFN}c : $w > 0$;\label{FVFN}}
\end{center}
\end{figure}
%
%
%
%
%
%

 When $c=0$ there is no effective dissipation and
the phase portrait of the dynamical system (\ref{FNPropv}) is sketched in Fig. \ref{PPFNPH}a.
In particular, there is an homoclinic trajectory connecting $V^+$ to itself.
When $c$ is large enough, the phase portrait has the shape depicted in Fig.  \ref{PPFNPH}b. 
Consequently, by continuity, there is an intermediate value of $c$, $c_0(w)$ depending on $w$, where
there is an \textit{heteroclinic orbit} connecting the point $V^-$ and $V^+$. This heteroclinic orbit corresponds
to a moving transition layer, travelling with a speed $c_0(w)$. More precisely, the heteroclinic orbit corresponds
to an ``ascending''  front connecting  neurons  where $v$ belongs to the $-$ branch and with a coordinate  $\xi \to -\infty$
to neurons where $v$ belongs to the $+$ branch  and with a coordinate  $\xi \to +\infty$ (see the Fig. \ref{Front}).
 Note that for each $w$ there is a \textit{unique} such $c_0$:
this is the dissipation rate required to reach asymptotically the lower bump of $V$ ($V^-$ in the
case $w >0$) with an orbit starting from an arbitrary small  neighbourhood of the higher bump ($V^+$ in the
case $w >0$) with a zero initial speed.

Obviously the same argument can be done when $w$ is negative. One obtains then a descending front connecting
 connecting  neurons  where $v$ belongs to the $+$ branch and with a coordinate  $\xi \to -\infty$
to neurons where $v$ belongs to the $-$ branch  and with a coordinate  $\xi \to +\infty$.

\begin{figure}[ht]
\begin{center}
\includegraphics[height=5cm,width=5cm,clip=false]{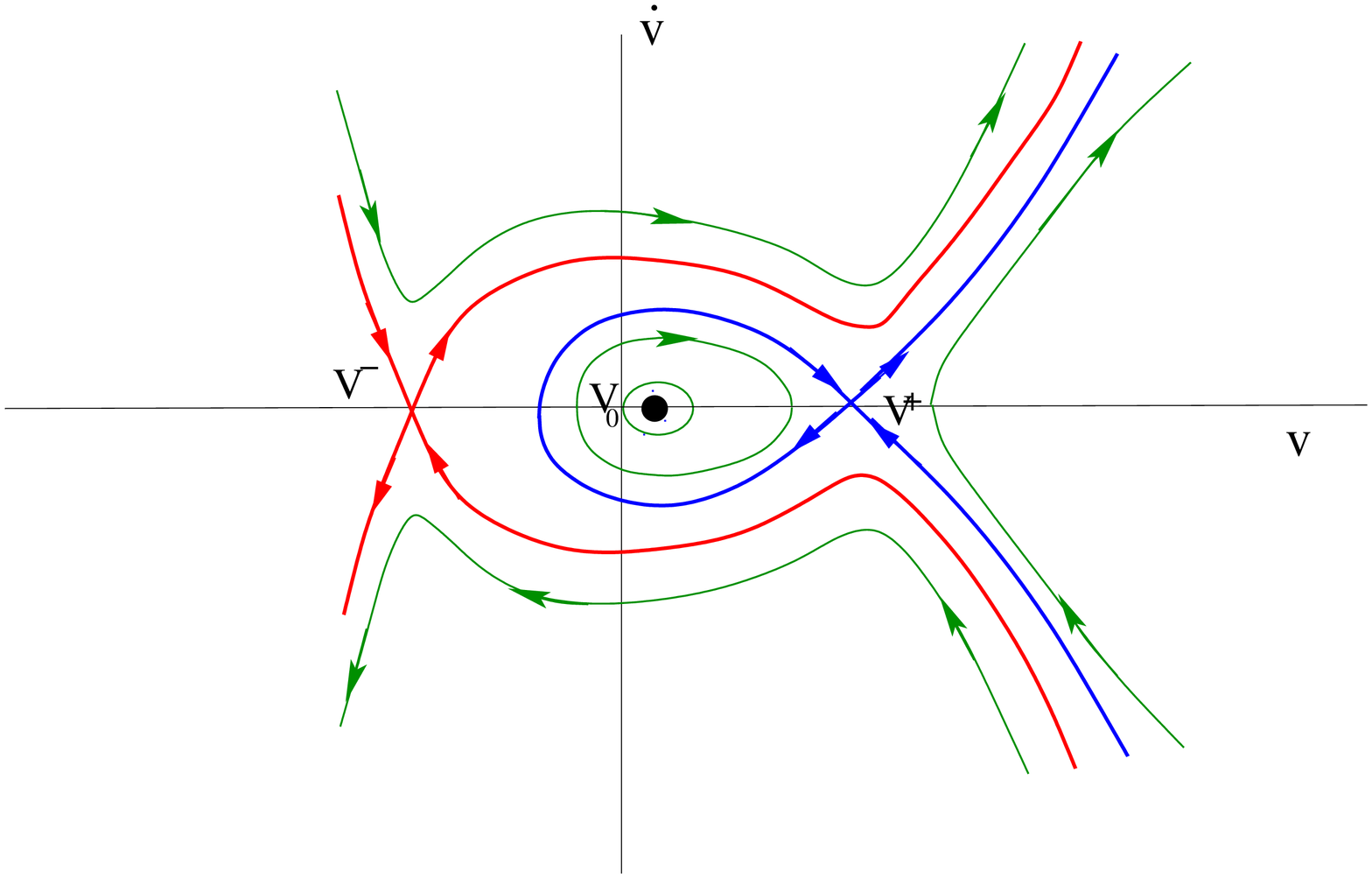}
\hspace{0.5cm}
\includegraphics[height=5cm,width=5cm,clip=false]{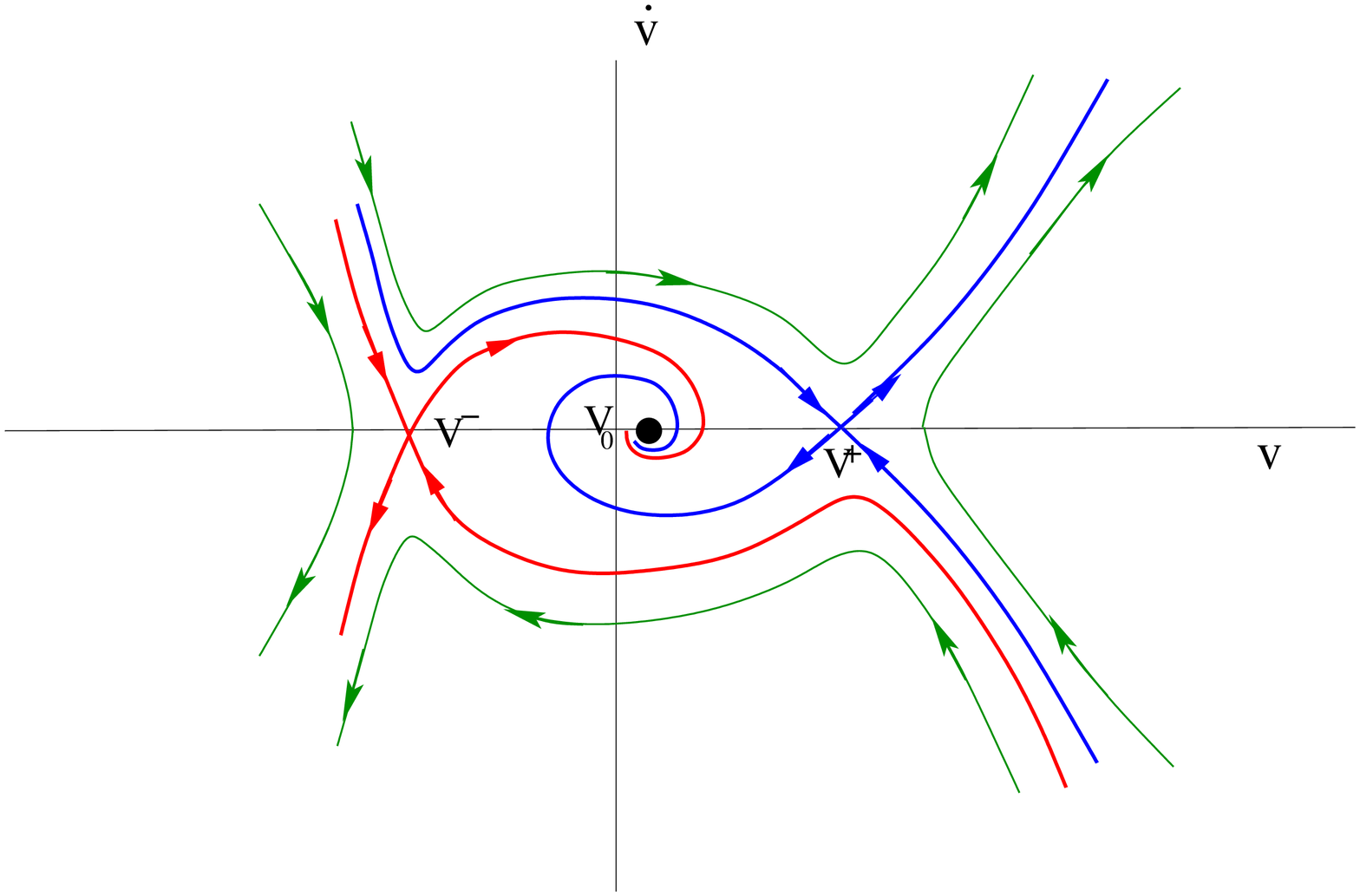}
\hspace{0.5cm}
\includegraphics[height=5cm,width=5cm,clip=false]{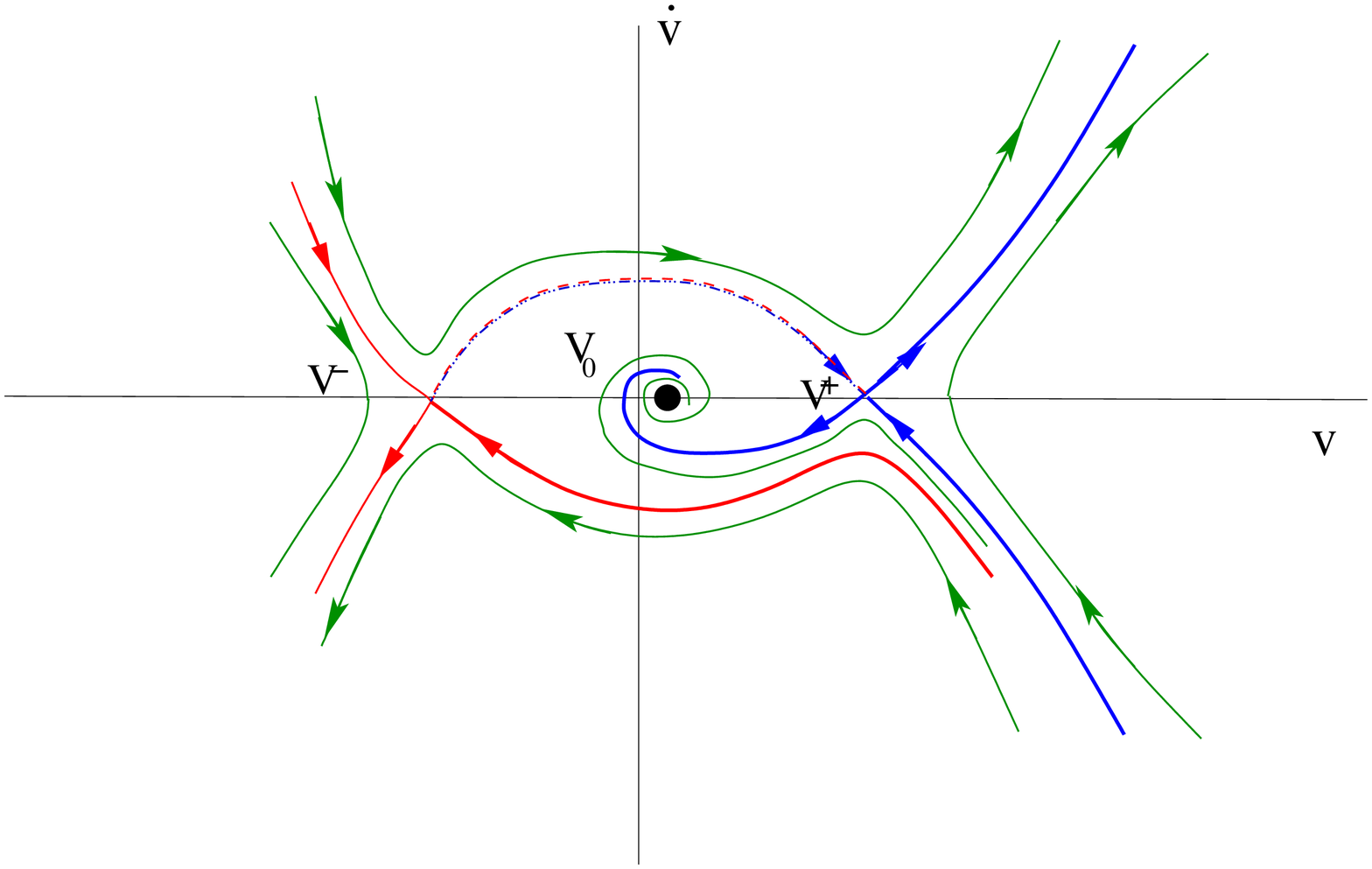}
\vspace{0.5cm}
\caption{Phase portrait of eq. (\ref{FNPropuV}) for : Fig. \ref{PPFNPH}a : $c=0$;
Fig. \ref{PPFNPH}b : $c>0$ ;Fig. \ref{PPFNPH}c : $c=c_0$. The situation corresponds to $w >0$.\label{PPFNPH}}
\end{center}
\end{figure}
%
%
%
%
%
%
The complete picture is the following\footnote{Strictly speaking, one has still to show that this picture,
obtained for $\epsilon=0$,
persists when  $\epsilon >0$.  One can indeed show that the heteroclinic orbit persists
by using perturbation theory and Fredholm arguments.}. In most space the outer equations (\ref{Outv},\ref{Outw}) are satisfied.
When a transition between the two branches occurs, there is a sharp transition in $v$, travelling at a speed
$c(w,\epsilon)$ connecting the two branches (and $w$ is essentially a constant during the transition). 
This corresponds to a travelling pulse consisting in an excitation front followed by a recovery back (see Fig. \ref{FPropPulse}). 
Note however that the medium needs to be sufficiently excitable to maintain a propagation. This corresponds
to the mathematical condition: $\int_{V^-(w^+)}^{V^+(w^+)} f(v,w^+)dv >0$ ensuring that there is a positive speed
of propagation..

%
%
%
%
\begin{figure}[ht]
\begin{center}
\includegraphics[height=6cm,width=6cm,clip=false]{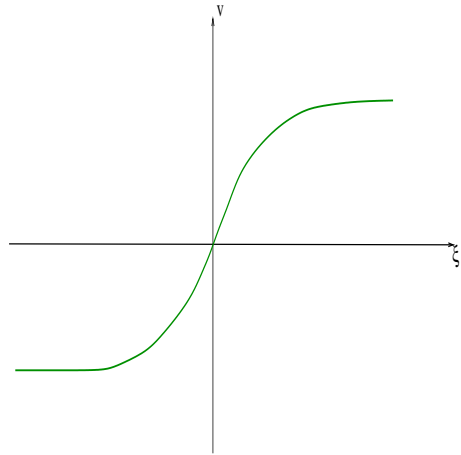}
\caption{\label{Front}  Front corresponding to the heteroclinic connection represented in Fig. \ref{PPFNPH}.}
\end{center}
\end{figure}
%
%
%
%
%
%
%
%
\begin{figure}[ht]
\begin{center}
\includegraphics[height=6cm,width=10cm,clip=false]{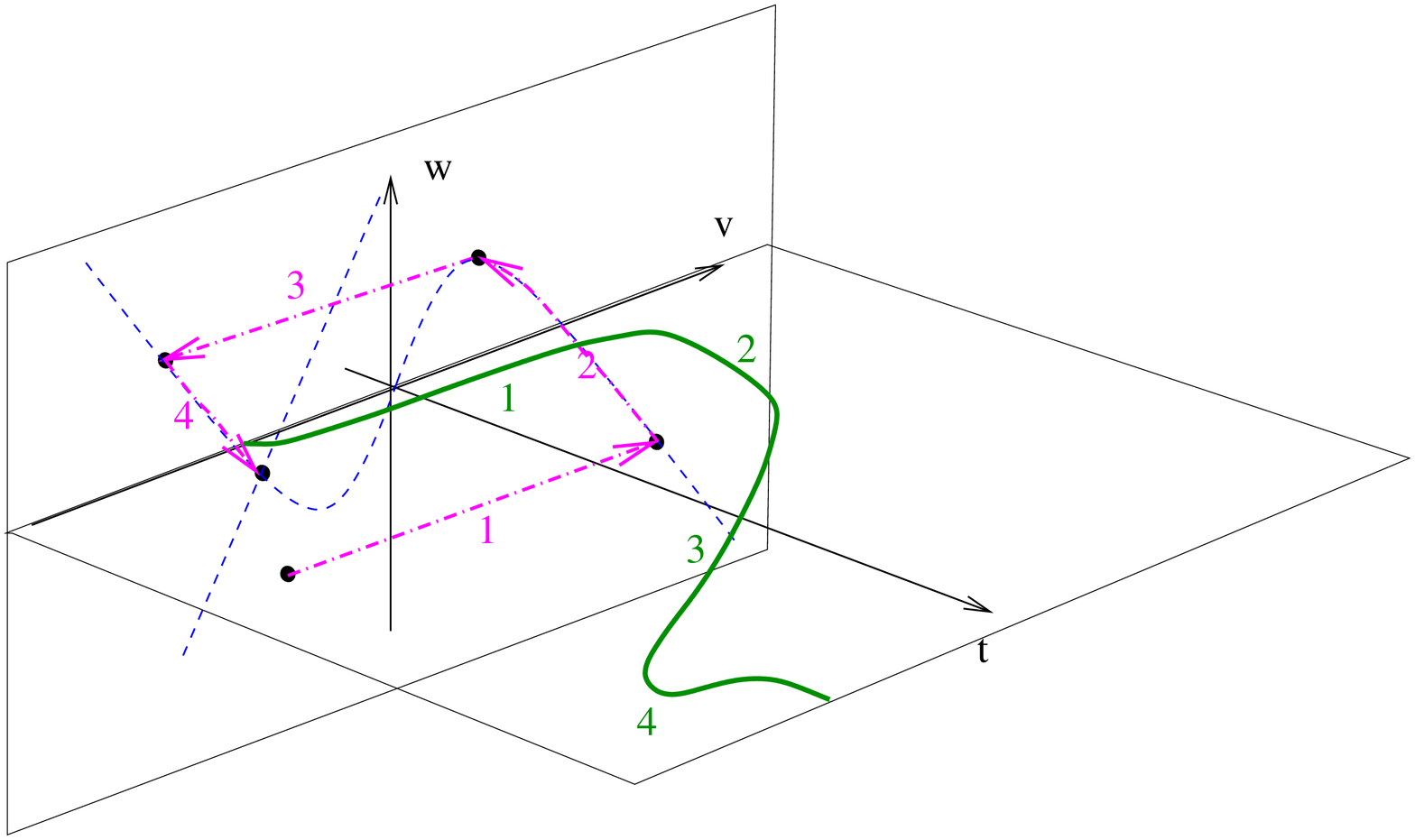}
\caption{\label{FPropPulse} Schematic sketch of spike propagation in the spatially extended Fitzhugh-Nagumo model.}
\end{center}
\end{figure}
%
%
%
%
%
%

This picture has therefore allowed us to understand the mechanism of spike propagation in neurons,
by using simple dynamical systems arguments. It is important to note the role of the refractory period.
If the
action potential reaches a given point, the neighboring points that have not been yet reached
by the spike are depolarized to the threshold, while the  neighboring points that have just been  reached
by the spike are in the refractory period and cannot emit a new spike. This imposes
a propagation direction.    

Finally, note that the existence of travelling spike in the Hodgkin-Huxley model can also be shown  rigorously \cite{Hastings,Carpenter}
For the typical values for squid axon one finds a speed value $c=21 mm/ms$ very close to the experimental value
found by Hodgkin and Huxley ($21.2 mm/ms$).

\su{Neural coupling.}\label{GenNN}

Up to now we have only considered the behavior of individuals
neurons described more or less accurately by a set of differential
equations. But neurons are not isolated entities and
it is absolutely clear that the brain functions are the result of \textit{collective}
effects. If  formal Neural Networks are (more or less rough) 
models for the brain, the emergent collective dynamics resulting from the coupling of
individual (formal) neurons should  exhibit properties such as information storage, recognition tasks, learning, that a lone
neuron should not able to perform. If we stay at the level of mathematical models, then dynamical systems
theory should be able to provide us some hints about the collective evolution when
parameters are varied, external inputs are presented, learning is performed, \etc.
This aspect are further addressed in the next sections.

\ssu{Synapses and synaptic plasticity} \label{Synapses}

The main function of neurons is to propagate informations via
electric signals. This is reflected in their structure. They have two types of specific extensions:
dendrites and axons. The dendrites form a tree like structure. They collect
signals coming from other neurons and transmit them to the neural cell nucleus.
The axon transmit spikes towards other neurons via connections
called ``synapses'' (from Greek " syn " (together) et " haptein " (join)).
There exists two type of synapses: electrical and chemical.
In the first case (electric synapses) neurons are touching and the neural flux can directly go from one neuron
to the other. In the second case (chemical synapses), the neurons are not touching and
the neural flux is transmitted vi neurotransmitters (Acetylcholin, Dopamin,
Gamma-Aminobutyric Acid, Glutamat etc...).  The action potential opens ion
channels producing an influx of $Ca^{2+}$, leading to the release of a neurotransmitter into the synaptic cleft. The transmitter
diffuses then to the other side of the cleft and binds to receptors, causing ion-conducting channels to open. This results in a excitatory or
inhibitory post synaptic current, depending on the nature of the ion flow. 
Most synapses are
chemical.

When two neurons are connected via  synapses the emission of spikes from the pre-synaptic
neurons may evoke spikes in the post-synaptic neuron.  These spikes have a variable
height depending on the \textit{synaptic efficiency}. Synaptic efficiency evolves with time
via different mechanisms. \textit{Long Term Potentiation} (LTP) is a synaptic reinforcement mechanism
 involved in memory. It corresponds to an increase in the post-synaptic response after
an intensive presynaptic excitation, applied on a short time scale ($\sim 1$s),
 but with a high frequency ($> 100$ Hz), inducing a strong depolarisation in the post synaptic neuron.
\textit{Long Term Depression} (LTD) is  complementary to LTP. This mechanism arises when the pre-synaptic neuron
has a low frequency activity (1-5 Hz) but the post-synaptic neuron essentially does not fire. This lack
of synchrony between the two neurons has the  effect of reducing the synaptic efficiency.
It is believed that LTD is used in structures such as hippocampus, to bring back to a normal level of efficiency
synapses whose efficiency has increased via LTP,
rendering them available for new informations storage. 
A last mechanism, called  \textit{Spike Timing Dependent Plasticity} (STDP) has recently attracted 
much efforts. One can experimentally show that 
LTP and LTD can be elicited by carefully adjusting the timing 
of the pre- and post- synaptic activity. If the post-synaptic spike fires just before the pre-synaptic cell 
then the association between the two neurons weakens. On the opposite
this association is reinforced if  the post-synaptic spike fires just after the pre-synaptic cell. 
Important references for STDP studies were published in  \cite{Fitzsimonds,Bi}. 
However, there seem to be a wide  variety of different rules which may have different functionalities  for dynamical neural networks. 
\cite{Abbott}

\ssu{Modeling neural networks.}\label{MNN}

Synapses are complex objects, as neurons are. However,
the more accurate one desires to model the evolution of a neural
assembly, the less it is possible to handle analytically the dynamics.
Consequently, one has to simplify the neurons and/or synapses description
in order to obtain tractable models. 
Therefore, in many models 
 synapses are roughly 
 represented by a ``wire'' connecting the pre- and post-synaptic
neuron and weighted by a number $J_{ij}$ modeling the efficiency of the synaptic
connection from neuron $j$ to neuron $i$.  
This number
can be positive (excitatory synapse) or negative (inhibitory synapse).  
 It can be random or constant, and may evolve in time 
(via learning for example, see sections \ref{Hebb} and \ref{App}). 
Although the synapses are asymmetric in general
(the influence of $j$ on $i$ is not the same as the influence of $i$ on $j$), some models consider
symmetric synapses (sections \ref{Hebb}, \ref{SCG}). Indeed, the symmetry in the interactions lead, for some models,
to convergence properties, useful for performing tasks (see section \ref{SCG}).

Obviously,
representing the synaptic connections between
two neurons by an edge between two nodes 
is certainly a very rough way of sketching
a neural network structure. Nevertheless, it
is widely used in this community. We would however
like to point out the following remark.
Since synapses are used to transmit
neural fluxes (spikes) from a neuron
to another one, the existence
of synapses between a neuron (A) and another
one (B) is implicitly 
attached to a notion of  ``influence'' or
causal and directed action\footnote{Note that the notion of influence
roughly sketched here is very close to the definition of synaptic weights
discussed by Hebb in \cite{Hebb}.}. 
However, a neural network is a highly dynamical
object and its behavior is the result of complex
interplays between the neurons dynamics and the
synaptic network structure. Moreover, the neuron
$B$ receives usually synapses from many other
neurons, each them being ``influenced'' by
many other neurons, possibly acting on $A$, etc...
Thus the actual ``influence''
or action of A on B has to be considered dynamically
and in a global sense, by considering $A$ and $B$
not as isolated objects, but, instead, as entities
embedded in a system with a complex interwoven dynamical
evolution. Thus the mere analysis of the synaptic graph 
topology is in general not sufficient to handle the
neural dynamics. A prominent example of this
is given in the section \ref{RepLin}.

On mathematical grounds this aspect can be addressed
as follows. Assume that the coupled neurons evolution is
described by a dynamical system:

\beq\label{NNG}
\frac{du_i}{dt}=F_i(u_1, \dots, u_N; \Gamma)
\eeq

\nid where $u_i$ is a variable  describing
the ``state'' of neuron $i$ (e.g. its membrane potential). $N$
is the total number of neurons. $\Gamma$ is a set of parameters
accounting for neurons characteristics, external stimuli,
and also including synaptic couplings (more specific examples
will be given throughout this paper). In the sequel we shall
use the notation $\bu$ for the vector $\left\{u_i \right\}$

Assume now that we weakly modify the state of neuron $j$,
for example by adding an external stimulus, such that the new
neuron state at time $t$ is $u_j(t) + \delta_j(t)$. The change
induced on neuron $i$ at time $t+dt$ can be formally computed by writing
a Taylor expansion of $F_i$ in powers of $\delta_j(t)$.
At the lowest order the change will be proportional to
the Jacobian matrix element $\frac{\partial F_i}{\partial u_j}(\bu)$.
This element measures in some sense the linear ``influence'' of the neuron $j$ on the neuron $i$,
when the system is in the state $\bu$. More precisely, it characterizes,  to the first order in a Taylor expansion, the modification induced
on $u_i$ when $u_j$ has a small variation. 

Although (\ref{NNG}) is generally a non linear system,
this Jacobian matrix can provide useful insight in the dynamical
properties as discussed in the sections \ref{Coop} and \ref{RepLin}. 
It is in particular possible to construct a graph from
the Jacobian matrix such that there is an oriented edge $j \to i$ iff $\frac{\partial F_i}{\partial u_j}(\bu) \neq 0$. The
edge is positive if $\frac{\partial F_i}{\partial u_j}(\bu) > 0$ and negative if $\frac{\partial F_i}{\partial u_j}(\bu) < 0$.
(Obviously, this graph depends in general on the state $\bu$). This graph has \textit{circuits} or \textit{feedback loops} 
If $e$ is an edge denote by $o(e)$
the origin of the edge and $t(e)$ its end. Then a circuit is a sequence of edges $e_1, \dots, e_k$ such that $o(e_{i+1})=t(e_i)$,
$\forall i =1 \dots k-1$, and $t(e_k)=o(e_1)$. A circuit is positive (negative) if the product of its edges is positive (negative).
A positive feedback loop basically induces (to the linear order) a positive feedback inducing an increase in the activity of the neurons
in this loop. Obviously, there is no exponential increase since rapidly non linear terms will saturate this effect.\\

The graph induced by the Jacobian matrix is usually distinct from the synaptic
graph. In particular, it depends on the state $\bu$ of the set of neurons. However,
in models such as the recurrent neural networks discussed in
the section \ref{SCG} and \ref{NotreModele} $\frac{\partial F_i}{\partial u_j}(\bu)$
 is proportional to $J_{ij}$ with a positive ($\bu$ dependent) coefficient.
Thus this graph preserves the excitation/inhibition nature of the synapse.
Nevertheless, even in this case, the mere fact that the graph of linear
influence depends on  the state of the system may have dramatic
effects e.g. on signal propagation. 
As discussed in section \ref{RepLin}, the notion of linear influence (and
more generally linear response) allows to handle to some extent 
the interplay between the network topology and neurons dynamics
and rather unexpected effects will be exhibited.

\ssu{Synaptic plasticity and learning.}\label{Hebb}

Synaptic plasticity occurs at many levels of organization and time scales in the brain.
It alters excitability of the brain and regulates behavioural states (e.g. transition between sleep and wakeful activity).
It is also involved in short and long term memory and learning. In this section
and in this paper we shall only focus on this last issue.

The synaptic weights are
evolving in time during learning. In formal neural network learning is thus represented by evolution schemes for the
synapses, called \textit{learning rules}.   Although learning rules can be proposed using
precise description of LTD, LTP and STDP, most of them rely on
some fundamental recipes inspired from D. Hebb's work.
One speaks then of \textit{Hebbian learning}. We shall focus on Hebbian learning
in this paper.

 D. Hebb has proposed in \cite{Hebb} a theory of behavior based on the physiology of the nervous system.
The most important concept to emerge from Hebb's work was his formal statement (known as Hebb's rule) 
of how learning could occur. \\

\textit{When an axon of cell A is near enough to excite a cell B and repeatedly or persistently takes part in firing it, some growth 
process or metabolic change takes place in one or both cells such that A's efficiency, as one of the cells firing B, is increased. 
}\\

Most of the learning rules in neural networks are based on Hebb's observations plus a few well established facts.
They rely upon a few recipes that can summarized as \cite{HoIz}:

\bit

\item Learning results from modifying synaptic connections between neurons.

\item Learning is local i.e. the synaptic modification depends only upon the pre- and post- synaptic
neurons activity and does not depend upon the activity of the other neurons.

\item The modification of synapses is slow compared with characteristic times of neuron dynamics.

\item If either pre- or post- synaptic neurons or both are silent then no synaptic change takes place
except for (exponential) decay which corresponds to forgetting.  
\eit

The first item
implies that learning results in a modification of the $J_{ij}$'s. The
second one basically says that the synaptic modification of $J_{ij}$ writes
$J_{ij}'=\epsilon h(J_{ij}^T,m_j,m_i)$ where $J_{ij}'$ is the value of the 
synapses $j \to i$ after the learning rule has been applied. The parameter $\epsilon$
has been added for convenience and will be discussed below.
The numbers $m_i$ ($m_j$) 
denotes the ``state'' or ``activity''
of the neuron $i$ ($j$). We do not precise yet what is this ``state'' since it can vary according
to the model.  Several examples will be discussed below.
 The third item  implies then that 
$\epsilon$ is small parameter, whose inverse corresponds to the characteristic time for a significant
change of $J_{ij}$.  The fourth item may lead to different forms according to the model
(see below).
But if one assumes that the changes in the $J_{ij}$'s are slow (item 3) and if $h$ is a smooth function
then one may simply consider a Taylor expansion of a generic regular function $h$. This gives, up
to the second order in $m_i,mj$.

$$J_{ij}'=\epsilon \left(a_{000}+ a_{100}J_{ij}+ a_{010}m_j + a_{001}m_i + 
a_{011}m_ i m_j + h.o.t. \right)$$

\nid where h.o.t. means ``higher order terms'' such as $J_{ij}m_i m_j$, etc....
In this chapter we shall focus on this form, forgetting the other terms.
Note that the terms $a_{100},a_{010},a_{001},a_{011}$ have all a ``biological''
interpretation. We shall not consider the term $a_{000}$. 
Writing $\lambda=\epsilon a_{100}$
the corresponding term models passive ``forgetting'': if a synapse is not solicited 
its intensity decreases with a decay rate $\frac{1}{\lambda}$ (we shall assume that
$1 \geq \lambda > 0$). On biological grounds, the situation is a little bit more complicated.
The decay of the synapse and more generally its evolution  
depend on the activity of the pre synaptic
($j$) and post synaptic ($i$) neuron,as we saw. These activities determines the production of $Ca^{2+}$
ions, which acts in turn on the width of ionic channels involved in the synapse activity.
 The production of $Ca^{2+}$
increases whenever $i$ \textit{and} $j$ are ``active'' increasing the synaptic efficiency.
On the other hand, when $x_i$ \textit{or} $x_j$ are active then the concentration
$[Ca^{2+}]$ stays constant, and enzymatic phenomena result in an effective decay of the
synapse (Long Term Depression). This gives an interpretation of the $3$ terms
$a_{010},a_{001},a_{011}$. \\

Thus, setting $\epsilon a_{100}=\lambda, \epsilon a_{011}=\alpha,
\epsilon a_{010}=-\beta, \epsilon a_{001}=-\gamma$
 we obtain a synaptic evolution having the form:

\beq\label{HebbRuleGen}
J_{ij}'=\lambda J_{ij} +\alpha \Gamma_{ij} -\beta m_i - \gamma m_j
\eeq

\nid$\Gamma_{ij}^T$ is a function of the activity of the pre- and post-
synaptic neurons. In most case $\Gamma_{ij}^T \sim m_i^T m_j^T$ but the form
(\ref{Synapses}) affords natural generalization that we shall
briefly discuss.  Note that all the coefficients $\alpha,\beta,\gamma,\lambda$
are proportional to $\epsilon$, which fix somehow the characteristic time scale
of the synaptic dynamics. 

Some examples of learning rules will be presented in this chapter but we shall
focus on situations where $\beta=\gamma=0$. A more
detailed discussion can be found in chapter III.

\su{Weakly connected neurons.}\label{WCNN}

What happens when neurons, having their own dynamics, are coupled via synapses ?
Though this question is too general to have a precise answer, it is possible
to address it when considering a \textit{weak coupling limit} with some
additional assumptions discussed below.  In a nutshell, the basic idea is to consider the situation where
a collection of neurons is coupled as a \textit{perturbation} of the uncoupled case, where each neuron
evolve independently from the other. The perturbation resulting from the coupling can however be either
irrelevant, when the coupled and the uncoupled systems are essentially equivalent from the dynamical
point of view (section \ref{WCNNSS}), or it can have a drastic effect. As argued below, this is basically
the case when some neurons are close to a bifurcation point.
In this case a rather detailed
analysis can be made by using standard tools from bifurcations theory
theory such as center manifold reduction and normal forms (sections  \ref{WCNNSU} and  \ref{GNF})
provided one restricts the overwhelming possibilities of bifurcations, that may potentially occur in 
a collection of coupled neurons, to some canonical ``scenarios'' (sections \ref{WCNNP}, \ref{WCNNH}).
 This is, of course, an important restriction, but the results obtained are quite illuminating from many aspects, 
especially with respect to the ability
of such Neural Networks to perform task, such as pattern recognition, manifested by  changes in the dynamics
when a pattern is presented to the network (section \ref{ExHebb}). We present here a short review of results
mainly due to Hoppensteadt and Izhikevich (see \cite{HoIz}).

\ssu{General setting.}

From now on, we consider therefore 
an assembly $\cR$ of $N$ coupled neurons.
Summarizing the previous section,  the dynamics of individual neuron is governed
by an equation of the form:

\beq\label{1N}
\frac{dX_i}{dt}=F_i(X_i;\lambda), \quad i=1 \dots N
\eeq

\nid where $X_i$ is a vector in $\bbbr^m$ describing the state
of the neuron. This form is quite general and includes in particular
the Hodgkin-Huxley equations (\ref{HHV}-\ref{HHh}) and the general form of excitable membrane equations (\ref{NGen}).
 $\lambda$ is a set of parameters
on which the neurons dynamics depends. An example is the applied
current $I$. We may assume that $\lambda$
belongs to a  space $\cE_\lambda$.
 Without loss of generality, and
for technical reasons we shall assume from now on that $X_i \in \cM$
where $\cM$ is a compact $m$ dimensional manifold.

 The basic requirement of the theory of weakly connected neural
networks is that the contribution of activity of one neuron to the activity
of another one is very small. More precisely, following \cite{HoIz} we call
\textit{Weakly Connected Neural Network} (WCNN) a dynamical system of the form:

\beq \label{DWCNN}
\frac{dX_i}{dt}=F_i(X_i;\lambda)+\epsilon G_i(X_1, \dots, X_n;\lambda,\rho,\epsilon); 
\quad X_i \in \cM, \quad i = 1 \dots N
\eeq

\nid or, in a more compact form:

\beq \label{DWCNN2}
\frac{d\bX}{dt}=\bF(\bX;\lambda)+\epsilon \bG(\bX;\lambda,\rho,\epsilon)
\eeq

\nid where we note in boldface  the $n \deq N \times m$ dimensional vectors
$\bX=\left(X_i \right)_{i=1}^N, \ \bF=\left(F_i \right)_{i=1}^N, 
\ \bG=\left(G_i \right)_{i=1}^N$. Note therefore that $\bF$ as a diagonal structure.
In equation (\ref{DWCNN})
$G_i$ is a smooth, bounded function of $(X_1, \dots, X_N;\lambda,\rho,\epsilon)$
that models the synaptic connections
between the other neurons $X_1, \dots, X_N$ and the neuron $i$. 
It depends on the set of parameters $\lambda \in \cE_\lambda \subset \bbbr^p$
 describing the state of individual neurons,
on the coupling parameter $\epsilon$, and on an additional set of parameters 
$\rho \in \cE_\rho \subset \bbbr^r$
corresponding to external constraints (for example the external environment
influence,  a static input, \etc).
Finally, it is assumed that $\epsilon$ is small, namely $\epsilon \ll 1$. 
This  purely ``mathematical 
assumption'' is required to perform the analysis presented below.

Of course, one may wonder whether this restriction still provides models for
biologically realistic situations. Note however that the mathematical condition $\epsilon \ll 1$
is abstract and refers to the  particular set of differential equations (\ref{DWCNN})
which attempts to model \textit{some} aspects of neuronal dynamics.
Henceforth, questions
such as : "How small is $\epsilon$ in the real brain" are essentially meaningless.
There are nevertheless different and non equivalent ways
 to estimate the strength of connexions between neurons,
One of them is based on the analysis of cross correllograms from pairs of neurons.
Performing such an analysis Abeles \cite{Abeles1} concluded that interactions between
adjacent neurons in the cortex are weak and the interactions between distant neurons
is even weaker. Another way to characterize weakness of synaptic connections
is to measure the amplitude of post synaptic potentials (PSP)
in the soma while the neuron membrane is far below the threshold value.
Indeed, in this state the size of PSP reflects the weakness of connections.
A detailed discussion can be found in \cite{HoIz}.

\ssu{Structurally stable case.}\label{WCNNSS}

In spite of the restriction to weak couplings, the dynamics of (\ref{DWCNN}) can be very rich
since, as we have seen in the previous sections, the behavior of individuals neurons can already be quite complex. Consequently,
it is impossible to analyze (\ref{DWCNN}) without further restrictions or specifications.
A  starting point is to consider first the case when each neuron has an
equilibrium point and is in a rest state when $\epsilon=0$. 
 In order to simplify the computations, and without loss of generality, one may
assume that this point is the origin when $\lambda=0$, namely $\bF(0,0)=0$.
Denote by: 

\beq
\bL \deq  D_{\bX} \bF(0,0)
= 
\left(\baR{cccccc} 
L_1 & 0 & \dots &0 \\
0 & L_2 & \dots & 0\\
. & . &. & .&\\
. & . & \ . & .&\\
. & . &  \ \ .& .&\\
0 & 0 & \dots & L_N  
\eaR\right) 
\eeq

\nid the Jacobian matrix of $\bF$ at the point $\bX=0, \lambda=0$, where
$L_i \deq \left(\frac{\partial F_{ik}}{\partial X_{ij}} \right)_{k,j=1 \dots m}$. 
A fixed point may be hyperbolic or not. The following result derived by E. Izhikevich
\cite{Iz}
is an example of application of the Hartmann-Grobman theorem
(see appendix) in the context of Neural Networks dynamics.

\bth\label{ThSSWCNN}
If the dynamics of each neuron is near an hyperbolic equilibrium
when $\epsilon=0$ then the uncoupled network, the weakly coupled 
one (\ref{DWCNN2}) and the linear system :

\beq
\frac{d\bX}{dt}=\bL\bX
\eeq

\nid are {\underline topologically conjugated}.
\enth
 
\bigskip

This means that the entire coupled neural network is essentially
a \textit{linear system} and is not more complex. If we admit that non linearity
plays a fundamental role in neural dynamics, the corollary leads to
the following conclusion. \textit{A WCNN with an hyperbolic fixed point is not
interesting as a ``brain'' model \footnote{More modestly, one may consider,
instead of the brain, small functional units such as cortical columns 
or simple nervous systems (worms). Fortunately, the same conclusion holds.}}. The corresponding dynamical
system being structurally
stable, we are led to the bewildering conclusion that a neural network model
where each neuron has an
equilibrium point and is in a rest state when $\epsilon=0$ needs
(at least) to be \textit{structurally unstable} in order to exhibit (real) non-linear effects 
and to be relevant for brain dynamics. 

At this point we may pose.  Indeed, it might be difficult to find out
a biologically realistic situation where the neural network
is in a state corresponding to a  \textit{fixed point}. There is at least an evident
one :  \textit{death} (this is indeed a structurally stable
situation). But this is not a very interesting example. This led us to several remarks.
First, as shown below, even if the uncoupled system consists of neurons in a rest state,
the coupled dynamics can be quite a bit more complex, even if the coupling are weak
(see section  \ref{KO}). As soon
as neurons are coupled, many different situations and dynamical regimes may occur.
We shall see several examples in this chapter, from periodic to chaotic regimes, with one
or several attractors etc... Consequently, coupling neurons with a rest state corresponding to a fixed point does not
necessarily mean that the coupled dynamics will be at rest. 

A more important issue concerns hyperbolicity. Though the theorem \ref{ThSSWCNN} deals
with hyperbolic fixed points, the notion of hyperbolicity extends
to quite more general attractors than fixed point, such as strange attractors (see appendix). 
 If we are interested in the ability of a Neural Network model to perform
tasks such as recognition of an external pattern, and if we agree that this recognition corresponds to some dramatic change
in the dynamics, then one can, in principle, extend the wisdom coming from theorem 
\ref{ThSSWCNN}: to be dynamically reactive to solicitations from the outside
the neural network needs to be close to a point in the parameter space
where the dynamics is structurally unstable with respect to perturbations corresponding e.g. to a
specific (learned) pattern. To be efficient and  adaptable the
system needs to be close to a critical point in order to display 
punctuated response to external world changes. We shall actually propose an example in section \ref{App}
exhibiting a behavior that can be related to this statement.
 It might well be that this conclusion extends more generally to biological systems and to living systems (see \cite{BakH},\cite{BCh}).
 
\ssu{Central manifold reduction.}\label{WCNNSU}

As we shall see right now, the situation is already quite a bit richer 
when some of the neurons
have a rest state (corresponding to a fixed point) which is \textit{not}
hyperbolic. Assume therefore that there is a subset $\cB \subset \cR$
of neurons such that the Jacobian matrix $L_i$ of the
uncoupled neuron $i$, $i \in \cB$ has eigenvalues with a zero
real part. We shall call these neurons \textit{critical} since they are close
to a bifurcation point. This means that the slightest change in the parameters,
induced either by the inputs of other neurons, or by an external stimulus,
\etc, may provoke a non linear dynamical response of the neuron such as spike, train of
spikes, etc... Moreover we order the neurons such that the $k$ first neurons are critical.

 Even when focusing on this situation, 
the analytical study of the changes occurring when the coupling $\epsilon$
is switched
on, and when the parameters $\rho,\lambda$ are modified is not tractable
in general. Indeed, in the most general situation, each matrix
$L_i$ have a number $n_i \leq m$ of neutral eigenvalues, some neurons
may be at the threshold for a Hopf bifurcation, some others close
to a saddle-node bifurcation, \etc. Moreover, if $m$, the number of control parameters, is sufficiently large (depending
on the accuracy of the underlying model) one may have bifurcations
of codimension larger than one or two. It is therefore
natural to start from the simplest situations, namely the case where all neurons undergo
the same (codimension one) bifurcation. Therefore, we focus now on the case where either
the matrix $L_i$ of each critical neuron has a simple zero eigenvalue,
or a pair of complex conjugate imaginary eigenvalues.
In these cases the techniques of central manifold projection and normal form reduction
(see appendix) allows us to reduce the dynamical system (\ref{DWCNN}) to a canonical form, 
close to the bifurcation point, and provided $\epsilon$ is sufficiently small.

The result presented below is rather abstract (though it is a direct application of the center manifold theory \cite{Carr} adapted to the present
context). But it has interesting consequences discussed in the next subsection. 
Basically, this results shows that the dynamics of the coupled system is \textit{locally governed by the critical neurons}.

%
%
%
%
%
%
%
%
%
%
%
\bth\label{TWC} (Izhikevich \cite{Iz})
Suppose that each of the first $k$ Jacobian matrices $L_i, \  i =1 \dots k$
is non hyperbolic. Then the 
 dynamics of (\ref{DWCNN}) is locally governed by a dynamical system
of the form :
\beq \label{DWC}
\frac{dx_i}{dt}=f_i(x_i;\lambda)+\epsilon g_i(x;\lambda,\rho,\epsilon), \quad i=1 \dots k
\eeq
\nid where\footnote{ The tangent space of each Jacobian matrix $L_i$ can be decomposed as:
$$\bbbr^m=E_i^s \oplus E_i^u \oplus E_i^c = E_i^h \oplus E_i^c$$
\nid where $E_i^s$, $E_i^u$, $E_i^c$ are
respectively the \textit{stable, unstable} and \textit{central} space.
} $x_i \in E_c^i$ and $J_i=D_{x_i}f_i(z,0)=L_{i|E_i^c}$. Moreover, if
$E^u =\lbrace 0 \rbrace$, there is a function $Z : E^c \times \cE_\lambda \times
\cE_\rho \times \bbbr \to \cM^n$ such that any local solution $\bX(t)$ of (\ref{DWCNN}) close
to $\bX=0$ tends exponentially to $Z(\bx(t),\lambda,\rho,\epsilon)$ where $\bx(t)$ is
a solution of (\ref{DWC}).
\enth

%
%
%
%
%
%
%
%

\ssu{General normal form} \label{GNF}

Once the center manifold reduction has been done and once one has
identified the type of instability occurring for the uncoupled neurons
one can further reduce the dynamical
equation (\ref{DWC}) to a canonical form or \textit{normal} form, allowing
somehow to classify the models into equivalence classes. For this, 
one needs however to provide some additional restrictions on the type of couplings
(function $G_i$),
and on the parameter dependence  (transversality conditions).

In the more general case, $\epsilon,\rho,\lambda$ are independent parameters
and the situation is quite complex. A way to simplify it is to
assume that $\lambda,\rho$ have the following form:

\bea \label{lambda}
\lambda \sim \lambda(\epsilon)&=& 0 + \epsilon \lambda_1 + \epsilon^2 \lambda_2 + 
O(\epsilon^2) \ \label{TaylLambda};
 \lambda_1,\lambda_2 \in \cE_\lambda\\
\rho \sim \rho(\epsilon)&=& \rho_0 + \epsilon \rho_1  + o(\epsilon^2) \ \label{Taylrho};
 \rho_0,\rho_1 \in \cE_\rho
\eea

\nid such that $\lambda(0)=0$ and $\rho(0)=\rho_0$.
 This form is convenient to handle
but it is not a loss of generality since $\lambda_1,\lambda_2$ (resp. $\rho_0,\rho_1$)
are still independent control parameters and can assume any value in $\cE_\lambda$ (resp.
$\cE_\rho$). As discussed above, $\lambda$ corresponds to control parameters allowing for example to tune
the neuron characteristics, while $\rho$ mimics ``external constraints''. In the form (\ref{Taylrho}) 
$\rho_0$ may correspond to some ``background'' influence $\rho_0$ while $\rho_1$ models 
an external input\footnote{A similar description is made in the model of section \ref{NotreModele}.
The microscopic parameters of the model (\ref{DNN}) corresponds to $\lambda$ while the 
external input $\bxi$ of section \ref{App} corresponds to $\rho_1$. Note however that the analysis performed
in section \ref{NotreModele} does not require the assumption of weak coupling and closeness to a rest state.}.

For $\epsilon=0$
the uncoupled neurons have a control parameter $\lambda=0$,  the critical
neurons are located at the bifurcation point, and the influence of the external
environment is modelled by the parameter $\rho_0$. When the coupling is switched
on, the physiological parameter $\lambda$ of individual neurons is
modified unless $\lambda_1=0$. Similarly, the external world influence
is manifested by an additional term $\epsilon\rho_1$ superimposed upon
the ``background'' influence $\rho_0$. A particularly interesting issue 
is to study the reactivity of the 
set of coupled neurons with respect to an external input, modeled
by the parameter $\rho_1$. If one thinks for example of recognition task
one expects a non trivial sensitivity to the input pattern $\rho_1$,
manifested by qualitative dynamical changes. 
E. Izhikevich analyzed this situation in great details
and made a classification of the normal form for the coupled system
according to the type of bifurcations the individual neurons are close to.\\

Let us give the main ideas and results.
Assume that we are in the situation of the theorem \ref{TWC}
and that the center manifold reduction (\ref{DWC}) has been performed.
Using the form (\ref{TaylLambda})(\ref{Taylrho}) for
$\lambda,\rho$ one rewrites the dynamical system (\ref{DWC}) in the form :
$$\frac{dx_i}{dt}=
f_i\left[x_i;\epsilon \lambda_1 + \epsilon^2 \lambda_2+O(\epsilon^2)\right]+
\epsilon g_i\left[\bx;\epsilon \lambda_1 + \epsilon^2 \lambda_2+O(\epsilon^2),
\rho_0 + \epsilon \rho_1  + o(\epsilon^2),\epsilon\right]=$$
$$=f_i(x_i;0)  +\epsilon \left[D_\lambda f_i(x_i,0).\lambda_1 + g_i(\bx;0,\rho_0,0) 
\right]+$$
$$+\epsilon^2 \left[D^2_\lambda f_i(x_i,0).(\lambda_1,\lambda_1) + D_\lambda f_i(x_i,0).\lambda_2
+ D_\lambda g_i(\bx;0,\rho_0,0).\lambda_1 + D_\rho g_i(\bx;0,\rho_0,0).\rho_1
+ \frac{\partial}{\partial \epsilon}g_i\right]+ o(\epsilon^3)$$

Close to the fixed point $\bx=0$ this reduces to:

\beq\label{NFG}
\frac{dx_i}{dt}=
\frac{\partial f_i}{\partial x_i}(0;0)x_i+ h_i x_i^2+
\epsilon a_i 
+\epsilon^2 d_i+
\epsilon \sum_{j=1}^n s_{ij} x_j +o(\epsilon^3)
\eeq

\nid where :

\bea
a_i&=&D_\lambda f_i(0;0).\lambda_1 + g_i(0;0,\rho_0,0) \label{ai}\\
h_i&=&\frac{1}{2} \frac{\partial^2 f_i}{\partial x_i^2}(0;0)\\
d_i&=&D^2_\lambda f_i(0;0).(\lambda_1,\lambda_1) + D_\lambda f_i(0;0).\lambda_2
+ D_\lambda g_i(0;0,\rho_0,0).\lambda_1 + D_\rho g_i(0;0,\rho_0,0).\rho_1
+ \frac{\partial}{\partial \epsilon} g_i\\
s_{ij}&=&\frac{\partial g_i}{\partial x_j}(0;0,\rho_0,0)
\eea

It is particularly remarkable that the coefficient $s_{ij}$ acts as a ``formal synapse'' coupling
the neuron $j$ to the neuron $i$. But, contrarily to the usual synapses, which establish a ``wired''
link between two neurons, the coefficient $s_{ij}$ is essentially generated by the \textit{(nonlinear) dynamics}.
Note that it is given by the Jacobian matrix of $g$. This is thus a first illustration
of the concept introduced in the section \ref{MNN}.
It corresponds to an \textit{effective link that is not necessarily supported by a wired link}. This fundamental aspect
is discussed below and in more details in the section \ref{RepLin} (see in particular the equation (\ref{chiij})).\\

Up to now we have written general equations without consideration about the (codimension one)
bifurcations of the critical neurons. To each type of bifurcations is associated
a set of transversality conditions (see appendix) that allows  to reduce further
the equations (\ref{NFG}). 

The result (\ref{NFG}) is rather abstract but it has interesting development in particular in
 the following case.
Assume that we slightly perturb the dynamical system with an external input corresponding to
the parameter $\rho_1$. What is the effect of this perturbation on the dynamics ? In particular, assuming
that some neurons are close to a bifurcation point, does a perturbation of the form (\ref{lambda})
have an effect on the global dynamics ? Such a change can be considered as an effective reaction of the system
to the external input, that can be used, for example, to perform recognition tasks. Consequently,
it can be useful to have analytical results on the normal form of the coupled dynamics near
the bifurcation point.
We shall not discuss all the cases investigated by Hoppensteadt
and Izhikevich.
 We shall focus instead on two examples that we find particularly rich and enlightening.

\ssu{Saddle-node and pitchfork bifurcations.} \label{WCNNP}

Let us first  analyze the case when the $k$ critical neurons are close
to a \textit{saddle-node bifurcation}. Then, for each critical
neuron $i=1 \dots k$ the uncoupled
vector field $f_i(x_i,\lambda)$ satisfies the transversality conditions
(see appendix) $\frac{\partial f_i(0;0)}{\partial x_i}=0; \ \frac{\partial^2}{\partial
x_i^2}f_i(0;0) \neq 0$ and $D_\lambda f_i(0;0) \neq 0$ (which means
that $D_\lambda f_i(0;0)$ has no zero eigenvalue). Therefore, it follows from (\ref{NFG}) that the \textit{normal form}
of a WCNN where $k$ neurons undergo a saddle-node bifurcation is:

\beq \label{NFS}
\frac{dx_i}{dt}= h_i x_i^2+
\epsilon a_i 
+\epsilon^2 d_i+
\epsilon \sum_{j=1}^n s_{ij} x_j +o(\epsilon^3)
\eeq

The variable change $x_i \to \epsilon^\frac{1}{2} x_i$ and the time change
$t \to \epsilon^{-\frac{1}{2}} t$ transforms these equations into:

\beq\label{NFSa}
\frac{dx_i}{dt}= a_i + h_i x_i^2+
\sqrt{\epsilon} \sum_{j=1}^n s_{ij} x_j +\epsilon d_i+
o(\epsilon^3)
\eeq

 We now want to analyze the effect of 
an external input $\rho_1$ on (\ref{DWCNN2}). From eq. (\ref{ai}) $a_i$ depends only on $\rho_0,\lambda$
while $h_i$ is independent on these parameters.  Moreover, if $a_i \neq 0$,
the dynamical system (\ref{NFSa}) admits, to the zeroth order in $\epsilon$,
the fixed points $\pm \sqrt{\frac{a_i}{h_i}}$, which are \textit{both hyperbolic}.
Since hyperbolicity is structurally stable, when $a_i \neq 0$ the dynamical system (\ref{NFSa}) \textit{is insensitive to the input pattern $\rho_1$}
(and, at least in this sense, cannot perform recognition
task).  It can react only
to $\rho_0$ and this reaction is trivial (by the implicit function theorem).

Consider now the case when $a_i=0$. Then the variable change $x_i = \epsilon h_i^{-1}y_i$
and the time change $\tau=\epsilon t$ in (\ref{NFS}) lead to:

\beq\label{NFSb}
\frac{dy_i}{d\tau}=r_i + x_i^2 + \sum_{j=1}^n c_{ij}x_j + O(\epsilon)
\eeq

\nid where:

\bea
r_i &=& h_i d_i\\
c_{ij}&=&h_i s_{ij}h_j^{-1}
\eea

The dynamical system (\ref{NFSb}) is the normal form of (\ref{NFSa}) under
the condition $a_i=0$. It depends now on the external input $\rho_1$ via
$d_i$. The fixed points of ((\ref{NFSb}) are now determined
by :

\beq
r_i + x_i^2 + \sum_{j=1}^n c_{ij}x_j=0
\eeq 
 
\nid Consequently, they depend on the structure of the matrix
$\cC \deq \lbrace c_{ij}\rbrace_{i,j=1}^n$ (corresponding to the effective links induced by the 
nonlinear dynamics).
The condition $a_i=0$ writes :

\beq\label{Adapt}
D_\lambda f_i(0;0).\lambda_1 + g_i(0;0,\rho_0,0)=0; \ \forall i= 1 \dots n.
\eeq

This specific condition is called by E. Izhikevich, the \textit{adaptation condition}.
He showed in particular this interesting result: in order for the reduced dynamical
system (\ref{DWCNN2}) to exhibit a non trivial sensitivity to the input pattern $\rho_1$,
one needs that the neurons \textit{adapt to the pattern}, via \textit{the internal parameter $\lambda$}.
In other words, $\lambda$ is not independent on $\rho$ but it must satisfy
(\ref{Adapt}) which means that the internal parameter $\lambda$ counterbalances (up to order $\epsilon$)
the steady state input from the entire network onto each neurons. This notion has deep implications,
because it suggests that a suitable training of a Neural Network such as (\ref{DWCNN2}) requires
\textit{an evolution of the control parameters $\lambda$, under the influence of the external input $\rho_1$,
with the constraint that the condition (\ref{Adapt}) is achieved}. In some sense, learning is efficient
if the training leads the system into a very specific part of the parameters space, corresponding to the condition
(\ref{Adapt}), where the system is close to a bifurcation point and where this bifurcation is only induced by the presentation of the
learned pattern (or a weakly perturbed version of it). This has interesting echoes
with the discussion of the effect of hebbian learning in the model described in the section \ref{App}.  

The adaptability  condition is also important if one considers other types
of bifurcations such has the pitchfork (in this case the transversality
conditions impose that (\ref{Adapt}) is automatically satisfied) and the cusp bifurcation.

In the case of pitchfork bifurcations, for example then the normal form is :

\beq\label{NFCP}
\frac{dx_i}{dt}=b x_i \pm x_i^3 + \sum_{j=1}^n c_{ij}x_j 
\eeq

We shall return back to this form in the section \ref{ExHebb} where we shall consider the effect
of having a synaptic matrix  constructed via Hebbian learning.
 
\ssu{Hopf bifurcations.} \label{WCNNH} 

Assume now that the uncoupled system has $k$ neurons close to a Hopf bifurcation.
This means that the Jacobian matrix of each corresponding critical has a pair of purely
imaginary eigenvalues $\pm \Omega_i$ at the critical value of the parameters. Call $v_i,\bar{v}_i$ (resp.
$w_i,\bar{w}_i$) the corresponding right (resp. left) eigenvectors.
Moreover, the vector
field satisfies the transversality conditions TH1 in the appendix.  Using
similar techniques as in the previous example one can prove the following \cite{HoIz}:
%
%
%
%
%
%
\bth \label{TNFBHopf}
If the dynamical system (\ref{DWC}) is near a multiple Hopf bifurcation then
there is a variable change and a time rescaling $\tau = \epsilon t$ reducing it to the normal form:
\beq
\frac{d z_i}{d\tau} = b_i z_i + d_i z_i |z_i|^2 + \sum_{j \neq i}^k C_{ij} z_j + O(\sqrt{\epsilon})
\eeq
\nid where $b_i,d_i$ are complex coefficients and where the coefficients
$C_{ij}$ are given by :
\beq\label{CijHopfWCNN}
C_{ij}= \left\{
\baR{ccc}
w_i . D_{x_j}g_i . v_j &\qquad& \mbox{if $\Omega_i = \Omega_j$}\\
0  &\qquad& \mbox{if $\Omega_i \neq \Omega_j$}
\eaR
\right.
\eeq
\enth
This results is quite interesting. It shows that close to the bifurcation point the oscillating
neurons can be divided into pools according to their natural frequency. There is an effective
coupling $C_{ij}$ between the neurons in the same pool, while the coupling between 
neuron from different pools is negligible (namely weaker than $O(\sqrt{\epsilon}$)) \textit{even if there is a synaptic connexion
between them in the global system. } This strongly suggests that \textit{resonances} appear in such system.
Indeed, one can exhibit Arnold like tongues in this situation (see \cite{HoIz} page 173). Also, 
a periodic input with a frequency $\omega$ one can establish interactions between oscillators transmitting
on different frequencies or it can disrupt interactions between oscillators transmitting on the same frequency.

This result as strong implications going far beyond the field of Neural Networks. Indeed, it
suggests that non linearity may induce effective paths among units that are not directly
connected to the graph of interactions. Moreover this type of behavior is not
specific to weakly coupled neural networks with a rest state close to a Hopf bifurcation.
In the section \ref{RepLin} we shall exhibit a similar behavior for a \textit{chaotic Neural Network}
and we shall show that a linear response theory based on recent results by D. Ruelle \cite{Ruelle2}     
might be used to locate these resonances.

\ssu{An example of ``Hebbian'' learning.}\label{ExHebb}

 In this section we give a first example of Hebbian learning  rule allowing
the neural network to perform
tasks such as pattern recognition. We start
from the ``generic'' form of Hebb's rule (\ref{HebbRuleGen}) with $\beta=\gamma=0$ and
where the ``state'' of the neuron $i$ ($m_i$ in eq. (\ref{HebbRuleGen}))
is  given by $x_i$ and where $\Gamma_{ij}=x_i x_j$.  

\beq\label{HebbEvol}
J_{ij}' = \lambda J_{ij} + \alpha x_i x_j + higher \ order \ terms.
\eeq

Suppose now that we have several ``patterns'' or ``images'' $\bxi^1, \dots, \bxi^p$ to be ``memorized'' by our neural
network. What means ``memorized'' ?
Many definitions are possible but, in the context of this chapter, we shall consider that a pattern is memorized
if  the neural networks has \textit{acquired}, via learning, the capacity to dynamically evolve towards a ``state'' ``associated
to the pattern'', provided that it was ``suitably prepared''.
We insist on the fact that this property must be \textit{acquired}, namely, it should not exist when no learning is performed.
This definition is however still rather ambiguous. What is a ``state'' ? What means ``can be associated'' to the pattern,
``suitably prepared'' ?
Again, there are many possible interpretations. But let us start with a very simple one. Assume that the ``patterns''
are vectors corresponding to points in the phase space of our neural network.
Assume then that, after learning, the dynamic evolution admits all patterns as \textit{stable fixed points}.
Then, starting from an initial condition in the attraction basin of the pattern $k$, say, the dynamics will
converge to this pattern. Here ``suitably prepared'' means that we start from an initial condition in the attraction basin
and the ``state associated to the pattern'' is the pattern itself and this is fixed point of the dynamics. 
Since an initial condition in the attraction
basin of the pattern is a (possibly small) perturbation of it, one may interpret the convergence
 as a \textit{recognition} of the pattern by the neural network when a perturbed version of it is presented.

How to manage such a ``learning ability'' ? Consider the equation (\ref{HebbEvol}) and assume first that  
the coupled neural network admits only a stable fixed point corresponding to the pattern $\bxi^1$. Then the Hebbian
rule implies that  $J_{ij} \to -\frac{\alpha}{\lambda} \xi_i^1 \xi_j^1$. A possible generalisation to $p$ pattern
is then:

\beq \label{HebbHopfield}
J_{ij}=\frac{1}{N}\sum_{k=1}^p \beta_k \xi_i^k \xi_j^k
\eeq 

We shall actually see, in the section  \ref{SCG}, the effect of this rule in a recurrent neural network.
We shall also show another application of Hebbian learning in a situation where the recognition of a pattern
does not correspond to the convergence to a fixed point, but to a more complicated object in the phase space
(section  \ref{App}). For the moment and to stay in the spirit of the section \ref{WCNN}
 we will address the following question.
What is the effect of the hebbian synapses (\ref{HebbHopfield}) in a neural network where some neurons
are close to a bifurcation ?\\

Let us for example consider the case of a pitchfork bifurcation and assume therefore that  the dynamics
is given by:

\beq\label{ExHebbP}
\frac{dx_i}{dt}=b x_i - x_i^3 + \sum_{j=1}^n J_{ij}x_j 
\eeq

\nid corresponding to having a subset of neurons close to a pitchfork bifurcation 
(see section \ref{WCNNP}). It is easy to see that $\bxs=0$ is always a fixed point and that 
its Jacobian matrix is given by $D\bF_\bxs= b I + \cJ$, where $I$ is the identity matrix and
 $\cJ$ the matrix
of synaptic couplings. Assume now that $\cJ$ is given by eq. (\ref{HebbHopfield}). 
Note that, according to this equation, $\cJ$ can be written  in the form
$\cJ=\frac{1}{N} \sum_{k=1}^p\bxi^k \tilde\bxi^k$ where $\tilde \bxi$ denotes the transpose of $\bxi$.
Assume now that the patterns $\bxi^k$ are \textit{orthogonal} and take binary values $\xi_i^k = \pm 1$ 
(thus $(\bxi^k,\bxi^l)=N\delta_{k,l}$). The mutual orthogonality of the patterns imposes that the number $p$ of patterns is lower than the dimension
of the phase space ($p \leq N$). Then it is easy to see that $\cJ \bxi^k = \beta_k \bxi^k$. Thus 
$\beta_1, \dots, \beta_p$ are eigenvalues of $\cJ$ with corresponding right 
eigenvectors $\bxi^1, \dots, \bxi^p$. The remaining $N-p$ eigenvalues are all zero and the corresponding
eigenvectors belong to the orthogonal of $span\left\{\bxi^1, \dots,\bxi^p \right\}$ (the kernel of $\cJ$).
If we finally assume that $\beta_1 \geq \beta_2 \geq   \dots \beta_p$ then we see that $DF_0$ has 
$p$ eigenvalues $b+\beta_1 \geq b+\beta_2  \dots \geq b+\beta_p$.
 Consequently, $0$ looses its stability when
$b \geq - \beta_1$. This destabilization arises via a pitchfork bifurcation occurring in the direction $\bxi^1$. Indeed, set $x_i=y_1\xi^1_i$ then $\frac{dy_1}{dt}=(b+\beta_1) y_1 - y_1^3$. Hence,
 two symmetric fixed points $\bx_\pm^1= \pm\sqrt{b + \beta_1}\bxi^1$ appear, \textit{proportional to the pattern}
 $\bxi^1$. If we further increase $b$ we have a sequence of similar pitchfork bifurcations, in the
direction $\bxi^k$, whenever
$b \geq - \beta_k$. Finally, a stability analysis of these new fixed points show that if
$b > - \beta_m + \frac{\beta_1 - \beta_m}{2}$, there are $m$ pairs of \textit{stable nodes corresponding to
the patterns} $\bxi^1, \dots, \bxi^m$. In this sense, the Hebbian rule (\ref{HebbHopfield}) gives
to a neural network of type (\ref{ExHebbP}) the capability to retrieve memorize patterns corresponding
to stable fixed points. We shall return on a more general version of eq. (\ref{ExHebbP}) in the section
\ref{SCG}. Note finally that if one increases $b$ beyond a positive value then ``spurious'' memories appear, that is, stable fixed points that do not correspond to any of the memorized patterns. These new patterns
belong to $Ker \cJ$ (see \cite{HoIz} for details).

\su{Recurrent models.}  \label{Rec}

In the previous section we have considered the effects of weakly coupling neurons, in a situation
where uncoupled neurons are close to a bifurcation point. We have presented some rigorous results
obtained from bifurcations and normal form theory. They reveal some illuminating aspects of the emergent
dynamics, such has the adaptation principle or the existence of an effective network induced by the dynamics
and not necessarily identical to the synaptic network.

We depart now from this setting. We want to analyze the collective dynamics of a neural network where the couplings
are not necessarily weak and where neurons are not necessarily close to a bifurcation point. For this, we consider
a recurrent neural network whose dynamics is given by the equations (\ref{ECG}) (continuous time)
or (\ref{DNN}) (discrete time). One can indeed go quite a bit deep in the dynamics description.
Moreover, the model presents an overwhelming richness and it can be partially analytically studied. 
This is also a good benchmark for developing tools in non linear networks analysis (see section \ref{RepLin}). 

We first show how this recurrent model can be derived by switching from a spiking description of the neuron
to a frequency rate description (section \ref{SpikeFire}). We then discuss  the dynamics of the model
when the synaptic weights are symmetric (section \ref{SCG}). General results from dynamical systems theory allow one
to prove a convergence property and to exhibit a Lyapunov function (see appendix for a definition).
This function has some analogies with the magnetic energy in a system of interacting spins. Actually, the
``energy'' landscape presents a structure similar to the rich and complex energy landscape of spin glasses
models. There exist a large number of minima (fixed points), whose number increases exponentially with
the system size. The existence of these many minima is closely related to the competition excitation/inhibition
induced by synapses, and corresponding to \textit{frustration} in spin glasses. Actually, the techniques developed
in statistical mechanics of spin glasses can be adapted to estimate the number of minima \cite{MW} and
to have an good description of energy landscape. The convergence to minima can then be used to store informations
if one uses the Hebbian rule  (\ref{HebbHopfield})  \cite{HopfieldTank}. 

In section  \ref{Coop} we present briefly cooperative systems, where one still have a convergent
dynamical system even if the synapses are not symmetric. We discuss in particular shortly a fundamental result
by Hirsch having recent extensions in the field of genetic networks \cite{Gouze}, \cite{Soule}. But usually, when symmetry is broken,
the dynamics is not convergent and can be, for example, chaotic. The section \ref{NotreModele} is entirely devoted to
a chaotic model, which has, furthermore, nice properties when submitted to an Hebbian like learning.

\ssu{From spiking neurons to firing rate neurons.}\label{SpikeFire}

In section \ref{WCNN} we have discussed the effect of coupling neurons without giving
a detailed description of the individual neuron dynamics. For further development we have now to specify 
it. For this purpose and in the spirit of table 1 we switch now from
neurons having an activity described in terms of spikes to neurons described  in terms of
firing  rate. \\

Assume therefore that each neuron is emitting spike trains and call $x_j(t)$ the firing rate of 
the neuron $j$ at time $t$. Note that the definition of firing rate requires an integration over a certain
time window, (see the introduction), that we shall assume to be short compared to the time scale
for the evolution of the variables considered in this description. Call $\tu_i(t)$
be the time average of the membrane potential on this time window. In standard models the
firing rate is a function of  $\tu_i$: $x_i(t)=f_i(\tu_i(t))$
where $f_i$ is a sigmoidal function such as\footnote{One also finds in the literature
the case where $f_i(u)=\tanh(g_iu)$. Hence $f$ takes its values in $[-1,1]$.}
 $f_i(x)=Erf(g_i u)$ or $\frac{1+\tanh(g_i u)}{2}$. $g_i$ is called the 'gain' of the transfer
function $f_i$. Since the slope at the inflexion point of $f_i$ is proportional to $g_i$ and since $f_i(-\infty)=0$ and $f_i(+\infty)=1$
  this parameter measures the level of non linearity of the function.
The sigmoidal shape of $f_i$ can be understood by the following argument.

A spike is emitted each times the average membrane potential exceeds the neuron threshold $\theta_i$.
This threshold, as we saw, depends on time. In particular, after emission, it increases to infinity during
a time $\tau_a$, assumed here to be identical for all neurons, corresponding to the absolute refractory period.
Then it decreases to reach its initial value. Hence, if $\theta(\tau)$ is the threshold value at time $\tau$,
the initial time being the time where a spike is emitted, one has:

\beq
\theta(\tau) = 
\left\{
\baR{ccc}
\infty \qquad \mbox{if} \quad &0 < \tau < \tau_a&\\
\mbox{decreasing function if} \quad &\tau > \tau_a&
\eaR
\right.
\eeq  

At each time $\tau_i$ such that $\theta_i(\tau_i)=\tu_i$ there is an emission of a spike, therefore
the corresponding average time of firing (the initial time being the time where a spike is emitted)
 is given  by $\tau_i=\theta_i^{-1}(\tu_i)$ and the (normalized) frequency rate is :

$$x_i=\frac{\tau_a}{\tau}=\frac{\tau_a}{\theta_i^{-1}(\tu_i)} \leq 1$$

Consequently, $x_i$ is an monotonously increasing function of $\tu_i$ with values
 in $[0,1]$. In the case of  integrate and fire neurons driven by an external stochastic
current $I(t)$ one has has an explicit
equation for $f_i$. The membrane potential
evolution is given by:

$$\tau_m \frac{du}{dt}=-\gamma u(t)+RI(t)$$

\nid (eq. \ref{EIF}, section \ref{SIF}).
Assume that $I(t)$ is random (e.g. Poisson process) with
a (stationary) probability distribution $\cP$. Then the probability
that the neuron fires at a time $t+dt$ is given
by 

$$\cP\left[u(t+dt)  \geq \theta \right]=
\cP\left[(1-\frac{\gamma}{\tau_m}dt)u(t)+ \frac{RI(t)dt}{\tau_m} \geq \theta\right]=$$

$$=\cP\left[I(t)dt \geq \frac{\tau_m}{R}(\theta-u(t)) + u(t) \frac{\gamma}{R}dt\right]
\sim 1-\cF\left[\frac{\tau_m}{R}(\theta-u(t)) \right]=f(u)$$

\nid where $\cF$ is the repartition function of $I$.\\

Consider now an assembly of such neurons. The neuron $i$ receives
the spikes coming from other neurons, and the total current $I_i(t)$ is the sum of  spikes arriving from each neuron $j$,
weighted by $J_{ij}$. The membrane potential $\tu_i$ of the neuron $i$
 depends on  the frequency rates of the spikes trains emitted by the neurons connected to $i$.
The current received by $i$ is therefore
$\sum_{j=1}^{N} J_{ij} x_j(t)$ where $x_j$ can be viewed as an integration over a small time window of
the current appearing in eq. (\ref{CIF}). 
Then the analog to the evolution equation  (\ref{EIF}), section \ref{SIF}, 
is :

\beq
\frac{d\tu_i}{dt}=-\tu_i(t) + \sum_{j=1}^{N} J_{ij} x_j(t)
\eeq 

\nid 
where we have dropped the time constant in front of $-\tu_i$, incorporating it into the time scale $dt$.

  Fixing some threshold reference (e.g. $\theta_i^0=\theta(2\tau_a)$, setting
$u_i(t)=\tu_i(t)-\theta_i^0$ and $x_i=f_i(u_i)$ one finally obtains:

\beq\label{ECG}
\frac{du_i}{dt}=-u_i(t)+\sum_{j=1}^N J_{ij}f_j(u_j(t))-\theta_i; \quad i=1 \dots N.
\eeq

 The model (\ref{ECG})
displays a wide variety of dynamical behavior according to the form of $\cJ$ the matrix of synaptic couplings.

The equation (\ref{ECG}) is a particular form of the Cohen-Grossberg model \cite{CG}.
The general form is:

\beq\label{GECG}
\frac{du_i}{dt}=a_i(\bu)\left[b_i(u_i(t))+\sum_{j=1}^N J_{ij}f_j(u_j(t)) \right]; \quad i=1 \dots N.
\eeq 

\nid where $a_i,b_i$ are mild functions (e.g. $a_i$ is bounded, positive and locally Lipschitz continuous and
$b_i,b_i^{-1}$ are locally Lipschitz continuous). In the sequel we shall however restrict to the model (\ref{ECG}).

\ssu{Symmetric synapses.}  \label{SCG}

Consider first the case with symmetric synapses $J_{ij}=J_{ji}$.
One shows then that (\ref{ECG}) is \textit{convergent} whenever $\cJ$ is symmetric.

Here is an elegant proof due to M. Benaim\cite{Benaim}. It has the  advantage to hold
in more general case than for symmetric synapses. M. Benaim proved indeed the following theorem:

\bth\cite{Benaim} Consider the differential system:
\beq\label{EBen}
\frac{du_i}{dt}=b_i(\bu)G_i(\bu)=F_i(\bu), \qquad i=1 \dots N
\eeq
\nid where $b_i : \bbbr^N \to \bbbr^{\ast +}$ are strictly positive $\cC^1$ functions and assume that there exist
a family of strictly positive $\cC^1$ functions $\psi_i : \bbbr \to \bbbr^{\ast +}$ such that the following
 holds (\textit{``detailed balance''} condition)\footnote{The name ``detailed balance'' comes from the evident analogy with
the equilibrium conditions for the stochastic evolutions, such as Glauber dynamics, used in statistical physics.}:
\beq\label{DB}
\frac{\frac{\partial G_j}{\partial u_i}}{\frac{\partial G_i}{\partial u_j}}=\frac{\psi_j(u_j)}{\psi_i(u_i)}
\eeq
\nid  Then :

\ben

\item (\ref{EBen}) admits a strict Lyapunov function.

\item The isolated equilibria of (\ref{EBen}) are generically hyperbolic.
\een

\enth

The proof relies on the remark that the differential form $\omega= \sum_{i=1}^N \psi_i(u_i) G_i(\bu)  du_i$
is exact ($d\omega=0$) since:

$$d\omega=\sum_{i < j} \left(\psi_i(u_i)\frac{\partial G_i}{\partial u_j} - \psi_j(u_j)\frac{\partial G_j}{\partial u_i}\right) du_i  \wedge du_j=0$$

It follows that there exists a function $V$ such that $\omega=-dV$. Consider now
the Riemmanian metric  defined by:

\beq \label{scal}
\left<\bx,\by \right>_\bu = \sum_{i=1}^N \frac{\psi_i(u_i)}{b_i(\bu)}x_i y_i
\eeq

Then $dV=-<\bF,d\bu>_\bu$ and consequently $\frac{dV}{dt} = -<\bF,\bF>_\bu$. Hence $V$  is strictly decreasing along
the trajectories of the dynamical system (\ref{EBen}). Thus (\ref{EBen}) is a gradient field for the previous metric
and  $V$ is  a strict Lyapunov function. Consequently, from the Lasalle
invariance principle (see appendix) the dynamical system is convergent and the fixed point are generically isolated.
 The Jacobian matrix being self adjoint for this scalar
product its eigenvalues are real and are generically non zero.\\

In the case of the dynamical system (\ref{ECG}) one can write:

\beq\label{PCCG}
\frac{du_i}{dt}=\frac{1}{f'_i(u_i)}\left[
-u_i(t)f'_i(u_i)+\sum_{j=1}^N J_{ij}f_j(u_j)f'_i(u_i)-\theta_if'_i(u_i)\right]
\deq \frac{G_i(\bu)}{f'_i(u_i)} 
\eeq

\nid  (since $f'_i$ vanishes only at infinity, one may assume that the initial conditions are
chosen in some compact set of $\bbbr^N$). Then $b_i(u_i)=\frac{1}{f'_i(u_i)}$ and :

$$\frac{\partial G_i}{\partial u_j}=-\delta_{ij}\left((u_i+\theta_i)f''(u_i) +f'(u_i)\right)
+J_{ij}f'_j(u_j)f'_i(u_i)$$

The detailed balance condition holds if $J_{ij}=J_{ji}$ and it writes 
$\frac{\partial G_i}{\partial u_j}=\frac{\partial G_j}{\partial u_i}$ (hence one may take
$\psi_i=1$ in the scalar product (\ref{scal})).
Thus, (\ref{ECG}) is a gradient system with a Lyapunov function $V$ given by $dV = 
-\sum_{i=1}^N G_i(\bu) du_i = \sum_{i=1}^N (u_i(t)f'_i(u_i) -\sum_{j=1}^N J_{ij}f_j(u_j)f'_i(u_i)+\theta_i)du_i$.
This gives, up to an irrelevant additive constant:

\beq\label{LyapCG}
V(\bu)= \sum_i \int_{0}^{u_i} (u+\theta_i) f'_i(u)du - \sum_{j=1}^N J_{ij}f_j(u_j)f_i(u_i)
\eeq

\bigskip

This result has several important consequences. 
First it shows that (\ref{ECG}) is convergent \cite{CG}. But it gives substantially more. The equilibria are the minima
of the function $V$. Actually, this function looks very much like the \textit{magnetic energy} in physical systems. Assume indeed, 
that the slope of the sigmoid tends to infinity. Then $f$ becomes a binary function. If $f(u)=tanh(gu)$ then $x_i=f(u_i)$ takes
value in $\left\{-1,1 \right\}$  as the binary
spins of the Ising model. The Lyapunov function (\ref{LyapCG}) writes, in this limit,
$-\sum_{j=1}^N J_{ij}x_ix_j$ (since $f'(u) \to 0$ everywhere but at $u=0$ where it becomes infinite).
The Lyapunov function as therefore exactly the form of the 
energy resulting from the magnetic interaction between Ising ``spins''.   

The  structure of the energy landscape of  magnetic system with ferromagnetics ($J_{ij}>0$) and antiferromagnetics
($J_{ij}<0$) magnetic interactions is astonishingly complex. Actually, when the $J_{ij}$'s are randomly distributed
(modeling the presence of impurities in a magnetic sample) one obtains a model for a \textit{spin glass} (for
a review see for example \cite{BY},\cite{Mezard},\cite{Sherrington}). Spin glasses have
extremely rich properties and the canonical models (such as the Edwards-Anderson model \cite{EA} or the Sherrington-Kirckpatrick model
\cite{SK}) are not yet entirely understood.
 This analogy with magnetic spin glasses has been very fruitful and in particular Hopfield made
a breakthrough in the field of formal Neural Network by developing the analogy between a Neural Network
with binary neurons and a spin glass. He showed that information can be stored in the minima of
$V$ and he proposed a method, inspired from the Hebb's rule to construct the interactions
in a way such that the patterns to be learned correspond to the minima of $V$. 

Indeed, assume that the $J_{ij}$'s are now given by equation (\ref{HebbHopfield}), where the $\bxi^k$'s are some patterns 
that we want to store and retrieve from the dynamics (\ref{ECG}). For simplicity we take $\beta_k=1$ in 
(\ref{HebbHopfield}). We assume that the number of patterns , $p$,
is $\leq N$. Since, in the limit $g \to \infty$ the phase space becomes $\left\{-1,1 \right\}^N$,
the $\xi^k_i$ takes binary values $\left\{-1,1 \right\}$. Assume moreover that the vectors $\bxi^k$ are mutually orthogonal,
 i.e. $(\bxi^k,\bxi^l)=N\delta_{(k,l)}$. The Lyapunov function writes then:

\beq
V(\bu)= -\frac{1}{N}\sum_{i,j=1}^N \sum_{k=1}^p \xi_i^k \xi_j^k x_i x_j
\eeq

\nid Note that in the case $g \to \infty$ $x_i =sgn(u_i)$. 
If $\bx = \bxi^1$ then $V=-\frac{1}{N} \sum_{i,j=1}^N \left(\xi_i^1\right)^2\left(\xi_j^1\right)^2 - 
\frac{1}{N}\sum_{i,j=1}^N \sum_{k=2}^p \xi_i^1\xi_i^k \xi_j^1\xi_j^k=
- \frac{1}{N} \|\bxi^1\|^4 -\frac{1}{N} \sum_{k=2}^p \left(\bxi^1,\bxi^k \right)^2$. 
Since the $\bxi^k$'s are orthogonal one gets finally
$V=-N$ which is the absolute minimum of the magnetic energy. This results holds obviously for all patterns.
Consequently, all patterns are absolute minima of $V$ and they are stable\footnote{Actually, we have still to define the ``limit''
of the dynamical equations (\ref{ECG}) when $g \to \infty$. The Hopfield model with binary states uses a discrete time sequential
dynamics (see \cite{Hopfield1} for details)}. When the number of patterns is larger than $N$, the patterns
can not be all orthogonal. Therefore, the second term $\frac{1}{N} \sum_{k=1}^p \left(\xi^1,\xi^k \right)^2$ plays
an important role, since it generates spurious minima of $V$. The exact analysis of the Hopfield model for a finite
and infinite numbers of patterns have been performed by Amit, Gutfreund and Sompolinsky \cite{AGS}, in the thermodynamic
limit $N \to \infty$. For this, they uses spin glasses techniques such as the replica methods (in the case $p \to \infty$).
Their results have been rigorously proved in \cite{Gayrard}.\\

When $g$ is finite the neural network (\ref{ECG}) has still the capacity to store and retrieve patterns
via Hebb's rule \cite{HopfieldTank}. Moreover, for random symmetric $J_{ij}$'s 
the minima of $V$ can still be computed by using techniques coming
also from the spin glasses literature. For example, using a method  developed by Bray and Moore \cite{BM}
for spin glasses, Marcus \& Westrevelt have been able to compute the number of fixed points
for (\ref{ECG}), when the size $N \to \infty$ (thermodynamic limit), in the case where 
$\cJ$ is a (symmetric) random matrices with independent entries and such that 
$E[J_{ij}=0]$ and $Var[J_{ij}=\frac{1}{N}]$. They found equations for the fixed points which are similar to
the Thouless-Anderson-Palmer equations \cite{TAP} giving the mean-field solutions in the Sherrington-Kirckpatrick
model, and the Bray \& Moore techniques gave a similar result: 
the number of fixed points increases exponentially
with the system size.

\ssu{Cooperative systems.}\label{Coop}

When the synaptic couplings $J_{ij}$ are not symmetric, (as  in biological systems)   there is in general no Lyapunov function,
and many kinds of dynamics are possible.
 However, for some systems called \textit{cooperative systems} one has still convergence properties, 
without Lyapunov function, but relying on a specific property of the flow, that preserves some pseudo
order on the phase space. The results presented here are due to Hirsch
\cite{Hirsch}.

A dynamical system 

\beq
\frac{du_i}{dt}=F_i(\bu)
\eeq

\nid is called \textit{cooperative} if:

\beq
\frac{\partial F_i}{\partial u_j}(\bu) \geq 0, \qquad \forall i \neq j
\eeq

\nid and it is called \textit{competitive} if:

\beq
\frac{\partial F_i}{\partial u_j}(\bu) \leq 0, \qquad \forall i \neq j
\eeq

This has the following interpretation. As discussed in the section \ref{MNN} 
the Jacobian matrix element $\frac{\partial F_i}{\partial u_j}(\bu)$
measures in some sense the ``influence'' of the neuron $j$ on the neuron $i$,
when the system is in the state $\bu$. More precisely, it characterizes,  to the first order in a Taylor expansion, the modification induced
on $u_i$ when $u_j$ has a small variation. In cooperative systems the corresponding
interaction graph has therefore only positive edges, whatever the state of the neural network and
consequently, only positive circuits. They have moreover the following property. 
Assume that the phase space is convex and define the partial order $\bu \leq \bv \Leftrightarrow u_i \leq v_i, \ i =1 \dots N$.
A cooperative flow preserves this order. Thus $\bu(0) \leq \bv(0) \Rightarrow \bu(t) \leq \bv(t), \ \forall t >0$ (this corresponds
to the positive feedback discussed above). Note that if $\bF$
is competitive a reversal of the time arrow leads to $\bu(0) \nleq \bv(0) \Rightarrow \bu(t) \nleq \bv(t) \ \forall t >0$. From these
inequalities, Hirsch \cite{Hirsch} was able to prove that for a two dimensional cooperative dynamical system, any bounded
trajectory converges to a fixed point.
In larger dimension, one needs moreover a technical condition on the Jacobian matrix of $\bF$ : it must be irreducible. 
Then Hirsch proved that the $\omega$ limit set of almost every bounded trajectory is made of fixed points. 
One does not have the same property for  competitive systems \cite{Smale}. 

Some extension of these results have been recently made \cite{Gouze,Soule}. Though these works where intended
to obtain mathematical results about metastability in the context of genetic networks, they hold in a very
general context, and, in particular, in the context of neural networks. 
In 1981, R. Thomas made the following conjectures
\cite{Thomas}.\textit{ 1) A positive feedback loop  in the graph of interactions of a differential dynamical system is a necessary
condition for the existence of several equilibria. 2) A negative loop is a necessary condition for a stable periodic 
behavior.} J.L. Gouz\'e proved these conjectures under the hypothesis that the sign of the Jacobian matrix elements
do not depend on the state. Consequently, the graph is the same everywhere in the phase space. The proof
of the conjecture 1  has been extended by Soul\'e in \cite{Soule}.
The main idea in the proof
of conjecture $1$ is that if the dynamical system has several fixed points then $\bF$ has several zeroes
and thus cannot be injective. Thus knowing sufficient conditions for $\bF$ to be injective, their negation give necessary
conditions for $\bF$ to have several zeros. The injectivity is ensured by conditions on the determinant of the Jacobian
matrix. The proof of the second conjecture uses the fact that if all semi circuits (closed path in the non oriented graph) 
of length $2 \leq p \leq N$ are nonnegative then the dynamical system is equivalent, up to change of sign of some variables,
to a cooperative system. Then there is no attracting periodic trajectory. \\

As a conclusion, let us remark is that the notion of negative circuits is related
to a notion of \textit{frustration} introduced
in the context of spin glasses. A loop is frustrated if 
the magnetic energy cannot be minimized for all the edges of this loop. 
This implies  that the magnetic energy of this loop
cannot reach its absolute minima and that severals spin configurations lead to the same
local minimum. The loop is frustrated because there are always ``unsatisfied'' edges where  the spins are not in a configuration allowing 
them to minimize their energy.
Flipping one spin may satisfy them but then others link will become unsatisfied.  
The notion of positive  circuits is similar
to the notion of non frustrated loop.  Actually, one can obtain
convergence results for symmetrically signed networks ($sgn(J_{ij}=sgn(J_{ji})$)
 provided that the corresponding graph is not frustrated  \cite{Smith,Benaim}. The notion of frustration has therefore
nicely been adapted here.

Finally, note that feedback effects can generate very complex situations, \textit{even if the dynamics is convergent}.
For example, in the case of symmetric synapses where a Lyapunov function exists, the presence 
of feedback terms (and frustration) induces a multiplicity of stable fixed points. This effect is analogous
to the multiplicity of solutions for the Thouless-Anderson-Palmer \cite{TAP} equations giving the various
phases in the Sherrington-Kirckpatrick spin glass.

\ssu{Neural oscillators.} \label{NO}

What happens now if the synapses have no particular symmetry ? Actually, there are many
possibilities including chaos. An example is presented below. But to end up the section
\ref{Rec} we would like to point out an easy way to generate oscillations in a system of two neurons
having the dynamics (\ref{ECG}). Consider therefore the dynamical system (\ref{ECG}) with two neurons
where, for simplicity, $\theta_1=\theta_2=0$:

\beq\label{EOsc}
\left\{
\baR{ccc}
\dot u_1 &=& -u_1 + J_{11}\tanh(gu_1)+ J_{12}\tanh(gu_2)\\
\dot u_2 &=& -u_2 + J_{21}\tanh(gu_1)+ J_{22}\tanh(gu_2)\\
\eaR
\right.
\eeq

\nid where $g$ controls the non linearity. It is easy to see that $\bu=0$
is always a fixed point. Moreover the Jacobian matrix at
$\bu=0$ is:

\beq\label{JOsc}
DF_0=
\left(
\baR{cccc}
-1 + gJ_{11}& g J_{12}\\
g J_{21}& -1+gJ_{22}\\
\eaR
\right)
=-I + g\cJ
\eeq

\nid where $I$ is the identity matrix and $\cJ$ the matrix of synapses.
Therefore, the eigenvalues of $DF_0$ are $\lambda_k = -1 + g s_k$
where $s_1,s_2$ are the eigenvalues of $\cJ$. We note therefore that the stability of the origin
is determined, in this case, by the eigenvalues of $\cJ$. We shall return back to this point
in the section \ref{NotreModele}. The eigenvalues of $\cJ$ are straightforward to compute
and the eigenvalues of $DF_0$ are given by
$\lambda_{1,2}= -1 - g\frac{J_{11}+J_{22}}{2} \pm \frac{g}{2} \sqrt{(J_{11}-J_{22})^2 + 4 J_{12}J_{21}}$.
Consequently, it is possible to have \textit{oscillations} in the system, provided that
$(J_{11}-J_{22})^2 + 4 J_{12}J_{21} \leq 0$. This imposes that $J_{12}J_{21}<0$. Namely  one neuron (say the first one)
excites the second one while the second neuron inhibits the first one (see Fig. \ref{FOscill}).
 Note however that this is only a \textit{necessary} condition.
Actually, a sufficient condition corresponds to having a Hopf bifurcation destabilizing $\bu=0$ and generating stable oscillations.
This happens whenever the two following conditions are fulfilled.

\bea
g\frac{J_{11}+J_{22}}{2}\geq 1 \\
(J_{11}-J_{22})^2 + 4 J_{12}J_{21} \leq 0
\eea
 
For example, the following matrix (corresponding to the diagram drawn in Fig. \ref{FOscill})
generates oscillations whenever $g \geq 1$: $\cJ=\left(
\baR{cccc}
1 & 1\\
-1 & 1\\
\eaR
\right)$.

%
%
%
%
%
%
%
%
\begin{figure}[ht]
\begin{center}
\includegraphics[height=4cm,width=3cm,clip=false]{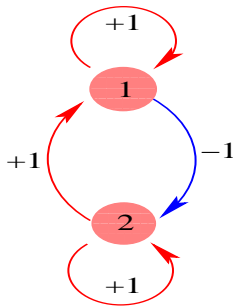}
\caption{\label{FOscill} Example of a system of two oscillating neurons (provided that $g >1$).}
\end{center}
\end{figure}
%
%
%
%
%
%

This example shows that once the synaptic symmetry is broken the dynamics usually does not settle onto a fixed
point but on a more complex attractors. Even with two neurons  the competition of excitation/inhibition can generate periodic oscillations.
When the number of neurons increases, one can have pools of synchronized and oscillating neurons, as discussed in the section
\ref{WCNNH}. But one can also have more complex situations ranging from oscillations to chaos. An example is given in the next section.

\su{A complete example.} \label{NotreModele}

In the previous section we have discussed various aspects of a recurrent model
with convergent dynamics.  Actually, in the early eighties
a large effort was devoted to the study of convergent recurrent neural networks. Indeed,
convergence was interpreted as a retrieval of a stored pattern \cite{Hopfield1}.
As shown above the symmetry of the synaptic connexions ensures, in the Cohen-Grossberg,
model, the existence of a Lyapunov function and, consequently, the model is convergent.
When the synapses are not symmetric, the dynamics can be quite a bit more complex,
exhibiting a wide variety of dynamical regimes such as periodicity, quasi periodicity,
chaos, existence of several complex attractors, etc...

In the end of the eighties, some attention have been paid to recurrent neural networks
exhibiting such complex dynamics. Indeed, the real brain is clearly a highly dynamical
system and the convergence of the EEG to a fixed point is not a sign of a good health. 
From the biological
point of view, accurate models have been designed
to modelize temporal phenomena in the brain:
synchronization of oscillations for feature linking \cite{GS},\cite{EB},\cite{Ga},
transition between coherent states
\cite{KG}, and chaos \cite{Bablo1},\cite{Bablo2},\cite{Bablo3},\cite{Bablo4},\cite{GB}. 
Relying on recent neurophysiological results, the study of chaos in neural
networks seemed a very promising way in at least two directions: the
comprehension of the cognitive processes in the brain \cite{SF},
\cite{BP}, \cite{CD} and the development
of new technologies involving the control of chaos
and the massively
parallel computability of neural networks. However, due to their complexity, they were very
difficult to treat onto a mathematical ground and
they lacked a theoretical background explaining the
behavior of the networks in function of a few relevant control parameters.

In this setting   particularly astonishing experiments were made by Freeman and his collaborators
on the olfactory bulb of the rabbit \cite{Freeman1},\cite{Freeman2}.
 They  suggested that the \textit{spontaneous} dynamics of
the olfactory bulb could be \textit{chaotic}. But they also lead the authors to conclude that the  recognition of a previously learned smell
is manifested by a \textit{temporary reduction of the chaotic activity}. On the basis of these experiments, Skarda and 
Freeman \cite{SF}  proposed an interpretation and a modeling 
scheme of the learning and recognition processes. 
In this scheme, the spontaneous dynamics of the neurons is chaotic
and the retrieval of some previously learned pattern corresponds to a
reduction of the chaotic attractor towards an attractor of lower dimension.
During the alert waiting state, the network explores a large region 
of its phase space through chaotic dynamics. When the learned stimulus
is presented, the dynamics is reduced and the system follows the lower dimensional
attractor which has been created during the learning process.

This idea is exciting but quite controversial, since, in particular, it is
extremely difficult to measure quantities, such as fractal dimensions,
on the basis of time series which are, intrinsically, non stationary. Nevertheless,
this paradigm merits to be explored. For this, and to escape from the inherent
limitations of data measurements in biological experiments, one possibility is to
propose a model, that can be an oversimplification of the biological reality,
but which captures some important features. The advantage of such a model is that
one can simulate it numerically and have a better control on the time series.
Also, sometimes, it is possible to obtain analytical results.\\

This section is entirely devoted to such a model. Its structure is directly inspired by
the Cohen-Grossberg model (but with a discrete time). Despite its simplicity, it displays
an astonishing variety of dynamical behaviors and it has quite unexpected properties.
Moreover, a relatively deep mathematical analysis can be performed  combining concepts and 
methods from dynamical systems theory, statistical physics and ergodic theory.
The subsections \ref{Mod},\ref{ResGen},\ref{KO},\ref{MFT} are devoted to a preliminary analysis
of the spontaneous dynamics of the model, namely without learning. In the subsection \ref{App}
we discuss its behavior when an Hebbian learning is applied and we show that a behavior similar
to Freeman's paradigm is exhibited. Namely, this Neural Network has the ability to store
information and to retrieve it by reduction of chaos. 
Finally, the subsection \ref{RepLin} explores an important
aspect and propose a new analytical tool to partially answer a basic question, arising naturally from the discussion
above. Assuming that the model presented here as some relevance for brain dynamics, 
how can a \textit{chaotic} network process information ?   

\ssu{Model description.} \label{Mod}

Let us  consider  a discrete time version of the equations (\ref{ECG}):
\begin{equation} \label{DNN}
 \left\{
\begin{array}{lc}
x_i(t+1) = f(u_i(t+1)) \\
\\
u_i(t+1) = \sum_{j} J_{ij} x_j(t) + \theta_i \qquad i=1..N
\end{array}
\right.
\end{equation}
\nid where $f$ is a sigmoidal function such as 
$f(x)= tanh( gx)$ or $f(x)= \frac{1 + tanh(gx)}{2}$. 
Henceforth, $f$ maps $\bbbr$ to an interval $[a,b]$
and the dynamics(\ref{DNN}) occurs in a compact space $\Omega=[a,b]^N$.
 The  parameter $g$ controls
the non linearity of  $f$. This nonlinearity plays an important role in many aspects.
Firstly, it is directly related to the transition to chaos described in section \ref{KO}.
But it has also an important influence when discussing the amplification/saturation 
effects on the propagation of a signal transiting via a neuron (see section \ref{RepLin}).

In this model, each neuron interact with each other (fully connected model). 
The ``output'' state $x_i(t+1)$ is a function of the weighted sum 
of the signals arriving at $i$ at time $t$, $u_i(t) = \sum_{j} J_{ij} x_j(t)$.
We call $u_i(t)$  the \textit{local field}.
Moreover, the ``synapses'' $J_{ij}$ are independent, identically distributed random variables,
with expectation $E[J_{ij}]$ and variance $Var[J_{ij}]$ given by:
\begin{equation}\label{Jij}
 E[J_{ij}] = \frac{\bar{J}}{N} \quad ; \quad Var[J_{ij}] =
\frac{J^2}{N} 
\end{equation}
\nid such that the expectation and the variance of the ``synaptic potential'' $ \sum_{j} J_{ij} x_j(t)$
remains bounded as $N \to \infty$.
 For technical reasons we shall furthermore assume that the probability density of the $J_{ij}$'s, 
$\rho$, obeys:
\begin{equation} \label{HypGirko}
 \left\{
\begin{array}{lc}
(i) \ \exists \beta > 1 \ \mbox{s.t.} \ \int \rho^\beta(x)dx < \infty \\
(ii) \ \exists  \delta > 0 \ \mbox{s.t.} \ E\left[|J_{ij}^{2+\delta}|\right] < \infty \\
(iii) \ \exists \alpha > 0 \ \mbox{s.t.} \ E\left[|J_{ij}^n|\right] \leq n^{\alpha n}, \forall n \geq 2.
\end{array}
\right.
\eeq
\nid Note that these conditions hold for a Gaussian or a uniform distribution. In the sequel
the matrix of synaptic couplings will be denoted by $\cJ$.

In the section \ref{App} we shall discuss the effect 
of an Hebbian learning on the synapses and on the dynamical evolution.
But at the present stage, assuming  independence between the $J_{ij}$'s may be viewed as an attempt
to model a neural network initially ``empty'' of any information encoded in its
synaptic structure (\textit{tabula rasa}). 
Note that the synapses are therefore (almost-surely) asymmetric in this model. The situation is thus
different from the previous section, where the symmetry allowed us to exhibit a Lyapunov function,
ensuring convergence to fixed points. In the present model, the attractors of the dynamics are in general
not fixed points, but complex objects (e.g. strange attractors). \\

In eq. (\ref{DNN}) the quantities $\theta_i$ play two different roles. 
In the present case (without learning) they correspond to a threshold
in the neuron response. To take into account neuron diversity we assume that 
the $\theta_i$'s are Gaussian, independent, and identically distributed random variables,
such that:
\begin{equation} \label{theta}
E[\theta_{i}] = \bt;
\qquad Var[\theta_i] =
\sdt. 
\end{equation}
\nid We call ${\Bth}$ the vector $\left\{\theta_i\right\}_{i=1}^N$.

In section \ref{App} we shall consider the effect of an input on the dynamics.
In this cases, the $\theta_i$'s will correspond to the input submitted to the neuron
$i$. Also, we shall discuss in section  \ref{RepLin} the
case where the input is time dependent. 
Finally, note that in eq.  (\ref{DNN}) the synapses and thresholds do not evolve in time
(quenched disorder). \\

The dynamical system (\ref{DNN}) depends a priori 
on $\cN \deq N^2+N+1$ parameters ($N^2$ synapses, $N$ thresholds, and $g$).
We call the vector $\lambda=(g,\cJ, \Bth)$ the vector of \textit{microscopic parameters}.
 $\lambda$  has therefore $N^2+N$ random entries.
In the sequel, it will be useful to write the dynamical system (\ref{DNN}) in the
form:
\beq \label{SD}
\bu(t+1) = \bF\left[\bu(t);\lambda\right]
\eeq
Note that the Jacobian matrix as a simple form:
\beq\label{DF}
D\bF\left[\bu;\lambda\right]=\cJ\Lambda(\bu)
\eeq
\nid where $\Lambda(\bu)$ is the diagonal matrix such that $\Lambda(\bu)_{ij}=f'(u_j)\delta_{ij}$.

Let us now denote by  $P^{(N)}_{\cJ,\Bth}$ the joint probability distribution for
couplings and thresholds, in a $N$ dimensional system.  This probability distribution
determines therefore the actual realization of the $J_{ij}$'s and $\theta_i$'s.
 Moreover,  
 $P^{(N)}_{\cJ,\Bth}$ is determined by the parameters $\bar{J},J,\bt,\sigma_\theta$.
Hence, these parameters fix the statistical properties of the microscopic parameters
$J_{ij}$ and $\theta_i$. 
we shall call the parameters\footnote{In this particular case $g$ is both a microscopic and a macroscopic
parameter. This is simply because all neurons have the same $g$. One can imagine a generalized
version where the nonlinearity of the neuron $i$, $g_i$, depends on the neuron and where the $g_i$'s
are randomly distributed.
In this case the $g_i$ would be additional microscopic parameters, while the
 parameters controlling their probability distribution would be additional macroscopic parameters.}
 $g,\bar{J},J,\bt,\sigma_\theta$ the \textit{macroscopic parameters}. Note that we have  
only  four independent macroscopic parameters because the dynamical system (\ref{DNN}) is invariant under the transformation.
\beq\label{RenJ}
g \to gJ; \quad J_{ij} \to \frac{J_{ij}}{J}; \quad \theta_i \to \frac{\theta_i}{J}; 
\eeq
%
%
\nid Hence $J$ is somewhat irrelevant.

From the dynamical system point of view widely developed in the previous sections,
 it is natural to seek the generic (in a \textit{topological} sense) behavior of (\ref{DNN}) 
when the \textit{microscopic} parameters are varied. However, this is a formidable task and
one may argue that, since these parameters are random, it is certainly more
useful to investigate the generic behavior (in a \textit{probabilistic} sense) when the
\textit{macroscopic} parameters are varying. In some sense, we substitute the analysis of 
the dynamical system (\ref{DNN}), with uncountably infinitely many possible realizations
of the microscopic parameters, by an ``averaged'' dynamical system depending on
four independent deterministic macroscopic parameters. In this spirit, a few results are given in the next section, 
obtained by combining dynamical systems theory and probabilistic results about random matrices (section \ref{ResGen}).
But, essentially, this approach  is the core of the 
dynamic mean field theory,
that will be fully developed in the chapter II. In the present chapter (section \ref{MFT}), we 
derive the mean field equations by an heuristic argument, and discuss  their
dynamical properties in relation with the dynamical system (\ref{DNN}).

Before entering into the detailed analysis let us make a last remark. The dynamics of the uncoupled neurons
in the dynamical system (\ref{DNN}) is rather poor. It indeed writes $x_i(t+1)=f(gJ_{ii}x_i(t)+\theta_i)$. This dynamical
system exhibit either a stable fixed point  or bistability (appearing by a saddle node bifurcation). Contrarily
to the examples studied in the section \ref{Spiking} there is no Hopf bifurcation, no homoclinic loops, \etc.
Nevertheless, the \textit{coupled} system, as we shall see, has a rather rich dynamics. This provides
a prominent example of \textit{emergent} collective behavior. 

\ssu{Preliminary results}\label{ResGen}

Let us first establish a few preliminary results.
Firstly, it is easy to show that, for each realization  $\cJ(\omega)$ of $\cJ$,
there exists a  $g$ value, $g_{as}(\omega)$,
independent of  $\Bth$, and given by:  

\beq\label{gas}
g_{as}(\omega)=\frac{1}{\alpha\|\cJ(\omega)\|}
\eeq

\nid such that $\bF$ is a contraction whenever $g<g_{as}$ \cite{JPA}. In (\ref{gas}) $\| \ \|$ is the operator norm
induced by the euclidean norm and $\alpha$ is such that $\alpha  =f'(0)$. This result is straightforward since
$$
\|\bF(\bu;\lambda)-\bF(\bv;\lambda)\| \leq 
\sup_{\bw \in \Omega} \|D\bF(\bw;\lambda)\| \|\bu-\bv\| = 
 \|\cJ\| \|\bu-\bv\|\sup_{\bw \in \Omega}\|\Lambda(\bw)\| 
$$
\nid where the last inequality holds from eq. (\ref{DF}). Since $\Lambda(\bu)$ is a diagonal matrix 
such that $\Lambda_{ij}(\bu)=f'(u_j)\delta_{ij}$ and since $f$ is a sigmoidal function where the maximal
slope is equal to $\alpha g$, $\bF$ is a contraction provided that $\alpha g \|\cJ\|<1$. The result follows.  

When $\bF$ is a contraction the dynamical system 
(\ref{DNN}) is absolutely stable  i.e. \textit{it admits  a unique fixed point, attracting all
trajectories}.  
The matrix $\cJ$ is random and the result (\ref{gas}) holds for each realisation. One
can obtain from this a statistical result by using a theorem proved by Geman
\cite{Geman1}. Provided that the $J_{ij}$'s obey the condition (\ref{HypGirko}(iii)),
 $\cJ$ converges almost surely, when $N \to \infty$ ,
 to a finite value that can be explicitly computed \cite{These,JPA}, depending on
the parameters  $\bar{J}, J$. From this, one obtains the asymptotic limit of $g_{as}$ when $N \to \infty$.
This provides, for finite $N$, an estimate of the $g$ parameter values where the system
is absolutely stable with a high probability.\\

When $g$ increases, one expects  bifurcations leading to dynamical changes.
When $f(x) = tanh(gx)$ and when there are no thresholds, the function $\bF(\bu;\lambda)$
has the symmetry $\bF(\bu;\lambda) = - \bF(-\bu;\lambda)$. Thus, $\bu=0$ is 
always a fixed point. Also, in this case $D\bF(0)=g\cJ$. Consequently,
the stability of this fixed point is determined by the spectrum of the random
matrix $g\cJ$. Obviously, the eigenvalues of $g\cJ$  are proportional to
the eigenvalues of $\cJ$ with a coefficient $g$. $\cJ$ being real, the eigenvalues
are either real or complex conjugated.
When $g$ increases, the spectrum
is dilated and for sufficiently large $g$ some eigenvalues are crossing the stability
circle  $\left\{z \in \bbbc| |z|=1 \right\}$ (see Fig. \ref{FCirclestab}).
However, the probability that several eigenvalues cross simultaneously this circle
is zero\footnote{Having several eigenvalues crossing simultaneously the stability circle corresponds
to impose algebraic relations of codimension larger than one  between the coefficients of the characteristic
polynomial of $\cJ$. Since the $J_{ij}$'s are selected randomly 
the probability to fulfill these algebraic conditions is zero.} if one excepts the case of  a pair
of complex conjugated eigenvalues. We expect therefore a destabilization
of $0$ by \textit{a codimension one bifurcation}.
%
%
%
%
\begin{figure}[ht]
\begin{center}
\includegraphics[height=6cm,width=10cm,clip=false]{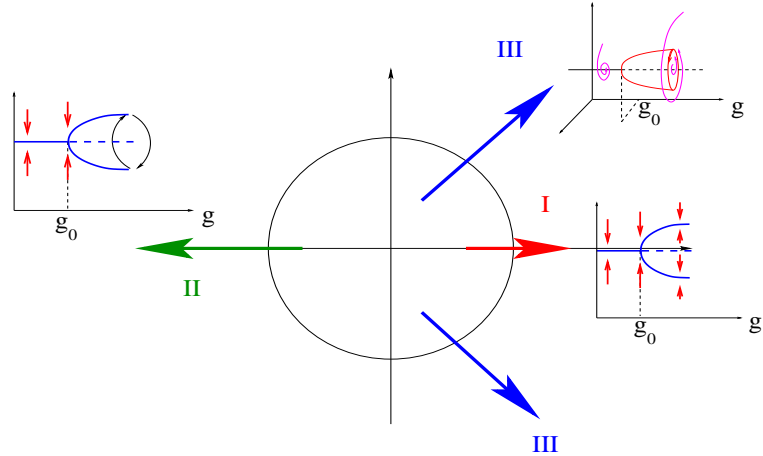}
\vspace{0.5cm}
\caption{Three possibilities of codimension one bifurcation occurring in (\ref{DNN}) when
the non linearity $g$ increases.  \label{FCirclestab}}
\end{center}
\end{figure}
%
%
%
%
%
%
The possible codimension one bifurcations for a map having the symmetry $\bF(\bu;\lambda) = - \bF(-\bu;\lambda)$
are described in Fig. \ref{FCirclestab}.
They are (see the appendix for details):

\bit
\item Case I. Pitchfork bifurcation. The fixed point destabilizes and two symmetric stable fixed points appear.
\item Case II. Period doubling bifurcation. The fixed point destabilizes and stable periodic orbit of period two appears. 
\item Case III. (Discrete time) Hopf bifurcation. 
The fixed point destabilizes and stable periodic orbit  appears. Note that, \textit{stricto-sensu}, orbits of period 
$2$,$3$ and $4$ \textit{do not} correspond to a Hopf bifurcation (the normal form is different, see \cite{Arnold2} for details).
Orbits of period $3$ and $4$ are observed for small $N$'s \cite{IJBC}. 
\eit

Call $\rho(\cJ)$ the spectral radius of $\cJ$ (the value of the largest modulus of the eigenvalues).
Then the destabilization occurs when:

\beq\label{g0}
g_0=\frac{1}{\rho(\cJ)}
\eeq  

\nid This is a random variable. However,
the statistical behavior of random matrices obeying the conditions (\ref{HypGirko} (i), (ii))
is well known when the size tends to infinity \cite{Girko}. The limiting
spectral density converges almost surely to the uniform density in the disc of center $0$ and radius
$J$ in $\bbbc$. Consequently,
$g_0$ converges almost surely to $\frac{1}{J}$ and the destabilization value is given by
$g_0J=1$. Note that the same result can be obtained from the dynamical mean field theory (see \cite{Sompo1} and \cite{Crisanti}
for a continuous time version of (\ref{DNN})). 

The repartition of eigenvalues is also known in the finite size case \cite{Edelman}.
One can show that there is an over density of real eigenvalues that disappear in the
limit $N \to \infty$. Consequently, for finite size, one observes destabilization by pitchfork and flip
bifurcations, but the Hopf bifurcation becomes more and more frequent when $N$ increases \cite{IJBC,PD,JP}.
Finally, in the infinite system an infinite number of eigenvalues cross simultaneously
the unit circle. This corresponds to a sharp transition from \textit{fixed point}
to \textit{white noise} discussed in section \ref{MFT} (see also \cite{JP}).

Let us now make a remark about the Hopf bifurcation. As we somehow anticipated in the sections 
\ref{WCNNH},\ref{NO} oscillations arise because there is a competition between excitation/inhibition effects
among the neurons. Actually, one expects from the study performed in \ref{WCNNH} to have, near the bifurcation, pools
of almost synchronized neurons oscillating coherently. This is revealed in the study of the correlation function
which has usually a bloc structure as revealed from example in \cite{Dauce0}. Note also that the period
of oscillations is generically irrational. Finally, the results above the spectrum of $\cJ$
imply that the phase $\nu$ of the largest eigenvalue, generating the Hopf 
bifurcation, is uniformly distributed between $[0^+,\pi]$. Hence $Prob[0^+ < \nu < \theta] = \frac{\theta}{\pi}$ and,
since the period is $T = \frac{2\pi}{\nu}$, $Prob[T > \tau=\frac{2\pi}{\theta}] = \frac{2}{\tau}$.
Therefore the probability density of the period is $\rho_T(\tau)=\frac{2}{\tau^2}$. Thus, there is a high
probability to have oscillations with \textit{a low period}.  \\

These results have been obtained by combining elementary results from dynamical system
theory, holding for each realization of the disorder, and convergence results in random matrices theory.
The convergence results, holding when $N \to \infty$,  are then used as a guideline for
a typical realization of the finite dynamical system. 
They are however quite restricted. For example, we have assumed that
the system has the symmetry $\bF(\bu;\lambda) = - \bF(-\bu;\lambda)$. But when we consider the equation (\ref{DNN})
with thresholds, this symmetry disappears. Then, the fixed point of the absolutely stable regime
is a random variable. Moreover, when $g > g_{as}$ new fixed points can appear by saddle-node bifurcations:
they are also random. Finally, we have been able to analyze the first bifurcation relatively easily but, after the destabilization
the usual techniques (central manifold reduction, normal forms) are difficult to handle since the 
coefficients are random (hence, for example, the eigenvectors of $\cJ$ are random).

One has therefore to develop an alternative statistical approach. This is done in the  section \ref{MFT}.
Before this, we discuss in the next section the typical behavior of the dynamical system (\ref{DNN})
when $g$ further increases. The results presented are a combination of genericity results in dynamical
systems theory and numerical simulations.

\ssu{Transition to chaos} \label{KO} 

Numerical simulation is a fundamental tool for the exploration
of the wide dynamical richness of the model (\ref{DNN}). But clearly,
exploring the parameters space of this system at ``random'', without
any preliminary idea of what is going on is like ``searching a needle
in a straw pile'' (it is in fact a bit more tricky since a straw is only a three dimensional
object).
Indeed, basically, dynamical systems are structurally stable on wide
ranges of parameter values and only the points where structural stability
fails (bifurcations points) matter.
But bifurcations occur for parameter values usually located on manifold
of smaller dimension than the ambient space. For example the codimension
one bifurcations discussed above correspond to a $\cN-1 = N^2+N$ manifold in the microscopic
parameters space. Since we select the $J_{ij}$'s and $\theta_i$'s with an absolutely continuous
probability distribution (i.e. having a density), the probability to fall on a bifurcation
point is zero. 
Obviously, since we are seeking statistical properties, we are rather interested in locating
 the bifurcation points in the \textit{macroscopic} parameter space. Having
a bifurcation map in this space would correspond to having statements such as:
``If you fix the parameters $\bJ,\bt,\st$ in this region of the macroscopic parameters space,
and if you vary $g$ between such and such value, then, typically you will observe this type
of bifurcation''. 

For this, we need to have theoretical guidelines. The preliminary results given above
are an example of such guidelines. The mean field approach briefly discussed below
provides additional hints. Consequently, the numerical simulations described in the
present section have been made with the informations given by these theoretical results,
plus a few standard and generic facts in bifurcation theory. These facts are:

\bit
\item Breaking the symmetry $\bF(\bu;\lambda) = - \bF(-\bu;\lambda)$ transforms 
pitchfork bifurcations into saddle-node bifurcations as depicted in Fig. \ref{PitchSN}.
We observe indeed such bifurcations and we have an analytical way to locate them
(see the next section).
%
%
%
%
\begin{figure}[ht]
\begin{center}
\includegraphics[height=3cm,width=6cm,clip=false]{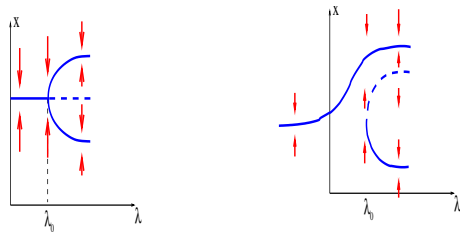}
\vspace{0.5cm}
\caption{Effect of breaking the symmetry $\bF(\bu;\lambda) = - \bF(-\bu;\lambda)$ on a pitchfork bifurcation.  \label{PitchSN}}
\end{center}
\end{figure}
%
%
%
%
%
%
\item The fixed points can be destabilized when $g$ increases. They generically do it by
Hopf bifurcation (namely with an increasing probability as $N$ increases).

\item As shown in Fig. \ref{RouteKO}, after the first Hopf bifurcation, the standard scenario is 
the ``Ruelle-Takens-Newhouse'' transition to chaos by quasi periodicity \cite{RT},
(though our system is a discrete time system).
As $g$ increases the limit cycle generated from the first Hopf bifurcation destabilizes
by a second Hopf bifurcation giving rise to a two dimensional ($\cT 2$) torus.
 Near the bifurcation, the trajectories densely fill 
the torus since the frequencies corresponding to the first and second Hopf bifurcation
are, in general, irrational. However, a further increase of $g$ leads to a frequency locking:
the frequencies corresponding to the first and second Hopf bifurcation
synchronize in a rational fashion and  the trajectories are periodic orbits on the torus.
Though frequency locking is structurally stable, increasing enough $g$ finally lead
to chaos, by different ways (for a detailed explanation in general models see \cite{GT,MT};
for a detailed description of this model see \cite{These}). Note however that there may exist
``re-stabilisation phases'' when $g$ further increases. This corresponds usually to the crossing
of ``Arnold tongues'' where the dynamics locks on a quasi periodic orbit. An example is given in Fig. \ref{SpectreLyap}a
where have plotted the first and the second Lyapunov exponents. The first Lyapunov exponent increases
with $g$ except  at some points where it takes a zero value. Since the second Lyapunov exponent is
also zero this corresponding to a reduction of the chaotic dynamics on a T2 torus. 
If one excepts these points, the positive Lyapunov exponents and the fractal dimension of the strange attractor increases 
 as $g$ increases.

In fig. \ref{SpectreLyap}b we have plotted the Lyapunov spectrum for $g=3.5$. One notes that, in the example chosen, there is only
one positive exponent. Thus the corresponding (Kaplan-Yorke) dimension is low ($D_{KY}=1.967$). More generally, one observes that the strange
attractor is usually a low dimensional object (compared to the dimension of the embedding space). One consequence is that
an arbitrary perturbation of a point on the attractor has almost all its components outside the attractor. 
Note finally that this transition to chaos generates resonances peaks in the power spectrum (Fig. \ref{SpectreRouteKO})
some of them resulting from the Hopf bifurcations. Thus, even if in the chaotic regime the power
spectrum is continuous, it is not flat, like white noise, but it has peaks or resonances. These remarks lead
to important issues discussed in the section \ref{RepLin}.

\item As $N$ increases the transition to chaos occurs on a $g$ range becoming
more and more narrow. This leads to the conjecture that a sharp transition 
from fixed point to infinite dimensional chaos occurs in the thermodynamic limit.
This conjecture is related to the observation above that the eigenvalues of the Jacobian
matrix accumulate on the stability circle as $N \to \infty$. Exact equations and
analytical description are discussed in the section \ref{MFT}.

\eit

%
%
%
%
\begin{figure}[ht]
\begin{center}
\includegraphics[width=5cm]{Attracteur_g1N8.eps}
\includegraphics[width=5cm]{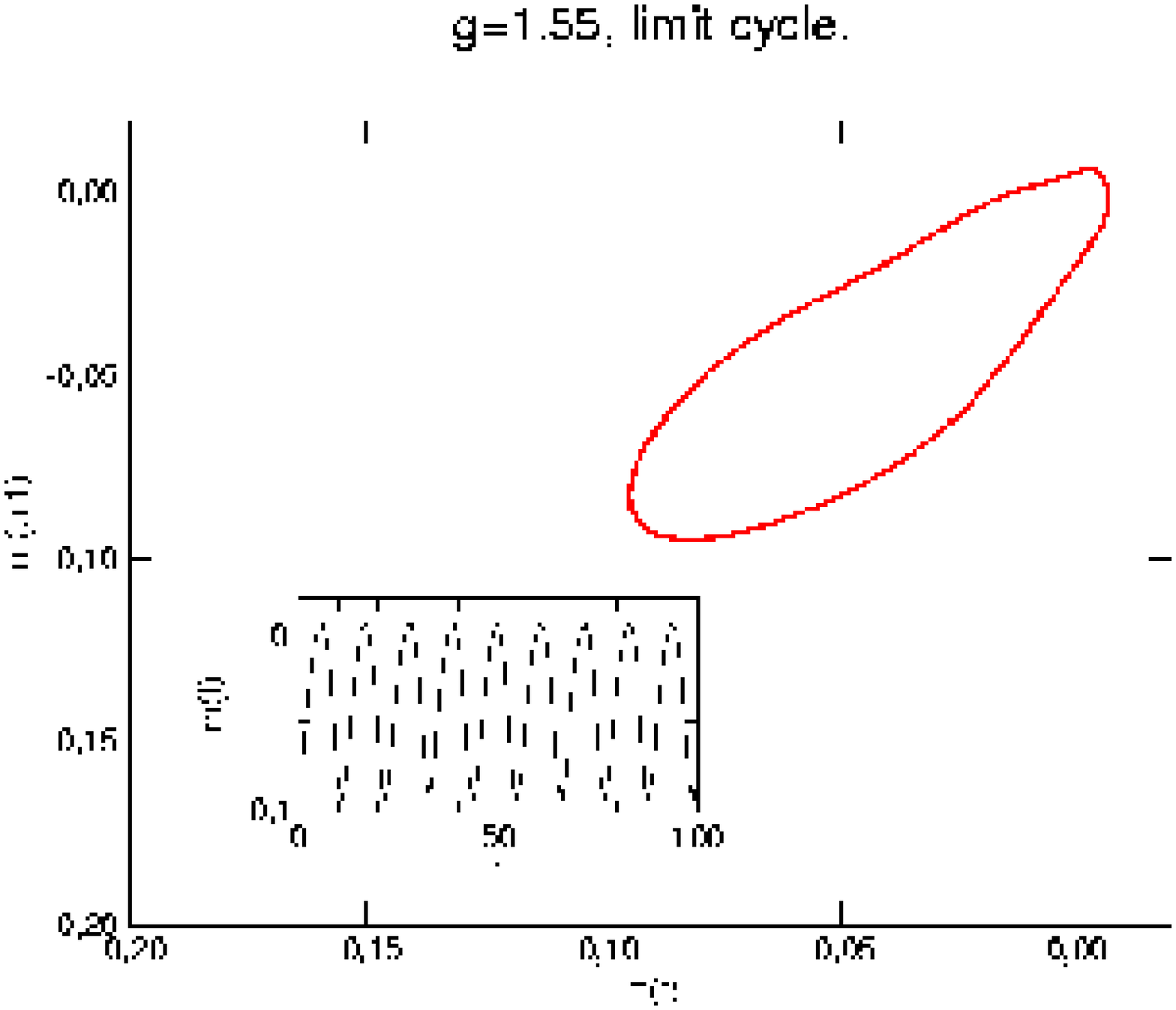}\\
\vspace{1cm}
\includegraphics[width=5cm]{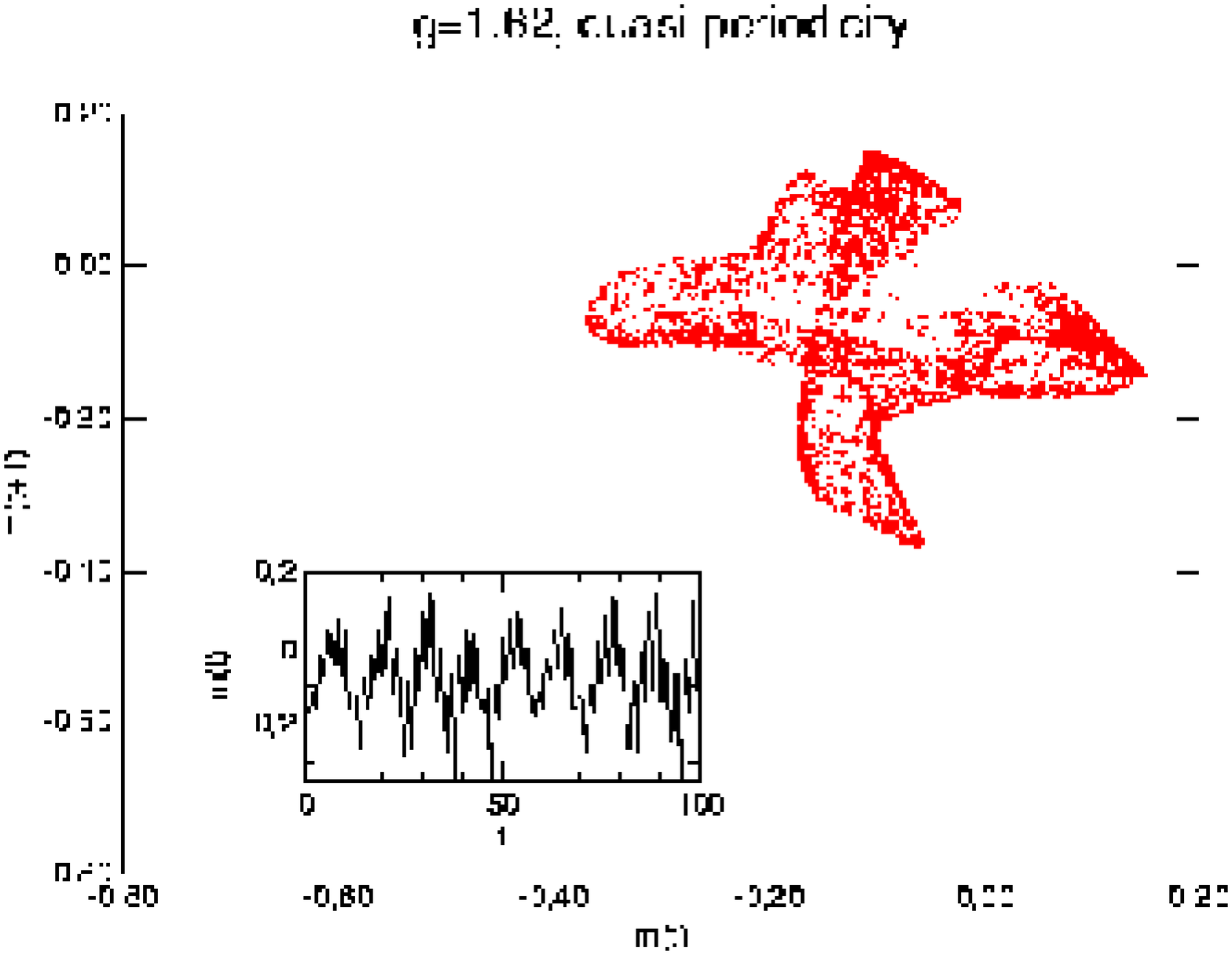}
\includegraphics[width=5cm]{Attracteur_g1.7N8.eps}\\
\vspace{1cm}
\includegraphics[width=5cm]{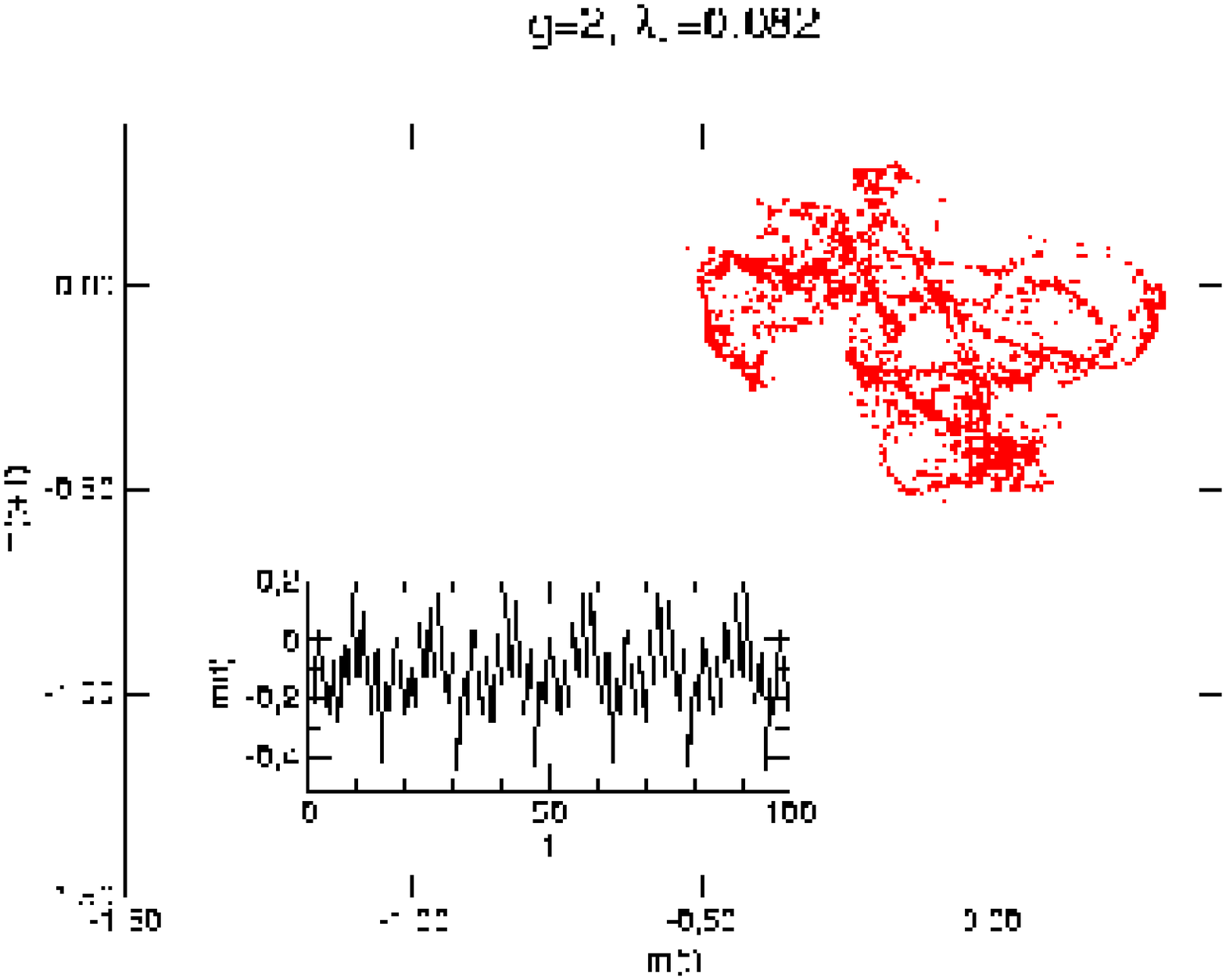}
\includegraphics[width=5cm]{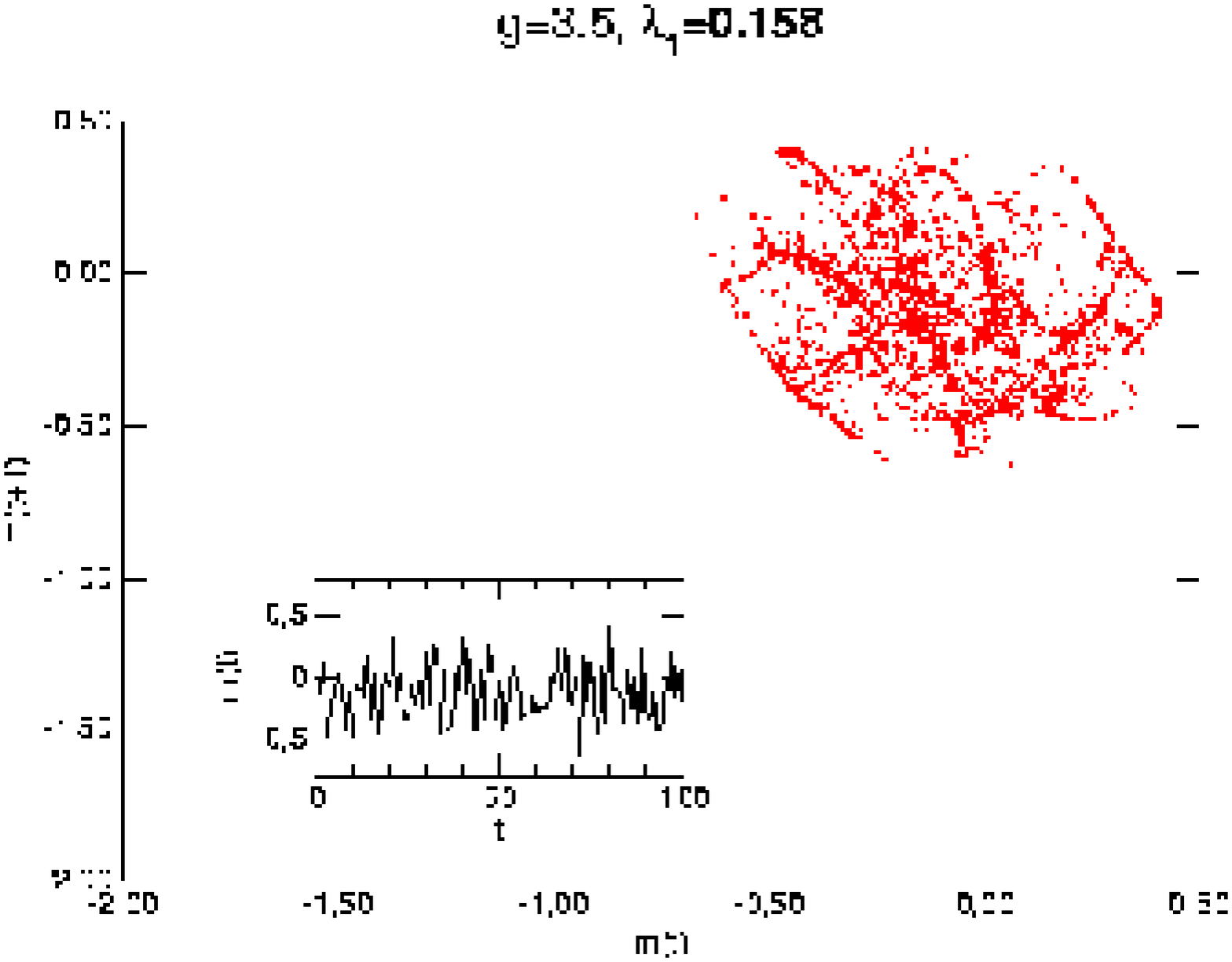}
\vspace{1cm}
\caption{\footnotesize \label{RouteKO}
An example of transition to chaos by quasi periodicity in the model (\ref{DNN}).
We used the representation $m(t+1)$ versus $m(t)$ where $m(t)=\frac{1}{N}\sum_{i=1}^Nx_i(t)$
is the empirical average of the neurons state at time $t$. This representation provides
a projection of the dynamics of $m(t)$ in a two dimensional phase space. The insets represent the
evolution of $m(t)$ versus $t$. In the two last figure
$\lambda_1$ is the largest Lyapunov exponent.}
\end{center}
\end{figure}

%
%
%
%
%
%
%
%
%
\begin{figure}
\begin{center}
\includegraphics[width=6cm]{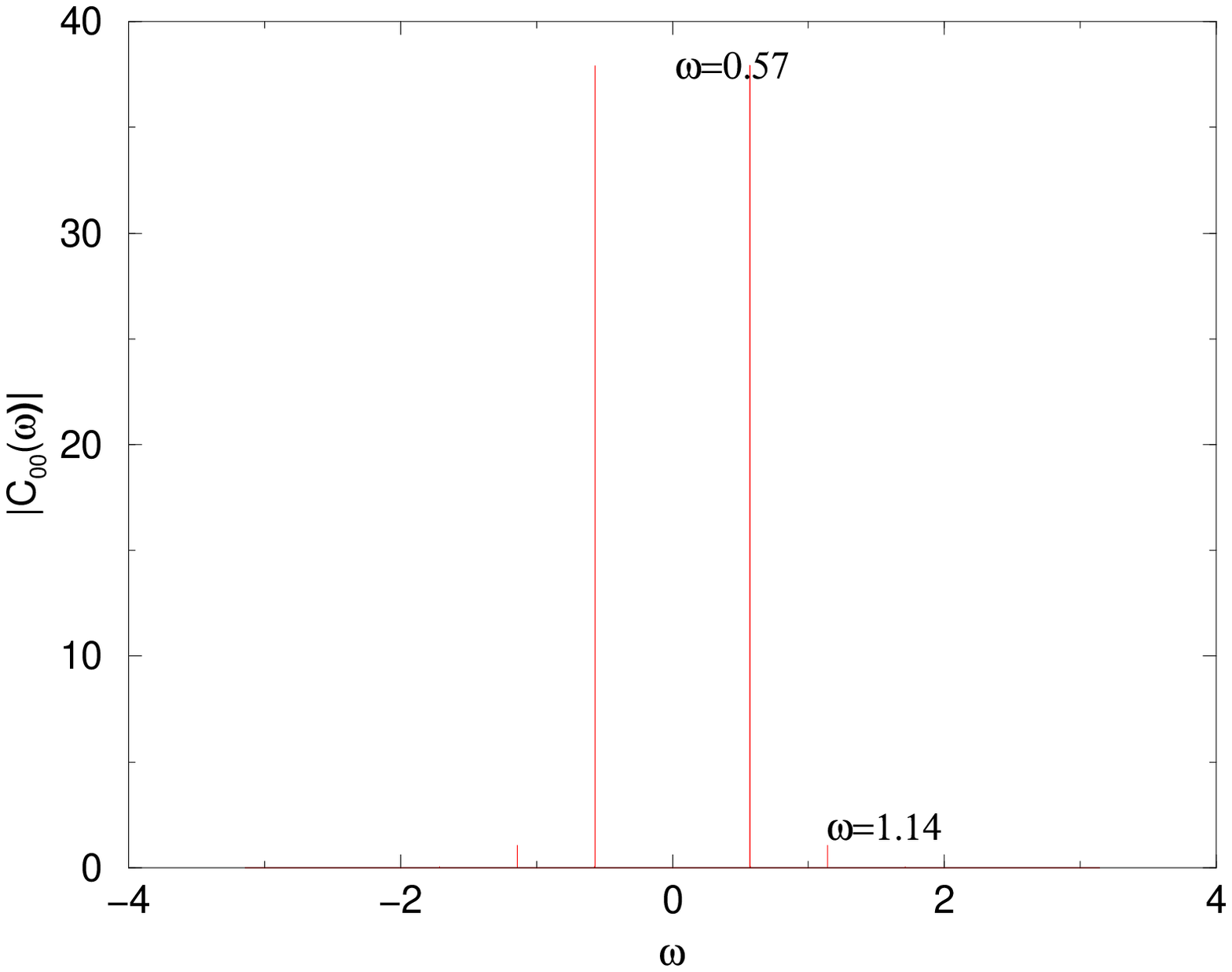}
\vspace{1cm}
\includegraphics[width=6cm]{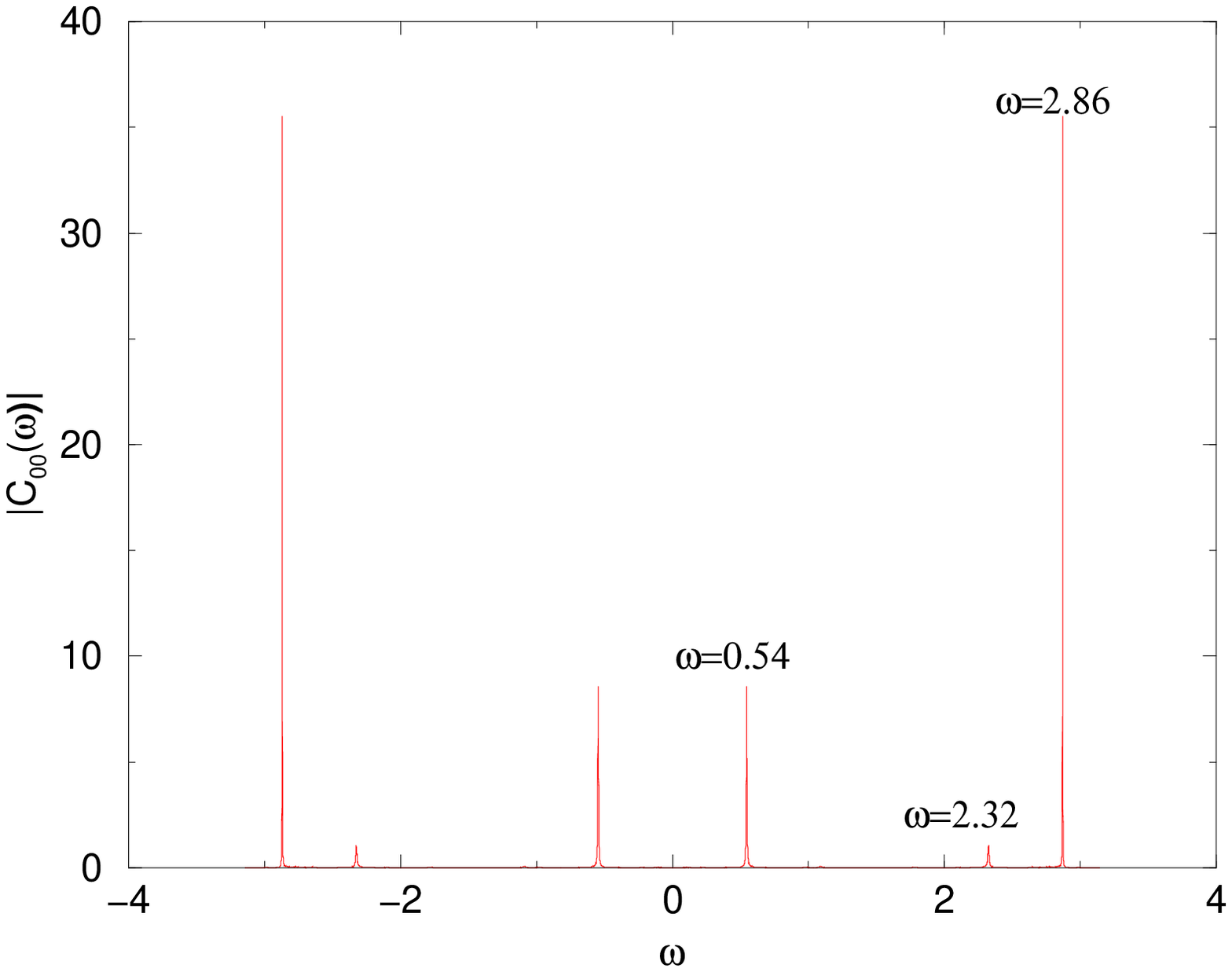}
\includegraphics[width=6cm]{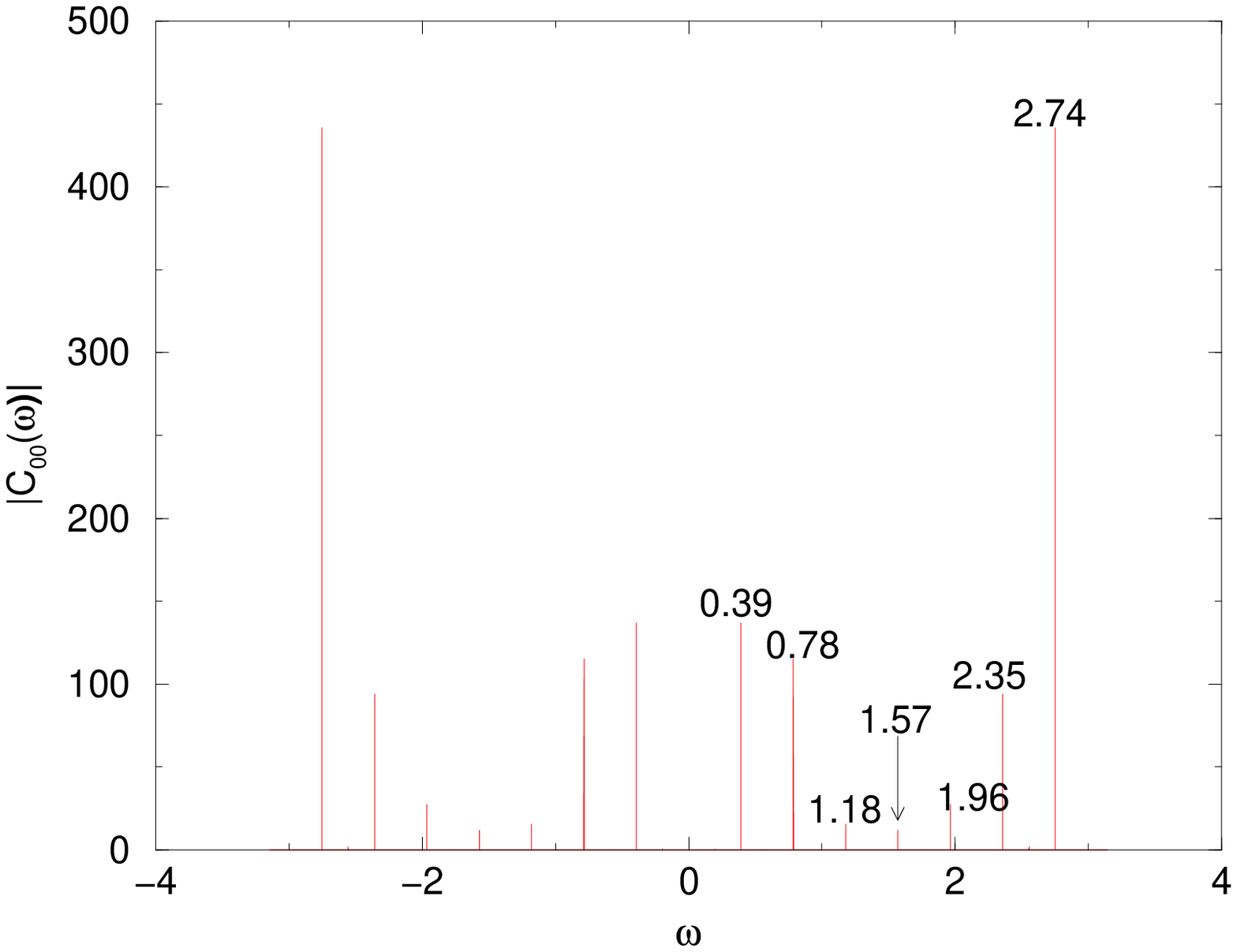}
\vspace{1cm}
\includegraphics[width=6cm]{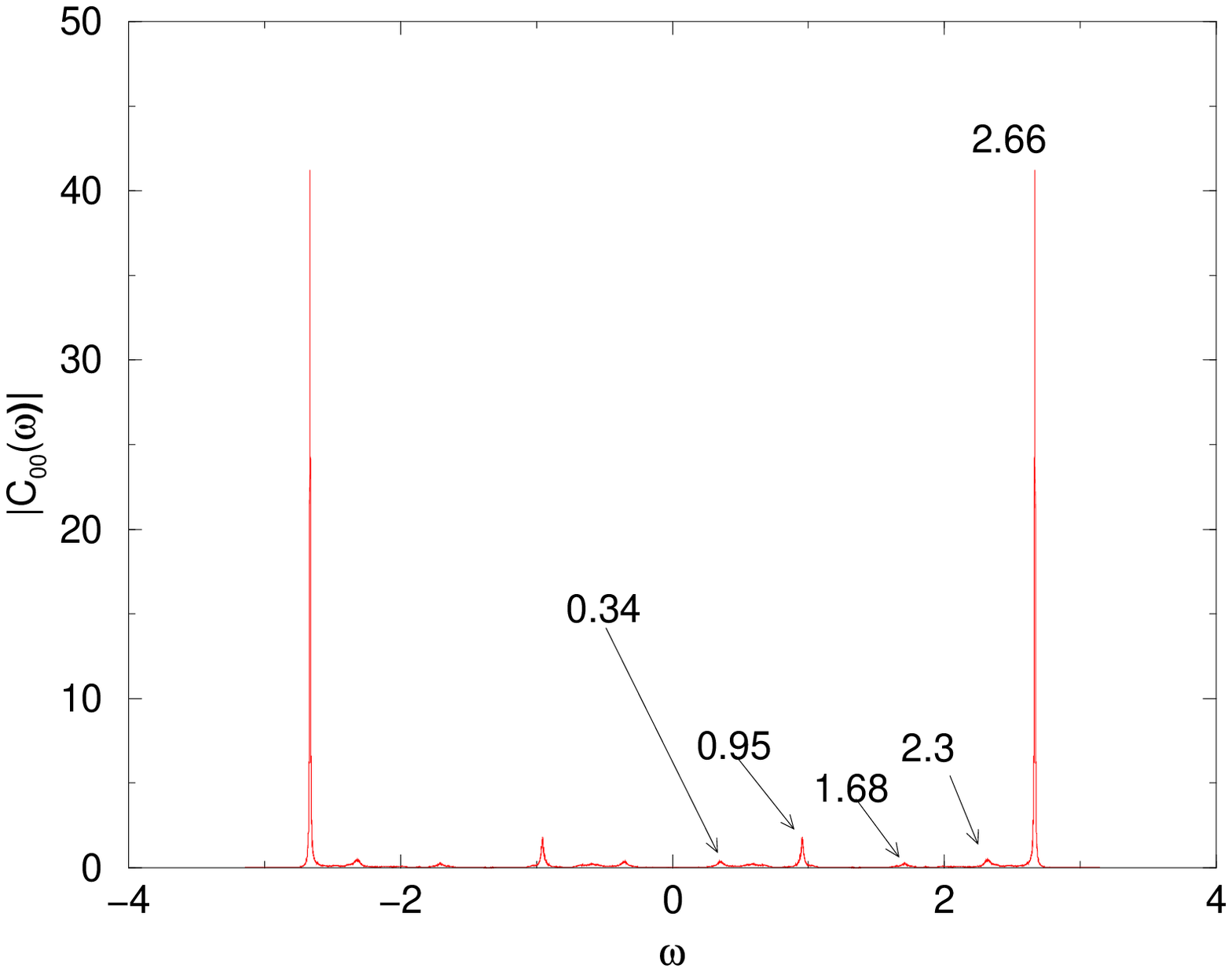}
\includegraphics[width=6cm]{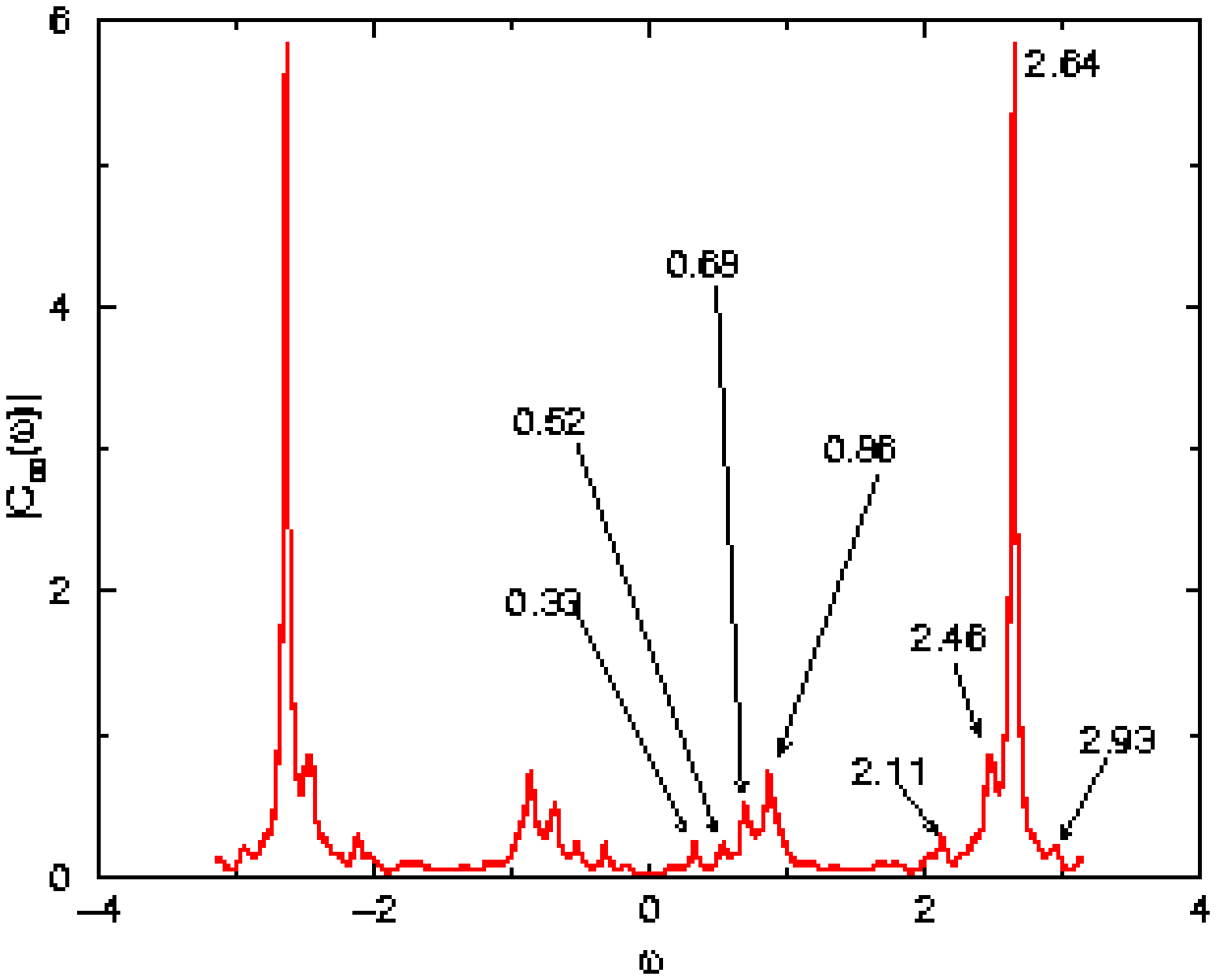}
\caption{\footnotesize \label{SpectreRouteKO}
Power spectrum of the neuron $0$ in the transition to chaos corresponding to the previous figure.}
\end{center}
\end{figure}
%
%
%
%

%
%
%
%
%
%
%
%
%
\begin{figure}
\begin{center}
\includegraphics[width=6cm]{evol_SpectrevsgN8.eps}
\includegraphics[width=6cm]{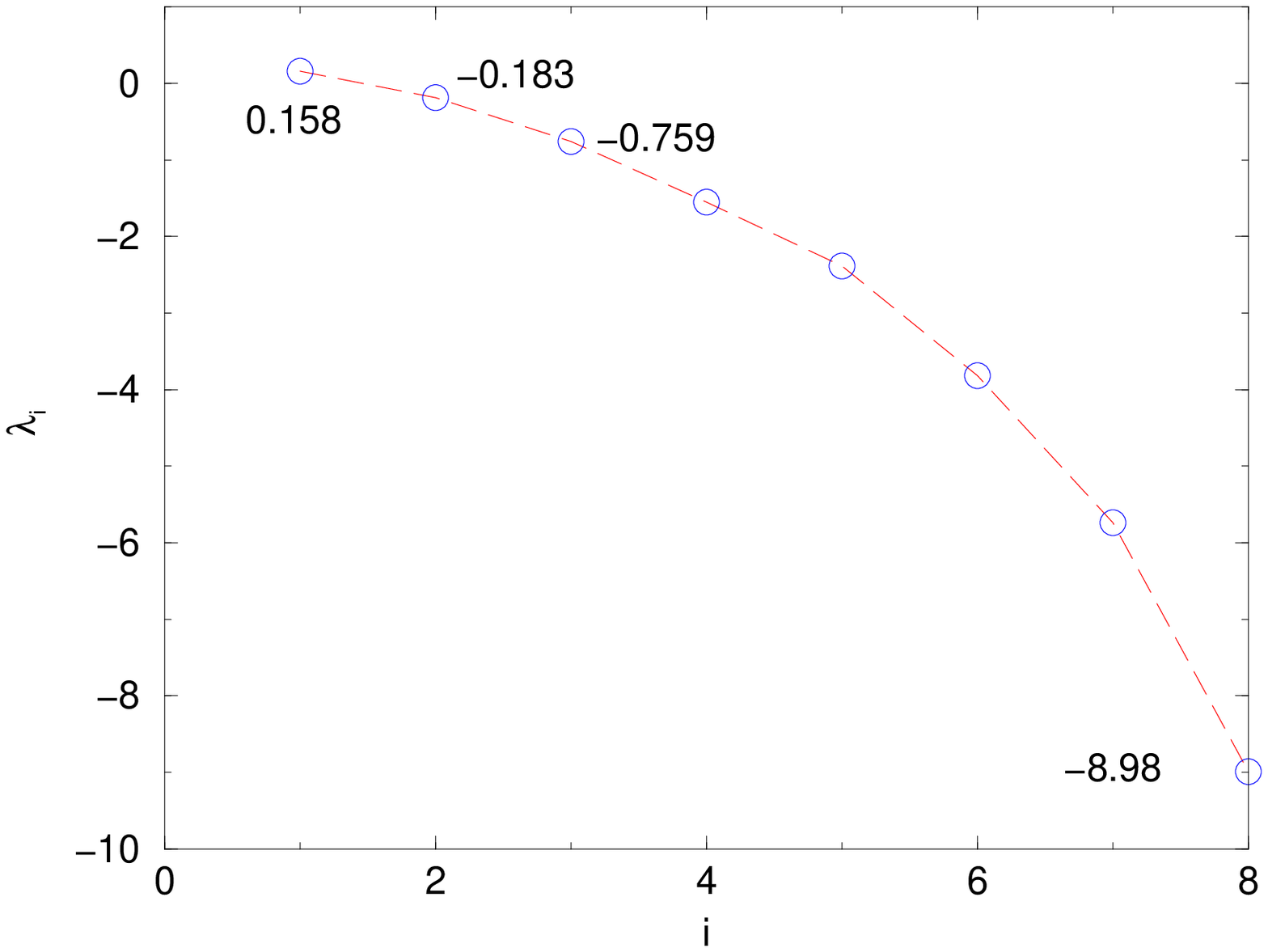}
\caption{\footnotesize \label{SpectreLyap}
\ref{SpectreLyap}a Evolution of the first two Lyapunov exponents when $g$ increases.
\ref{SpectreLyap}b Lyapunov spectrum in the chaotic regime when  $g=3.5$.}
\end{center}
\end{figure}

\ssu{The mean-field dynamical system.}\label{MFT}

The mean field ``approximation'' is quite well known in statistical physics. Though it is \textit{stricto-sensu wrong}
in many cases (for example it gives a wrong critical temperature for  the Ising model), it provides
often an astonishingly good qualitative insight in the description of many models of phase transitions in statistical physics.
Also, for some models (such as the Curie-Weiss model) it is exact in the thermodynamic limit.
In the field of neural networks the use of mean field approaches has a long history,
for analogic networks \cite{Sompo1} but also for spiking networks (see for example \cite{Brunel}).
The chapter II is entirely devoted to mean field approaches, and in the present chapter,
we shall focus only on the model (\ref{DNN}).
 
Basically, the mean field approach applied to this model consists in assuming that the $x_i(t)$'s are independent from
each others and independent from the $J_{ij}$'s !! Though this looks very rough, this approach leads to 
\textit{exact results} that can be rigorously proved (see chapter II). It can also be justified at an heuristic
level \cite{JP}: one easily shows that the key ingredients ensuring the success of this approach
are the independence of the $J_{ij}$'s and the fully connected structure of the model. Hence, the mean field approach
breaks down as soon as some correlation between the $J_{ij}$'s exist (e.g. after learning). Note also that it breaks
down when the $J_{ij}$'s are symmetric. More precisely, one has to correct the mean field equations derived below
by adding a feedback term corresponding to the delayed action that a neuron has on itself \cite{TAP,Mezard,CriSom,JP}
(this action cancels, on average, when the $J_{ij}$'s are independent \cite{JP}).\\  

Assume therefore that   $x_i(t)$'s are independent from
each others and independent from the $J_{ij}$'s. Then the central limit theorem 
states that the ``local fields''  $u_i(t+1) = \sum_{j} J_{ij} x_j(t) + \theta_i$
are Gaussian processes in the limit $N \to \infty$.
 Moreover, they are independent and identically distributed.
Hence the joint probability of the $u_i(t)$'s factorizes in an (infinite) product of
identical Gaussian distributions.
To characterize the Gaussian distribution at time $t$ one needs the average value $\mu(t)=E[u_i(t)]$,
the variance $v(t)=E[u^2_i(t)]-\mu^2(t)$ and the time covariance $\Delta(t,t')=E[u_i(t)u_i(t')]-\mu(t)\mu(t')$
(note that the left hand side terms are independent of $i$ since all the local fields have the same
distribution). It is straightforward to see that these quantities are functions of $m(t)=E[x(t)]$,
$q(t)=E[x^2(t)$, and $C(t,t') =E[x(t)x(t')]$ (see eq. (\ref{SKd1}),(\ref{SKd2}),(\ref{SKd3})) below.
Finally, since $x(t)=f(u(t))$ and since $u(t)$ is Gaussian one obtains:
\bea \label{SKd1}
\mu(t+1)&=&\bJ m(t) + \bt \\
m(t)&=&
\int_{-\infty}^{+\infty} \frac{e^{-\frac{h^2}{2}}}{\sqrt{2\pi}}f(h\sqrt{v(t)}+\mu(t))dh \nonumber
\eea
\bea\label{SKd2}
v(t+1)&=&J^2q(t)+\sigma_\theta \\
q(t)&=&
\int_{-\infty}^{+\infty} \frac{e^{-\frac{h^2}{2}}}{\sqrt{2\pi}}f^2(h\sqrt{v(t)}+\mu(t))dh \nonumber
\eea
\bea\label{SKd3}
&&\Delta(t+1,t'+1)=J^2 C(t,t')+\sigma^2_\theta \nonumber\\
C(t,t')
&=&\\
&=&\int_{-\infty}^{+\infty}\int_{-\infty}^{+\infty}
 \frac{e^{-\frac{h^2}{2}}}{\sqrt{2\pi}}
\frac{e^{-\frac{h'^2}{2}}}{\sqrt{2\pi}}
f\left(
\frac{\sqrt{v(t)v(t')-\Delta^2(t,t')}}{\sqrt{v(t')}}h+
\frac{\Delta(t,t')}{\sqrt{v(t')}}h'+ \mu(t)\right)
f\left(h'\sqrt{v(t')}+\mu(t') \right)
dhdh' \nonumber
\eea
These are the dynamic mean field equations of the model (\ref{DNN}). The parameter $m,q$ are called 
``order parameters'' in the statistical physics literature. They characterize the emergent behavior
of a system  with a large  number of degree of freedom and they exhibit drastic changes
corresponding, in statistical physics, to phase transitions, and in our context
to a macroscopic bifurcation. 

 Let us now
make a few remarks. Firstly, these equations  can also be derived from the computation
of a generating functional for the probability distribution of the trajectories $\left\{\bu(t)\right\}_{t=1}^\infty$.
This has been developed by Crisanti et al. \cite{Crisanti} for a continuous time version (without thresholds) and
by Molgedey et al. \cite{Molgedey} for a discrete time version. Both approaches lead obviously to the same equations.
But they also deal with the \textit{same type of convergence} namely \textit{weak convergence}. As said
above, the idea below the mean field approach is to have informations about the ``average'' behavior
of the dynamical system (\ref{DNN}). This is what we have obtained, but in a very rough sense. The equations
(\ref{SKd1}),(\ref{SKd2}),(\ref{SKd3}) tell us about the evolution of the average value of $u(t)$ when
the average is performed over \textit{infinitely many realizations of the disorder}. But, weak convergence
\textit{does not give any information about \underline{one} typical system whose size tends to infinity}.
For this, one needs a stronger convergence, the almost-sure convergence\footnote{Almost sure convergence
corresponds to the statistical physics notion of \textit{self averaging}. The empirical average
of a quantity in one realization of the disorder converges with probability one to the average
of this quantity over the disorder}. The large deviations
approach developed in chapter II will allow us to show the almost sure convergence
(and incidentally that the equations (\ref{SKd1}),(\ref{SKd2}),(\ref{SKd3}), derived from a
``questionable'' Ansatz, are correct).\\

Let us now discuss these equations, their solutions and their interpretation.
One remarks firstly that,  for $t=t'$, the equation of $\Delta(t,t)$ is the equation of (\ref{SKd2}) for
the variance $v(t)$ (as expected since  $\Delta(t,t)=v(t)$). Also, $\Delta(t,t') \leq v(t)$.
 One can therefore write the equation
(\ref{SKd3}) in the form
\beq\label{SKd4}
\Delta(t+1,t'+1)=H_{g,\bJ,J,\bt,\st}\left[\Delta(t,t')\right] \deq J^2 C(t,t')+\sigma^2
\eeq
\nid where  $H$ is defined only when $\Delta^2(t,t') \leq v(t)v(t')$ and is given by (\ref{SKd3}).\\

Having these equations in hand, the idea is now to study the reduced dynamical system  (\ref{SKd1}),(\ref{SKd2}),(\ref{SKd3})
and to infer informations about the typical dynamics of (\ref{DNN}). More precisely, we are interested
in the time asymptotic that corresponds to a stationary regime of (\ref{SKd1}),(\ref{SKd2}),(\ref{SKd3}).
The stationary equations are given by:
\bea\label{SK1}
\mu&=&\bJ m+\bt\\
m&=&
\int_{-\infty}^{+\infty} \frac{e^{-\frac{h^2}{2}}}{\sqrt{2\pi}}f(h\sqrt{J^2q+\sigma^2_\theta}
+\bJ m + \bt)dh \nonumber
\eea
\bea\label{SK2}
v&=&J^2q+\sigma^2_\theta
 \\
q&=&
\int_{-\infty}^{+\infty} \frac{e^{-\frac{h^2}{2}}}{\sqrt{2\pi}}
f^2(h\sqrt{J^2q+\sigma^2_\theta}
+\bJ m + \bt)dh \nonumber
\eea
\bea\label{SK3}
\Delta(t-t')&=&J^2 C(t-t')+\sigma^2_\theta=H_{g,\bJ,J,\bt,\st}\left[\Delta(t-t')\right]\\
C(t-t')&=&
\int_{-\infty}^{+\infty}\int_{-\infty}^{+\infty}
 \frac{e^{-\frac{h^2}{2}}}{\sqrt{2\pi}}
\frac{e^{-\frac{h'^2}{2}}}{\sqrt{2\pi}}
f\left(
\frac{\sqrt{v^2-\Delta(t-t')}}{\sqrt{v}}h+
\frac{\Delta(t-t')}{\sqrt{v}}h'+ \mu\right)
f\left(h'\sqrt{v}+\mu \right)
dhdh' \nonumber
\eea
These equations give important informations about the statistical
behavior of the model (\ref{DNN}) with an increasing accuracy
when the size increases. Moreover,  $m,q,\Delta$ act
as ``order parameters'' allowing us to distinguish different
dynamical regimes and to draw an effective bifurcation map in
the space of the macroscopic parameters. Let us list a few
results \cite{PD},\cite{EPL},\cite{These},\cite{JP}.

\ben

\item In the absolutely stable regime the fixed point is a random variable.
One checks numerically that the corresponding distribution of the local
fields is Gaussian and the equations (\ref{SK1}), (\ref{SK2})
give the mean and variance of this distribution with a very good accuracy.
For fixed $N$ the empirical mean and variance computed over a large
number of networks is close to the theoretical values. Moreover, 
the statistical dispersion of these empirical values decreases as
$N$ grows (as expected from the almost-sure convergence proved in Chapter II).

\item The equations (\ref{SK1}), (\ref{SK2}), have \textit{several solutions} 
in some regions of the macroscopic parameter space. More precisely,
they exhibit \textit{saddle-node} bifurcations. The critical values
where saddle-node bifurcations occur in the space of macroscopic parameters
can be computed from equations (\ref{SK1}), (\ref{SK2}).
It is remarkable that
these bifurcations have a direct correspondence with the saddle-node
bifurcations observed in the system (\ref{DNN}) in the following sense.

On one hand if one fixes the parameters $\bt,\st,\bJ$ in a region where
the mean field equations predict a saddle-node bifurcation as $g$ increases 
one observes indeed (in general) saddle-node bifurcations in the system (\ref{DNN}).
Of course, the exact $g$ value where the bifurcation occurs is random
and depends of the actual realization of the disorder. However, if one performs
a statistical analysis of these values, one finds an average value close to
the value predicted by the mean field equations. Moreover, the empirical variance
decreases with the system size.

On the other hand, the various fixed points appearing from saddle-node bifurcations in
the dynamical system (\ref{DNN}) are also random. But the coordinates of
these points (in the $u_i$ space)  are distributed according to a Gaussian distribution
whose mean and variance are in good agreement with the value obtained from
the fixed points of (\ref{SK1}), (\ref{SK2}).

As a conclusion, the analysis of the fixed points of (\ref{SK1}), (\ref{SK2}) allow
us to draw a bifurcation map in the macroscopic parameter space giving
the average $g$ value where saddle-node bifurcations occur, for a given value
of the parameters $\bt,\st,\bJ$. It gives also the probability distribution of
the corresponding fixed point in the dynamical system (\ref{DNN}).

\item Once we know the statistical distribution of the fixed points,
one can compute a destabilization condition by estimating the spectral radius of the
Jacobian matrix in the same way as we did above (but now the distribution
of eigenvalues depends on the distribution of the fixed points \cite{EPL}).
This condition is given by:

\beq\label{g0gen}
g_0\rho(\cJ \Lambda(\bxs))=1
\eeq  

\nid where $\rho()$ is the spectral radius, $\bxs$ the fixed point, and $\Lambda$ the diagonal
matrix $\Lambda_{ij}(\bxs)=f'(u^\ast_i)\delta_{ij}$. This condition generalizes (\ref{g0})
since for $f(u)=tanh(gu)$, $\bxs=0$ and $\Lambda(0)$ is the identity matrix.
From this one obtains the average $g$ value where destabilization occurs, for a given value
of the parameters $\bt,\st,\bJ$. Moreover, the Jacobian matrix as similar spectral 
properties as in the case $\bF(\bu;\lambda) = - \bF(-\bu;\lambda)$ and 
when a fixed point destabilizes it does this (generically) by a Hopf bifurcation.
Therefore the mean field equations allow us to draw a bifurcation map in the macroscopic parameter space giving
the average $g$ value where a Hopf bifurcations occurs (see ref. \cite{PD}).

\item A finer analysis of the complete set of equations ((\ref{SK1}), (\ref{SK2}),(\ref{SK3})),
 and especially of the equation for the time covariance (\ref{SKd4})), reveals that there are in fact two regimes.
The equation (\ref{SK3}) admits  always the solution $\Delta=v$. This solution
is stable for the map  (\ref{SKd4}) if $\frac{dH}{d\Delta}<1$. The destabilization condition is therefore given 
by\footnote{The equations ((\ref{SK1}), (\ref{SK2})) are similar to the Sherrington-Kirkpatrick equations describing
the Sherrington-Kirkpatrick spin glass model\cite{SK} at high temperature, while the equation (\cite{AT})
corresponds to the De Almeida-Thouless line. A detailed discussion of this aspect has been done in\cite{EPL,JP}.}:
\beq \label{AT}
\frac{dH}{d\Delta}(\Delta=v)=J^2
\int_{-\infty}^{+\infty} \frac{e^{-\frac{h^2}{2}}}{\sqrt{2\pi}}f'^2(h\sqrt{J^2q+\sigma^2_\theta}
+\bJ m + \bt)dh=1
\eeq
This equation defines in the space of macroscopic parameters $\left(g,\bJ,\bt,\st\right)$
a codimension 1 manifold which separates this space into two regions.

In the region corresponding to
$\frac{dH}{d\Delta}<1$ the  solution $\Delta=v$  is stable and it is the only solution of (\ref{SK3}).
The asymptotic stochastic process described by the mean field equations is then
stochastically equivalent to the Gaussian process given by:
\beq \label{ProcStat}
\left\{
\baR{ccc}
Y(0)&=&X \\
Y(t+1)&=&Y(t)
\eaR
\right.
\eeq
\nid where $X$ is a Gaussian random variable $\cN(\mu,v)$. Henceforth, $Y$
is a process with almost-surely constant trajectories. Its interpretation is easy:
it corresponds to a regime of (\ref{DNN}) where we have only fixed points.

In the other region, one can write $ v\delta(t,t')+\Delta^*(1-\delta(t,t'))=(v-\Delta^*)\delta(t,t')+\Delta^*$.
Consequently, the asymptotic stochastic process described by the mean field equation 
is stochastically equivalent to the Gaussian process given by: 
\beq\label{ProcDyn}
\left\{
\baR{ccc}
Z(0)&=&X \\
Z(t+1)&=&Z(t)+B(t)
\eaR
\right.
\eeq
\nid where $X$ is a Gaussian random variable $\cN(\mu,\Delta^\ast)$
and where $B(t)$ is  a \textit{white noise} with zero mean and variance $(v-\Delta^*)$.
$Z(t)$ is therefore the superimposition of a \textit{process with almost-surely constant
trajectories plus a white noise}.

It is also remarkable that the equation (\ref{AT}) corresponds exactly to the equation
of destabilization of the fixed point. We conclude therefore that the crossing
of the manifold (\ref{AT}) corresponds, in the infinite system, to a sharp transition from fixed point to
infinite dimensional chaos. Note however that this ``manifold`` is a very rough representation of the edge of chaos for finite size systems.
Indeed, it is known \cite{GT,MT} that in the transition to chaos by quasi periodicity, the edge of chaos
has a fractal structure corresponding to the intersections of Arnold tongues.

\een

The theoretical results described in the sections give
a fairly good description of the various dynamical regimes generically exhibited
by (\ref{DNN}). However, mean field equations have the drawback to hold only when the size
of the system tends to infinity. And we have just seen that this limit is rather poor
(either fixed points or white noise). Therefore, though  mean field equations are 
a good guideline for describing the statistical behavior of (\ref{DNN}) they miss
a lot of important aspects: intermediate regimes between fixed points and chaos,
dynamical properties of a \textit{given realization} of the network, etc.
In the next section we depart from the rough vision provided by the mean field theory and
develop two aspects drastically related to the \textit{finite size} system.  

%
%
%
%
%
%
%
%
%
\begin{figure}
\begin{center}
\includegraphics[height=10cm,width=12cm]{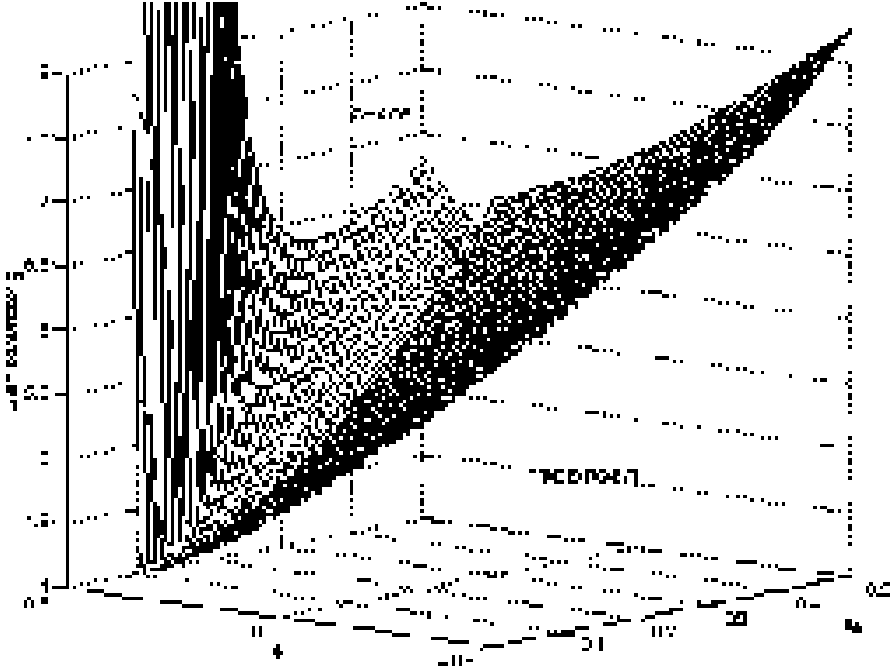}
\caption{\footnotesize \label{Fbifmap}
An example of bifurcation map. The  surface drawn in the parameter space $\bt,\st,g$ corresponds
to the sharp transition from fixed point to chaos, obtained from the mean field equations in the thermodynamic limit. 
 }
\end{center}
\end{figure}

\ssu{Hebbian learning effects.}\label{App}

Let us now  consider the effect of \textit{Hebbian learning} on the dynamical 
system (\ref{DNN}). For this, we return back to the recipes discussed in the section \ref{Hebb}.
Learning is based on the modification of synaptic connections between neurons. 
In the present context, this is interpreted as a a slow
evolution of the synapses  $J_{ij}$ when the network is submitted to 
a pattern  that one would like to ``teach'' to the network. In our model, a pattern 
is a vector  $\bxi = \lbrace \xi_i \rbrace_{i=1..N}$ and a presentation
consists in adding the vector $\bxi$ to
the vector of thresholds $\Bth$ (i.e. $\theta_i \to \theta_i + \xi_i, \ i=1 \dots N$).

Several (many) learning rules can be proposed, based on the recipes presented in the section
\ref{Hebb} (see for example \cite{Dauce0}). A straightforward implementation, very
similar to equation (\ref{HebbEvol}) is:

\begin{equation} \label{Papp}
J_{ij}(t+1) = \lambda J_{ij}(t) +\frac{\alpha}{N}(x_i(t+1)-\eta) \times (x_j(t)-\eta) \times H(x_j(t)-\eta)\qquad i,j = 1..N ; \qquad T>1 
\end{equation}

The parameter $0 < \lambda \leq 1$ corresponds to a decay of the synapse when it is not used.
$H$ is the Heaviside function ($H(x)=1$ if $x>0$ and $0$ otherwise). $\eta$ defines a level
of activity allowing us to decide whether a neuron is ``active'' at time $t$ ($x(t) \geq \eta$)
or ``silent'' ($x(t) < \eta$). Consequently, the term $(x_i(t+1)-\eta)(x_j(t)-\eta)H(x_j(t)-\eta)$ 
 corresponds to modifying the synapse only if the  post- synaptic neuron is active.
This corresponds to the  fourth\footnote{We have removed the condition that the
pre- synaptic neuron is active. Indeed, in this naive model, a term 
$(x_i(t+1)-\eta) \times H(x_i(t+1)-\eta) \times (x_j(t)-\eta) \times H(x_j(t)-\eta)$,
always positive or zero, would lead to an increase of the synapses linking active neurons and to an exponential  decay of the other synapses
toward $0$.
Hence we would rapidly have a network composed by positive synapses only, with a value increasing in time. Thus, rapidly,
all active neurons would saturate.} ``recipe'' in the section \ref{Hebb}. Finally,  
a weight $J_{ij}$ cannot change its sign
(this corresponds to demanding that a synapse
cannot switch from excitatory to inhibitory or vice-versa). 

On biological grounds, the learning rule (\ref{Papp}) can be interpreted as follows.
The synaptic weight $J_{ij}$ connects the neuron $j$ to the neuron $i$ and the output 
signal emitted by $j$ at time $t$ is transmitted to the neuron $i$ at the next time step
with the weight  $J_{ij}$. Let us assume that $\lambda=1$ (no forgetting). Then,
the learning rule has the effect of enhancing the synaptic strength $J_{ij}$ 
if the  neuron $j$ is active at time $t$ and if the  neuron $i$ is active at time  $t+1$.
On the other hand, if $j$ is active at time $t$ and the neuron $i$ is inactive then the synapse decreases. \\

The joint evolution of  (\ref{DNN}) and (\ref{Papp}) occurs as follows. 
The initial couplings and thresholds $J^0_{ij}, \theta^0_i$
are fixed to an initial random value determined by the probability distributions
 (\ref{Jij}), (\ref{theta}), for a determined value of the macroscopic parameters
$\left(\bar{J},\bar{\theta}, \sigma_{\rm \theta}^2\right)$. These parameters
and $g$ are fixed such that the corresponding dynamics is {\it chaotic}.
The values of these parameters can be roughly determined from the bifurcation
map described above (see Fig. \ref{Fbifmap}).

After a sufficiently long time such that the neurons dynamics 
has ``reached'' its chaotic attractor, one presents a pattern $\bxi$.
This means that one
modifies the thresholds: $\theta^1_i = \theta^0_i + \xi_i$. 
The weight $J_{ij}$ are not modified at this stage. The pattern is
a random vector $\bxi$ whose entries are independent, identically distributed,
Gaussian, with a mean $\bxi$ and a  variance $\sxi$. Henceforth, each
neuron feels an effective threshold $\theta^1_i = \theta^0_i + \xi_i$.
This modifies the dynamics. However, from the macroscopic
parameters point of view, this amounts to have the transformation  $\bt \to \bt+\bxi$, $\st \to \st+\sxi$.
Hence, it is still possible to know the average behavior of the perturbed system
by using the bifurcation map. In the experiments described below, the pattern
is chosen such that the perturbed dynamics \textit{remains chaotic}.
Then one iterates the learning procedure (\ref{Papp}). The stimulus
is always present. Once the learning phase is finished
one removes the stimulus $\bxi$ (i.e. the thresholds are reset to their initial value
$\theta^0_i$). \\

What is the effect of the learning rule (\ref{Papp}) (1) on the neurons dynamics ? (2) on the synapses ?
The typical effects on the neurons is depicted in Fig. \ref{Fapp}. One observes generically an inverse quasi periodicity route.
Namely, starting from a chaotic attractor,
the modification of the $J_{ij}$'s by the Hebbian rule (\ref{Papp}) 
leads first to a  ${\cal T}^2$ torus, then to a limit cycle and, finally,
to a fixed point (with possibly a crossing of several Arnold tongues leading to temporary synchronizations).
 Thus, too long a learning phase basically ``kills'' the dynamical
activity (see Fig. \ref{Fapp}).

%
%
%
\begin{figure}[ht]
\begin{center}
\includegraphics[width=8cm]{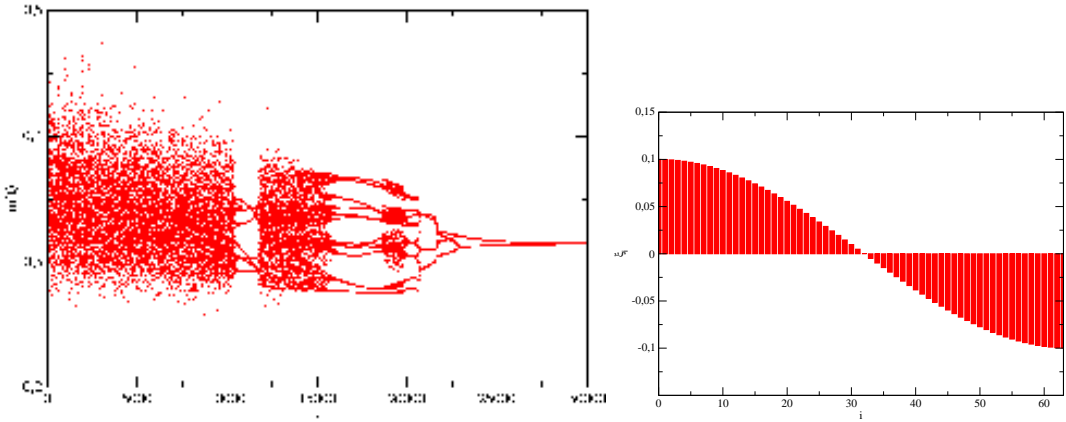}
\includegraphics[width=6cm]{Motif.eps}
\vspace{1cm}
\caption{\footnotesize \label{Fapp}
Fig. \ref{Fapp}a Inverse quasi periodicity route induced by learning. In this simulation $N=64$, $g=8$
$\alpha=10^{-4}$, $\eta=0.5$, $\lambda=1$ (no forgetting). $30000$ learning steps are represented. The plotted quantity is the average value
of the output states $m(t)=\frac{1}{N}\sum_{i=1}^N x_i(t)$. Fig. \ref{Fapp}b Graphical representation of the learned pattern. 
}
\end{center}
\end{figure}
%
%
%
%

Now, assume that we stop learning when the systems is in an intermediate phase (e.g. quasi periodic
or periodic). Different results are possible depending on the time where we stop learning but also
on the pattern, the spontaneous dynamics, the learning rule \etc. Nevertheless, it is possible to observe
the following phenomenon, reminiscent of Freeman's paradigm.  In some cases,  {\it removing the pattern when the activity of the network+pattern is periodic gives back a strange attractor.}
 Then, {\it a new presentation of
the pattern leads back to the limit cycle}. 
An example is given in Fig. \ref{Fapp2}. The initial regime is chaotic (Fig. \ref{Fapp2}a I)
and presenting the pattern does not change the chaotic nature of the dynamics (Fig. \ref{Fapp2}a II). Obviously, this changes the 
attractor, but nothing significant happens. In particular a  glance to Fig. \ref{Fapp2}b does not reveal any
clear cut effect induced by the pattern presentation, \textit{before} learning. The situation is drastically different \textit{after} learning.
If one stops the learning phase corresponding to the figure  \ref{Fapp} after 11.000 learning
steps one ends with a periodic attractor (Fig. \ref{Fapp2}a III). Then, removing the pattern leads back to chaos (Fig. \ref{Fapp2}a IV).
Again, the form of the attractor is different from (Fig. \ref{Fapp2}a I,II) but observing the dynamics does not tell us
that learning has been performed. However, a new presentation of the pattern induces a sharp reduction of the dynamics
onto the limit cycle (Fig. \ref{Fapp2}b). Since, the pattern presentation didn't lead to this effect before learning, this property
has been acquired \textit{via learning}. 

Consequently, in this situation,
the learning process associates to a given pattern a {\it dynamical} pattern,
and recognition is manifested by a dynamical reduction from chaos to
the associated dynamical pattern. We have therefore another possible interpretation for
the loose statement given in the section \ref{Hebb}: ``a pattern is memorized
if  the neural networks has \textit{acquired}, via learning, the capacity to dynamically evolve towards a ``state'' ``associated
to the pattern'', provided that it was ``suitably prepared'' ''. Here the ``state'' is an attractor\footnote{To be more
precise the state is an ergodic  probability measure with support on the attractor. A natural choice is the SRB
measure introduced in the appendix. Thus, in the present context, the notion of state
is closer to statistical mechanics framework where a (macro)state is a probability measure on
the phase space.}, and ``suitably prepared''
means that we add the pattern $\bxi$ to the vector of thresholds $\Bth$ (pattern presentation). Hence, the effect of learning
is quite different from the Hebb-Hopfield learning where a pattern is associated to a fixed point and ``suitably prepared'' means
choosing an initial condition in the attraction basin of the pattern.  

%
%
%
\begin{figure}[ht]
\begin{center}
\includegraphics[width=7cm]{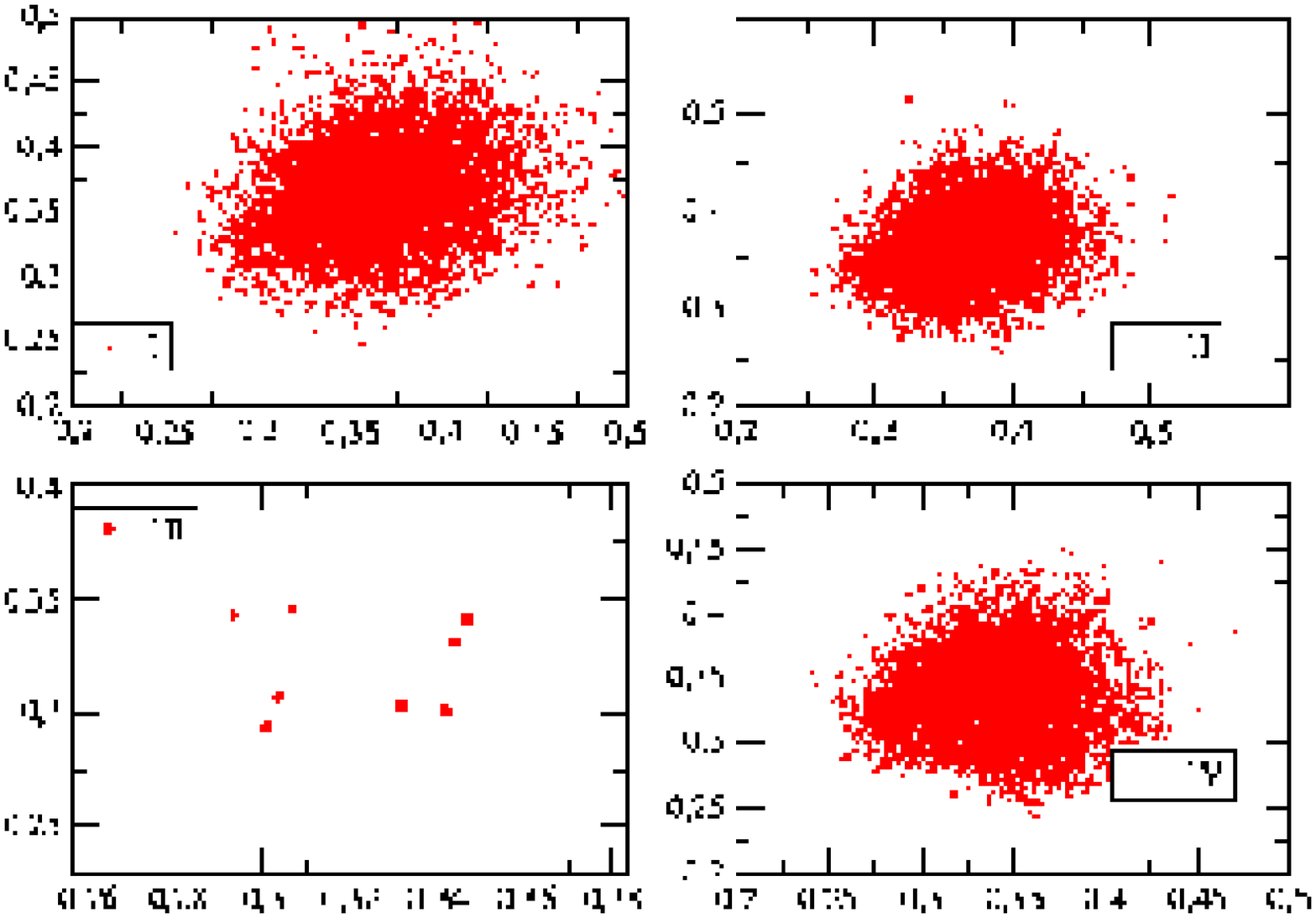}
\hspace{1cm}
\includegraphics[width=7cm]{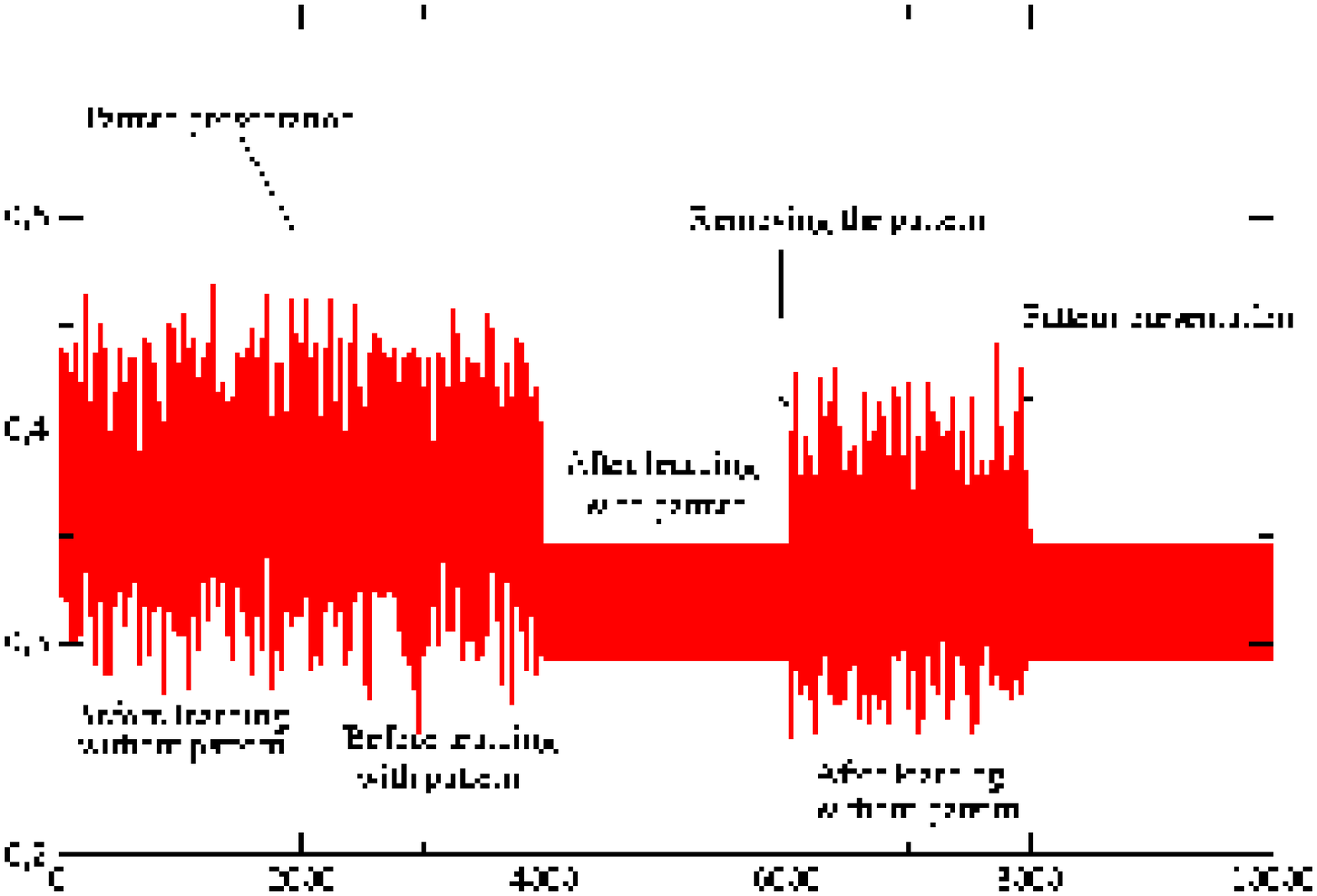}
\vspace{1cm}
\caption{\footnotesize \label{Fapp2}
 Learning and effect of a pattern presentation after learning. Fig. \ref{Fapp2}a Attractors. I. Attractor before learning and before the pattern presentation.
II . Attractor before learning when the pattern is presented. III. Attractor after 11000 learning steps with the pattern.
IV. Attractor  after 3000 learning steps without the pattern. Fig. \ref{Fapp2}b Time trajectories.
 }
\end{center}
\end{figure}
%
%
%
%

The remarkable fact is that the learning dynamics has lead the system in a
state different from the initial one. Without excitation by the stimulus,
the neuron dynamics is chaotic and there is no apparent difference
between this case and the situation before learning. More precisely,
certainly the learning phase has changed the characteristics of the strange 
attractor, but this change
does not tell us anything about the fact that \textit{an information has been
encoded in the network}. This fact is revealed only if one presents the stimulus
and its manifestation is drastic (remember that the presentation of the pattern
\textit{before} learning didn't reduce the dynamic).

This observation raises however many questions in particular with respect to the robustness
of this behavior, and the mechanisms leading to it. We postpone these questions to the end
of the section and we investigate now the second point listed above. What is the effect of the learning rule (\ref{Papp})  on the synapses ?\\

The remarkable fact is that no clear cut changes are observed even if the learning phase is  long.
Obviously the $J_{ij}$'s are modified by the learning rule (\ref{Papp}) but there is no
striking change in the structure of the matrix or in the histogram of the $J_{ij}$'s, 
even if these infinitesimal changes in the $J_{ij}$'s
are \textit{sufficient to modify the dynamics}. An example is given
in Fig. \ref{FappJij}. After 11000 times steps, the dynamics settle onto a limit cycle
but the histograms and the matrix $\cJ$ looks very much like the initial one. To observe
significant changes one has to iterate the learning phase far beyond the time where the dynamics
has died. The  Fig. \ref{FappJij} shows the distribution of the $J_{ij}$'s and the matrix
after $10^6$ time steps. Here a clear modification is revealed. The weights emitted by some neurons
have increased, while the others have not been modified. But the time scale to observe a significant change
is substantially larger than the time necessary to have a dynamical reduction.

%
%
%
\begin{figure}[ht]
\begin{center}
\includegraphics[width=7cm]{HistoJijvsapp.eps}
\hspace{1cm}
\includegraphics[width=7cm]{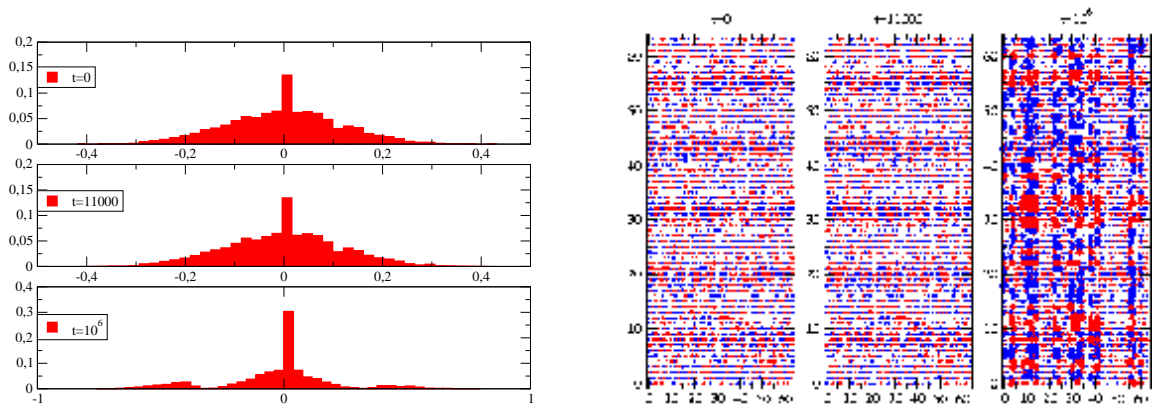}
\vspace{1cm}
\caption
{\footnotesize \label{FappJij}
Fig. \ref{FappJij} Effect of learning on the synapses. 
\ref{FappJij}a Histogram of the $J_{ij}$'s.
I Initial. II After 11000 learning steps. III. After $10^6$ learning steps.
 \ref{FappJij}b Matrix $\cJ$ I Initial. II After 11000 learning steps. III. After $10^6$ learning steps.
The radius of the circle is proportional to the absolute value of the synapses. Blue circles correspond to inhibitory synapses
and red circles to excitatory synapses.  }
\end{center}
\end{figure}
%
%
%
%

From the theoretical point of view, one is far from the degree of understanding of the
model without learning and there is no quantitative theory allowing to predict and control the effect of learning.
  As a matter of fact, the dynamic mean field cannot be applied,
since the learning  dynamics (\ref{Papp}) creates correlations between the weights.
However, one can give the following heuristic explanation of the phenomenon.
First, the mean field approach developed in the section \ref{MFT} has left us with a somewhat misleading
picture of the neural network. Indeed, in the mean field treatment all neurons are equivalent and thus
they have all the same level of activity. This is correct if one considers the activity of the neurons \textit{averaged}
over a large number of networks. But the situation is different when one considers \textit{one particular realisation
of the $J_{ij}$'s}. In the figure \ref{FAct} we have represented the time averaged (see the appendix)  value of $u_i$ and $x_i$,
in the various phases of the learning procedure. In the first row we have represented the average output activity $\left<x_i\right>$
only for the neurons such that $\left<x_i\right> < \eta = \frac{1}{2}$. Thus these neurons are (on average) active neurons.
Though the learning rule (\ref{Papp}) uses the instantaneous activity of the neurons and not the average (for a variant of this
rule, see eq. (\ref{app}) below), this representation gives us an indication of the repartition of ``active'' and ``silent'' neurons.
This repartition is clearly not uniform, since it results from the interplay of the neuronal connections $J_{ij}$'s and the non linearity
of the transfer function (this interplay and the resulting properties are discussed in more details in the section  \ref{RepLin}). 

%
%
%
\begin{figure}[ht]
\begin{center}
\includegraphics[width=7cm]{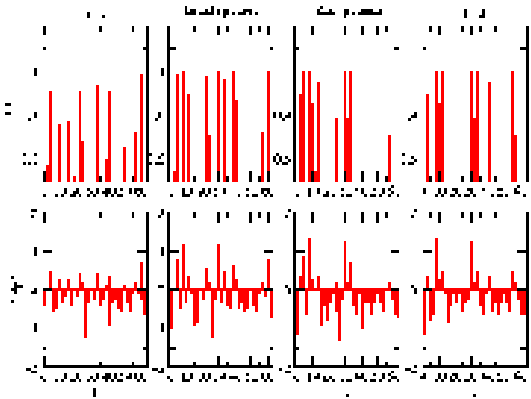}
\vspace{1cm}
\caption
{\footnotesize \label{FAct}
Fig. \ref{FAct} Effect of learning on the activity of the neurons. First row. Average output activity $\left<x_i\right>$ of the neurons 
Second row. Average value of the local field $\left<u_i\right>$. }
\end{center}
\end{figure}
%
%
%
%

The pattern presentation as a direct but weak effect on the local fields and an even weaker effect on the activity
(the pattern is represented in Fig. \ref{Papp}b; note that obviously $\left<x_i\right>=\left<f(u_i)\right> \neq f(\left<u_i\right>)$).
The learning rule selects then the active neurons and modify their outgoing synapses in the following way. Assume that $j$
is an active neuron. Then if $i$ is active $J_{ij}$ increases.  Thus $J_{ij}$ becomes more excitatory if it  positive and it
becomes less inhibitory if it is negative. On the opposite, if  $i$ is ``silent'' then $J_{ij}$ decreases. Thus 
$J_{ij}$ becomes less excitatory if it  positive and it becomes more inhibitory if it is negative. In all other cases
$J_{ij}$ stays constant (for $\lambda=1$). If we admit that one step of learning has a small influence on the level
of activity of the neurons\footnote{This can be assumed away from bifurcations point (see section \ref{WCNN}) but it is incorrect near a bifurcation.}
then the picture remains essentially the same at the next learning step. Thus, in this rough picture, we have a set of active neurons whose outgoing
synapse gradually evolves. The excitatory links towards active neurons become more and more excitatory, the  
inhibitory links towards silent neurons become more and more inhibitory; in the same time 
the excitatory (inhibitory) links towards silent (active) neurons decay to zero and eventually vanish since a weight cannot change
its sign. Consequently, a very long learning phase leads to an histogram such has Fig. \ref{FappJij}a III, with a high peak at zero,
two bumps corresponding to excitatory and inhibitory synapses with a large\footnote{Note that there is 
no upper or lower bound on the synapses in the learning rule (for a variant see eq. (\ref{app})). Thus, the
modified synapses diverge asymptotically.}
 absolute value, and finally a background of synapses
that have essentially not been modified during learning. The active neurons become ``hubs'' for the dynamics, in the sense
that they have a relatively large connectivity and some weights with big absolute values. This corresponds to
the vertical bands with big circles revealed in
Fig. \ref{FappJij}c. One also has horizontal bands with mainly either red or blue big circles. The ``red lines'' corresponds to the links received
by active neurons coming from active neurons, while the blue ones corresponds to the links received by silent neurons also coming from active neurons.
 It is thus remarkable that the Hebbian like learning 
rule (\ref{Papp}) leads to a structuration of the network \footnote{The Hebbian learning generates in fact  small word structures, as shown
in \cite{Berry}. This is basically because Hebbian learning builds ``shortcuts''.
If two neurons are not wired (thus far apart from the synaptic graph point of view) but if they are ``synchronized'',
(e.g. $i$ is active at time $t+1$ whenever $j$ is active at time $t$)
then the learning rule will construct a synapse between them. Note that, as discussed in the section \ref{WCNN} and
in the next section, 2 neurons can be synchronized even i they are \textit{not} wired, by the mere effect of the non linear dynamics
(see section \ref{RepLin} for a discussion of this aspect in chaotic networks).} into ``pools'' of neurons. 
Finally, from the dynamical point of view, since active neurons become more and saturated the dynamics converges to a fixed point.\\

This pictures gives us a fair understanding of the (somewhat trivial) behavior of the system when learning is performed on long time
scales. But, what about the small time scales and what about the inverse quasi periodicity route ?    
For this, let us use the wisdom acquired in the preceding subsections.
 The dynamical system (\ref{DNN}) can be represented by a (randomly) chosen point in a space of parameters with $\cN=N^2 + N + 1 $ dimensions.
In this space, many ``critical'' manifolds exist, whose crossing corresponds to various type of bifurcations.
As discussed above a complete investigations of this space is impossible but standard results in dynamical systems 
theory, completed with numerical simulations and mean field theory have allowed us to roughly locate the ``boundary
of chaos'' as a function of the macroscopic parameters. Note however that the ``bifurcation manifold`` obtained
from the mean field approach in the figure \ref{Fbifmap} is a very rough representation of the edge of chaos.
Indeed, it is known \cite{GT,MT} that in the transition to chaos by quasi periodicity, the edge of chaos
has a fractal structure corresponding to the intersections of Arnold tongues. Thus the transition is usually not
sharp when one modifies the parameters but one has succession of phase locking with various rotation numbers
and chaos (see e.g. Fig. \ref{SpectreLyap}a and \ref{Fapp}). 

On the other hand, the manifold corresponding to the destabilization of
the fixed point has a nicer structure. It is indeed given by eq. (\ref{g0gen}) $g_0\rho(\cJ \Lambda(\bxs))=1$.
 Now,  the learning dynamics corresponds to a motion of the representative point 
of the dynamical system in the subspace of synaptic weights, while the presentation or removal of the
pattern  correspond to a translation in the subspace of thresholds. 
 These motions  lead to bifurcations when crossing critical manifolds.
Consider now the destabilization condition  (\ref{g0gen}). Since learning has the effect
of slowly increasing the level of activity of active neurons (and inhibit more and more the silent ones),
the derivatives of the  transfer function of the neurons has a tendency (on average) to become smaller.
Thus, the entry of $\Lambda(\bx(t))$ become smaller on average. If we (roughly!) replace the condition
$g_0\rho(\cJ \Lambda(\bxs))=1$ by $g_0\rho(\cJ <\Lambda(\bx)>)=1$ and if we neglect the modifications
of the $J_{ij}$'s induced by learning, one sees that the  $g$ value to destabilize the network
\textit{increases} while learning is performed. Thus, the effective  motion induced by learning in the parameter space
corresponds to get closer and closer from the destabilization manifold, with an eventual crossing
when learning is to long. Finally, since for large $N$, the destabilization manifold and the edge of chaos are very close
one concludes that learning lead the system closer and closer from the edge of chaos.

What about presentation or removal of a pattern ? The learning rule (\ref{Papp}) depends
on the activity of the neurons.  Since the initial presentation 
of the pattern leads to changes in the distribution of the neuronal local fields $u_i$, this activity is (possibly slightly) modified
by a pattern presentation. From the parameter space point of view the pattern presentation
corresponds to a translation in the subspace of thresholds, in the direction of the vector
$\bxi$.  The whole learning phase is conditioned by the presence of the pattern.
It has the effect of increasing the numbers of saturated, $x_i=1$ or silent, $x_i=0$ neurons.
  The global effect is similar to having an effective threshold whose value grows
 during the learning phase, leading to the observed dynamical collapse.
Removing the pattern has in general the effect of reducing the width of the distribution 
of neural local fields and the number of saturated/silent neuron decreases. 
If the system is close to the edge of chaos (this happens when we stop learning slightly after the
reduction of chaos to a periodic or quasi periodic attractor)
 this can induce the drastic dynamical change observed. Thus, this scenario lead us to conclude that
 the learning dynamics leads the system "to the border of the chaos", 
{\it in a state where it is sensitive to the learned pattern} (i.e. a translation in the direction of the pattern (presentation, resp. removal)
induce the crossing of the border of the chaos). In some sense, the network has adapted itself to the pattern, via learning.
This has an interesting echo  with the adaptation condition (\ref{Adapt}) of the section
\ref{WCNN}. In particular, there should exist transversality conditions ensuring that the presentation/removal of
the pattern leads to a ``transverse crossing of the edge of chaos''.  \\

This discussion gives us interesting hints but is not entirely satisfactory. Firstly,
as already said, we don't have a real theory to validate this scenario. Also, we didn't discuss the effect
of presenting \textit{another} pattern, after the learning phase. More generally
all the discussion above dealt with a specific example of a specific rule. What about the genericity
of this result ? What about its robustness ? What happens if one changes the learning rule ? 

Actually, the rules (\ref{Papp}) is rather rough and not really robust. It has been introduced
as a straightforward implementation of the recipes in section \ref{Hebb} providing an interesting
pedagogical example. However, to have robust effects one needs to consider more elaborated rules.
Systematic investigations have been performed in \cite{Quoy0},\cite{Dauce0},\cite{NN}.
Various learning rules have been proposed, having the general form (\ref{HebbRuleGen})

\begin{equation} \label{app}
J_{ij}' = J_{ij} +\frac{\alpha}{N}\Gamma_{ij}\qquad i,j = 1..N ; 
\end{equation}
\nid where $\Gamma_{ij}$ may either depend on the value of the ``instantaneous'' pre- ($t$) and post- synaptic ($t+1$)
neuron or on averages such has $\Gamma_{ij}=m_i m_j$ or $\Gamma_{ij}=C_{ij}(1)$, where $C_{ij}(1)$ is the time 1 correlation
function between $j$ and $i$. In the case where $\Gamma_{ij}$ depends on average values, one has to consider two coupled dynamical systems.
A \textit{fast} one corresponding to the neurons
evolution  and a slow one corresponding to the evolution of the $J_{ij}$'s.
In the joint evolution one has then to wait that the fast neurons dynamics settle onto
its attractor before performing one learning step. 

The main observations above remain \cite{Manna,NN}. Moreover,
it is possible to improve the learning rule so that the response of the system to the
pattern in terms of chaos reduction is \textit{selective and specific}. 
Presenting another, completely distinct stimulus, does not lead to
a dynamical reduction. However, a weakly noisy version of the stimulus 
has this effect. Henceforth, this mechanism \textit{is robust to noise}.
It is also possible to learn several stimuli but the storage capacity
of the learning rule (\ref{app}) is weak. More elaborated versions
can reinforce the storage capacity \cite{Dauce0}.\\

These results are fascinating for they are the demonstration of an effect 
similar to Freeman's paradigm
(even one should take care when drawing biological conclusions from this simple model).
To the best of our knowledge this model is the first example of a  formal neural network
exhibiting this effect. However, one may can ask what are the potential applications of this.
Actually, one may complain that to observe this dynamical reduction one needs to somehow
``assist'' learning since one has to stop it before  the dynamics irremediably die. 
Also, learning left us with an association pattern/attractor, but how can we use  this ?
In fact, the more interesting observations are on one hand that the spontaneous dynamics
is chaotic and on the other hand that learning a given stimulus leads to a repartition
of active neurons that depend on the stimulus \cite{QD3}. Chaos allows the spontaneous
dynamics of the neural network to explore a wide range of ``possibilities'' each them
corresponding to a state of the network, while having neurons selectively responding
to stimuli/patterns can be used to perform tasks or make decision. In this sense, the
Neural network (\ref{DNN}) can be used as a first layer of a complex neural architecture.
This has been for example used in the training of an autonomous robot designed to 
adapt its motion to a random environment \cite{QD3}.

This subsection leave us with an interesting conclusion. The hebbian learning
rule (\ref{app}) allows us to store some information in the \textit{chaotic} neural network (\ref{DNN})
and this information can be somehow retrieved. But this leads to several natural questions :
How can a \textit{chaotic} network store and treat information ? Where is the ``learned'' information stored ?
Is there a way to see that such a network has learned something without presenting the pattern ?
The collapse effect is clearly a collective effect, but  this does not mean that all neurons play
the same role in the dynamics ? These questions will not be answered in this paper but 
in the next subsection we present a new
analytical tool that may, in the long term, be used to tackle such problems.
Recent developments have been recently made in this direction in \cite{EffetApp}.

\ssu{Influence of a time dependent input: signal propagation and linear response theory.}\label{RepLin}

Let us first return to the point raised in section \ref{MNN}. Since
synapses are used to transmit neural fluxes (spikes) from a neuron
to another one, the existence of synapses between a neuron (A) and another
one (B) is implicitly  attached to a notion of  ``influence'' or
causal and directed action.  However, as we saw, a neural network is a highly dynamical
object and its behavior is the result of complex
interplays between the neurons dynamics and the
synaptic network structure. Moreover, the neuron
$B$ receives usually synapses from many other
neurons, each them being ``influenced'' by
many other neurons, possibly acting on $A$, etc...
Thus the actual ``influence''
or action of A on B has to be considered dynamically
and in a global sense, by considering $A$ and $B$
not as isolated objects, but, instead, as entities
embedded in a system with a complex interwoven dynamical
evolution. In this context it is
easy to imagine examples where  there is a synapse from $A$ to $B$
but no clear cut influence, or, in the opposite, no synapse and nevertheless
an effective action.

Consider indeed the figure \ref{Fnet}. Neuron $1$ excites neurons $3$, but in the same
time it excites neuron $5$, which inhibits neuron $3$. What is the effective
action of $1$ on $3$ ? This clearly depends not only on the synaptic weight,
but also on the state of the neurons $1$,$3$,$5$. More generally, the spikes or signals
emitted by a neuron can follow different paths, and its effective influence
results from the contribution of all these paths. Actually, one can easily
figure out by a simple glance at figure \ref{Fnet} that feedback loops (that
is closed circuits in the synaptic graph) play an important role. However,
as pointed out several times in this chapter one has to consider topological
aspects (such as the feedback circuits) \textit{and} dynamical aspects.\\

%
%
%
\begin{figure}[ht]
\begin{center}
\includegraphics[width=5cm]{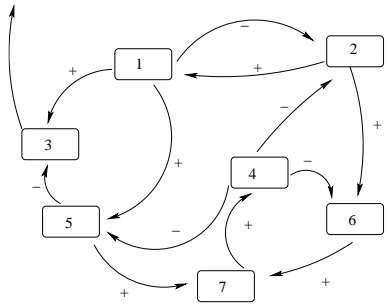}
\vspace{1cm}
\caption
{\footnotesize \label{Fnet} Example of network illustrating the effect
of feedback loops.}
\end{center}
\end{figure}
%
%
%
%
%
%

One way of doing this is to compute cross correllogramms. Indeed,
the time correlation function $C_{AB}(t)$ between the ``state''
of $A$ and the state of $B$ incorporates
the dynamical evolution and the effective
effects due to the neural network as a whole.
However, correlations functions do not really
provide causal information. Indeed, a strong
correlation between $A$ and $B$ at time
$t$ does not tell us if $A$
acts on $B$ or if $B$ acts on $A$ (note in particular
that $C_{AB}(t)=C_{BA}(-t)$).

Another way to measure a causal action consists in
exciting neuron $A$, say with a weak signal, and observe
the effects on $B$, e.g. by comparing its evolution
with and without the signal applied on $A$. We
shall give later on an explicit way to do this.
Nevertheless, there is a common wisdom, coming
from non equilibrium statistical mechanics, stating that the response of $B$ to a weak
perturbation on $A$ (linear response), \textit{if it exists}  should
 actually be a correlation function. This is
the celebrated fluctuation-dissipation theorem (FDT).
We shall however see below that the FDT may not hold
in simple neural networks models, due in particular
to saturation effects in the spike rate emission.

Finally, a natural choice for an excitatory signal
is a periodic signal, with a tunable frequency.
Thus, the response function, drawn versus frequency,
provides similar information as the complex 
susceptibility in physics. In particular, peaks
in the susceptibility corresponds to resonances,
that is a response of maximal amplitude. We shall
see below how these resonances can be used to provide
an effective, frequency dependent notion of network
structure. We shall also see how they incorporate
non linear effects in the dynamics even though
they are obtained in the context of linear response
theory. \\

With these ideas in mind consider the model (\ref{DNN})
in the \textit{chaotic regime} and assume that we superimpose upon the  state  $u_j(t)$ of the node $j$ a small external signal $\xi_j(t)$.
How does this signal propagate  inside the network ? Because of the sigmoidal shape of the transfer functions the answer 
depends crucially, not only on the connectivity of the network, but also on the value of the $u_k$'s.
 Assume, for the moment and for simplicity, that the time-dependent signal $\xi_j(t)$ has variations
substantially faster than the variations of $u_j$. Consider then the cases depicted in
Fig. \ref{FSat}.
 In the first case (a) the signal $\xi_j(t)$ is amplified by $f$, without distortion if  $\xi_j(t)$ is weak enough.
In the second case (Fig. \ref{FSat}b), it is damped and distorted by  the saturation of the sigmoid.
More generally, when considering the propagation of this signal from the node $j$ to some node $i$
one has to take into account the level of saturation of the nodes encountered in the path, but
the analysis is complicated by the fact that the nodes have their own dynamical evolution
 (Fig. \ref{chaine_relais}).
A mathematical formulation of this is given e.g. in eq. (\ref{chiij}) below. This shows once again that
the analysis of this signal propagation  must take into account   the topological structure
 of the graph as well as the nonlinear dynamics. 

 %
%
%
%
%
%
%
%
%
\begin{figure}[ht]
\includegraphics[height=5cm,width=5cm,clip=false]{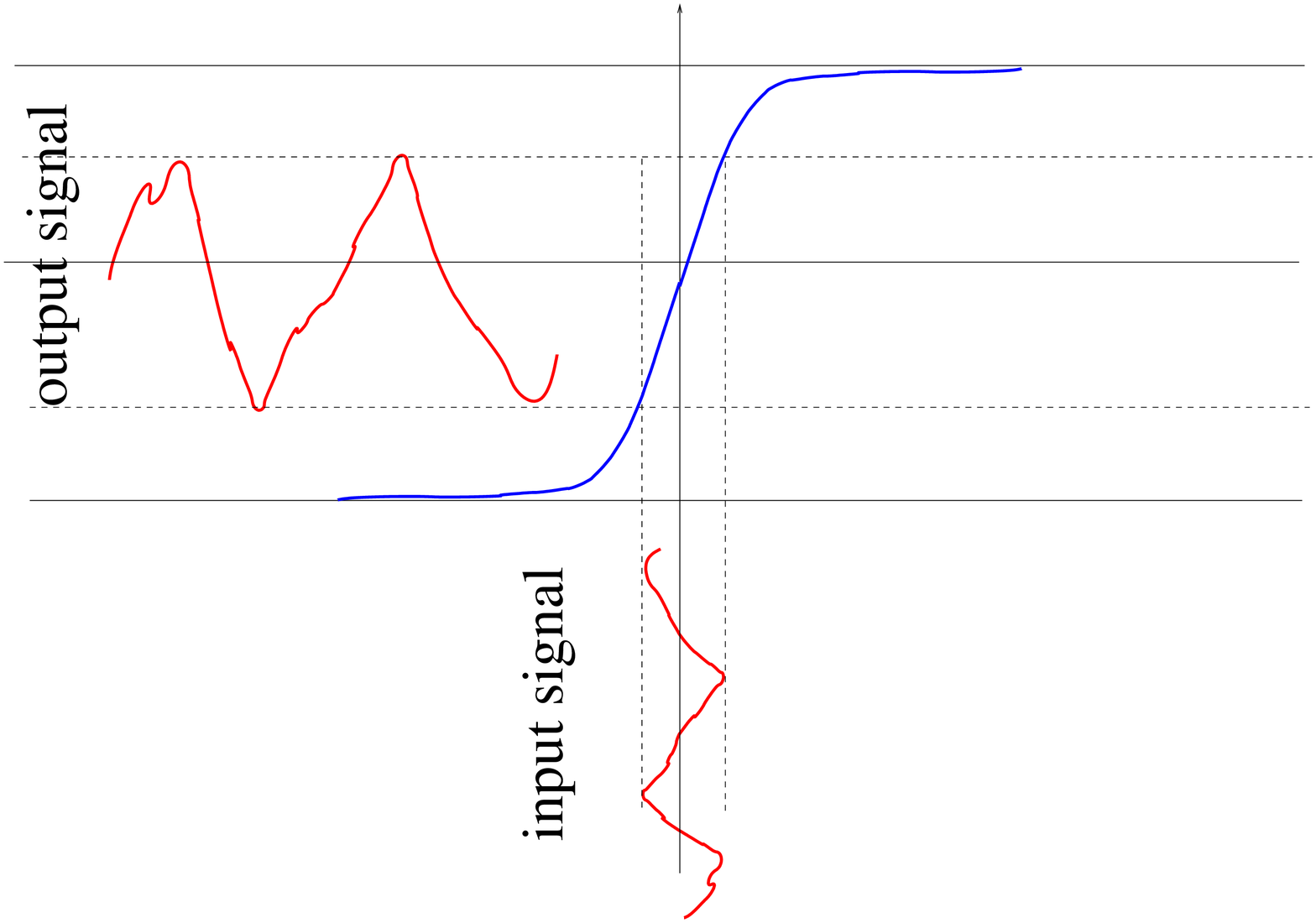}
\hspace{1cm}
\includegraphics[height=5cm,width=5cm,clip=false]{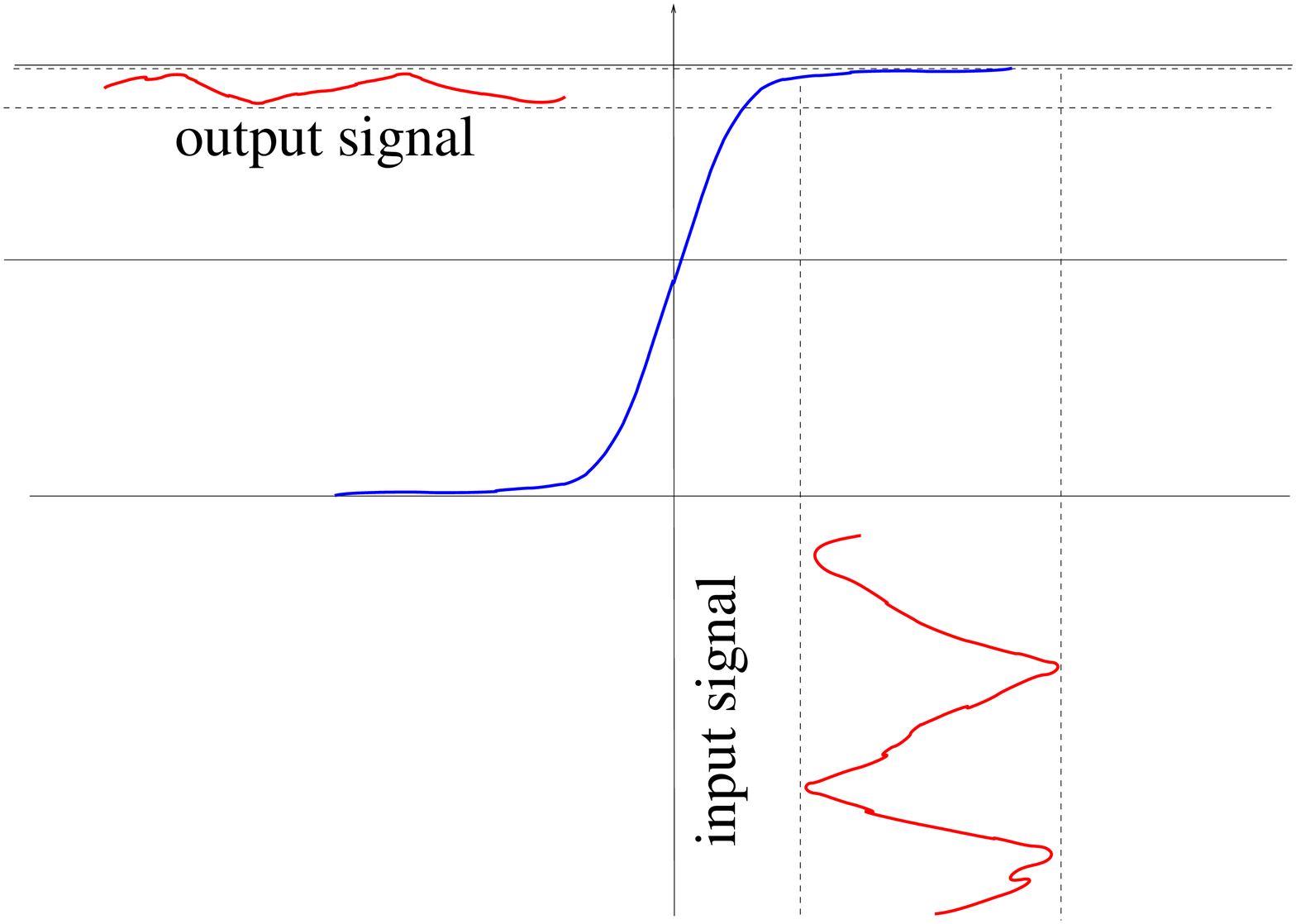}
\caption{\label{FSat} Nonlinear effects induced by a transfer function with a sigmoidal shape on signal
transmission. Fig. \ref{FSat}a. Amplification. Fig. \ref{FSat}b. Saturation.}
\end{figure}
%
%

%
%
%
\begin{figure}[ht]
\includegraphics[height=6cm,width=10cm,clip=false]{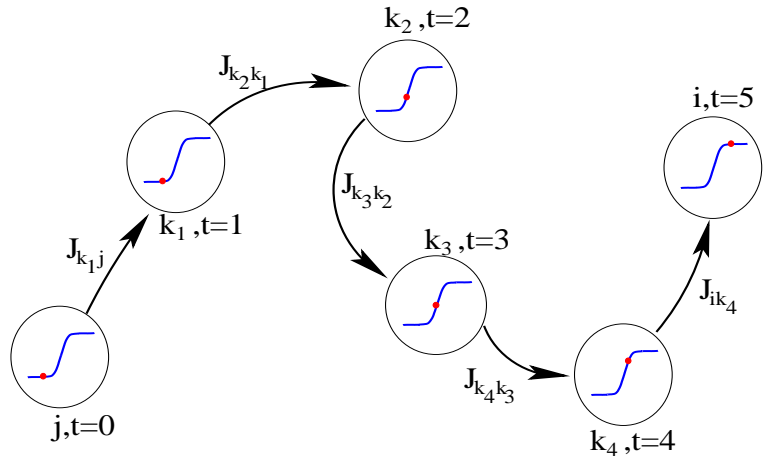}
\caption{\label{chaine_relais} The propagation of a signal along a path in the network
depends not only on the weights of the links but also on the level
of saturation of the nodes that the signal meets. The level of saturation
depends on the current state of the node (schematically represented as a
red point in the figure). This state evolves with time.}
\end{figure}

In this context we would like to measure the \textit{average} ``influence''
of neuron $A$ on neuron $B$ (namely how a weak
signal applied on $A$ perturbs on average the state of $B$), including
the effects of the non linear dynamics.
There is a natural notion of \underline{average} in chaotic systems 
such as (\ref{DNN}) related the so-called Sinai-Ruelle-Bowen measure $\rho$
(SRB) \cite{SRB} (see appendix)
 which is obtained as the (weak) limit of the Lebesgue measure $\mu$ under the dynamical evolution\footnote{
A crucial property is that a SRB measure
 has a density along the unstable manifolds,
 but it is singular in the directions transverse to the attractor.  
This feature is at the origin of the distinction between unstable 
and stable poles of the susceptibility (see below).}:
\beq \label{SRB}
\rho=\lim_{n \to +\infty} \bG^{n} \mu.
\eeq
In the following we will assume that all Lyapunov exponents are bounded away from zero.
  Then for each $\bu \in supp \rho$, 
where $supp \rho$ is the support of $\rho$, there exists a splitting 
$E^{(s)}_{\bu} \oplus E^{(u)}_\bu$ such that $E^{(u)}_\bu$,
 the unstable space, is locally tangent to the attractor (the local unstable manifold)
and $E^{(s)}_\bu$, the stable space, is transverse to the attractor (locally tangent to the local stable manifold).
 Let us emphasize that the stable and
unstable  spaces depend on $\bu$ (while the Lyapunov exponents are $\mu$ almost surely constant).
Let us consider a point $\bu$ on the attractor
and make a small perturbation $\delta_\bu$. This perturbation can be decomposed as
$\delta_\bu = \delta_\bu^u + \delta_\bu^s$ where $\delta_\bu^u \in E^{(u)}_{\bu}$ 
 and $\delta_\bu^s \in E^{(s)}_\bu$. 
$\delta_\bu^u$ is locally amplified with an exponential rate (given by the largest
positive Lyapunov exponent). On the other hand 
 $\delta_\bu^s $  is damped with an exponential speed (given by the smallest
negative Lyapunov exponent)

 Assume now that we superimpose a  signal of weak amplitude upon some of the ``membrane potentials'' ($u_k$) 
in such a way that the dynamics is still chaotic  (with only a tiny variation of the Lyapunov exponents). (This means
that the method of signal injection is intended to be non invasive).  
For simplicity,
we suppose  that the signal does not depend on the state of the system, but we can consider this generalisation without difficulty
(linear response still applies in this case, but the equations (\ref{dro},\ref{Chi}) do not hold anymore).
Denote by $\bxi$ the vector $\left\{\xi_i \right\}_{i=1}^N$. The new dynamical
system is described by the equation:
\beq \label{pert}
\tbu(t+1)=\bG\left[\tbu(t)\right]+\bxi(t)
\eeq 
The weak signal $\bxi(t)$ may be viewed as a small perturbation of the trajectories of the unperturbed system (\ref{DNN}).
  At each time this perturbation has a decomposition $\bxi(t) = \bxi^{(s)}(t)+\bxi^{(u)}(t) $ on the local stable and unstable spaces.
The stable component $\bxi^{(s)}(t)$ is exponentially damped. The unstable one $\bxi^{(u)}(t)$  is 
 amplified by the dynamics and then  scrambled by the nonlinear terms.  
Consequently, it is impossible to predict the long term effect of signal $\bxi(t)$ on the global dynamics.

This is true for  \textit{individual trajectories}.    However, the situation
is substantially different if one considers the \textit{average} effect of the signal, the average being performed with respect
to the SRB measure $\rho$ of the unperturbed system.  Indeed, as an application of the general theory~\cite{Ruelle2},  
 it has been established in \cite{CS1},\cite{CS2} that
the \textit{average} variation $\delta_{u_i}(t)$ of the membrane potential $u_i$ under the influence
of the signal is given, to the linear order, by:
\beq\label{dro}
\left< \tu_i(t) - u_i(t) \right>=\sum_{\sigma=0}^{\infty}\sum_{j}\chi_{ij}(\sigma)\xi_j(t-\sigma-1)
\eeq
\nid We used the shortened notation $< \ >$ for the average with respect
to $\rho$. In this expression  $\chi_{ij}(\sigma)$ are  the matrix elements of :
\beq\label{Chi}
\chi(\sigma) = \int \rho(d\bu)  D\bG^{\sigma}_{\bu} 
\eeq
Thus $\chi(\sigma) $ is a matrix representing the average value of the iterate $\sigma$ of the Jacobian.  
 Let us note that the 
 fact that $\chi(\sigma) $   stay bounded for $\sigma \to \infty$ is not a trivial result  because $D\bG^{\sigma}_{\bu} $ diverges
 exponentially with $\sigma$.    The convergence of  $\chi(\sigma) $ has been rigorously shown by Ruelle under the hypothesis of 
uniform hyperbolicity.
It results from the exponential correlation decay (mixing) in the unstable directions and on the exponential contraction In our framework, this means that, provided that $\bxi(t)$ is sufficiently small,
and for any smooth observable $\obs$, the  variation $< \obs >_t-< \obs >$ is \textit{proportional} to $\bxi(t)$ up to small 
non linear corrections.
In other words,  $\rho_t $ is \textit{differentiable}  with respect to the perturbation.
The derivative is called the  \textit{linear} response.\\

 It is interesting to note that the response
at time $1$ is $< D\bG (\bu)>$, namely this is the average value of the Jacobian matrix. Thus, at time $1$ we have
a complete correspondence between the notion of influence discussed in the section \ref{Coop} and the linear response.
This suggests us to construct circuits of influence as we did in the section \ref{Coop}.
Unfortunately, this correspondence does not hold for larger times. This is basically because the quantity $ < D\bG^{\tau} (\bu)>$
does not obey the chain rule (contrarily to $D\bG^{\tau} (\bu)$). Therefore, if
$j$ influences $i$ and  if $i$ influences  $k$, this does not imply that $j$ influences $k$.
In the case of dynamical system~(\ref{DNN}) one can decompose $\chi_{ij}(\tau)$ as :
\beq\label{chiij}
\chi_{ij}(\tau)=
\sum_{\gamma_{ij}(\tau)}
        \prod_{l=1}^{\tau}J_{k_l k_{l-1}}
\left< \prod_{l=1}^{\tau}f'(u_{k_{l-1}}(l-1))\right>,
\eeq
\nid  The sum holds on each possible paths
$\gamma_{ij}(\tau)$, of length $\tau$, connecting the
neuron $k_0=j$ to the neuron $k_\tau=i$, in $\tau$ steps.
One remarks that each
path is weighted by the product of a \textit{topological} contribution
depending only on the weight $J_{ij}$ and
a \textit{dynamical} contribution. Since, in the kind of systems we consider, functions $f$ are sigmoid, 
the weight of a path $\gamma_{ij}(\tau)$ depends crucially on
the state of saturation of the neurons $k_0, \dots, k_{\tau-1}$ at times $0, \dots, \tau-1$.
Especially, if $f'(u_{k_{l-1}}(l-1))>1$ a signal is amplified while it is damped if
$f'(u_{k_{l-1}}(l-1)) < 1$. Thus, though a signal has many possibilities for going from $j$ to $i$ in $\tau$ time steps,
some paths may be ``better'' than some others, in the sense
that their contribution to $\chi_{ij}(\tau)$ is higher.  Therefore eq.~(\ref{chiij})  underlines a key point.
  The analysis of signal transmission in a coupled network of 
dynamical neurons with non linear transfer functions
 requires to consider both the topology of the interaction graph {\em and}  the nonlinear dynamical regime of the system.\\

One can decompose the response function (\ref{Chi})  into two terms.
The first one is obtained by locally projecting the Jacobian matrix  on the unstable directions of the tangent space.  This term will be named  the``unstable"
 response function.
It corresponds to linear response of the system to  perturbations locally parallel to the local unstable manifold (roughly speaking perturbations ``parallel
 to'' the attractor).
One can show that the linear response associated with this type of perturbation is in fact  a  correlation function, 
as found in  standard fluctuation-dissipation theorems~\cite{Ruelle2}. 
Hence, as usual for correlation functions of a chaotic system,  \textit{it decays exponentially} (because of mixing) and the decay rates are associated
 with the poles of its Fourier transform.  More precisely, 
these exponential decay rates correspond to the imaginary part of the complex poles of the unstable part of the 
susceptibility~(\ref{chiij}).  Thus they will be called  ``unstable'' poles.
More generally, it can be shown that these poles  are also the eigenvalues of the operator governing the time-evolution of the probability
 densities (which we denoted above as $\bG^{t} \mu$), the so-called Perron-Frobenius
operator~\cite{RP}.  Therefore, these poles, whose signatures are visible in the peaks of the modulus of the correlation functions, do not depend on
 the observable, though some residues may accidentally
vanish for a given observable.   

The second term \footnote{Note that a linear response theory has also been proposed in \cite{Vulpiani}.
However, it requires  the invariant measure to have a density. This is only true for the conditional
measure along unstable manifolds. As a matter of fact, this theory does not contain the stable term.}
 is obtained by locally projecting the Jacobian matrix  on the stable directions of the tangent space. 
It corresponds to the response to perturbations  locally parallel to the local stable manifold 
(namely transverse to the attractor).  Therefore, it is \textit{exponentially damped} by the dynamical contraction. 
[Note that, according to the specific
form of the Jacobian matrix, this contraction is, in our case, mainly due to the saturation of the sigmoid
transfer function].
The corresponding exponential decay rates are given by the complex poles (``stable'' poles) of
the stable part of the complex susceptibility. But here the poles depend a priori on the observable.
One can easily figures this out if one decomposes the stable tangent space of a point in the
orthogonal basis of Oseledec modes (directions associated to each of the negative Lyapunov
exponent). The projection of the $i$-th canonical basis vector on the $k$-th Oseledec mode 
depends on $i$ and $k$. This
dependence persists even if one takes an average along the trajectory, as in (\ref{Chi}). 

Hence, both stable and unstable terms are exponentially damped, ensuring the convergence of the
series (\ref{dro}), but for completely different reasons. Moreover, the stable and unstable 
part of the linear response have drastically different properties. As a matter of fact, the stable part
\textit{is not a correlation function and it does not obey the fluctuation-dissipation
theorem}. In particular, the unstable poles and stable poles
are usually distinct. 
Moreover, the stable poles allow to distinguish the neurons in their
capacity to transmit a signal. \\

The existence of this linear response theory  opens up the way to  applications involving  chaotic networks 
 {\it used as a linear filter}.  Indeed  eq.~(\ref{dro})  describes a linear system which   transforms an input signal $\bxi(t)$ 
of small amplitude into  an output signal 
$ \left< \tu_i(t) - u_i(t) \right> $ according to a standard convolution product.    
In particular, if  the external signal is chosen as: 
\beq\label{xi_per}
\bxi(t) = \epsilon e^{-i\omega t} \, \hat{\be}_j
\eeq
 (where
$\hat{\be}_j$ is the unit vector in direction $j$), then the response of the system is also harmonic with :
\beq\label{harmo}
  \left< \tu_i(t) - u_i(t) \right>  =   \epsilon \hat{\chi}_{ij}(\omega)  e^{-i\omega( t-1)} 
\eeq
where the frequency-dependent amplitude:
\beq\label{suscep}
 \hat{\chi}_{ij}(\omega) =  \sum_{\sigma=0}^{\infty}  \chi_{ij}(\sigma)  e^{i\omega \sigma} 
\eeq
is called the {\it complex susceptibility}.  
In ref.\cite{CS1}  we have conceived and implemented a method to compute 
 $ \hat{\chi}_{ij}(\omega) $ numerically. 
   The knowledge of  the susceptibility matrix  is very useful as it enables 
one to detect resonances, i.e. frequencies for which  the amplitude response of the system to a periodic input signal  
is maximum.   In fact the existence of a linear response implies that $ \hat{\chi}_{ij}(\omega) $ is bounded for all
 $\omega \in [0,2\pi]$.  Moreover, in view  of eq.~(\ref{suscep}), it is analytic in the complex upper plane.   On the other 
hand, $ \hat{\chi}_{ij}(\omega) $ can have poles within a strip 
in the lower half plane, e.g. in $\omega_0 - i\lambda$, $\lambda >0$.  In this case,
 and if $\lambda$ is small, the amplitude $| \hat{\chi}_{ij}(\omega) | $ exhibits a peak of width $\lambda$  and  
height $| \hat{\chi}_{ij}(\omega_0) |$  which can be interpreted in the present context as follows:   when unit  $j$ 
 (whose state varies chaotically due to the global dynamics)   is subjected to a small periodic excitation  at 
frequency $\omega_0$ and amplitude $\epsilon$  then the \textit{average} response of  unit   $i$  behaves
 periodically with same frequency and amplitude $\epsilon | \hat{\chi}_{ij}(\omega_0) | $ which  is maximal 
in a frequency interval centred about $\omega_0$. \\

Let us numerically computes the susceptibility $\hchi(\omega)$ for real
values of $\omega$ (see \cite{CS1},\cite{CS2} for details) in the following example. 
This is a sparse network  where each unit receives connection
from exactly $K=4$ other units (sparse neural networks of type
(\ref{DNN}) exhibits also chaos via quasi periodicity \cite{IJBC}).
The number of units was fixed to $N=9$.  
The $J_{ij}$'s have been drawn at random according to a Gaussian distribution with
mean zero and a variance $\frac{J^2}{K}$. 
The corresponding network is drawn in Fig. \ref{FJij}. (Note that the corresponding graph is not decomposable).
Blue stars correspond to inhibitory links
and red crosses to excitatory links. In this example the unit $7$ is a ``hub'' in
the sense that it sends links to almost every units, while $0$, $2$, $3$ or $5$ send at most two links.

%
%
%
\begin{figure}[ht]
\includegraphics[height=6cm,width=6cm,clip=false]{Jij}
\hspace{2cm}
\includegraphics[height=6cm,width=8cm,clip=false]{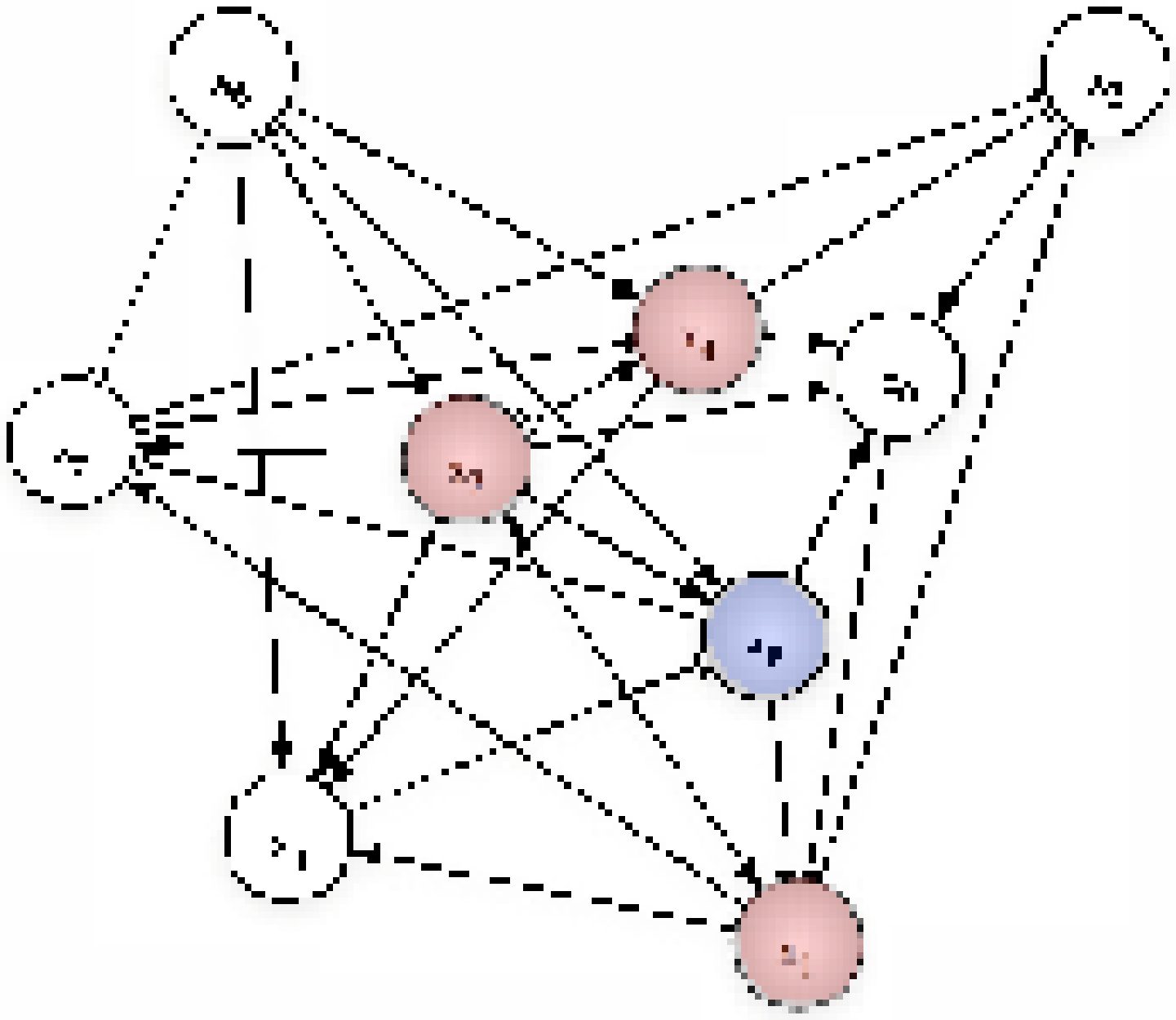}
\caption{\label{FJij} Connectivity matrix (Fig. \ref{FJij}a, on the left) and the corresponding network 
for the investigated system (Fig. \ref{FJij}b, on the right). In Fig. \ref{FJij}b each node
is represented by a circle. A filled circle means that there is a link from the corresponding
node to itself (red: self-excitation, blue: self-inhibition). Inhibitory links are terminated by a vertical bar while
excitatory links are terminated by an arrow.}
\end{figure}

A small constant $\theta_i$ has been added to each $u_i$ to break 
down the symmetry $\bu \to -\bu$ (i.e. $u_i(t)=\sum_j J_{ij}x_j(t)+\theta_i$).
As expected the corresponding dynamics exhibits a transition to chaos by quasi-periodicity. For $g=3$
the dynamics has one positive Lyapunov exponent ($\lambda_1=0.153$)
and $8$ negative Lyapunov exponents (with $\lambda_2=-0.427$). The Lyapunov exponents have been
computed with the Eckmann-Ruelle algorithm \cite{ER}. The chaotic regime is stable to small perturbations,
as we checked. 

Computing the susceptibility one obtains the curves  shown in Fig.\ref{FSuseps0.01}.
Several remarks can be made. First,  some resonance peaks are rather high ($\sim 20$) corresponding to 
an efficient
amplification of a signal with suitable frequency. It is also clear
 that the intensity of the resonance has no direct connection with the intensity or the sign
of the coupling and is mainly due to nonlinear effects. For example, there is no direct connection from $0$ to $3$ or $5$
but nevertheless these units react  strongly to a suitable signal injected at  unit $0$.\\

%
%
%
%
%
%
\begin{figure}[ht]
\includegraphics[height=5cm,width=14cm,clip=false]{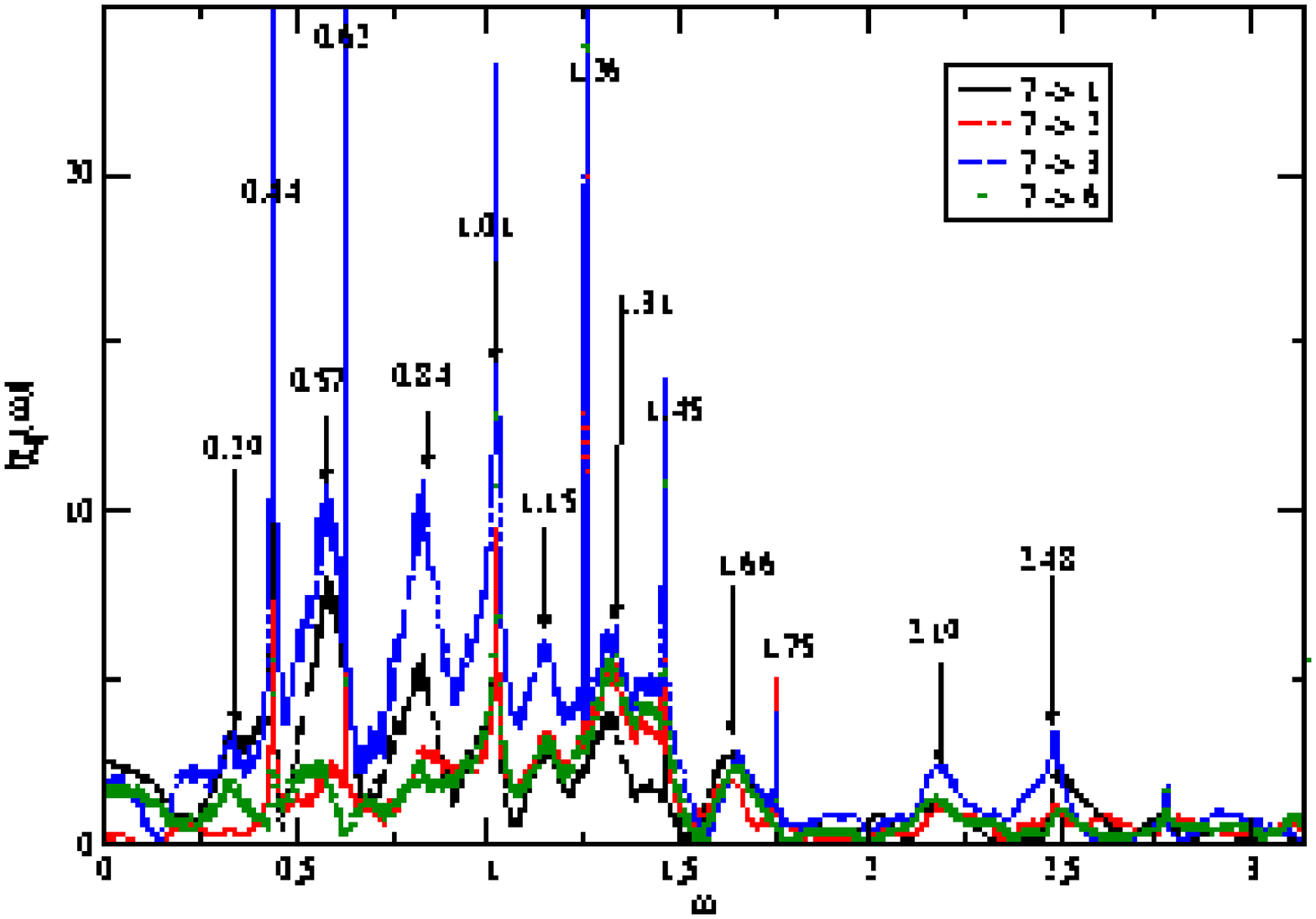}
\vspace{0.2cm}
\includegraphics[height=5cm,width=14cm,clip=false]{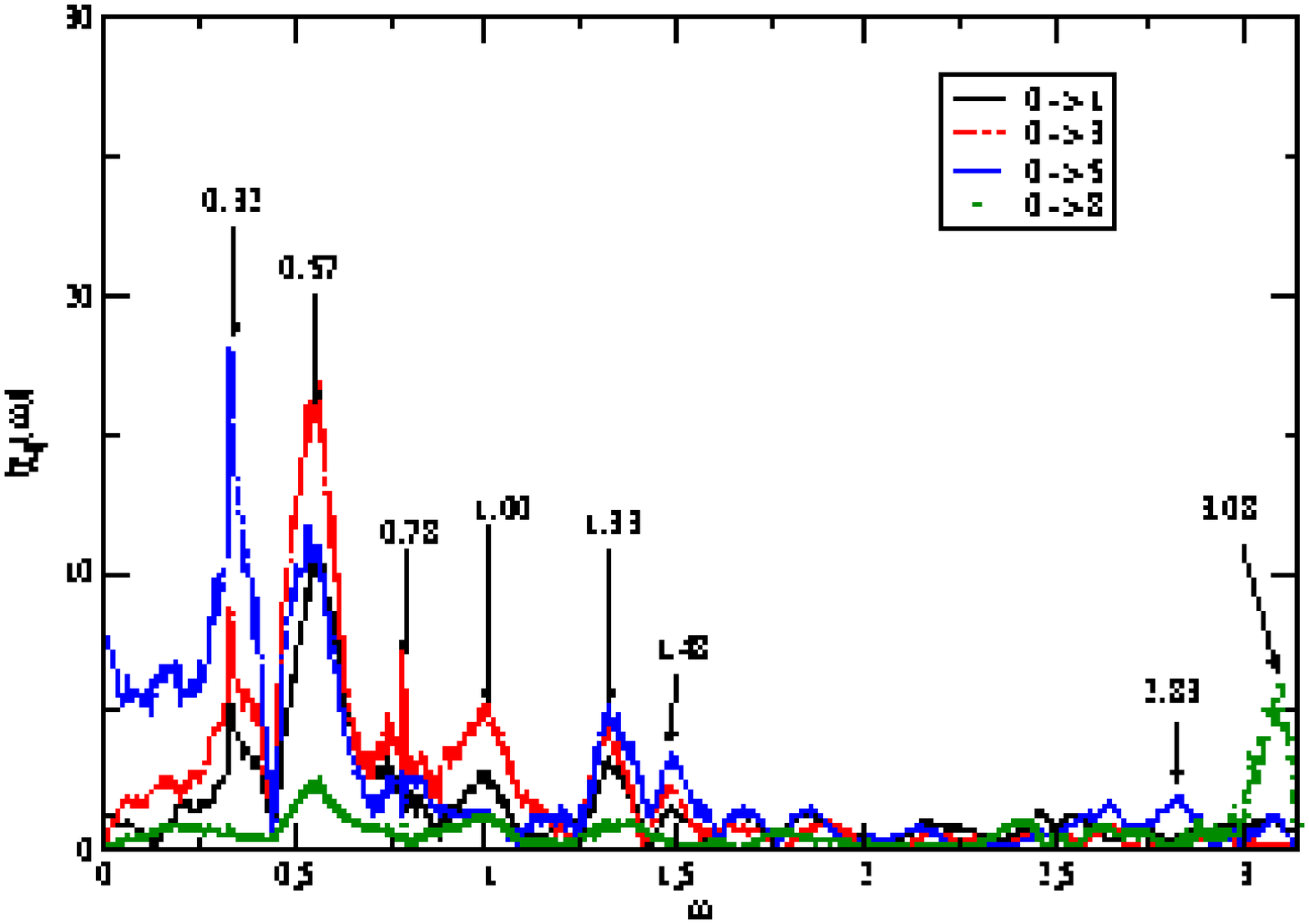}
\vspace{0.2cm}
\includegraphics[height=5cm,width=14cm,clip=false]{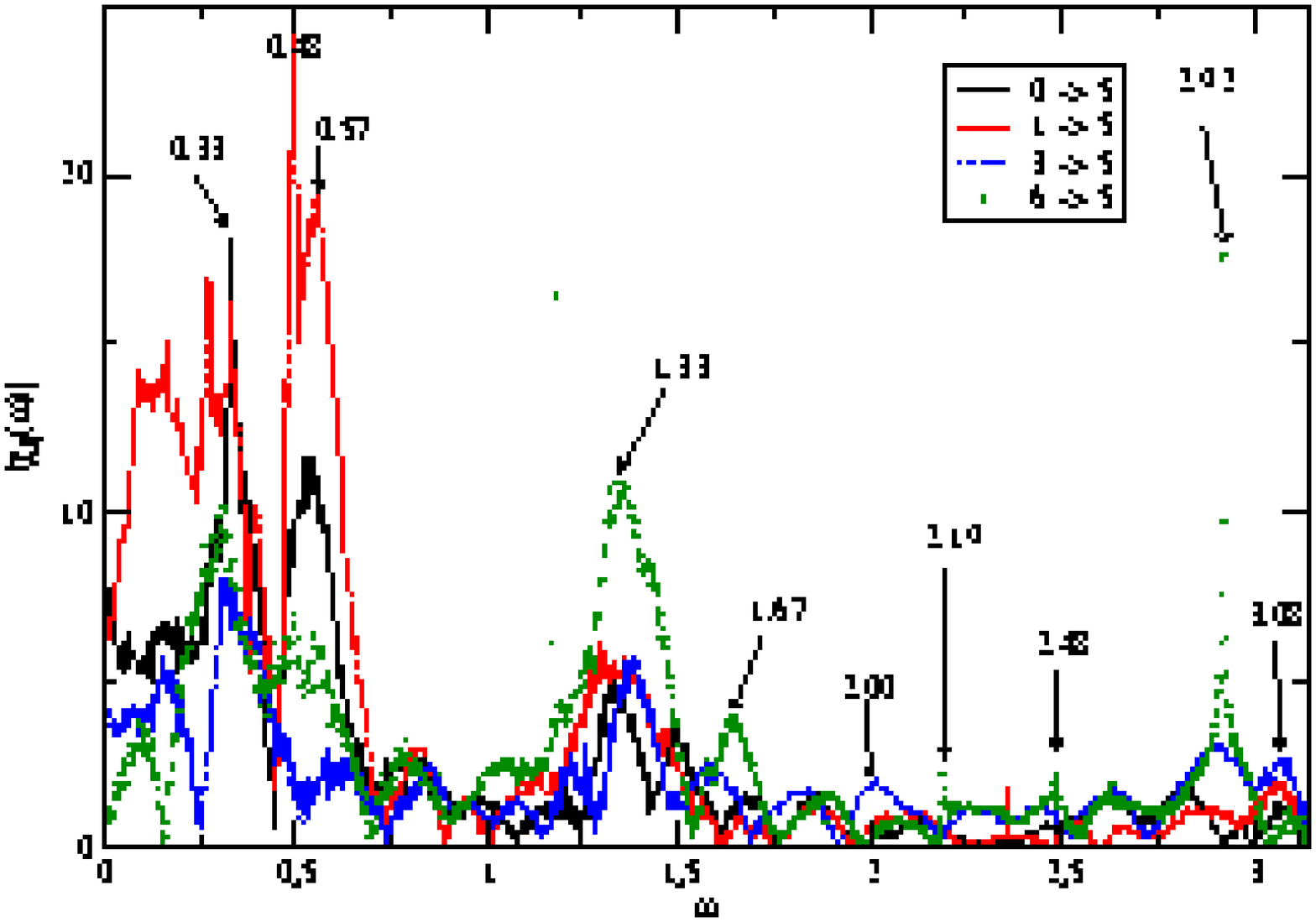}
\caption{\label{FSuseps0.01}. Modulus of some susceptibilities. Fig. \ref{FSuseps0.01}a (top). 7 (highly connected unit) excites the units:
1 (excitatory link with intensity $J_{17}=0.007$); 2 (no direct link); 3 (excitatory link with intensity $J_{37}=0.722$);
6 (inhibitory link with intensity $J_{67}=-0.041$). Fig. \ref{FSuseps0.01}b (middle). 0 (weakly connected unit) excites the units:
1 (inhibitory link with intensity $J_{10}=-1.131$); 3,5,8 (no direct link); Fig. \ref{FSuseps0.01}c (bottom). 5 receives the excitation
from the units:  0 (no direct link); 1 (excitatory link with intensity $J_{51}=1.015$); 5   (no direct link); 6 
(inhibitory link with intensity $J_{56}=-1.312$).}
\end{figure}

Let us now compare the Fourier transform of the correlations function $C_{ij}(t)$ for the same pairs
(Fig. \ref{FCorrelations}).  One remarks that  these  functions exhibit  less resonance peaks. This is expected since
the  Fourier transform of the correlation  function $C_{ij}(t)$ only contains unstable resonances while
the susceptibility contains stable and unstable resonances. Note also that the resolution in resonance
peaks is quite better in the susceptibility.

%
%
%
\begin{figure}[ht]
\includegraphics[height=5cm,width=14cm,clip=false]{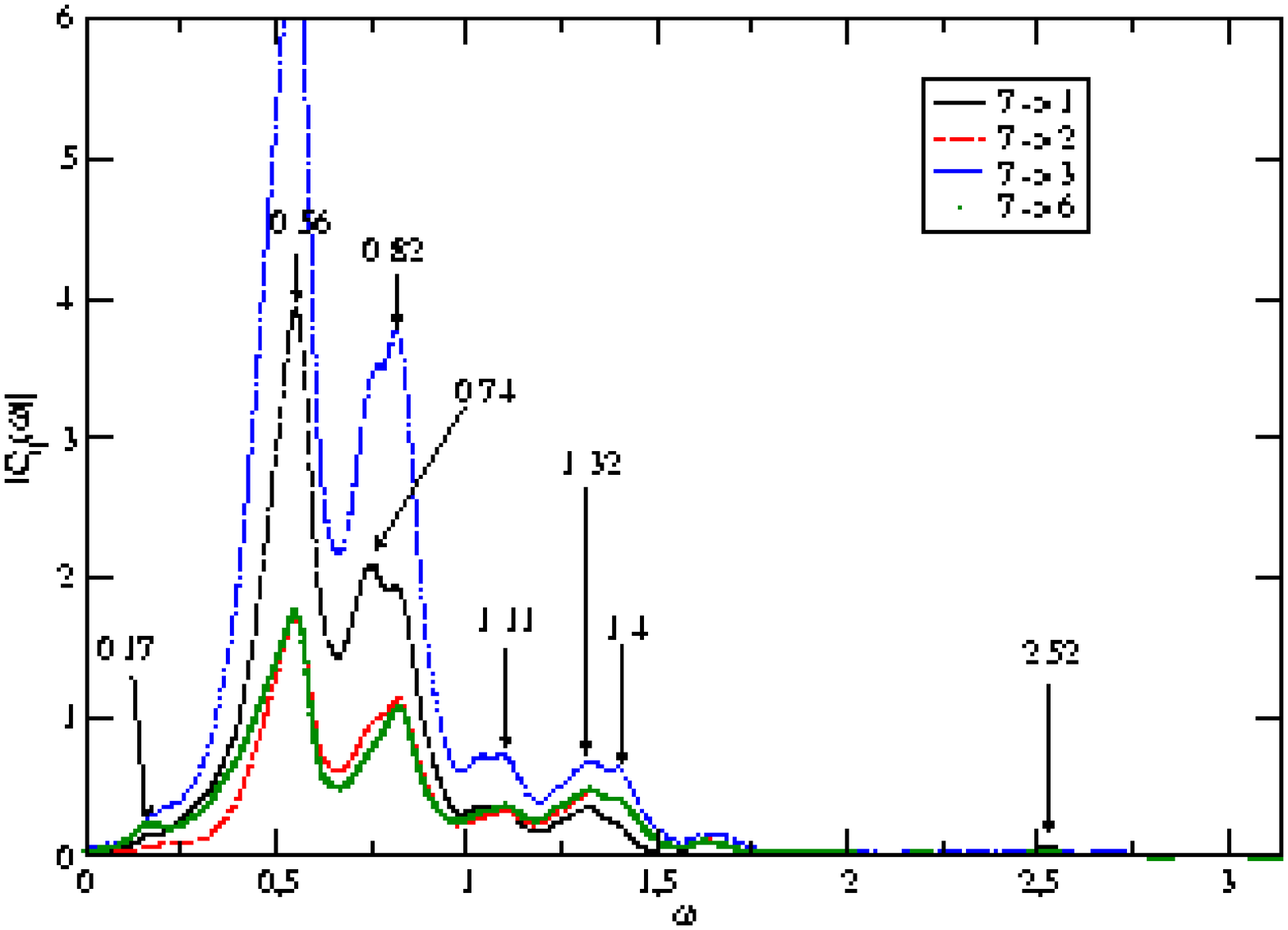}
\vspace{0.5cm}
\includegraphics[height=5cm,width=14cm,clip=false]{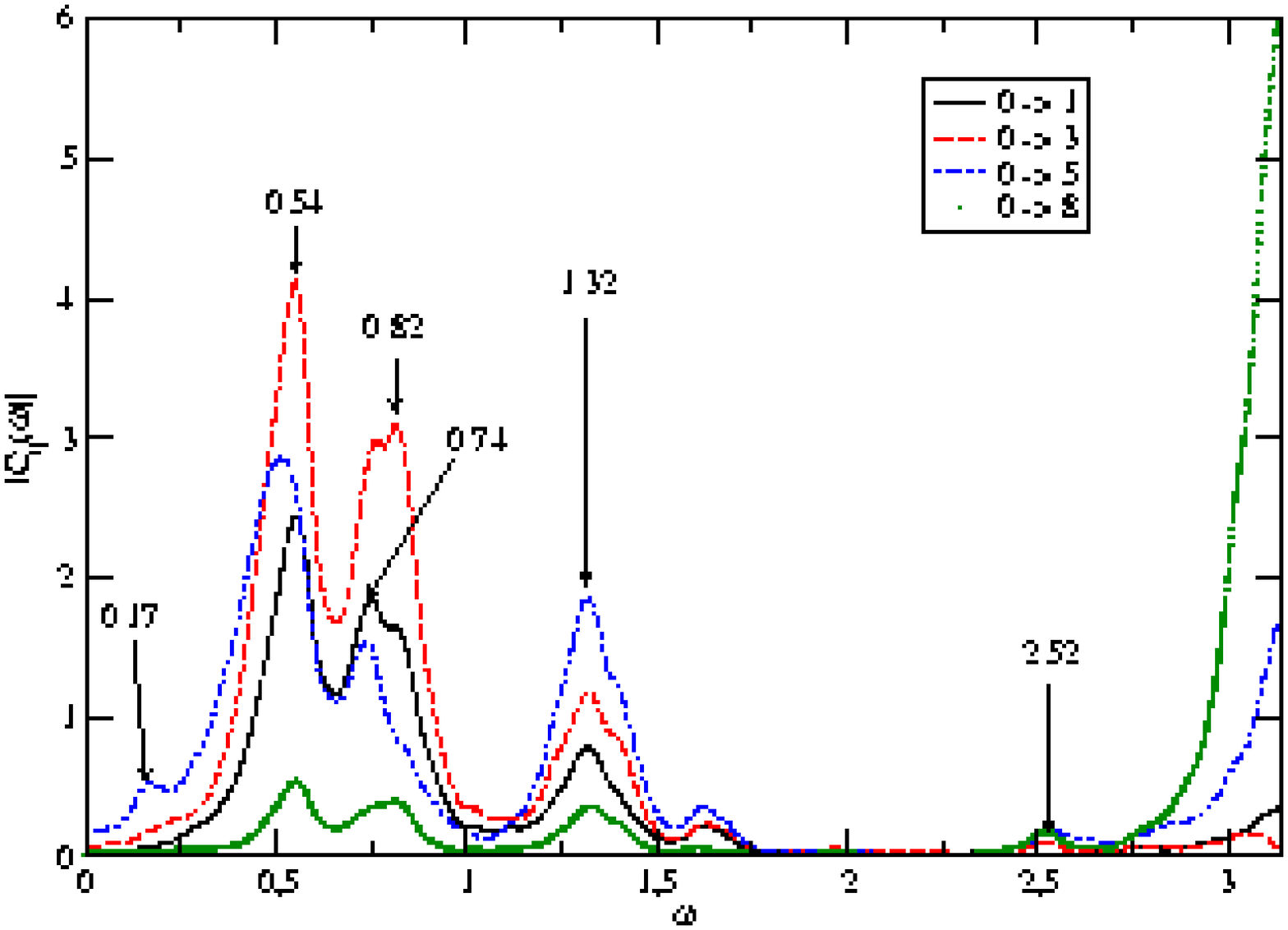}
\vspace{0.5cm}
\includegraphics[height=5cm,width=14cm,clip=false]{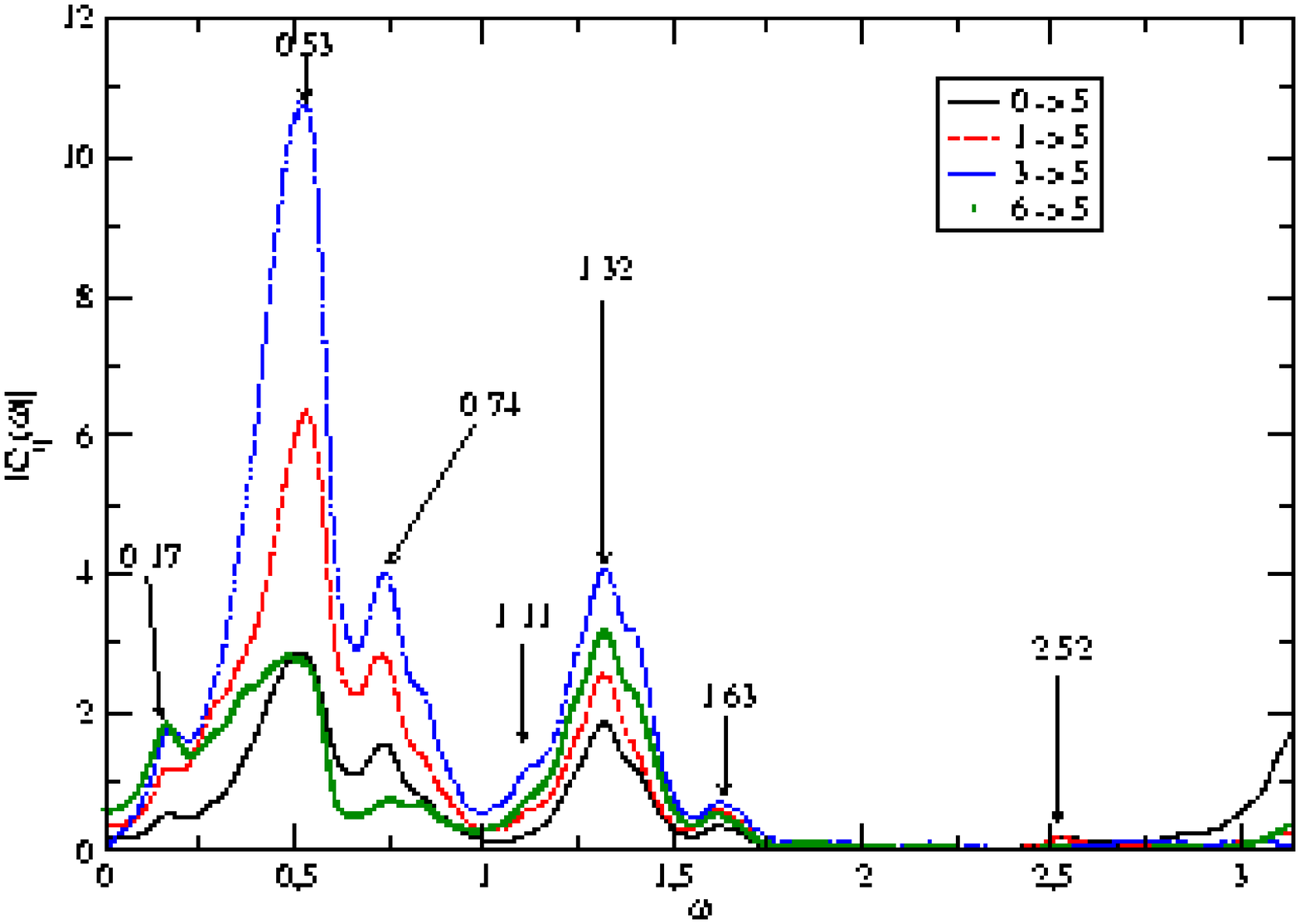}
\caption{\label{FCorrelations}. Modulus of the correlation functions corresponding to the susceptibilities
represented in the Fig. \ref{FSuseps0.01}a,b,c. }
\end{figure}

 The previous analysis leads then us to propose a notion of ``effective'', frequency dependent, connectivity
based on susceptibility curves. For a  given  frequency $\omega$, we plot the modulus of the susceptibility
$|\chi_{ij}(\omega)|$ with a representation assigning to each pair $i,j$ a circle whose size is proportional
to the modulus. Some examples are represented in Fig. \ref{FConn}. We clearly  see in this figure that changing the
 frequency changes the effective
network.

For example, with a frequency $\omega=0.125$ (Fig. \ref{FConn}a),
  the node $1$
has a strong ability to transmit signals towards the node $5$ (namely the response of this unit is high).
On the contrary, nodes $5,6$ and $7$ have weak performances in signal 
transmission at this frequency. Moreover, one sees that $7$ is a bad sender and a bad receiver. 
With a frequency $0.57$ the effective network has a rather symmetric
structure and basically all units respond to this excitation (however with a different amplitude).
Also, some units present a strong affinity with some others, at a specific frequency. Obviously, one also checks
that for frequencies that do not correspond to resonances (such as $\omega=2.33$ in
Fig.   \ref{FConn}f) the response is essentially inexistent whatever the pair.
Finally, this figure shows that it is possible to excite
any unit from any other one in such a way  that this unit (and possibly a few other but \textit{not all} the other units)
have a maximal response.

All these effects are due to a combination of topology and dynamics  and they
cannot be read in the connectivity matrix $\cJ$.

%
%
%

\begin{figure}[ht]
\includegraphics[height=8cm,width=10cm,clip=false]{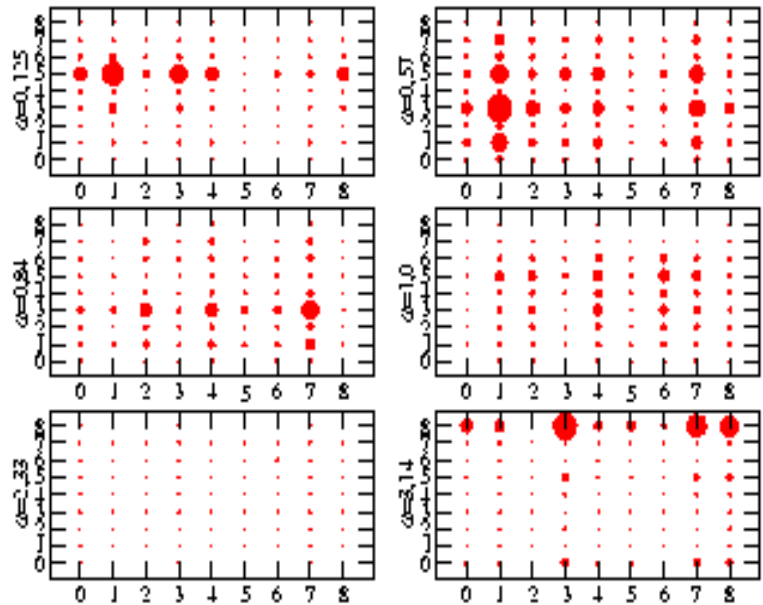}
\caption{\label{FConn} Effective connectivity for : Fig \ref{FConn}a (top left) $\omega=0.125$;
 Fig \ref{FConn}b (top right) $\omega=0.57$;  Fig \ref{FConn}c (middle left) $\omega=0.84$; 
 Fig \ref{FConn}d (middle right) $\omega=1.0$;  Fig \ref{FConn}e (bottom left) $\omega=2.3$;  
Fig \ref{FConn}e (bottom right) $\omega=3.14 $.}
\end{figure}

\ssu{Conclusion} This section was devoted to the analysis of some recurrent neural networks, which
are a particularly prominent example in this field. We have analyzed in some details the collective
dynamics and exhibit several important effects revealing the richness and complexity of the emergent
dynamics. Indeed, as noted in the begin of the section, the dynamics of the uncoupled neurons is rather poor.
This justifies somewhat the claim, made in the introduction, that one can make rather drastic simplifications
in the description of the neurons of a coupled system, and still get a complex and relevant model. However,
one must be cautious. Removing some characteristics and still get an interesting behavior does not mean
that the removed characteristics are irrelevant ``details''. Actually, the models presented here are quite
simplistic as ``brain'' models. To our opinion, there main interest is to provide ``benchmarks'' for
developing and testing tools that one may use, later on, to analyze more realistic models.

\su{Conclusion}

In this chapter, we have provided examples suggesting that
the mathematical analysis of neural networks dynamics can be
pushed relatively far, in some simple models.
However, a remaining question is: can we perform the same kind of analysis
for neural networks closer to biological systems ?
At the actual stage of research the techniques of cerebral imagery; brain
analysis and neurophysiology allows to go relatively deep in
the structure and dynamics of cerebral areas, but it allows also
to make an explicit cartography of the nervous system of primary
animals such as worms (e.g. Caenorhabditis elegans \cite{CElegans}).
Thus, it is in principle possible to write the explicit dynamical
system accounting for the evolution of small area containing
a relatively small number of neurons ($\sim 100-1000$).
However, the detailed analysis of these equations
is still intractable. This is also, in some cases, useless.
Indeed, often these area exhibit a relatively simple collective
behavior. It is thus profitable to define a phenomenological
model, described by a small set of differential equations and
a few parameters that one can adjust to fit experimental results.

A prominent example concerns
the cortical columns implicated in vision.     
A cortical column is a population of
pyramidal cells receiving excitatory and inhibitory
inputs from others cells in the same column but also excitatory
inputs coming from other columns, close or distant.
A celebrated dynamical model has been proposed by
 Lopes Da Silva \cite{Lopes} and Jansen \& Rit \cite{JR}
describing the activity of cortical columns.
A mathematical analysis of the bifurcations exhibited by this  model
has been performed in \cite{Grimbert}. 
It shows the existence of  oscillations generated by a Hopf  bifurcation, induced
by the variation of a parameter modeling the frequency of stimuli emitted
by an external source.  The value of the oscillations frequency
is about $10$ Hz corresponding to the $\alpha$ rhythm.
One can also exhibit spikes emission, looking very much like
epileptic activity, and related to a saddle-node bifurcations on a cycle. 
In the vicinity of this bifurcation, an excitation with an external stimulus with a specific frequency
induces a spike train emission. The techniques used by the Authors combine standard results
from dynamical systems theory and numerical analysis, in the spirit of the analysis
presented in the section \ref{SBifHH}. Note that 
rhythm $\delta,\theta,\beta,\gamma$ have also been
numerically exhibited in \cite{DavidFriston}, when varying the excitatory-inhibitory effects
in biologically realistic ranges.

This example shows that it is indeed possible to analyse neural networks
closer to biology, possibly after  some simplification of the initial system.
More details about vision and cortical columns are given in chapter IV.\\

We have made a trip in the world of Neural Network dynamics, following the path 
represented in Table 1. As said in the beginning many examples, models, \etc have been omitted.
However, we have tried to give an outlook of the various methods available for the study of the dynamics.
This excursion has also shown that, when going from a level of complexity (one neuron dynamics)
to another level (collective dynamics), it might be fruitful to adopt  different perspectives
(accurate description of a neuron versus emergent behavior of ``simplified'' neurons) and 
different (but complementary) methods (dynamical system theory versus probability theory and statistical
physics). It also shows us the necessity to develop accurate tools to handle neuronal dynamics
(this is well known and not new) and the possibility to do this by combining existing
theories and numerical analysis. This is a formidable task but the byproducts are on one hand
a better understanding of neuronal dynamics and on the other hand a possible insight
in other fields.

\pagebreak

\su{Appendix}

This appendix is mainly devoted to non-specialists. It gives a brief summary
of the concepts and techniques in dynamical systems theory 
used in this chapter. Our main references are \cite{Arnold1},\cite{Arnold2},\cite{GH},\cite{Katok},\cite{RuelleBif}.

\ssu{Elementary notions in dynamical systems theory.}

\sssu{Basic definitions.}

The dynamical systems studied in this chapter are either defined by a (finite) 
set of differential equations :
\beq\label{SD1}
\frac{d\bX}{dt} = \bH(\bX;\lambda)
\eeq
\nid or a set of recurrences\footnote{This implies that we do not consider the case of Neural Networks
with \textit{sequential} dynamics.}:
\beq\label{SD2}
\bX(t+1) = \bF(\bX;\lambda)
\eeq
\nid where $\bX \in \cm$, $\cm$ being a \textit{compact} set in $\bbbr^N$,
where $N$ is the number of degree of freedom and $\bX$ denotes the vector
$\lbrace x_i \rbrace_{i=1}^N$.  
The vector fields $\bH$ (resp. the recurrence $\bF$) in eq. (\ref{SD1}), (resp. (\ref{SD2}))
do not  depend
explicitly on time. The corresponding dynamical system is then called
 \textit{autonomous}. We mainly deal with the autonomous case in the paper and in this appendix.
 $\lambda \in \cE_\lambda \subset \bbbr^p$ refers to a set of  
$p$ (real) parameters on which the system depends. This might be an external current applied
to a neuron, an external input submitted to an assembly of neurons,
the set of synaptic weights, \etc. Therefore, $\lambda$ can have a large (though finite)
dimension. It can also be deterministic or \textit{random}. The last case requires 
however combinations of techniques from dynamical system theory and probability
theory. An example is developed in  section \ref{NotreModele}.

We assume that $\bH,\bF$ are smooth (at least $\cC^2$)
 functions of $\bX,\lambda$. In the continuous time case (\ref{SD1}) the Cauchy theorem
ensures the local unicity of the solutions  provided that $\bH$ is a Lipschitz function.
Namely,
if $\bX \in \cm$, there is a time interval $]-c,c[$ and a neighborhood
$\cU \ni \bX$ such that there is a unique solution  of (\ref{SD1}), $\bX(t) \in \cU, t \in ]-c,c[$
 and such that $\bX(0)=\bX$. Moreover,
when $\cm$ is compact, the solutions extend to $t \in [-\infty,+\infty[$ \cite{Chillingworth}.
Denote by $\tbx \deq \lbrace \bX(t) \rbrace_{t=0}^{+\infty}$
the (forward) \textit{orbit} or \textit{trajectory} such that $\bX(0)=\bX$ and
by $\tbx^- \deq \lbrace \bX(t) \rbrace_{t=-\infty}^{0}$ the backward trajectory. 
The unicity of trajectories implies that two trajectories cannot
cross (though they can accumulate on the same set, as shown below).
Also, the equations (\ref{SD1}) have the meaning that any trajectory
is locally tangent to the vector field $\bH$. In low dimensional cases (namely $N \leq 3$)
this is helpful to draw a qualitative sketch of the main dynamical system features (phase portrait), without any computation
(see for example the sections \ref{SFN},\ref{SProp}). 

In the case
of the recurrence (\ref{SD2}) the forward trajectory is simply
constructed by iterating the map $\bF$. Therefore it is always defined (provided that the initial condition
is in the domain of definition of $\bF$). The backward trajectory is uniquely
defined only if $\bF$ in invertible. In the sequel we shall assume that $\bF$ is a $\cC^2$
diffeomorphism.
For the dynamical system
(\ref{SD1}) one can prove the existence of a one parameter family
of diffeomorphisms $\phi^t$ (or \textit{flow}), such that $\phi^0=id$,
$\phi^t \circ \phi^s = \phi^{t+s}$ and $\bX(t)=\phi^t(\bX)$.
In the sequel, we shall use the notation $\bX(t)=\Bf^t(\bX)$
for both dynamical systems (\ref{SD1}),(\ref{SD2}). Consequently,
$\Bf$ will refer to the flow in the case (\ref{SD1}) and
to the map $\bF$ in the case (\ref{SD2}).\\

The dynamical systems (\ref{SD1}), (\ref{SD2}) may exhibit a wide
variety of dynamics, from very simple (rest state attracting all trajectories),
to complex (chaotic behavior) and even more complex (coexistence of 
many chaotic attractors, \etc). Consequently, in most cases the explicit
solution of (\ref{SD1}),(\ref{SD2}) are not known. The 
current philosophy in dynamical systems theory, initiated by H. Poincar\'e \cite{Poincare},
is that  finding a general solution is not only impossible, but
also useless. Indeed, in many cases, a qualitative study of the dynamical
system is enough to extract quite a large amount of informations
which often allows us to capture the main features of the dynamics.
In particular, one can extract characteristic ensembles
such has attractors, repellors, periodic orbits, \etc,
which contains the main informations one needs.
In many cases, one is indeed interested in the asymptotic behavior of
the forward orbits. The \textit{$\omega$-limit} set of $\bX$ is the set
of accumulation points of the forward trajectory $\bX(t)$.
The $\omega$-limit set of $\Bf$ is the 
union of the $\omega$-limit sets for all $\bX \in \cm$.
It contains in particular the attractors of the dynamics (see the definition
below). 
The same notion ($\alpha$-limit) set can be defined for the backward
trajectory when it is defined. A more general and related notion is the \textit{non wandering} set. 
This is the set of points $\bX$ such that  for any open
neighborhood  $\cU \ni \bX$ there is a time  $t_0 >0$ 
such that $\Bf^{t_0}(\cU) \cap \cU \neq \emptyset$. 
This set contains the main elements of the dynamical system
such as the $\omega$ limit set.

 The $\omega$-limit set lay have a quite complex structure. 
However, it contains in general some characteristic objects such
as \textit{fixed points}, or \textit{periodic orbits}. $\bX^\ast$ is a fixed point
if its orbit consists of $\bX^\ast$ only. In other words, $\bH(\bX)=0$ [resp. $\bF(x)=\bX$].
$a$ is a periodic point if there is some $t>0$ such that $\Bf^ta=a$. The lower
bound of such $t$ is the \textit{period} of $a$, $T(a)$. The set $\Gamma=\left\{\Bf^t a \ ;
0 \leq t \leq T(a) \right\}$ is called a \textit{periodic orbit} or a \textit{closed orbit}.
For a discrete time dynamical system it is a finite set; for continuous time it
is continuously infinite.

\sssu{Fixed points and linear analysis.}

The first step of the analysis of (\ref{SD1}),(\ref{SD2}) is to seek for
\textit{equilibria} or \textit{fixed point}. A fixed point is \textit{stable} iff for any neighborhood
$\ \cU \ni \bX^\ast$, there exists a neighborhood $\ \cU_1 \subset \cU$
such that $\forall \bX_0 \in \cU_1$, $\forall t >0, \ \Bf^t(\bX_0) \in \cU$.
$\bX^\ast$ is asymptotically stable if there exists an open neighborhood
$\cU_1$ such that $\forall \bX_0 \in \cU_1$, $\Bf^t(\bX_0) \to \bX^\ast$
as $t \to \infty$. Asymptotically stable fixed points are called \textit{sink}.
A stable fixed point which is not asymptotically stable
is called a \textit{center} (see Fig. \ref{FFixPt}). A well known example is the stable 
equilibrium position of the undamped pendulum. A fixed point is \textit{unstable}
if it is not stable. Note that the notion of stability is a \textit{local} notion.
%
%
%
%
\begin{figure}[ht]
\begin{center}
\includegraphics[height=3cm,width=12cm,clip=false]{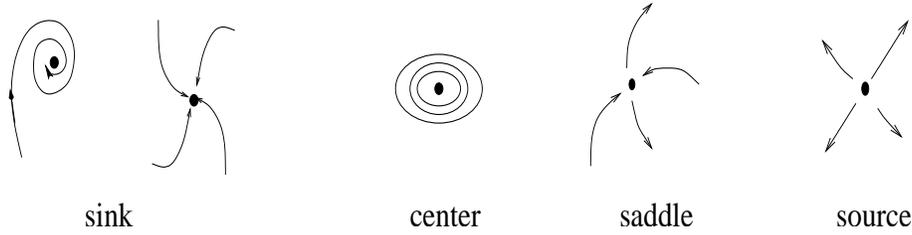}
\vspace{0.5cm}
\caption{Various kind of fixed points. \label{FFixPt}}
\end{center}
\end{figure}
%
%
%
%
%
%
Among the various kind of fixed points, the stability of \textit{hyperbolic}
 fixed points can be analyzed by linearization about $ \bXs$. Indeed,
call $D\bH_{ \bXs}$ (resp. $D\bF_{ \bXs}$) the Jacobian matrix of $\bH$ (resp. $\bF$)
at $ \bXs$. Since the coefficients of this matrix are real, the eigenvalues are
either real or complex conjugate. Call $Sp\left[\cA \right]$
the spectrum of a matrix $\cA$. One decomposes
$Sp\left[D\bH_{ \bXs} \right]$ [resp. $Sp\left[D\bF_{ \bXs} \right]$ ]into three parts : 
the \textit{stable} eigenvalues
are such that $\Re(\lambda) < 0$ [resp. $|\lambda| < 1$];
the \textit{neutral} eigenvalues are such that $\Re(\lambda) = 0$ [resp. $|\lambda| = 1$]
and the \textit{unstable} eigenvalues are such that $\Re(\lambda) > 0$ [resp. $|\lambda| > 1$].
Moreover, the Jacobian matrix can be reduced to a diagonal (or more generally to a Jordan normal form)
in a basis $\bv_1, \dots \bv_N$ corresponding to the (generalized) eigenvectors. The \textit{stable
space} $\cE^s( \bXs)$ is the subspace of $\bbbr^N$ generated by the eigenvectors corresponding
to the stable eigenvalues. In the same way one defines the  \textit{central space} $\cE^c( \bXs)$
and the \textit{unstable space} $\cE^u( \bXs)$.

Then $ \bXs$ is an \textit{hyperbolic fixed point} of
(\ref{SD1})  if there is no neutral eigenvalues (resp. $\cE^c( \bXs)=0$).
 $ \bXs$ is \textit{linearly stable}  if additionally $\cE^u( \bXs)=0$
(namely all eigenvalues are stable). A linearly stable equilibrium is asymptotically
stable and the rate of convergence is given by the largest real part of the eigenvalues
in the case (\ref{SD1}) (continuous time), and by the largest modulus of the eigenvalues in the case (\ref{SD2}) (discrete time).
Unstable hyperbolic fixed points are divided into \textit{saddle points} (there are stable and unstable
eigenvalues) and sources (all eigenvalues are unstable) (see Fig. \ref{FFixPt}).
Hyperbolic fixed point
have the following important properties\footnote{Note that the notion of hyperbolicity
extends to moving points (see section \ref{SHyp}) and that the result below can be generalized
(see \cite{Katok}) }. 

\ben

\item\textit{Hartman-Grobman linearization theorem.} If $ \bXs$ is hyperbolic then there exists
an homeomorphism $h$ preserving the sense of orbits, 
locally mapping the orbits of the flow of (\ref{SD1}) (resp. the map
(\ref{SD2})) to the orbits of the \textit{linear} flow $e^{tDH_\bXs}$ (resp. the linear
map $DF^t_ \bXs$). The Hartman-Grobman theorem implies that the dynamics near an hyperbolic fixed is 
essentially equivalent (up to a smooth variable change) to a linear system (for a nice
application to Neural Networks  see section \ref{WCNNSS}).

\item\textit{Invariant manifolds.}  Let $\cU$ be a neighborhood of $ \bXs$.
If $ \bXs$ is hyperbolic then there exists \textit{local} stable and unstable manifolds:

\bea
\cW^s_{loc}( \bXs) &=& \left\{\by \in \cU \ | \Bf^t(\by) \to  \bXs \ as \ t \to \infty \ and \ \Bf^t(\by) \in \cU, \ \forall
t \geq 0\right\}\\
\cW^u_{loc}( \bXs) &=& \left\{\by \in \cU \ | \Bf^t(\by) \to  \bXs \ as \ t \to -\infty \ and \ \Bf^t(\by) \in \cU, \ \forall
t \leq 0\right\}
\eea

\nid respectively with the same dimension $n_s, n_u$ as the eigenspace $E^s_ \bXs,E^u_ \bXs$
 of the linearized system, respectively  locally tangent to $E^s_ \bXs,E^u_ \bXs$ at $ \bXs$,
as smooth as the function $\bH$ (resp. $\bF$) and dynamically invariant.
 Moreover the angle between $E^s_ \bXs,E^u_ \bXs$ is bounded away from zero.  
The local stable and unstable manifold have global analogues :

\bea
\cW^s( \bXs) &=& \cup_{t \leq 0} \cW^s_{loc}( \bXs) =
\left\{\by \in \cm\ | \Bf^t(\by) \to  \bXs \ as \ t \to \infty \right\}\\
\cW^u( \bXs) &=& \cup_{t \leq 0} \cW^u_{loc}( \bXs) = \left\{\by \in \cm \ | \Bf^t(\by) \to  \bXs \ as \ t \to -\infty \right\}
\eea
\een

%
%
%
%
\begin{figure}[ht]
\begin{center}
\includegraphics[height=4cm,width=6cm,clip=false]{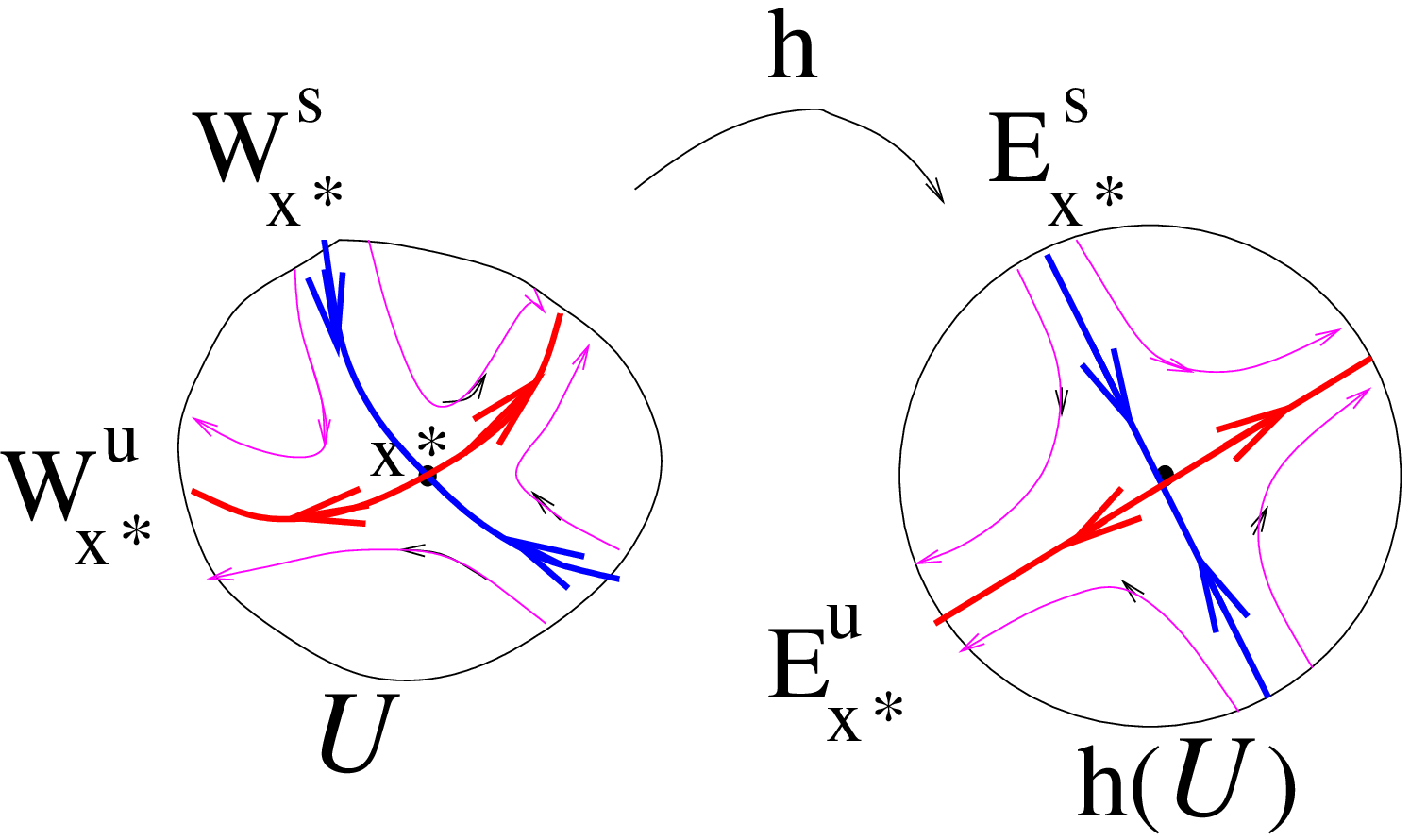}
\hspace{2cm}
\includegraphics[height=4cm,width=6cm,clip=false]{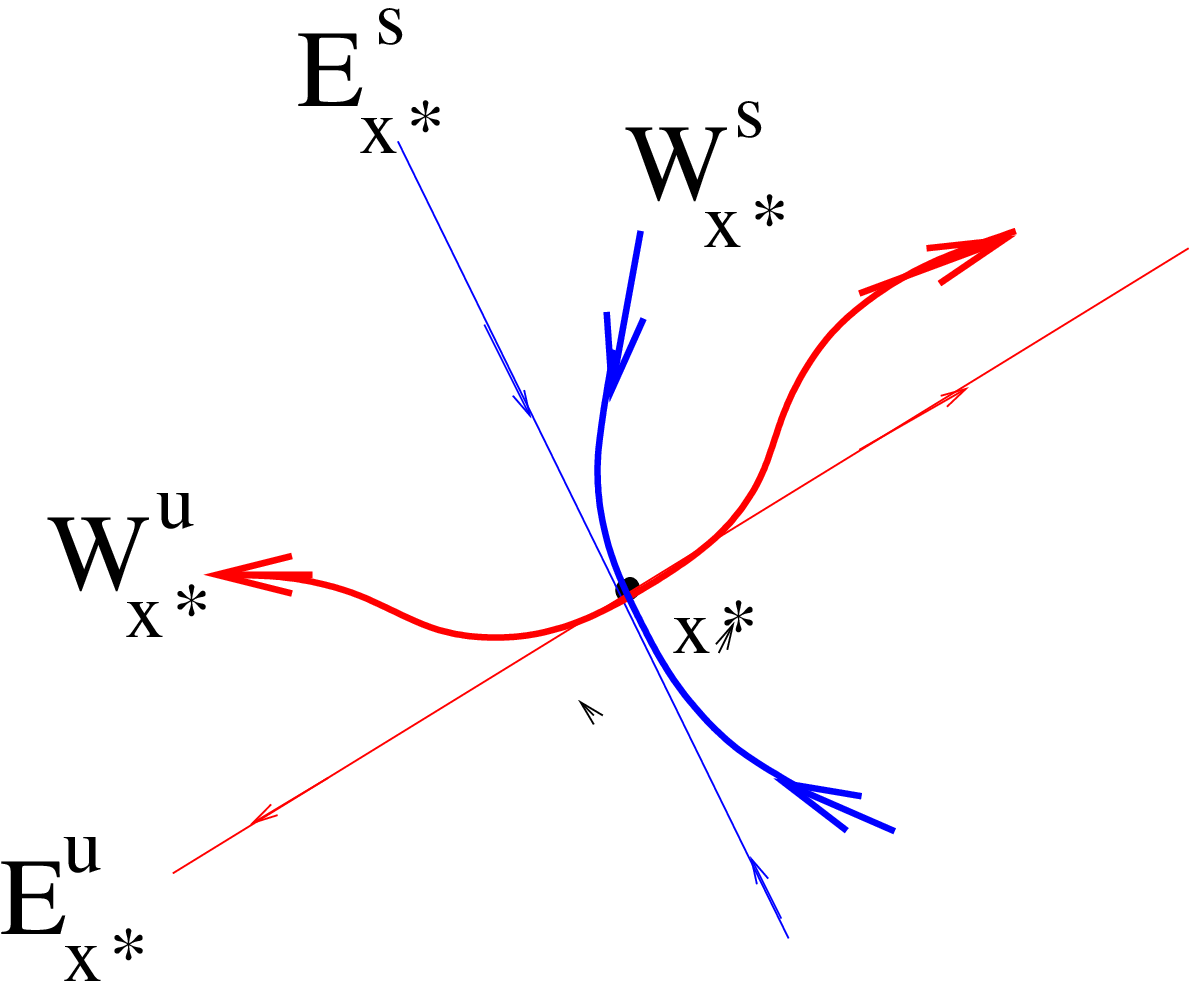}
\vspace{0.5cm}
\caption{Fig. \ref{FW} a. Hartman-Grobman theorem. Fig.\ref{FW} b. Local stable and unstable manifolds. \label{FW}}
\end{center}
\end{figure}
%
%
%
%
%
%

Stable and unstable manifolds may intersect in homoclinic (Fig. \ref{Fcline}a,b) or heteroclinic intersections
(Fig.\ref{Fcline}c) . 
This has important consequences. In particular, the global stable and unstable
manifolds of a fixed point may have strong influence to the global dynamics,
as in the case of transverse homoclinic intersections for maps (\cite{ArnoldAvez,GH}).

%
%
%
%
\begin{figure}[ht]
\begin{center}
\includegraphics[height=4cm,width=4cm,clip=false]{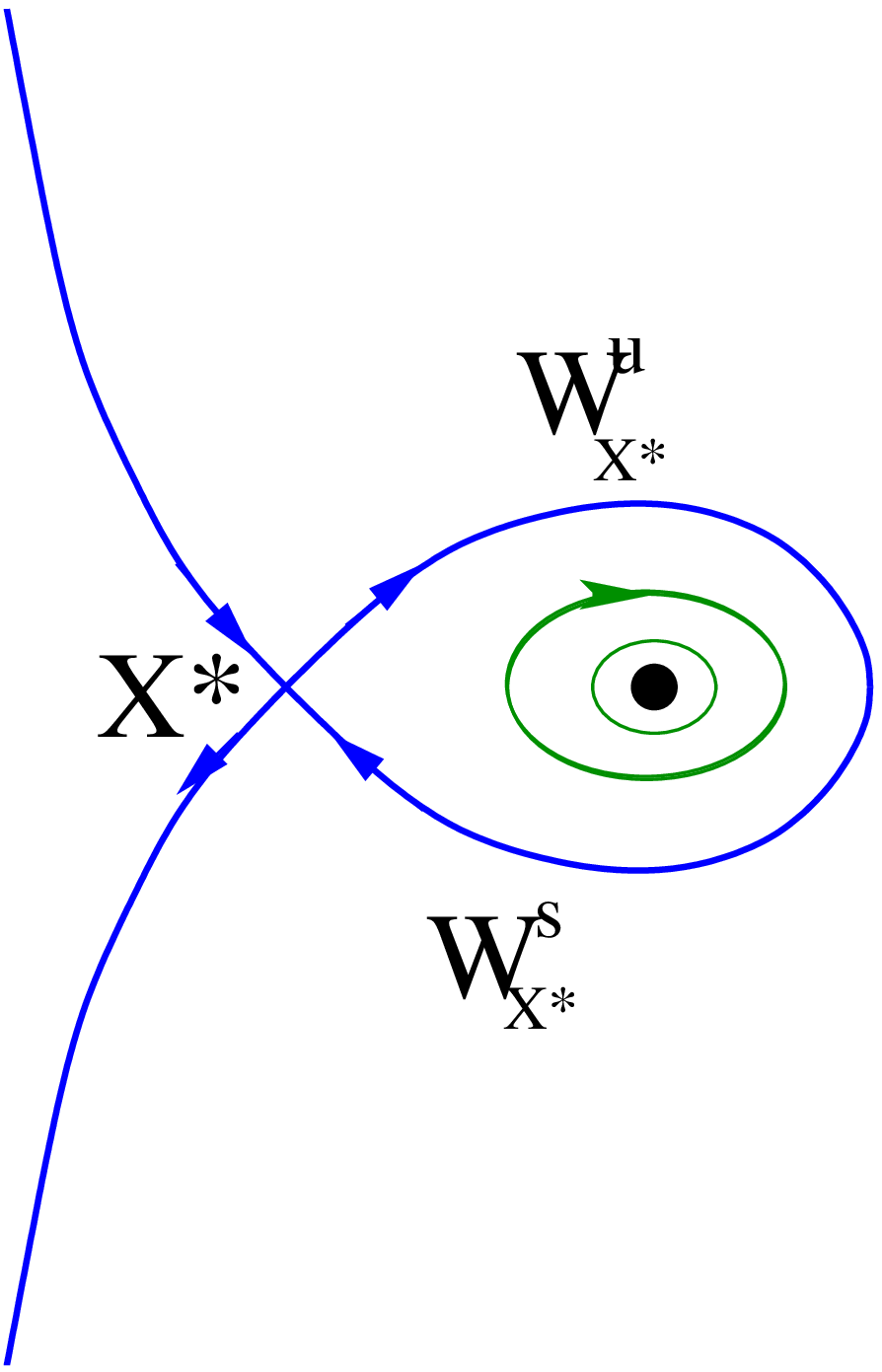}
\hspace{2cm}
\includegraphics[height=4cm,width=4cm,clip=false]{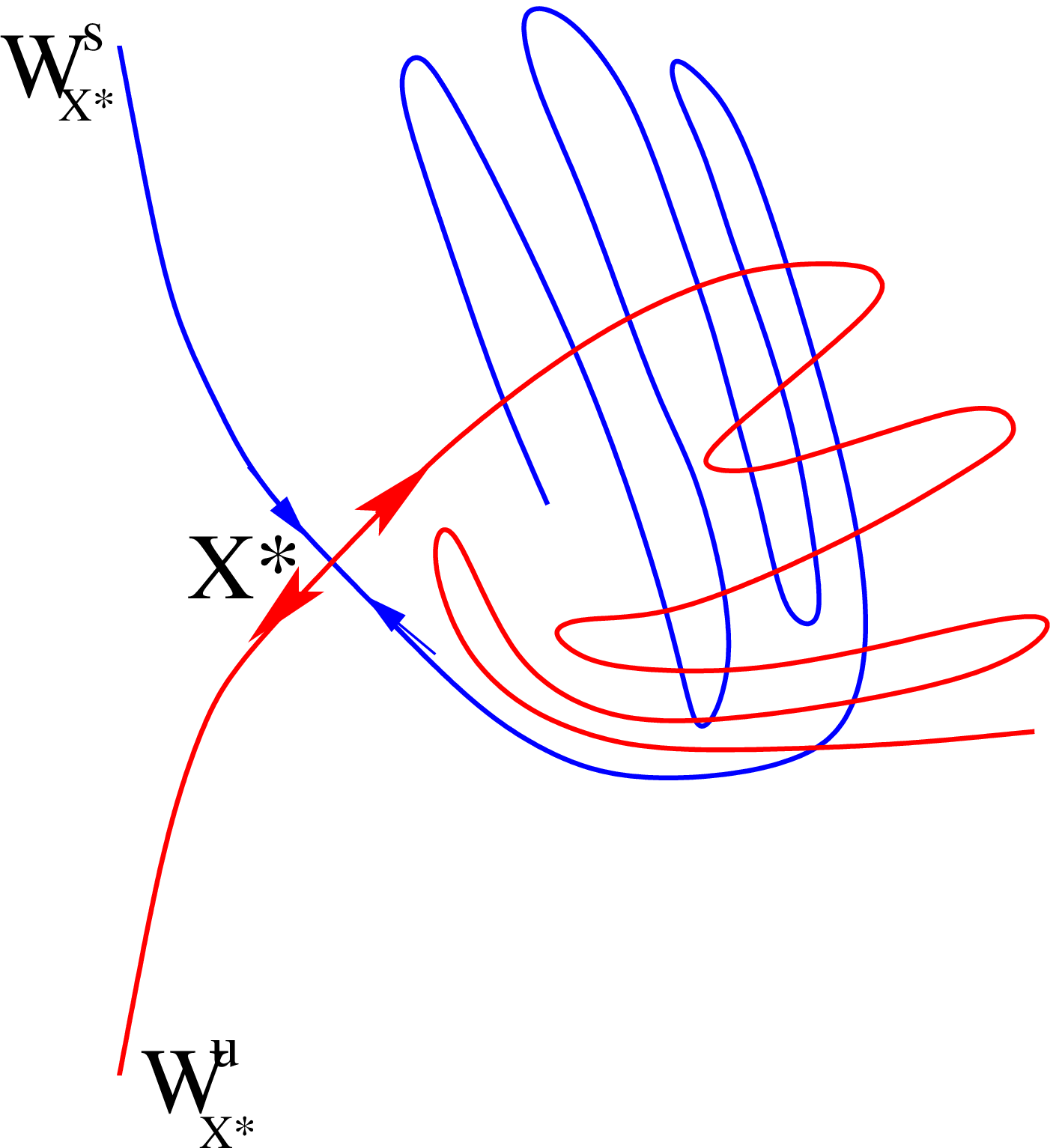}
\hspace{2cm}
\includegraphics[height=4cm,width=4cm,clip=false]{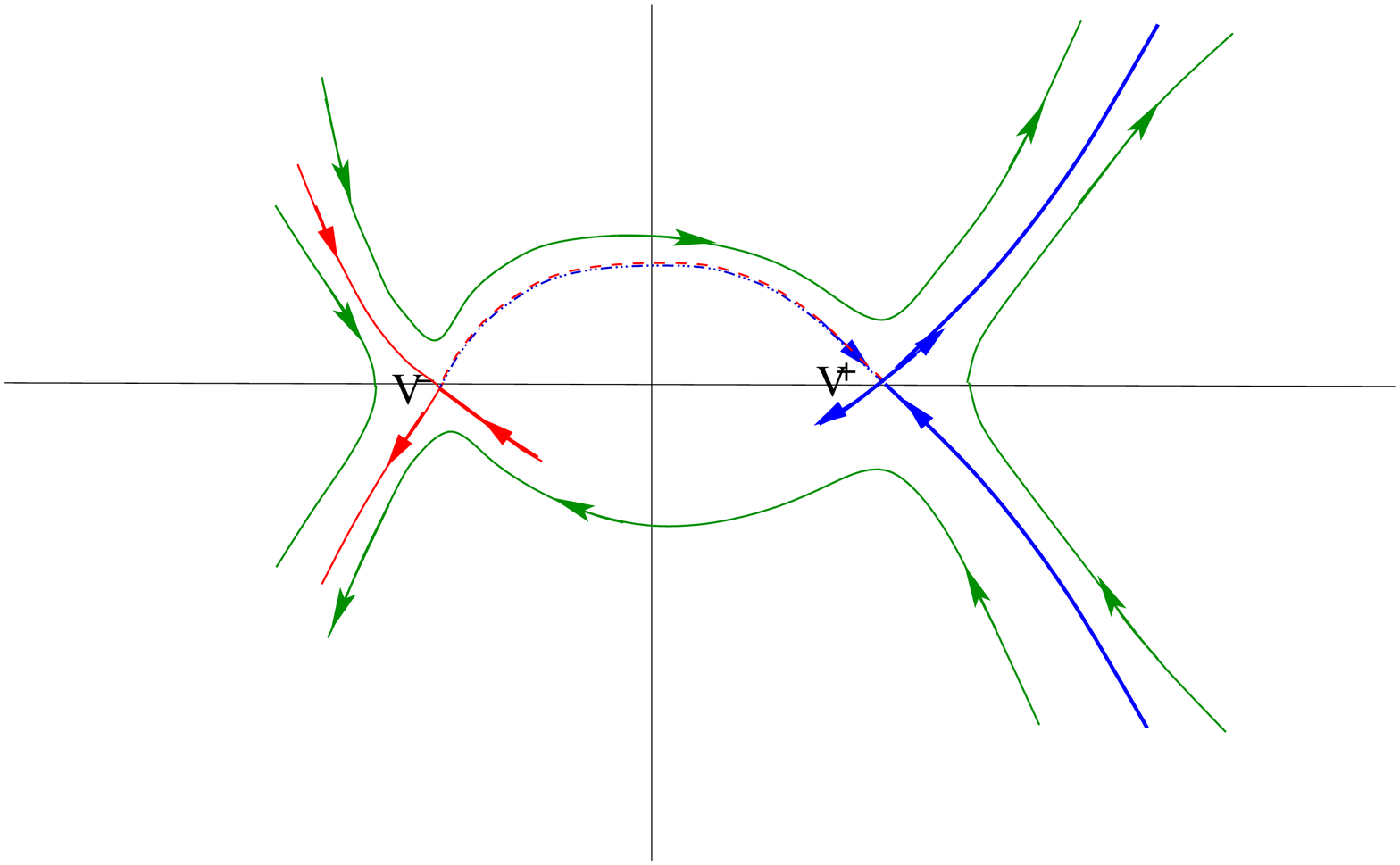}
\vspace{0.5cm}
\caption{Fig. \ref{Fcline} a. Homoclinic intersection (continuous time system).
\ref{Fcline} b. Transverse homoclinic intersection (discrete time systems). 
Fig.\ref{Fcline} c. Heteroclinic intersection. \label{Fcline}}
\end{center}
\end{figure}
%
%
%
%
%
%

The notion of fixed point is related to a more general notion called ``convergence''. Following
Hirsch \cite{Hirsch} we say that a dynamical system is

\bit
\item \textit{Convergent:} if all trajectories converge to equilibria.
\item \textit{Globally convergent or asymptotically stable:} if  all trajectories converge to a unique equilibrium.
\eit

\sssu{Lyapunov functions.}\label{SLyapFunc}

A Lyapunov function $\cV$ is a differentiable\footnote{
Note that the continuity is sufficient in the definition.
However, differentiability allows us to replace the condition $\frac{d\cV}{dt} \leq 0$ by $\left<\nabla \cV, \bH\right> \leq 0$,
where $<,>$ is a scalar product in $\bbbr^N$.
This means that the trajectories cross the level curves of $\cV$ ``inward''. Note that, reciprocally, if
there exists a metric such that $<\nabla \cV, \bH> \leq 0$, $\cV$ is a Lyapunov function for the corresponding
dynamical system. This allows one to show the convergence of some dynamical  systems under quite general conditions
(see section \ref{SCG} and  \cite{Benaim}.)   }
function which decreases along the trajectories and is bounded from below. In dissipative mechanical systems, the energy 
is a Lyapunov function. This notion is useful to locate fixed points (they are extrema of $\cV$) and
to analyze their stability. Indeed if $\frac{d\cV}{dt} \leq 0$ (resp. $\frac{d\cV}{dt} < 0$) in the neighborhood
of some fixed point $ \bXs$ then $ \bXs$ is stable (resp. asymptotically stable.) More generally the Lasalle invariance
principle \cite{Lasalle} asserts that the $\omega$-limit set of any point $\bX$ is included in the largest invariant
set where $\cV$ is a constant. An important corollary is that if $\cV$ is a strict Lyapunov function ($\frac{d\cV}{dt} < 0$)
on a compact set $\cm$ then the equilibria are isolated, and the system is convergent.
Lyapunov functions are used in section \ref{SCG}.
\ssu{Bifurcations.}
The dynamical systems (\ref{SD1}),(\ref{SD2}) depend smoothly on a set of parameters $\lambda \in \cE_\lambda$. 
When varying these parameters one modifies the dynamics. On open domains of parameters
the changes are essentially \textit{quantitative}, namely a variables change maps
the initial system to the modified one. One says that
two flows (or maps) $\Bf$,$\Bf'$ are \textit{topologically equivalent} if there
is an homeomorphism mapping the orbits of $\Bf$ to
the orbits of $\Bf'$ and preserving the ordering of points along the orbits. 
Two  topologically equivalent dynamical systems have therefore the same phase portrait (but quantitative
characteristics such as the convergence rate to a fixed point may differ).
A dynamical system $\Bf$ is  \textit{structurally stable} if  any sufficiently close $\Bf'$
\footnote{See \cite{RuelleBif} for a definition of a topology in a
space of flows.} is topologically conjugated to
$\Bf$. 

There exists in general a (closed) set of parameter values where the corresponding
dynamical system is not structurally stable. At these points, called \textit{bifurcations points},
the dynamics  changes \textit{qualitatively}. 
 The codimension
of the bifurcation is the number of independent parameters one has to adjust in order
to obtain the bifurcation. In this section
we focus on bifurcations occurring on \textit{fixed points}. Moreover, we only consider the case
where at most two independent parameters are varying. This is indeed the only cases where a complete
classification of fixed bifurcations is known \cite{GH}.  \\

Assume therefore that $ \bXs$ is a fixed point, namely this is
the zero of some function $\bG(\bX;\lambda)$ ($\bG(\bX;\lambda)=H(\bX;\lambda)$ in the continuous time case,
and $\bG(\bX;\lambda)=F(\bX;\lambda) - \bX$ in the discrete time case). 
When varying $\lambda$ the implicit function theorem guarantees that $ \bXs$
 moves along a regular curve $ \bXs(\lambda)$ provided that $D\bG(\bX;\lambda)$
is invertible. This also implies that the eigenvalues of $D\bG(\bX;\lambda)$ are moving continuously.
Note that, since $D\bG(\bX;\lambda)$ is real, the eigenvalues are either
real or complex conjugated. Then, at some parameter values, some eigenvalues can intersect the real axis (resp. the unit circle
in the discrete time case). There are two possibilities. Either they cross at the origin (resp. at $1$).
In this case the implicit function no more applies and several branches of solutions of the
equation  $\bG(\bX;\lambda)=0$ appear or disappear (see Fig. \ref{FSN},\ref{FPF}). Or they cross at imaginary values.
This induces in general a change of stability for $ \bXs(\lambda)$ and the appearance or disappearance of
a limit cycle (see Fig. \ref{FH}).\\

The initial dynamical system has $N$ degree of freedom. However, at the the bifurcation point, say $\lambda_c$, one expects 
that the only relevant information is contained in the eigendirections corresponding to the crossing
eigenvalues. This leads to a general method called the \textit{central manifold reduction}.
Let $E^c( \bXs)$ be the central space (it is non zero at the bifurcation point), $n_c=dim(E^c( \bXs))$
 the number of crossing eigenvalues and
call $E^h( \bXs)= E^s( \bXs) \oplus E^u( \bXs)$.
Then the central manifold theorem \cite{Carr} states that
 there is a function $\cH(\bX;\lambda) : 
E^c \times \cE_\lambda  \to E^h( \bXs)$ 
such that
$\cH( \bXs,\lambda_0)=0$, $D_x\cH( \bXs,\lambda_0)=0$ and such that the manifold:
$$\cW^c(\lambda) = \lbrace \bX + \cH(\bX;\lambda) \ | \ \bX \in E^c_ \bXs \rbrace$$  
\nid  contains $ \bXs$ and is tangent
to $E^c_ \bXs$ at this point. Moreover $\cW^c(\lambda)$ 
is \textit{locally invariant} for $\lambda$ sufficiently small and bounded.
This means that there is an open neighborhood $\cU$ of $ \bXs$ such that 
if $\bX(0) \in \cW^c(\lambda)
 \cap \cU$ then $\bX(t) \in \cW^c(\lambda)$ as long as $\bX(t) \in \cU$.
Finally $\cW^c(\lambda)$
 is \textit{locally attractive} if $E^u( \bXs) = 0$.
Therefore, in this case, all solutions staying in $\cU$ tend exponentially fast
to some trajectory on $\cW^c(\lambda)$.
$\cW^c(\lambda)$ is called the \textit{center manifold} (though it is not unique).

It is then possible to locally reduce the dynamics (\ref{DWCNN}) to the dynamics
on $\cW^c(\lambda)$ by projection. 
Denote by $\Pi^c : \bbbr^N \to E^c$ the projection
onto $E^c$, by $\Pi^h : \bbbr^N \to E^h( \bXs)$ the projection
onto $E^h( \bXs)$, and set $\bX^c(t)=\Pi^c\bX(t)$. Then if $\bX(t)$ is a solution of
(\ref{SD1}) such that $\bX(t) \in \cW^c(\lambda) \cap \cU, t \geq 0$ one has
$\bX^c(t)=\bX(t)+\cH(\bX(t);\lambda)$, namely, by a suitable (local) variable change
one can write down a smaller dynamically system leaving on $\cW^c(\lambda)$ and
characterizing the relevant part of the dynamics about $ \bXs$.\\

It is then possible to further reduce the dynamics by removing some non linear
terms with the appropriate variable changes. Actually, one cannot remove
all the non linear terms in this way (otherwise the dynamical system is basically
a linear system). Only the non linear terms satisfying \textit{non resonant} conditions (see
\cite{Arnold2,GH} for details) can be removed. Finally, one ends with a set of canonical
equations called a \textit{normal form}. In  some sense, the normal form reduction
for a dynamical system is a generalization of the diagonalisation for a matrix.
There are uncountably infinitely many matrices in $\bbbr^N$ but 
  many matrices  have the same diagonal (or Jordan) form. This means that
they are equivalent, up to a basis change, and the canonical form of their equivalence
class is the Jordan form. In the same way, an infinite number of dynamical system
undergoing a bifurcation at a fixed point can be represented under a canonical form
or normal form. 

It is remarkable that the different possible codimension one and two bifurcations 
are in fact only a few. Moreover, it is possible to write down general conditions on the dynamical
system, called transversality conditions, allowing to characterize the type of bifurcation
occurring. We now briefly describe these bifurcations. 

\sssu{Codimension one bifurcations.}

In this section, we assume that $\bXs=0$ is a fixed point, and that $\lambda$
is  one dimensional parameter. We review now the bifurcations arising generically
in this case. We denote by $\lambda_0$ the parameter value where the bifurcation arises.
We first consider the continuous case, and then the discrete time one.

\bit

\item\textbf{\textit{Saddle-node bifurcation.}}
The transversality
conditions, when written in a great generality, are rather abstract. However, it is
easy to understand them by taking a one dimensional example. Consider indeed
the system  $\dot{x}=f(x;\lambda)$ such that $x=0$ is a fixed point, and $\lambda_0=0$
is a bifurcation point. Performing a Taylor expansion about $(0;\lambda_0)$ gives:
\beq\label{Taylor}
f(x;\lambda)=f_{00}+f_{10}x+f_{01}\lambda+f_{11}x\lambda + f_{20}x^2 + \dots
\eeq
Since we want to characterize the dynamical system in a neighborhood of $(0,0)$
it is natural to consider the lowest order terms. Since $0$ is fixed point $f_{00}=0$.
Moreover, $\lambda_0=0$ is a bifurcation point where $\frac{\partial f}{\partial x}(0;0)=0$
which implies $f_{0,1}=0$. If we ask now that the  linear term in $\lambda$
does not vanish we get the first transversality condition for the saddle-node bifurcation (in one dimension):
$f_{01}=\frac{\partial f}{\partial \lambda}(0;0) \neq 0$.
 At the bifurcation point the implicit function theorem does not apply
and two branches of equilibria emerge (or disappear), with a vertical tangent (see Fig. \ref{FSN}).
 The second transversality condition, $f_{20} = \frac{\partial^2 f}{\partial x^2}(0;0) \neq 0$ 
ensures that these curves have a \textit{quadratic tangency}
at   $(0;0)$. One can then show that the term $f_{01}$ and all higher order terms in the expansion (\ref{Taylor})
can be removed. One can also use variable changes 
which allows us to eliminate $f_{10},f_{20}$.
 We finally end up with the normal form for the saddle-bifurcation in continuous time case.
\beq\label{NFSN}
\dot{x}=\lambda-x^2
\eeq

The corresponding bifurcation diagram is drawn in Fig. \ref{FSN}.
%
%
%
%
\begin{figure}[ht]
\begin{center}
\includegraphics[height=3cm,width=4cm,clip=false]{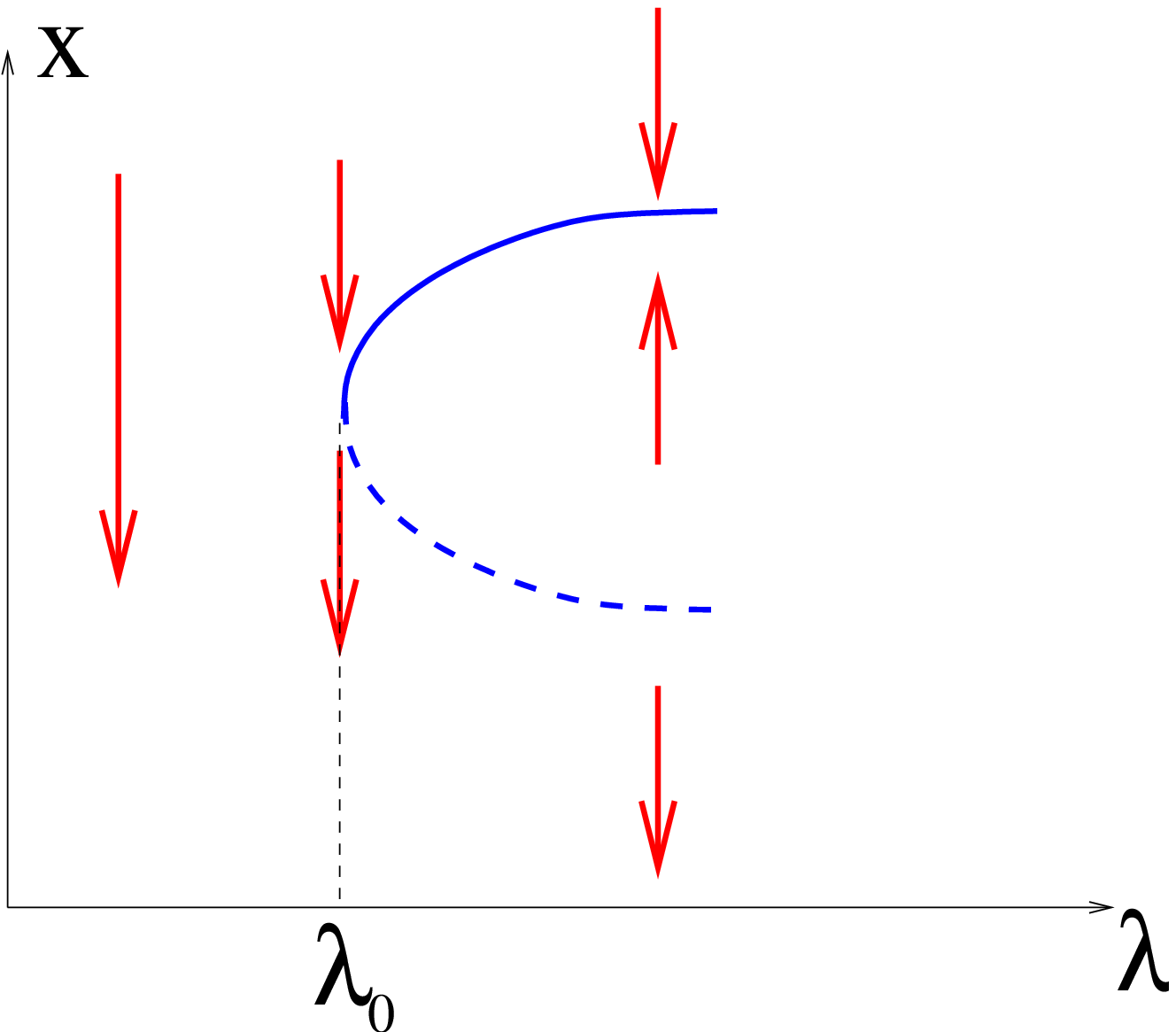}
\hspace{1cm}
\includegraphics[height=3cm,width=4cm,clip=false]{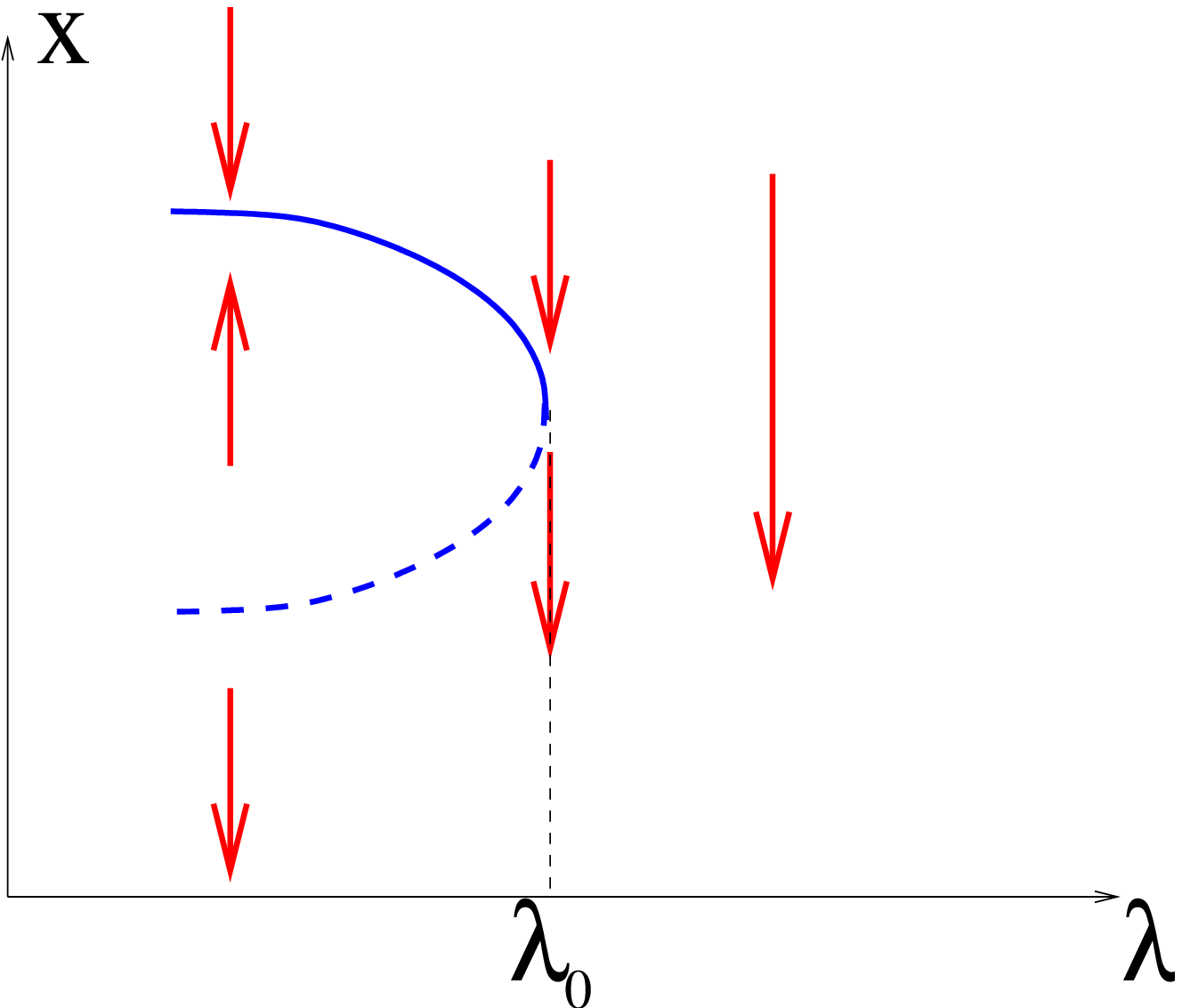}
\vspace{0.5cm}
\caption{Saddle node bifurcation. \label{FSN}}
\end{center}
\end{figure}
%
%
%
%
%
%

In a $N$ dimensional dynamical system the previous discussion generalizes as follows.
Assume that :

\ben

\item[(SN1)]  $D\bH(\bXs;\lambda_0)$ has a simple eigenvalue 0 with right eigenvector $\bv$
and left eigenvector $\bw$. It has also $k$ stable eigenvalues and $N-k-1$ unstable eigenvalues
(counting multiplicity).

\item[(SN2)] 

$$\bw.\frac{\partial \bH}{\partial \lambda}(\bXs,\lambda_0) \neq 0 $$

\item[(SN3)]

$$w_i.\frac{\partial^2 \bH_i}{\partial x_j \partial x_k}(\bXs,\lambda_0)v_j v_k \neq 0$$

\nid where we used the Einstein convention (sum over repeated indexes),
\een

\nid then the normal form of the bifurcation is (\ref{NFSN}). Namely 
the dynamical system behaves like eq. (\ref{NFSN}) in the direction of
the zero eigenvector, with hyperbolic behavior in the complementary directions.
This bifurcation is in some sense the ``most'' generic since the set of dynamical systems
which satisfy the transversality conditions (SN1),(SN2) is open and dense
in the space of $\cC^\infty$ one parameter families of vector fields with an equilibrium
with a zero eigenvalue.

For discrete time dynamical systems, the normal form writes :
\beq\label{NFSND}
x(t+1)=x(t)+\lambda-x^2
\eeq

\item\textbf{\textit{Transcritical bifurcation.}} Assume now that $\bXs$ is a fixed point before and after the bifurcation.
This implies that $f_{01}$ must be zero and the corresponding transversality (TSN2) condition 
cannot hold. If we replace it by the condition $f_{11}$ we obtain a normal form.
\beq\label{NFT}
\dot{x}=\lambda x-x^2
\eeq
\nid whose bifurcation diagram is depicted in Fig. \ref{FTrans}.
Two fixed points coexist and they exchange their stability at the bifurcation point.
%
%
%
%
\begin{figure}[ht]
\begin{center}
\includegraphics[height=3cm,width=4cm,clip=false]{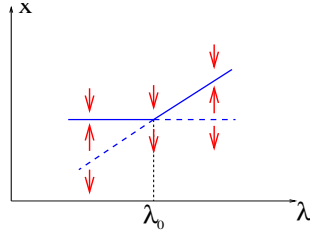}
\hspace{1cm}
\vspace{0.5cm}
\caption{Transcritical  bifurcation. \label{FTrans}}
\end{center}
\end{figure}
%
%
%
%
%
 
In the general case one has to replace  the transversality condition TSN2 by: 

\bit
\item[(TT2)] $$w_i \frac{\partial^2 \bH}{\partial \mu\partial \bX} \neq 0 $$
\eit
For discrete time dynamical systems, the normal form writes :
\beq\label{NFTD}
x(t+1)=x(t)+\lambda x-x^2
\eeq

\item\textbf{\textit{Pitchfork bifurcation.}} In some cases one has particular symmetries in the dynamical system. 
A particularly prominent example corresponds to the symmetry $\bX \to -\bX$.
Returning to our one dimensional example we see that $f_{1,1}$ and $f_{2,0}$
has to be zero. We must then consider higher order terms. It is clear
that the first remaining non linear term is $f_{3,3}x^3$ (and the next
one is $f_{5,5}$). All other terms of order $\leq 3$ vanish. Note that
according to the sign of $f_{3,3}$ one has \textit{supercritical} ($f_{3,3}>0$) or
\textit{subcritical} bifurcation ($f_{3,3}<0$). In this last case one has to take
into account the term $f_{5,5}$ in order to ``saturate the instability''.
 The normal form for the (supercritical) pitchfork bifurcation is
\beq\label{NFP}
\dot{x}=\lambda x-x^3
\eeq
The corresponding bifurcation diagram is drawn in Fig. \ref{FPF}. 
%
%
%
%
\begin{figure}[ht]
\begin{center}
\includegraphics[height=3cm,width=4cm,clip=false]{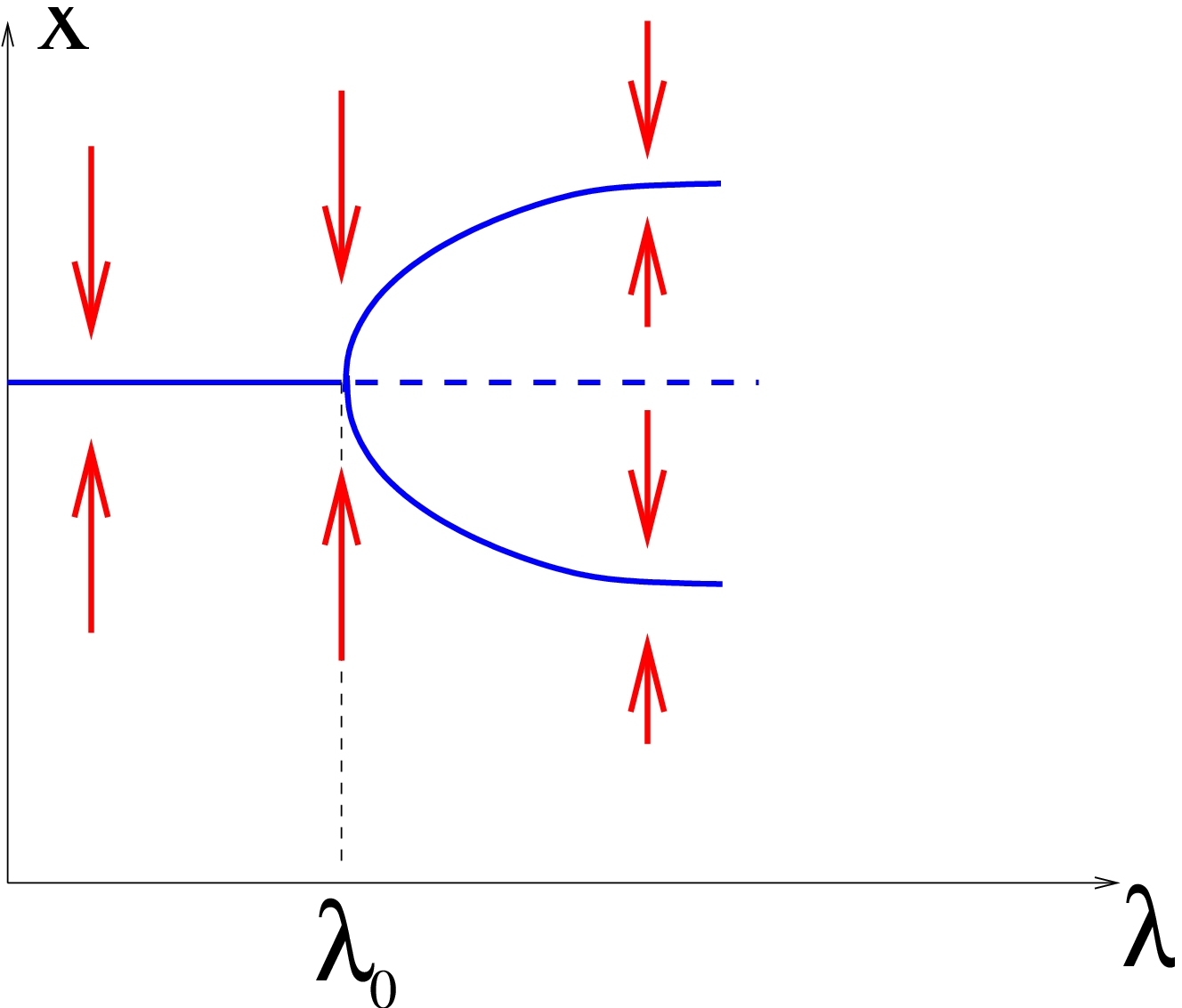}
\hspace{1cm}
\includegraphics[height=3cm,width=4cm,clip=false]{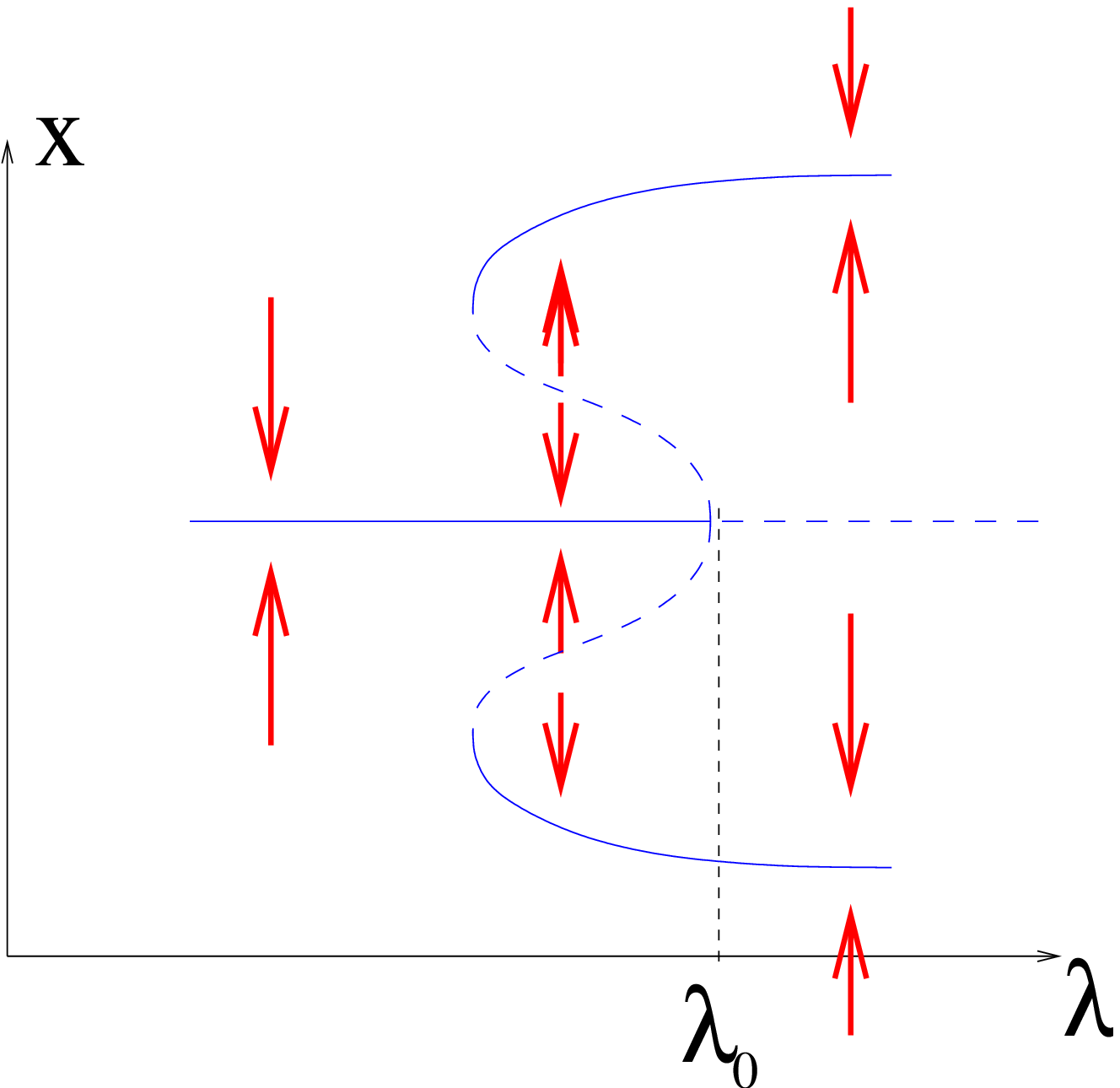}
\vspace{0.5cm}
\caption{Pitchfork  bifurcation. Fig. \ref{FPF} a. Supercritical. Fig. \ref{FPF} b. Subcritical  \label{FPF}}
\end{center}
\end{figure}
%
%
%
%
%
%

\item\textbf{\textit{Hopf bifurcation.}}
Assume now that there is a pair of complex conjugate eigenvalues of the Jacobian matrix crossing
the imaginary axis [resp. the unit circle] at the bifurcation point. Note that this requires
that the dynamical system has at least a dimension $2$. Having eigenvalues
with an imaginary part implies that the trajectories are locally oscillating around the fixed
point. When the eigenvalues cross from the left to the right the oscillations are exponentially
damped before the bifurcation point, and they are exponentially amplified after the bifurcation
(see Fig. \ref{FH}). The exponential amplification is obviously local. When moving
away from the fixed point the nonlinearities saturate the instability the trajectories
converge to a limit cycle. This corresponds to a  Hopf bifurcation. 

More generally,  \cite{Marsden,GH}
suppose that the dynamical system (\ref{SD1}) as an equilibrium $(\bXs;\lambda_0)$
such that

\bit
\item[(TH1)] $D\bH(\bXs;\lambda_0)$ has a simple pair of pure imaginary eigenvalues $\pm  i\omega$ and no other
eigenvalues with zero real parts.
\eit

Then there is a smooth curve of equilibria $(\bX(\lambda);\lambda)$ with $\bX(\lambda_0)=\bXs$.
The eigenvalues $\mu(\lambda),\bar{\mu}(\lambda)$ vary smoothly with $\lambda$. If, moreover

\bit
\item[(TH2)] 
$$ \frac{d \Re(\mu)}{d\lambda}(\lambda_0) =d \neq 0$$
\eit
\nid then the normal form is:
\beq\label{FNH}
\dot{z}=\gamma z + \alpha z^2 \bar{z} + O(|z|^5)
\eeq
\nid where $z$ is a complex variable corresponding to the reduction on the central manifold
and $\gamma=(\lambda-\lambda_0)+i\omega$.
Note that, written in polar coordinates $(r,\theta)$ the equations (\ref{FNH}) are:

\bea
\dot{r}&=&\left[d(\lambda-\lambda_0)+ar^2\right]r\\
\dot{\omega}&=&\omega+c(\lambda-\lambda_0)+br^2
\eea

\nid where $a,b,c$ are coefficients depending on the vector field (note that $a$ can be
positive or negative, corresponding to supercritical or subcritical bifurcation).
We remark that the amplitude of the limit cycle increases like the square root
of the difference $\lambda-\lambda_0$ and that the frequency depends on the parameter
and on the amplitude. Note also that, near the bifurcation point, the frequency is non zero
(see section \ref{SFN}).  
%
%
%
\begin{figure}[ht]
\begin{center}
\includegraphics[height=5cm,width=6cm,clip=false]{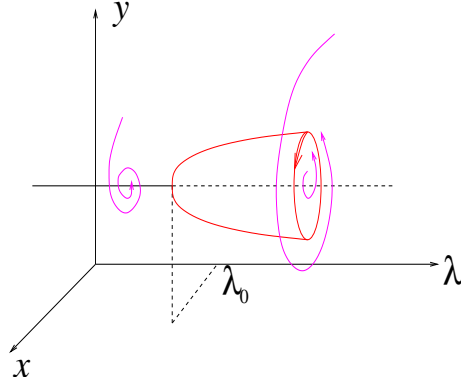}
\hspace{1cm}
\vspace{0.5cm}
\caption{Hopf bifurcation. \label{FH}}
\end{center}
\end{figure}
%
%
%
%
%
%
The discrete time case is substancially more complicated, with specific cases corresponding
to strong resonances. A detailed analysis can be found in \cite{Arnold2}. For a discussion
in the context of neural networks see also \cite{These}

\eit

\sssu{Codimension two bifurcations.}

We describe now one local codimension $2$ bifurcations, the Bogdanov-Takens bifurcation. (We only focus
on the examples found in this chapter). A complete description can be
found in \cite{GH}. Basically, codimension two bifurcations may arise
either if additional degeneracies in the non linear terms of the previous
bifurcations arise, or if the linear part of the vector field (the map)
is doubly degenerate. In this last case, the linear part  for flows takes the form
\beq\label{LinCod2}
\left(
\baR{ccc}
0 && 1\\
0 && 0
\eaR
\right);
\qquad
\left(
\baR{ccccc}
0 && -\omega && 0 \\
\omega && 0 && 0  \\
0 && 0 && 0
\eaR
\right);
\qquad
\left(
\baR{ccccccc}
0 && -\omega_1 && 0 && 0  \\
\omega_1 && 0 && 0 && 0  \\
0 && 0 && 0 && -\omega_2 \\
0 && 0 && \omega_2 && 0 
\eaR
\right);
\eeq
\bit
\item\textbf{\textit{Bogdanov-Takens bifurcation.}}
The Bogdanov-Takens corresponds to the first situation in eq. (\ref{LinCod2}).
The normal form is:
\bea \label{NFBT}
\dot{x} &=& y\\
\dot{y} &=& \lambda_1 + \lambda_2 y + x^2 + \sigma xy \nonumber
\eea
\nid where $\sigma = \pm 1$. In the sequel we shall consider the case $\sigma=1$. 
The second case can easily be obtained from the first one
by the substitution $t \to -t; \ y \to -y$. It is easy to show that  Hopf bifurcation 
occurs on the curve $\lambda_2 = \sqrt{-\lambda_1}$ (hence for $\lambda_1<0$) while
saddle-node bifurcations occur on $\lambda_1=0; \lambda_2 \neq 0$.
The complete bifurcation diagram is represented in Fig. \ref{FBT}.

\begin{figure}[ht]
\begin{center}
\includegraphics[height=6cm,width=8cm,clip=false]{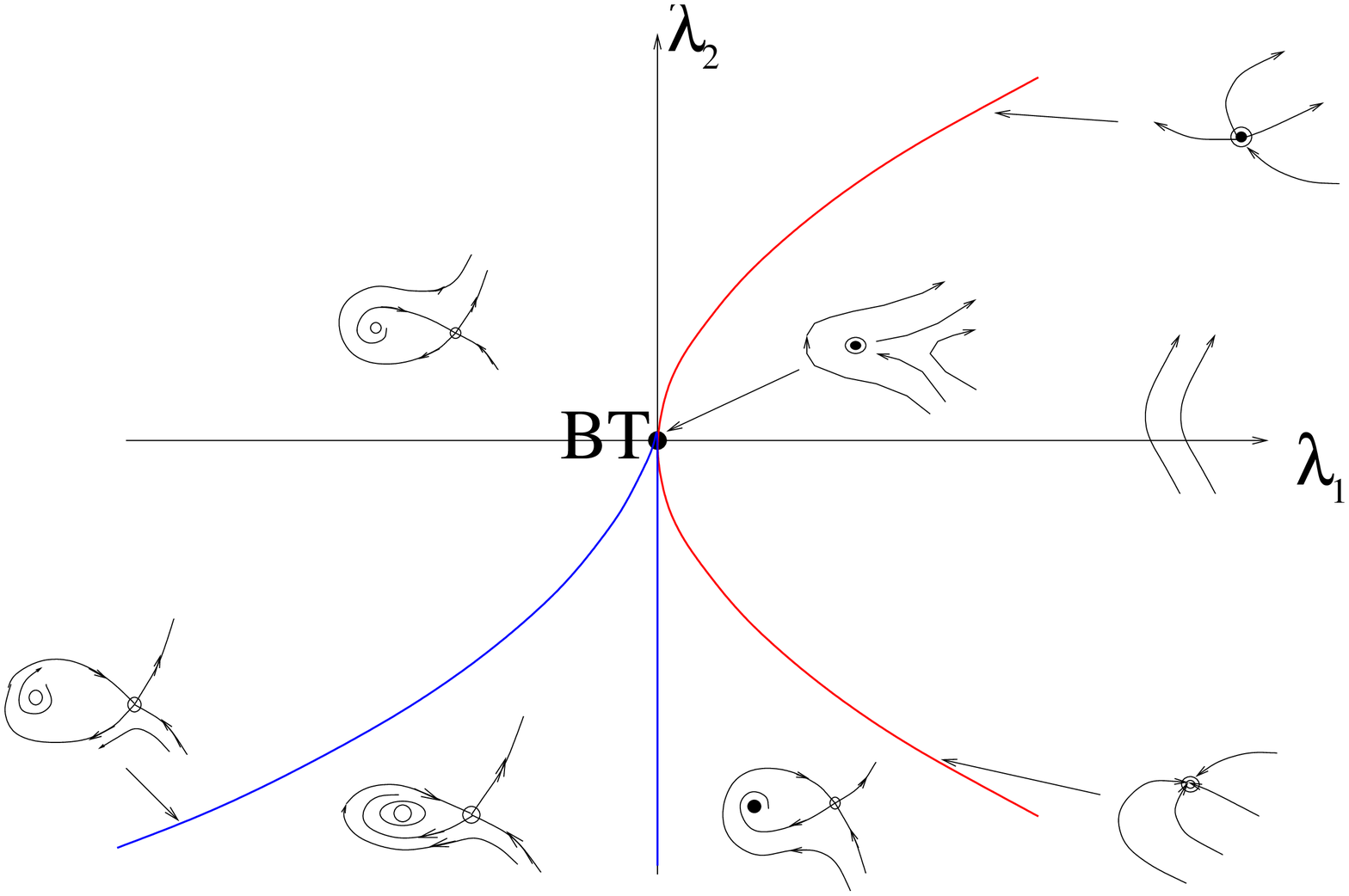}
\hspace{1cm}
\vspace{0.5cm}
\caption{Bogdanov-Takens bifurcation diagram. \label{FBT}}
\end{center}
\end{figure}
%
%
%
%
%
%
\eit

\ssu{Chaotic motion.}

\sssu{Attractor}

In the previous sections we have seen several example of topological objects
where the dynamics converges asymptotically: asymptotically stable fixed points
and stable limit cycle are examples of attractor. But attractor
can be quite a bit more complicated objects. Though there are many different
(and non equivalent) definitions of attractors \cite{ER,GH,Milnor,Williams,Cos}, they basically combine  
a notion of \textit{attractivity} and \textit{indecomposability}. 
Here is one definition \cite{Katok}.
An attractor is a compact set $\cA$ such that:

\bit
\item{(i) \underline{Attractivity}}. There exists an open set $\cU \supset \cA$ and a time
$t_0$ such that  $\bF^{t_0}(\cU) \subset \cU$ and $\cA= \cap_{t=0}^\infty \bF^t(\cU)$

\item{(ii) \underline{Indecomposability}}. $\forall \bX,\by \in \cA$ and $\forall 
\epsilon > 0$ there is a chain $\bX=\bX_0, \bX_1, \dots \bX_n=\by$ and a sequence of times
$t_1,t_2, \dots t_n \geq 1$ such that the distance between $\bF^{t_i}(\bX_{i-1})$ and
$x_i$ is $\leq \epsilon$.

\eit Note that (i) implies the dynamical invariance of $\cA$.

Attractors can have a simple topological structure (fixed points, cycles, tori)
or a complex one (strange attractors). Though there are several definitions of the 
``strangeness'' of an attractor, there is a general consensus about the necessity
to have initial condition sensitivity. This notion is in fact related to a more
general notions called hyperbolicity.

\sssu{Hyperbolic dynamical systems.} \label{SHyp}

A dynamical system is uniformly hyperbolic if  there exists 
$0 < \lambda < 1 < \mu$ and a constant $C$ such that:

\bit

\item{(i)} There exists two subspaces $\cE^s(\bX),\cE^u(\bX)$ respectively called
\textit{stable} and \textit{unstable}, forming an invariant  decomposition
of the tangent space at $\bX$:
$\cT_\bX=\cE^s(\bX) \oplus \cE^u(\bX)$ et $D\Bf_{\bX}^t\cE^s(\bX)
=\cE^s(\Bf^t(\bX))$ (resp. $D\Bf_{\bX}^t\cE^u(\bX)
=\cE^u(\Bf^t(\bX))$), $\forall t >0$, and such that the angle between the
two subspaces is bounded away from $0$.

\item{(ii)} $D\Bf_{\bX}$ is \textit{contracting} on $\cE^s(\bX)$:
If $\bv$ is a  vector in $\cE^s(\bX)$ :
$$\|D\Bf^t_{\bX}\bv\| \leq C\lambda^t \|\bv\|, \ \forall t > 0$$
\item{(iii)} $D\Bf_{\bX}$ is \textit{expanding} on $\cE^u(\bX)$:
If $\bv$ is a  vector in  $\cE^u(\bX)$ :
$$\|D\Bf^{-t}_{\bX}\bv\| \leq C\mu^{-t} \|\bv\|, \ \forall t > 0$$
\eit

(Note that the constant $C$ in the definition is independent of $\bX$. More generally
(non uniform case) this constant depends on $\bX$. 

(Uniformly) hyperbolic dynamical systems have several remarkable properties (see
\cite{Katok} for a wide description): existence of smooth local stable and unstable
manifolds locally tangent to  the spaces $\cE^s(\bX)$ (resp. $\cE^u(\bX)$);
 shadowing lemma;   density of periodic unstable orbits leading to trace formulas;
local product structure allowing the construction of Markov partition used
in symboling coding; structural stability; \etc.

The existence of an unstable direction implies initial conditions sensitivity while the existence
of contracting directions corresponds to asymptotic convergence onto an attractor.
Basically, a strange attractor is composed by the closure of the union of the unstable manifolds.
A perturbation ``parallel'' to the attractor (locally tangent to the unstable space)
is locally expanded at exponential speed (initial condition sensitivity) while
a perturbation transverse to the attractor   (locally tangent to the stable space)
is asymptotically damped. Parallel and transverse time dependent perturbations induce drastically
different effects on the dynamics (see section \ref{RepLin}) having interesting interpretation
in the context of Neural Networks.

\sssu{Statistical approach and ergodic theory.}

In chaotic systems, it is often useful to replace the study of individual trajectories
by a statistical analysis of the evolution in the phase space. The natural object
is then a probability measure. Actually, there is a close relationship between the (physically relevant)
probability measures and the notion of \textit{state} in statistical  physics.
The initial probability  $\mu^{(0)}$ corresponds to randomly selecting the initial conditions  
and the probability at time $t$, $\mu^{(t)} \deq \bF^{*t}\mu^{(0)}$ is the result of the action of the dynamics $\bF$ on $\mu^{(0)}$. 
It is given by:  
$$\mu^{(t)}\left[\cB\right] = \mu^{(0)}\left[\left\{\bX \ | \
\bF^t(\bX) \in \cB\right\}\right]=
\mu^{(0)}\left[\bF^{-t}(\cB)\right]$$
\nid where $\cB$ is a (measurable) set in $\Omega$.

The statistics of trajectories on the attractor is characterized by an invariant measure such that:
\beq\label{Defmuinv}
\mu(\bF^{-1}(\cB))= \mu(\cB)
\eeq
(The corresponding notion in statistical physics is the notion of \textit{phase}). 

Among all invariant measures the \textit{ergodic} measures play an important role (they correspond to \textit{pure phases}
in statistical physics). There are several equivalent definitions but the most known is certainly the identity
between time average and ensemble average. A  measure $\mu$ is ergodic if for \textit{$\mu$ almost every initial condition $\bX$} :

\beq\label{DefBirkhoff}
\lim_{T \to  \infty} \frac{1}{T} \sum_{t=1}^T \phi(\bX(t))
=\int \phi d\mu
\eeq

\nid where $\phi$ is a function in $L^1(d\mu)$.

The definition (\ref{DefBirkhoff}) is unfortunately rather poor since
one can show that a dynamical system in a compact space has often infinitely many such measures
\cite{ER,Katok}. A more useful notion is the \textit{Sinai-Ruelle-Bowen} (SRB)  measure.
A measure $\mu$ is a SRB measure (or natural, or physical measure) if the
 property (\ref{DefBirkhoff}) holds for a set of positive  \textit{Lebesgue} measure 
\cite{Viana,Ruelle2} of initial conditions. This means basically that
the time average and the ensemble average are equal for typical initial conditions.
Sinai, Ruelle and Bowen have shown that the SRB measure is a ``Gibbs like'' measure:
it has an exponential form, although the term  in the exponential is not the Hamiltonian encountered in statistical mechanics
but a dynamically relevant quantity. Usually, this the projection of the Jacobian along the unstable
fibers, which has direct connection with the regular part of the Perron-Frobenius operator   
Moreover the SRB measure maximizes some version 
of a free energy (topological pressure) : it has therefore the characteristics of an equilibrium state.

\pagebreak

 \ed
